# 面向多核架构的

# 操作系统可扩展性研究

（申请清华大学工学博士学位论文）

培 养 单 位： 计算机科学与技术系

学　　　科： 计算机科学与技术

研　究　生： 崔　岩

指 导 教 师： 史元春　　　教　授

二〇一二年十月

面向多核架构的操作系统可扩展性研究

崔　岩

# Research on Scalability of Operating Systems on Multicore Processors

Dissertation Submitted to

**Tsinghua University**

in partial fulfillment of the requirement

for the degree of

**Doctor of Engineering**

**in**
**Computer Science and Technology**

by

**Yan Cui**

Dissertation Supervisor: Professor Yuanchun Shi

**October, 2012**

# 关于学位论文使用授权的说明

本人完全了解清华大学有关保留、使用学位论文的规定，即：

清华大学拥有在著作权法规定范围内学位论文的使用权，其中包括：（1）已获学位的研究生必须按学校规定提交学位论文，学校可以采用影印、缩印或其他复制手段保存研究生上交的学位论文；（2）为教学和科研目的，学校可以将公开的学位论文作为资料在图书馆、资料室等场所供校内师生阅读，或在校园网上供校内师生浏览部分内容；（3）根据《中华人民共和国学位条例暂行实施办法》，向国家图书馆报送可以公开的学位论文。

本人保证遵守上述规定。

**（保密的论文在解密后遵守此规定）**

作者签名：______________ 导师签名：______________

日　　期：______________ 日　　期：______________

# 摘 要

多核系统上核数多、硬件资源共享的硬件特点给操作系统的设计带来了新的挑战。如何定位和分析操作系统在多核平台上对加速比的限制因素，如何模拟和避免操作系统锁竞争导致的加速比下降现象（锁颠簸）以及如何避免最后级缓存等共享资源的竞争问题是操作系统可扩展性研究的重要课题。本文工作包括：

1、分析获得影响可扩展性的关键限制因素。本文分析和比较了操作系统在多核平台上的可扩展性，设计了针对系统服务接口的微基准测试程序集，实际系统的测量发现操作系统的系统服务接口均有可扩展性问题。在此基础上的深入分析准确地还原了每种系统服务的可扩展性瓶颈的产生方式。综合测试实验分析，我们得出的重要结论是，多核系统中保护共享数据的同步互斥操作是限制服务接口加速比提升的主要因素。

2、提出了锁颠簸现象的模拟方法及避免机制与策略。基于离散事件仿真技术，对导致锁颠簸的软、硬件因素进行整合，设计了模拟锁颠簸现象的模拟器，能够准确、快速地重现锁颠簸。在此基础上，提出了基于等待者数目的可扩展锁机制，可利用等待者数目自适应地决定锁等待者的等锁方式，有效避免了锁颠簸现象；同时设计了锁竞争感知的调度策略检测每个任务的用锁程度，并利用模型驱动搜索将所有用锁密集任务集中在部分处理器上运行。测试结果表明，采用锁竞争感知的调度策略比相关工作提升加速比达 46%。

3、提出了资源竞争感知的调度策略。该策略利用离线搜集的资源竞争数据，选择出了度量任务资源需求的最佳启发式指标。选择过程不仅考虑了启发式指标的准确性，而且考虑了指标的稳定性。基于在线获取的任务资源需求信息，改进了任务到核的映射机制以及核上的任务选择机制，使得具有互补资源需求的任务被同时调度，显著降低了共享资源竞争程度。以科学计算应用为例，提升加速比达 13%，充分发挥出了资源竞争感知调度的潜力。

4、提出了基于函数可扩展性值的操作系统瓶颈定位方法。该方法定义一个函数的可扩展性值为该函数在多核和单核上每单位工作量的执行时间之差。通过分析具有较大可扩展性值的函数，得到影响加速比提升的因素。以在线事务处理应用为例，提升加速比达 49%，有效削减了操作系统对加速比的限制。

# Abstract

Large number of cores and hardware resource sharing are two characteristics on multicore processors, which bring new challenges for the design of operating systems. How to locate and analyze the speedup restrictive factors in operating systems, how to simulate and avoid the phenomenon that speedup decreases with the number of cores because of lock contention (i.e., lock thrashing) and how to avoid the contention of shared resources such as the last level cache are key challenges for the operating system scalability research on multicore systems. The contributions of this thesis include:

1. The key factors which affect scalability are acquired by analysing. In this thesis, a microbenchmark suite is designed and the speedups of system services on mainstream operating systems are analyzed and compared. Measurements on practical systems indicate that many system services of an operating system have problems in scalability. The in-depth analysis based on the measurement accurately reproduces the way of how bottlenecks are generated. Based on the analysis, our conclusion is that synchronization operations protecting the shared data in operating systems are the largest factors that limit the speedup of system services on multicores.

2. Methods are proposed to simulate and avoid the lock thrashing phenomenon. Based on the discrete event simulation technology, a simulator is proposed to simulate the lock thrashing phenomenon. This simulator integrates all hardware and software factors that cause the generation of lock thrashing and reproduces the lock thrashing phenomenon accurately and rapidly. Using the simulator to guide the avoidance of lock thrashing, a requester-based scalable lock is proposed. This lock adaptively determines the way of lock waiting according to the number of lock requesters and can avoid the lock thrashing phenomenon significantly. A lock contention aware scheduler is also proposed. By monitoring the lock usage in the per-task grain, this scheduler runs all lock-intensive tasks on a special set of cores and the optimal number of cores in the special core set is decided using model driven search. This scheduler achieves 46% better speedup when comparing with previous work.

3. A resource contention aware scheduler is proposed. This method selects the best heuristic metric using the offline collected resource contention data to represent a task's resource requirements. The metric selecting process not only considers a metric's accuracy, but also considers its stability. Based on a task's resource requirements, which is acquired online, the task-to-core mapping mechanism and task selection mechanism are both modified to co-run tasks with complementary resource requirements. Taking the scientific applications as an example, our proposal improves the speedup by 13%, achieving the full potential of the resource contention aware scheduler.

4. A method based on the function's scalability value is proposed to pinpoint the scalability bottlenecks in operating systems. This method defines the scalability value of a function as the execution time difference on multicore and single core environments, normalized by the number of completed unit works. By analysing functions with large scalability value, the scalability bottlenecks can be found. The effectiveness of this method is verified by the use of online tranaction processing applications. By completely removing the restrictive factors in operating systems, the speedup can be improved by 49%.

# 目 录

# 第1章 引言

本章介绍了多核处理器出现的背景及多核芯片的发展趋势，总结了操作系统在多核处理器上所面临的可扩展性挑战，即如何定位和分析操作系统在多核平台上对加速比的限制性因素、如何模拟和避免锁颠簸现象以及如何避免最后级缓存等硬件资源的竞争，同时概括了本文在操作系统可扩展性上的研究工作，给出了后文组织关系。

## 1.1 研究背景

### 1.1.1 处理器的变革

2005 年以前，主流的处理器均是单核心处理器。对于单核处理器而言，其性能提升的方式是靠不断地改进处理器的生产工艺。以 AMD Opteron 处理器系列为例，2003 年采用的是 100nm 工艺，而如今已经变成了 45nm[124]。处理器工艺的不断改进使得处理器的时钟频率以及单位面积上的晶体管数目不断地增加。增加的晶体管用来将处理器的指令流水线变得粒度更细、阶段更多。这种改进配合增加的时钟频率能够提升处理器的性能。对于处于这个时期的软件来说，最有效的性能优化方式就是将其执行在一个频率更高、集成度更高的处理器上，而软件本身不需要进行任何的改动。这种获取软件性能提升的方式被称为“免费的午餐”[1]。

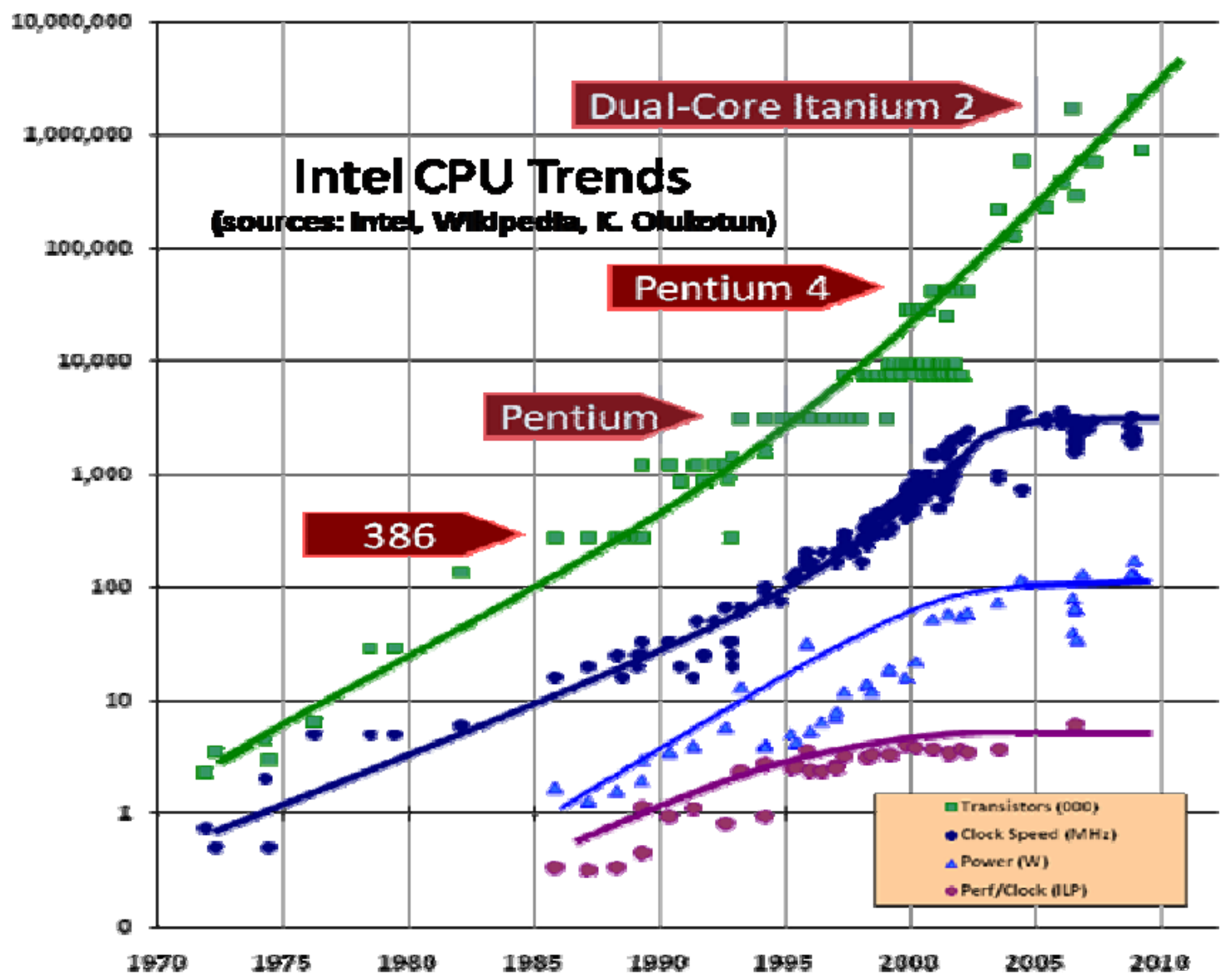

图 1.1 晶体管数目、时钟频率、功耗和指令级并行度随着时间的变化

例如，图 1.1 表示了 Intel 处理器上的晶体管数目、时钟频率、功耗和指令级的

并行程度随着时间而变化的曲线。可见, 2005 年以前, 芯片上晶体管的数目基本符合摩尔定律, 呈指数增长的趋势。而且, 随着时间的增加, 处理器时钟频率和指令级并行程度都以类似的趋势在上升。

然而, 一味地增加处理器的时钟频率和单一处理器上的晶体管数目却逐渐暴露出四方面的问题:1) 时钟频率的提升使得芯片功耗也有相应地提升(见图 1.1)。然而，芯片的功耗过大将产生大量的热量, 甚至能够熔化芯片本身。2) 制造粒度更细、阶段更多的指令流水线将极大地提升芯片的设计复杂度。3) 时钟频率的提升也导致处理器和内存的速度差距越来越大, 这使得访问内存的延时会极大地影响软件性能, 产生“内存墙”问题。4) 由于芯片集成度不断地提高, 处理器的线宽也越来越小, 已经接近了物理极限。

为了解决上述 4 个问题, 多核处理器应运而生。与单核处理器不同，多核处理器不再是一味地提升一个处理器的时钟频率和流水线深度, 而是将晶体管做成很多独立的核, 集成在同一个芯片内部。对于每个计算核心而言，其时钟频率不会很高, 流水线也不会被设计得很复杂。同时, 芯片上总体晶体管数目的增长趋势同样符合摩尔定律 (见图 1.1)。鉴于多核处理器的众多优势, Intel、AMD 和 IBM 等芯片制造厂商都已经采用了多核处理器。如今, 多核系统已经被广泛地应用在了各种各样的计算环境中, 如服务器领域、个人电脑领域甚至移动计算领域。不仅如此, 整合在同一芯片上的核数将会变得越来越多。如果根据摩尔定律进行预测, 那么, 不出十年, 每个计算机系统将拥有成百上千个核[2]。这种增长趋势已经在工业界中有所体现。例如, Intel 公司已经推出了具有 80 个核的原型芯片[3], 而 Tilera 公司已经发布了具有 100 个核的处理器[4]。作为一个例子, 图 1.2 表示了 Tilera 芯片的硬件架构以及每个处理器核的示意图。

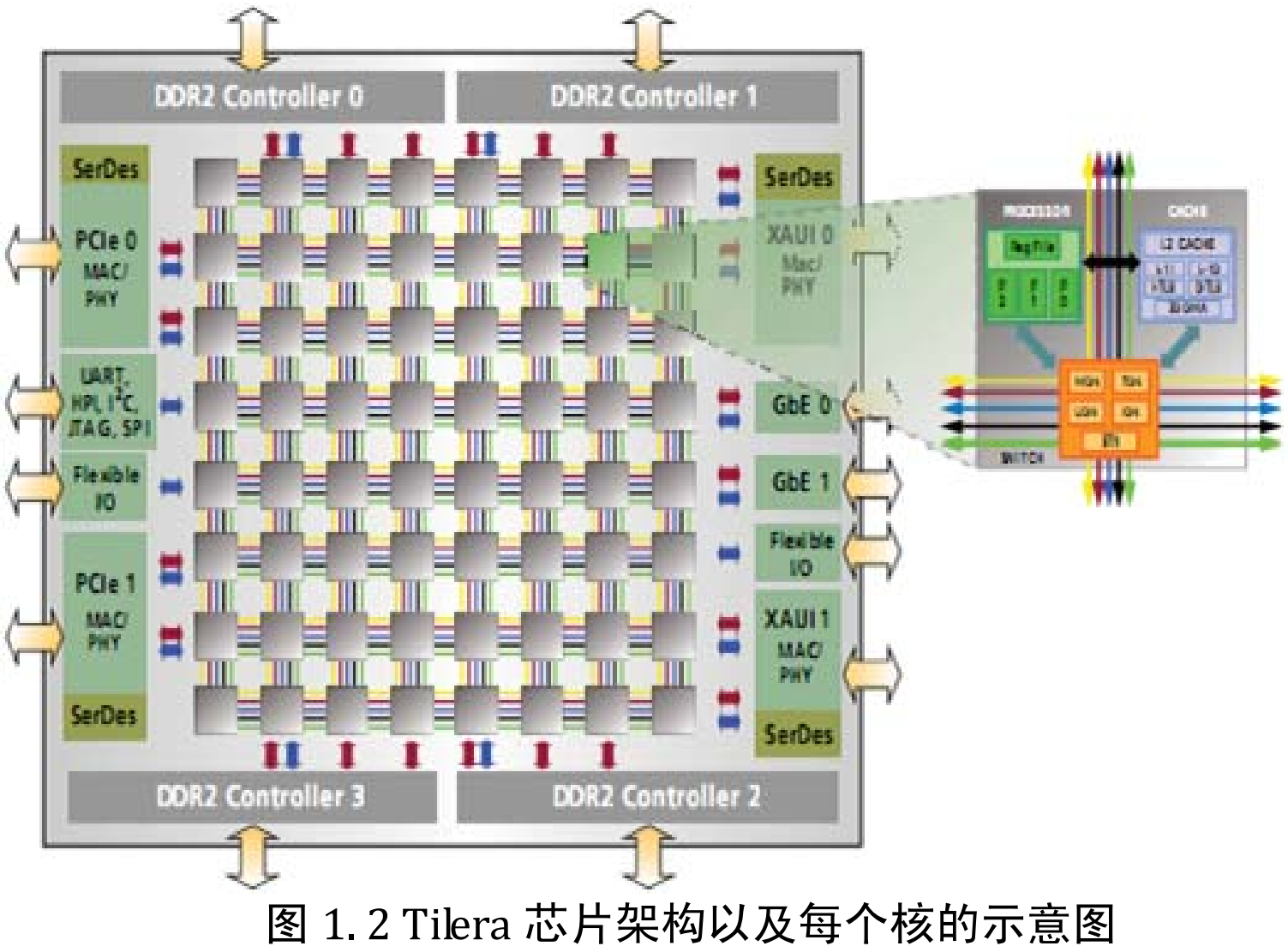

图 1. 2 Tilera 芯片架构以及每个核的示意图

### 1.1.2 可扩展性定义和量化

在多核时代，软件的性能主要取决于其利用并行的能力，而且这种能力的获取一定是通过编写并行代码显式实现的，不再对软件透明。举个例子，如果在拥有 20 个核的计算机系统中设计一个文件服务器应用，那么，提升该应用性能的最佳方法就是将其设计成并行的。这样，不同的任务可以被调度到不同的核上并行执行。否则，如果该应用是串行的，那么，在任何时刻系统中只有一个核处于工作状态，其它的 19 个核都处于空闲态。

本文利用“并行可扩展性”(简称“可扩展性”)来度量多核平台上软件的性能随着核数的增加而取得相应提升的能力。软件的可扩展性越好，说明它利用核数就越充分，那么向系统中添加更多的核就能够获得更好的性能。否则，它不能利用更多的核来提升性能，增加更多的核也是浪费的。本文中，可扩展性的好坏是通过程序的加速比(S)来量化的，其定义为 $S = T(N)/T(1)$。其中，T(N)表示程序在 N 核的系统上获得的总体吞吐量，而 T(1)表示程序在 1 个核的处理器上得到的吞吐量。根据这个定义不难发现，S 越接近 N，软件的加速比越大，可扩展性就越好。反之，S 越接近 0，说明软件的加速比越小，可扩展性就越差。针对面向吞吐量的应用，加速比 S 可以被固定时间的加速比模型（即 Gustafson 定律[6]，Amdahl 定律[7]的变型）描述为 $S=N\times(1-P)+P$，其中 P 为程序的串行执行成分占总体执行的百分比。当 P=1 时，表示该程序是串行程序，获得的加速比为 1。而当 P=0 时，表示该程序是完全并行的程序，其获得的加速比为 N。

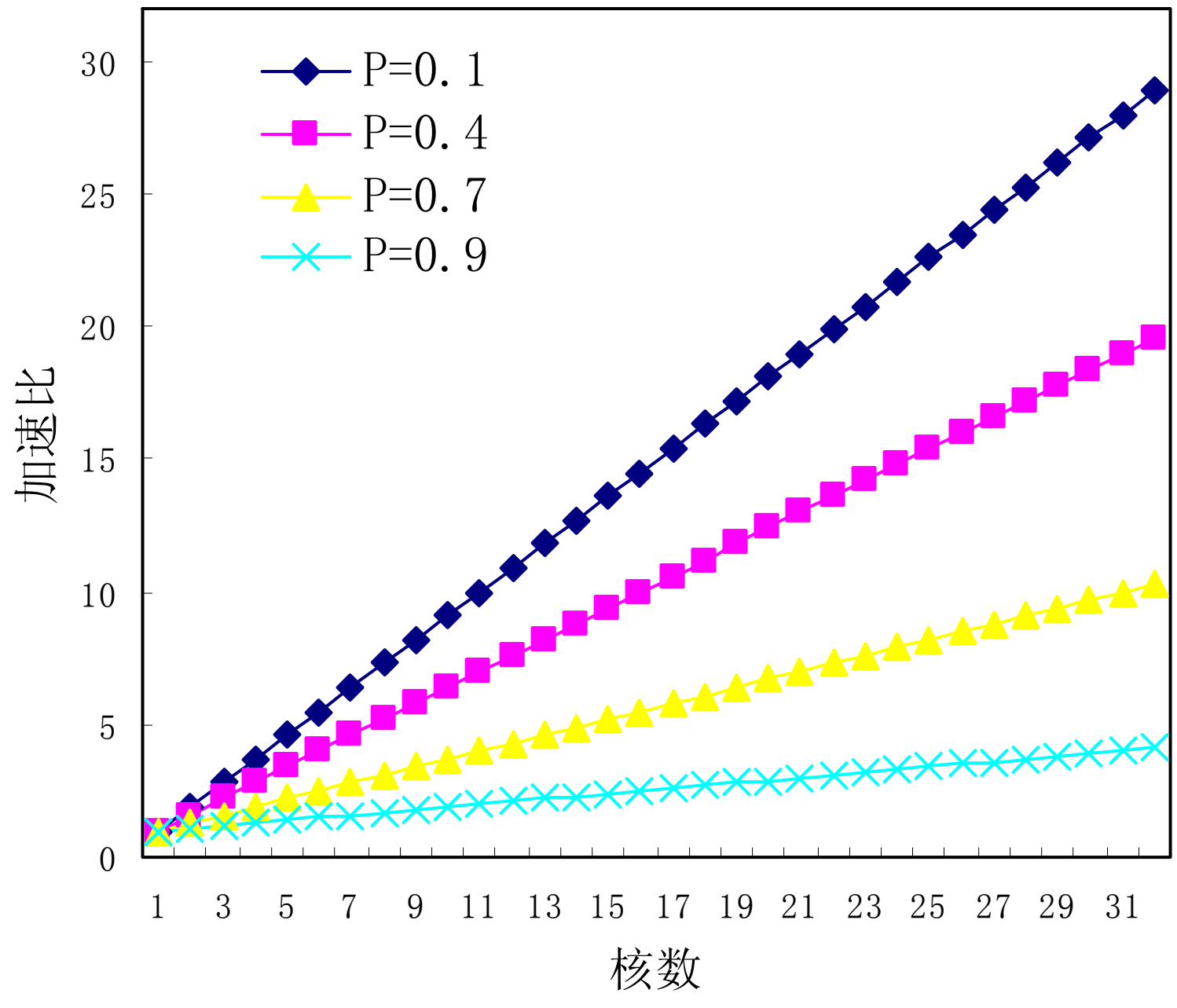

图 1.3 使用不同的 P 值时，获得的加速比随核数的变化

图 1. 3 表示了使用不同的 P 值时，获取的加速比随着核数的变化曲线。可见，P 的值越大，程序的行为越接近串行程序，获得的加速比也就越小，利用核数也就越不充分。从图中的结果可知，提高软件加速比的最有效方式是减少软件中串行执行成分占总体执行的比例。其途径可能是通过提高同步互斥代码的执行效率、减少临界区代码的长度或者降低硬件资源的竞争等。

### 1. 1. 3 操作系统的可扩展性挑战

本文的研究动机源于一个很自然的问题，即操作系统作为管理硬件、对应用层提供特权服务的软件，其性能是否能够随着核数线性地扩展。实际上，多核技术的成功很大程度依赖于在多核平台上运行的软件的可扩展性。然而，对于应用程序而言，其可扩展性不仅依赖于应用程序本身的并行性，而且依赖于应用程序运行在的操作系统。对于系统服务密集型应用(如网页服务器、文件服务器等)，如果操作系统的可扩展性不好，那么，这类应用将很难从硬件提供的并行上或者本身具备的并行上得到好处；对于计算密集型应用(如科学计算等)，尽管应用程序几乎不使用系统服务，但是可扩展性却受到操作系统的影响。因此，操作系统在多核平台上的可扩展性对利用多核处理器的计算能力是至关重要的。图 1.4 表示了本文的关注点在软、硬件体系结构中所处的位置。

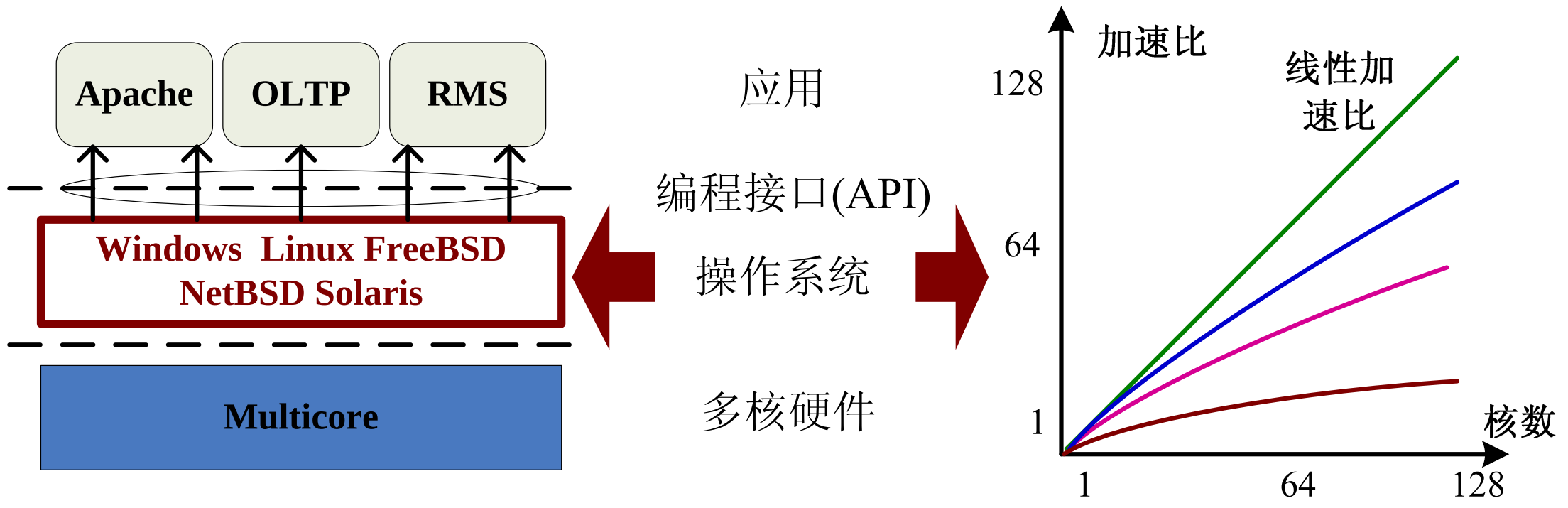

图 1. 4 本文的关注点在软、硬件体系结构中所处的位置

然而，当前的操作系统却是针对单核处理器以及小规模对称多处理器(SMP)[62]而设计的。相比与传统的对称多处理器，多核处理器至少具备两方面不同。1 核数的规模不同。对于对称多处理器而言，小规模的系统只有 2 个处理器，中等规模的系统会有 4 到 8 个处理器，而大规模系统的处理器数目才会多于 16 个。对于多核系统，目前具备 2 到 4 个核的系统已经相当普及，而且工业界已经发布了上百核的处理器。根据摩尔定律进行预测，假设 2005 年每个芯片上拥有 2 个核，那么，到了 2015 年，芯片上的核数将会达到 256 个，而到了 2018 年，核数将达到 1024 个。核数的增长趋势如图 1.5 所示。2 硬件资源共享。为了提高芯片资源的利用率，

很多关键的硬件资源如最后级缓存、内存控制器等在多核处理器上均是共享的。图 1.5 表示了对称多处理器和多核处理器的结构图。从图中对比可见，在多核系统上，每个核只拥有私有的第一级缓存(L1)，而所有核共享最后级缓存(L2)。然而，在对称多处理器上，不仅第一级缓存是私有的，最后级缓存也是私有的。

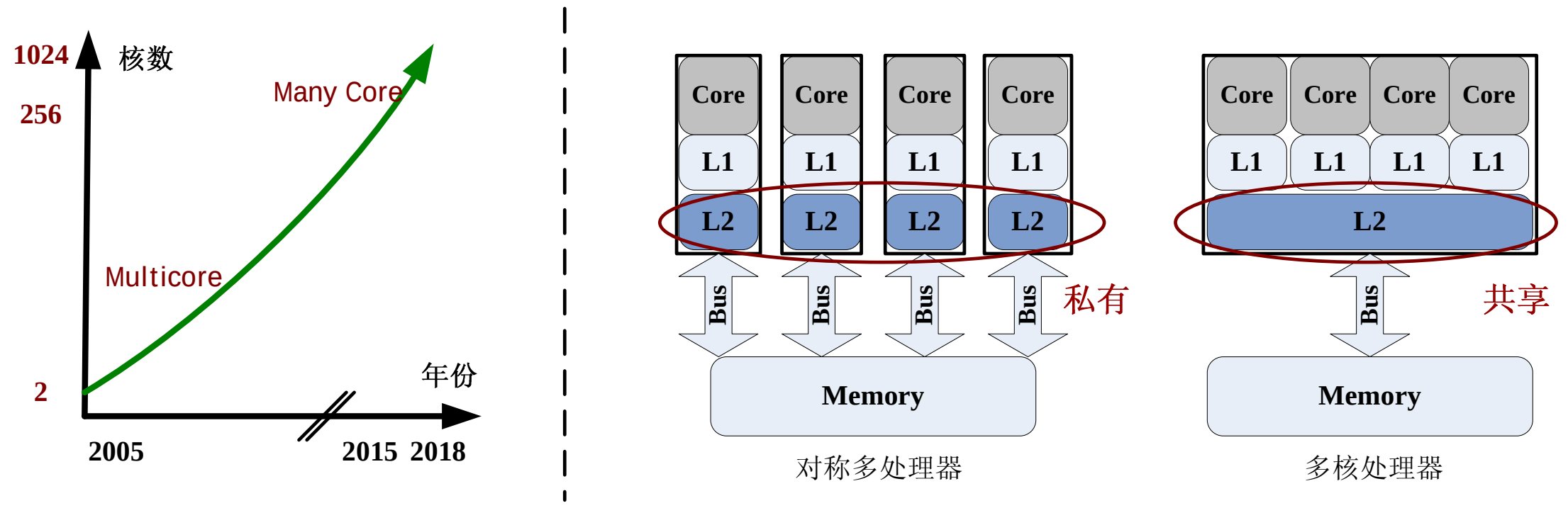

图 1.5 多核处理器和对称多处理器的两方面不同

由于操作系统并没有考虑到多核系统上核多、硬件资源共享的硬件特点，因此其可扩展性在多核这种新型的硬件平台上存在着挑战。本文针对如下 4 个具体的操作系统可扩展性问题进行了深入的研究。1) 针对系统服务接口，不同操作系统在多核平台上的可扩展性瓶颈是什么，是否相同，是如何产生的。2) 核数的增加导致操作系统中的锁激烈地竞争，进而导致锁密集型应用的吞吐量随着核数的增加反而下降(锁颠簸)。然而，我们需要回答锁竞争为什么会导致吞吐量下降、如何在操作系统中设计有效的机制或策略来避免该现象的发生等问题。3) 如何利用操作系统避免多核平台上共享最后级缓存等硬件资源的竞争。4) 如何有效地定位复杂系统服务密集应用中的操作系统的可扩展性瓶颈。

## 1.2 本文研究内容

操作系统在多核平台上的可扩展性研究的最终目标是要使得操作系统充分利用多核计算资源,对应用程序提供高效的编程接口。本文的具体研究内容包括:

1、针对系统服务接口的操作系统可扩展性分析和比较。任何一个软件层的可扩展性问题均会反映在最终的测试结果中。因此，如何设计测试程序可以只凸显操作系统层的可扩展性瓶颈是一个值得研究的问题。另外，每个操作系统的系统服务接口通常有上百个，如何在保证系统之间比较公平性的前提下从中选择最具代表性的系统服务接口也是一个必要的研究问题。而且，通常的操作系统内核代码有上百万行，操作系统中的分析工具也是种类繁多，如何对各种分析工具产生的输出进行提炼和结合以准确地还原出可扩展性瓶颈的产生方式也是一个挑战。

2、锁颠簸现象的模拟和消除。多核平台上核数的指数式增长导致操作系统中的锁被激烈地竞争。这种竞争会导致应用程序的吞吐量随着核数的增加反而下降(称为锁颠簸现象，如图 3.1 所示)。理解锁竞争导致吞吐量下降的软、硬件因素以及对这些因素进行整合以快速、准确地重现锁颠簸现象就是一大挑战。大部分模拟或者建模锁竞争的相关工作[48-52]不能重现实际系统上观察到的锁颠簸现象，文献[53]中提出了唯一一个可以观察到吞吐量下降的模型，但是该模型没有考虑非锁数据产生的缓存缺失，而且难于求解。另外，如何在操作系统内部设计有效的运行时机制来避免锁颠簸现象也是一个具有研究价值的问题。对相关领域的调研发现，解决该问题一般是通过改进的锁机制来达到。然而，该研究方向上仍有很大的挑战和进一步改进的空间。对于改进锁机制以避免锁颠簸现象的方法，最大的难点在于如何将锁实现引入的开销降至最低。尽管相关工作[55-56]已经通过使用复杂原子指令减少了这种开销，但是当锁竞争比较激烈时，这些方法并不能很好地避免锁颠簸现象，同时资源利用率和能耗有效性均较低。同时，由于调度策略会影响锁竞争的程度，因此通过改进的调度策略来避免锁颠簸现象以达到比基于改进锁机制更好的可扩展性也是可行的研究方向。对于通过改进调度策略来避免锁颠簸的方法，最大的挑战在于如何在线、准确地检测到吞吐量下降现象。为了达到这一目标需要解决如下的几个子问题:1) 如何在运行时估计应用程序的吞吐量。2) 如何避免系统中测量误差带来的负面影响。3) 如何动态适应程序用锁行为的变化。

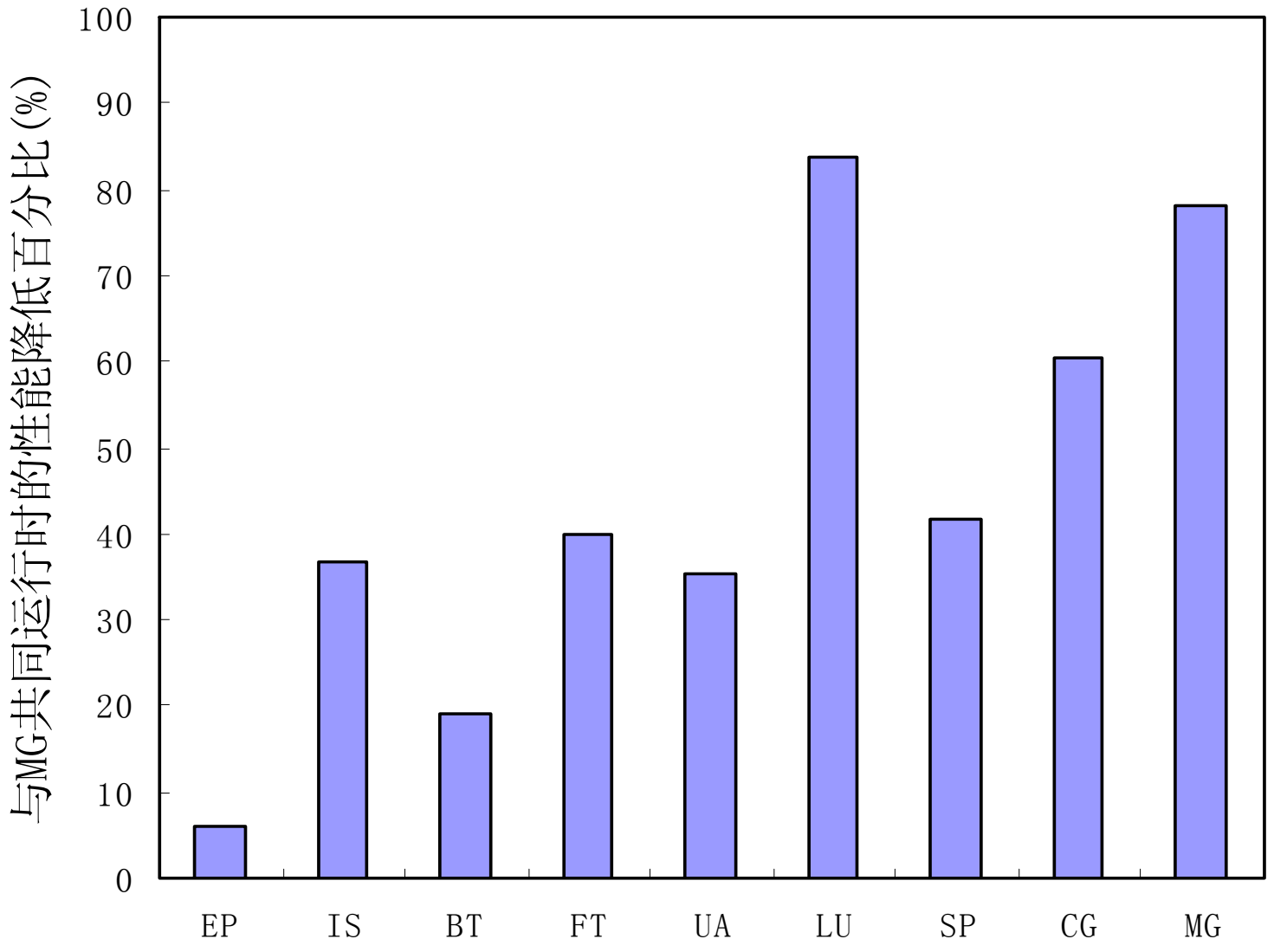

图 1.6 不同 NAS 程序和 MG 共同运行时导致的性能降级

3、 共享硬件资源竞争避免。多核平台上最后级缓存等硬件资源共享问题导致了系统的可扩展性差，同时影响独立程序的性能(如图 1.6 所示)。相关工作已经

证明调度策略的改进有助于提升系统总体的可扩展性和独立程序的性能[11-18]。但已有工作在如下 3 方面仍存在不足：首先，调度器需要感知每个任务的共享资源需求信息来进行调度。如何表示程序的共享资源需求是第一个不足。相关的工作[11-14]凭借经验来获取这种信息，因而不够准确。文献[15]提出了系统地选择合适度量指标的方法。但是该工作仅仅考虑了利用启发式指标代表程序资源需求时的准确性。而在实际的系统中，资源竞争可能会导致程序的资源需求大幅度地增加，如何将这种实际情况反映在指标的选择过程中是一个需要研究的问题。另外，相关工作[11-13, 15-18]仅利用负载均衡时任务到核的重新映射机制或者上下文切换时的调度顺序调整机制来避免硬件资源竞争。是否可以将两种机制结合以达到更好的系统可扩展性也是一个值得研究的问题。最后，很多相关工作[15-17]提出的调度策略在运行时均依赖于中心化的排序。然而，这种方法在任务相对较多时，可扩展性很差。能否在实际系统中实现纯分布式的竞争感知调度也是一个挑战。

4、 操作系统的瓶颈定位方法。许多复杂应用大量地使用操作系统的系统服务接口。对于这类应用，如何定位操作系统对加速比的限制性因素既具有挑战性又具有很大的研究价值。其挑战性表现在可扩展性瓶颈可能存在于整个软件架构中的任何一层，可能在应用本身，可能存在于操作系统中，也可能存在于硬件上。同时，虽然可用的分析工具种类繁多，但是均不直接针对操作系统可扩展性。如何从不同工具提供的大量信息中抽取出影响可扩展性的因素值得研究。

本文的研究工作可以概括为“一个中心，4 个研究点”。一个中心指的是本文所有的工作均是围绕着操作系统在多核处理器架构上的可扩展性问题展开的，重点开展上述 4 个研究点的工作。4 个研究点之间的关系如图 1.7 所示。从图 1.7 中，可以看出提升操作系统在多核平台上可扩展性的一般流程：1) 瓶颈定位。针对给定的应用，首先需要利用工具或者方法定位操作系统的可扩展性瓶颈。该步骤应当指明什么是可扩展性瓶颈，如锁竞争，最后级缓存竞争等。2) 操作系统分析和比较。对操作系统进行分析和比较，需要还原出瓶颈的具体产生方式。3) 瓶颈解决策略。基于瓶颈的产生方式信息，提出最有效的瓶颈解决方法，并在实际系统上验证该方案的有效性。需要说明的是，解决每个操作系统中的可扩展性瓶颈时，均需按照上述 3 个步骤。然而，不同的可扩展性瓶颈可能在不同的步骤中存在着挑战。例如，对于硬件资源竞争问题的研究，定位资源竞争瓶颈和对该瓶颈产生方式的分析相对容易，而提出瓶颈解决方案一步挑战性最大。又如，针对系统服务接口的研究中，定位瓶颈也很容易，但是如何在不同的服务接口上还原出瓶颈产生的方式以及如何针对出现锁颠簸现象的服务接口设计出通用的避免方法却具有挑战性。

图 1.7 中的 4 个研究点分别针对操作系统可扩展性研究中 4 个具体的挑战，虽

然位于解决操作系统可扩展性瓶颈流程中的不同步骤，但均是从多核系统的核多、硬件资源共享的硬件特点而引发出的：核数的大幅度增加而导致的操作系统可扩展性问题在研究点 1 中通过对系统服务接口的分析和比较被揭示；研究点 2 针对研究点 1 中发现的特殊现象(锁颠簸)，深入理解并提出了有效的解决方案；除了由软件资源(锁)而导致的竞争，多核平台上的硬件资源共享同样会导致竞争，该问题的解决由研究点 3 覆盖；解决操作系统可扩展性瓶颈的前提是能够有效地定位出瓶颈产生的位置。尽管某些可扩展性瓶颈的定位问题比较容易，但是达到对任何瓶颈均能有效地定位这一目标却具有很大的挑战性。研究点 4 针对多核这种新型的硬件平台解决了操作系统的瓶颈定位问题。由于这些研究问题均是操作系统可扩展性研究中非常核心和根本的问题，因此本文选择它们作为主要的研究内容。

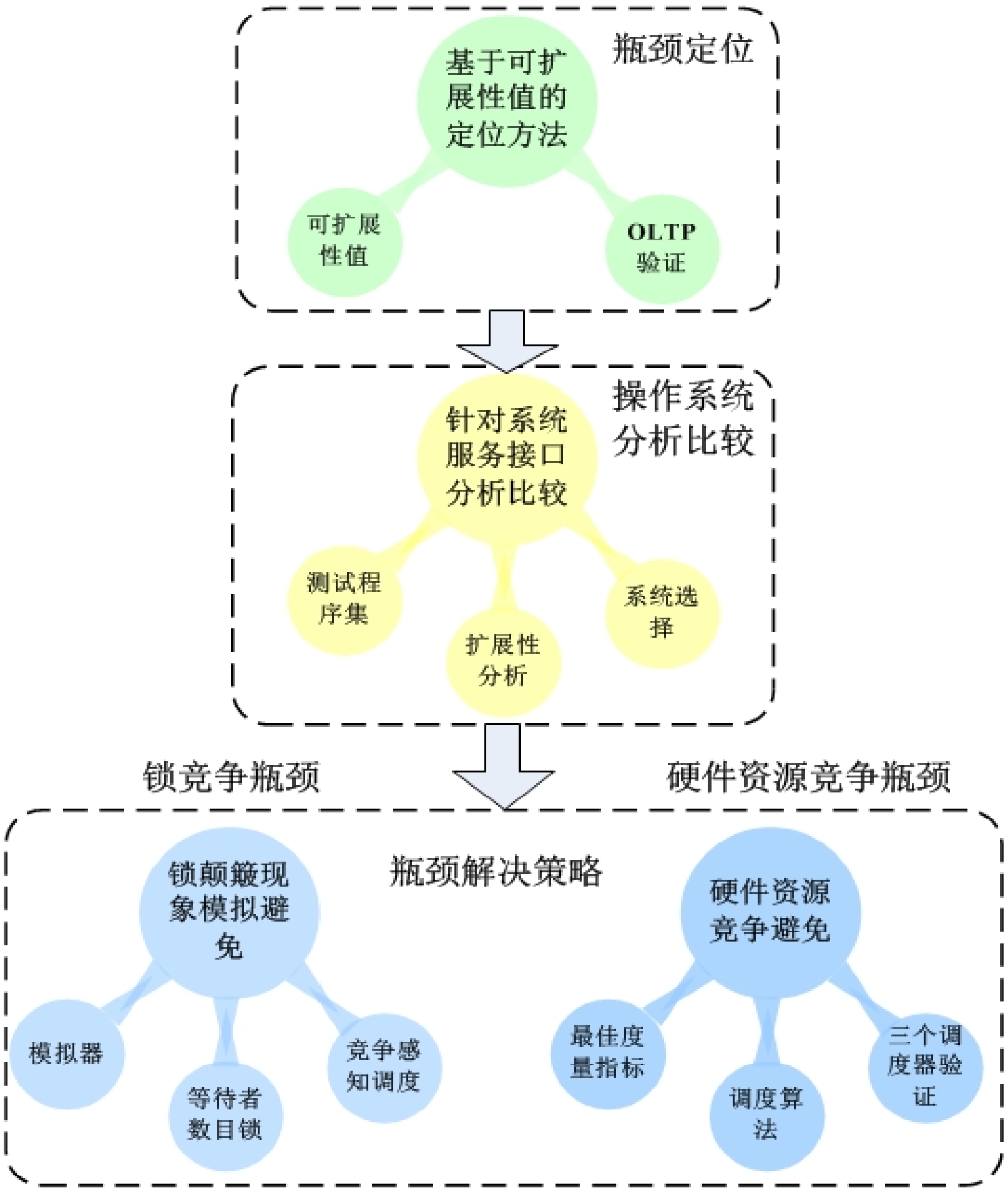

图 1.7 4 个研究点之间的关系

## 1.3 本文工作成果

针对 1.2 节列举的研究内容，展开了相应研究，取得了如下研究成果：

1、设计了微基准测试程序集，分析和比较了主流操作系统的服务接口在多核

平台上的加速比。测试程序集在保证系统之间公平比较的前提下，提出了准确的测试框架，测试了最具代表性的服务接口。在此基础上进行深入分析，准确地还原出了瓶颈的产生方式，并得出结论：多核系统中保护共享数据的同步互斥操作是限制服务接口加速比提升的主要因素。

2、提出了锁颠簸现象的模拟方法及避免机制与策略。基于离散事件仿真技术，对导致锁颠簸的软、硬件因素进行整合，设计了模拟锁颠簸现象的模拟器，能够准确、快速地重现锁颠簸。利用模拟器指导锁颠簸避免，提出了基于等待者数目的可扩展锁机制。该机制根据等待者的数目自适应地决定等锁方式，很好地避免了锁颠簸。同时设计了锁竞争感知的调度策略。该策略检测每个任务的用锁程度，利用模型驱动搜索将用锁密集任务集中在部分核上运行，采用锁竞争感知的调度策略比相关工作提升加速比达 46%。

3、提出了资源竞争感知的调度策略。该策略利用离线搜集的资源竞争数据，选择出了度量任务资源需求的最佳启发式指标。选择过程不仅考虑了启发式指标的准确性，而且考虑了指标的稳定性。基于在线获取的任务资源需求信息，改进了任务到核的映射机制以及核上的任务选择机制，使得具有互补资源需求的任务被同时调度，显著降低了共享资源竞争程度。以科学计算应用为例，提升加速比达 13%，充分发挥出了资源竞争感知调度的潜力。

4、提出了基于函数可扩展性值的操作系统瓶颈定位方法。该方法定义一个函数的可扩展性值为该函数在多核和单核上每单位工作量的执行时间之差。通过分析具有较大可扩展性值的函数，得到影响加速比提升的因素。以在线事务处理应用为例，提升加速比达 49%，有效削减了操作系统对加速比的限制。

## 1.4 论文组织

本文后续章节将按如下方式展开。

第 2 章，针对系统服务接口对主流操作系统在多核平台上的可扩展性进行了分析和比较，并介绍操作系统比较的相关工作。

第 3 章，介绍基于离散事件仿真技术的锁颠簸现象模拟器，并介绍基于等待者数目的可扩展锁机制以及锁竞争感知的调度策略以避免锁颠簸。

第 4 章，提出资源竞争感知的调度策略，并介绍资源竞争避免的相关工作。

第 5 章，提出基于函数可扩展性值的瓶颈定位方法，并介绍已有工作。

第 6 章，总结本文工作，并指出将来可能的研究方向。

# 第2章　针对系统服务接口的可扩展性分析和比较

## 2.1　本章引言

芯片上核数的增加意味着硬件可以提供的并行程度正逐渐增大。然而，一个不可避免的问题是应用程序是否可以充分地利用硬件提供的并行,或者应用程序是否可以随着核数的增加呈现较好的可扩展性。多核技术的成功很大程度上取决于运行其上的应用程序。但是，应用程序的可扩展性不仅取决于程序内在的并行，而且依赖于操作系统的可扩展性，尤其是使用系统服务比较多的应用程序(如网页服务器和文件服务器等)。

尽管操作系统的可扩展性在对称多处理器(SMP)平台上已经有了比较深入的研究[19-21]，然而操作系统在多核平台上的可扩展性研究却不尽相同。其原因有二。1）多核系统比多处理器系统的核数规模大，这使得可扩展性成为一个亟待解决的问题。2）很多对性能影响很大的硬件资源，例如最后级缓存和内存控制器等，都是共享的。因此，对于应用使用者和系统管理者而言，需要了解不同操作系统对不同系统密集型应用的可扩展性差别，以方便为应用选择合适的操作系统。对于打算提升操作系统可扩展性的系统开发者和打算避免操作系统可扩展性缺陷的应用设计者而言，明确内核可扩展性瓶颈产生的原因非常重要，并且依赖于对系统行为的深入分析。

鉴于开源操作系统在学术界和工业界的重要地位，本章设计并且实现了一套微基准测试程序集(microbenchmark suite)系统地评测和比较 3 个开源操作系统(Linux, Solaris 和 FreeBSD)在 AMD 32 核系统上的可扩展性。该套测试程序集测量了操作系统的 5 个重要系统服务接口，其中包括进程管理、内存管理、文件系统、网络和进程间通信。本文同时结合系统分析工具的输出和内核代码的分析还原出每个可扩展性瓶颈的产生方式。

实验结果表明尽管某个操作系统可能在某方面取胜，但没有一个操作系统在所有的子系统测试中均比其它操作系统表现出色。例如，Linux 系统在文件描述符的操作以及进程创建和删除密集的操作上比 Solaris 和 FreeBSD 的可扩展性好，但是，在套接字创建和删除频繁的操作上，Solaris 要好于 FreeBSD，而 FreeBSD 好于 Linux。通过使用分析工具以及源代码注入和分析，本文发现内核中的同步互斥操作是影响操作系统在多核平台上可扩展性的最重要因素。本文的实验结论可以用来为应用选择最佳的操作系统。而且，基于实验的结果以及在小规模平台上对目标

应用的分析，甚至可以预测该应用在大规模系统上的可扩展性。

## 2.2 实验环境搭建

### 2.2.1 实验平台架构和操作系统

本章选择了 3 个开源操作系统，分别是 Linux 2.6.26.8、Solaris (OpenSolaris 2008.11)和 FreeBSD 8.0-CURRENT。尽管每个操作系统都存在更加新的内核，但本文给出的实验结果仍然有意义，因为新版本的内核中仍然没有移除本章发现的大部分可扩展性瓶颈[8]。本文选用的 3 个操作系统均是 UNIX 或者类 UNIX 操作系统并且是 POSIX 语义兼容的。该特征使得 3 个操作系统对应用程序提供的编程接口(API)都是相同的。因此，相同的测试程序可以在不同的系统上不加修改的运行，这点也使得不同操作系统之间的可扩展性比较是公平的。

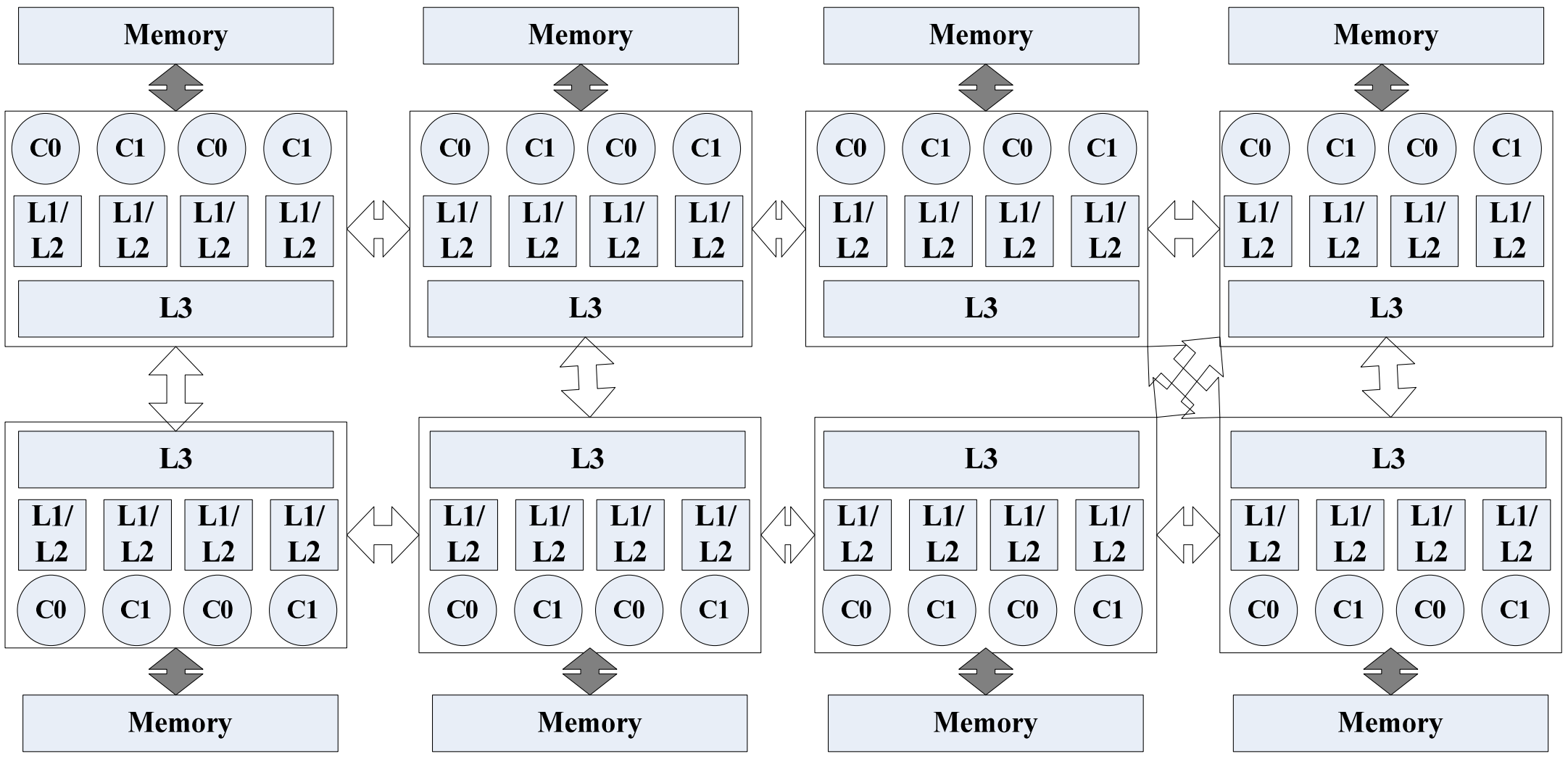

图 2.1 AMD 32 核系统硬件架构图

本章的实验均是在 AMD 32 核的 NUMA 系统上进行的。该系统由 8 个 AMD Opteron 8347HE 芯片组成，而每个芯片带有频率为 1.9GHZ 的 4 个核。每个核有 32KB 的 L1 数据缓存和指令缓存，2 个缓存均为 2 路组相连。同时，每个核有 512KB 的 8 路组相连的私有 L2 缓存。位于同一芯片上的 4 个核共享 2MB 的 L3 缓存。4 个芯片之间利用 Hypertransport 互联协议相连。系统中共有 32G 的内存，每个芯片连接 8G 的内存块(memory bank)。系统中共有 4 个 1TB 的硬盘，每个操作系统被安装到一个独立的硬盘中。该系统的硬件架构如图 2.1 所示。

### 2.2.2 测试程序

测试程序包括一套用于确定可扩展性瓶颈的微基准测试程序(microbenchmark)和用于可扩展性预测的实际应用。

1、微基准测试程序 本章设计的微基准测试程序集由 5 个测试程序组成，分别是 forkbench、mmapbench、dupbench、sembench 和 sockbench。测试程序分别对操作系统的进程管理、内存管理、文件描述符管理、SystemV IPC 以及套接字管理功能进行测试。所有的测试程序均使用相同的测试框架。在该框架中，有一个管理进程(master)负责创建工作进程(worker)和统计数据。每个工作进程被分配到一个核上产生负载(如图 2.2 所示)。这个框架确保用户层的测试代码不存在任何限制可扩展性的因素，如锁或者内存屏障。也就是说，在负载均衡的运行场景中，任何检测出的可扩展性问题肯定是由内核代码导致的。对于固定的核数，本文利用每个核的平均吞吐量度量程序的性能。本文中的吞吐量指的是单位时间里系统完成相关操作的次数。因为运行方式总是保持工作进程和测试的核数相同，而每个工作进程产生的工作量又都相同，因此，系统对于某个测试的可扩展性可以利用吞吐量随核数的变化趋势来刻画。在一个可扩展的系统中，吞吐量不应该随着核数的变化而变化，因为工作量和核数的增加程度是相同的。如果系统中存在可扩展性瓶颈，那么说明系统中的某些服务执行时受到某些串行因素(如锁，串行代码等)的影响，因此，任务不能像在单核系统上一样高效运行。进而，每个核的平均吞吐量将随着核数的增加而降低。降低的程度越大，其可扩展性越差。

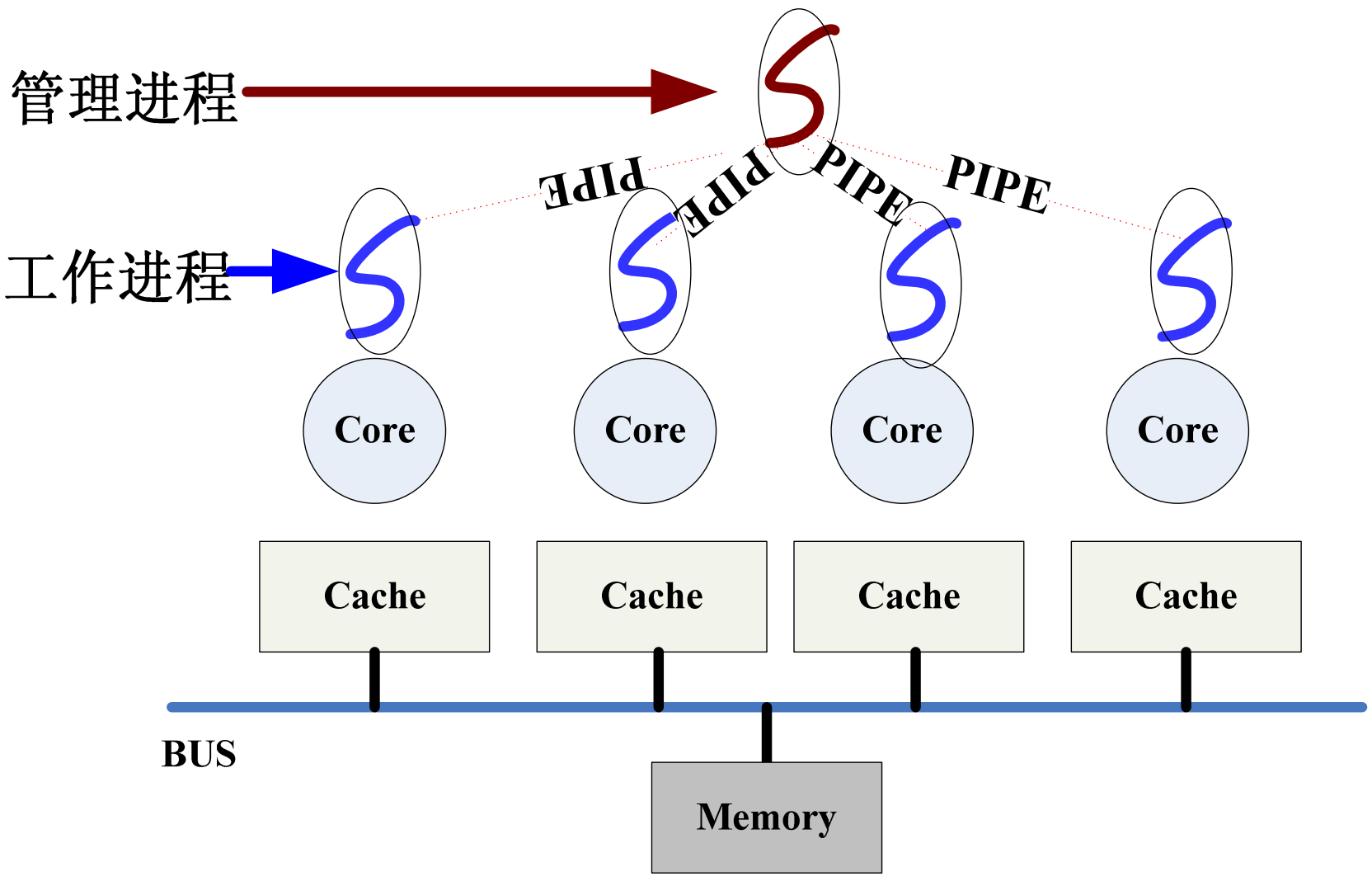

图 2.2 微基准测试程序的测试框架

2、实际应用 postmark 是被广泛使用的测试程序。它可以模拟电子邮件服务器，新闻服务器和电子商务应用服务器[22]。在程序开始时，postmark 产生一组随机大小的文件。在测试过程中，需要在创建的文件集上执行数目固定的一系列事务(transaction)。每个事务由两部分组成，首先是创建或者删除随机选择的文件，其次是读或者追加一个随机选择的文件。在程序结束时，所有的文件均被删除。在测试程序中，利用事务的吞吐量作为程序的性能。为了使用 postmark 研究系统的可扩

展性，本文将原来的串行程序改写成了多进程(multiprocess)程序。每个进程均有自己独立的文件集并且独立的执行一系列事务。在测试过程中，使用的进程数和测试的核数保持相同以发现可能的可扩展性问题。在实际的应用场景中，多个进程通常是工作在相同的文件集合上，比如 Apache 网页服务器[121]。尽管对相同文件的并发访问可能会造成竞争，进而使得应用本身具有可扩展性问题，但是本文侧重的是操作系统的可扩展性瓶颈，因此，我们让每个进程访问的文件集合互不相同。本章的实验结果可以认为是文件服务器的上限性能。另外，如果进程访问的是硬盘上的文件，那么 I/O 将会是系统的最主要瓶颈。这样，系统的可扩展性瓶颈对吞吐量的影响将变得微乎其微。因此，程序中的所有文件都被放在内存文件系统中(tmpfs)。

### 2.2.3 跟踪程序的执行

为了确定和分析可扩展性瓶颈，仅仅测量吞吐量随核数的变化是不够的。另外需要跟踪可扩展性瓶颈相关的系统行为，比如对锁的使用程度。对此，本文采用了各种性能分析工具。对于 Linux 系统，使用 Oprofile[9]获取程序执行时的函数调用图和每个关注函数的执行时间，并且使用/proc/lock_stat[23]获取激烈竞争的锁的信息。对于 Solaris 系统，使用 Dtrace[10]的工具 lockstat[24]来获取调用关系图，函数执行时间和锁的竞争程度。在 FreeBSD 系统上，我们的实验平台不支持 2 个广泛使用的内核调试工具 PMC[25]和 Kgmon[26]，而仅能够获取锁竞争的信息。为了避免调试工具对性能测量的影响，本文使用没有调试功能的内核来搜集测试数据，而使用带有调试和跟踪功能的内核进行可扩展性分析。

为了避免进程负载均衡带来的开销，每个进程被绑定在一个固定的核上。在 Linux 系统中，使用 sched_setaffinity()系统调用完成该功能。在 Solaris 和 FreeBSD 上，分别使用 pset_bind()和 cpuset_setaffinity()系统调用。在 Solaris 系统上，不能对所有的核均设置亲和性，至少有一个核是要留出来的[27]。因此，当在 Solaris 上使用 32 个进程时，留出一个核不做绑定。所有的测试程序均利用 GCC 进行编译，编译时开启优化选项 O3。

## 2.3 微基准测试程序的可扩展性瓶颈

### 2.3.1 forkbench

forkbench 是用于测量操作系统进程管理功能的微基准测试程序。在该测试程序中，每个工作进程不断的利用 fork()系统调用创建子进程然后利用 waitpid()等待

子进程结束。在子进程中，如果调度器调度其运行，则子进程立即调用 exit()。这样，系统的进程创建和删除操作将被频繁的执行。该测试程序在不同操作系统上随着核数的吞吐量如图 2.3 所示。

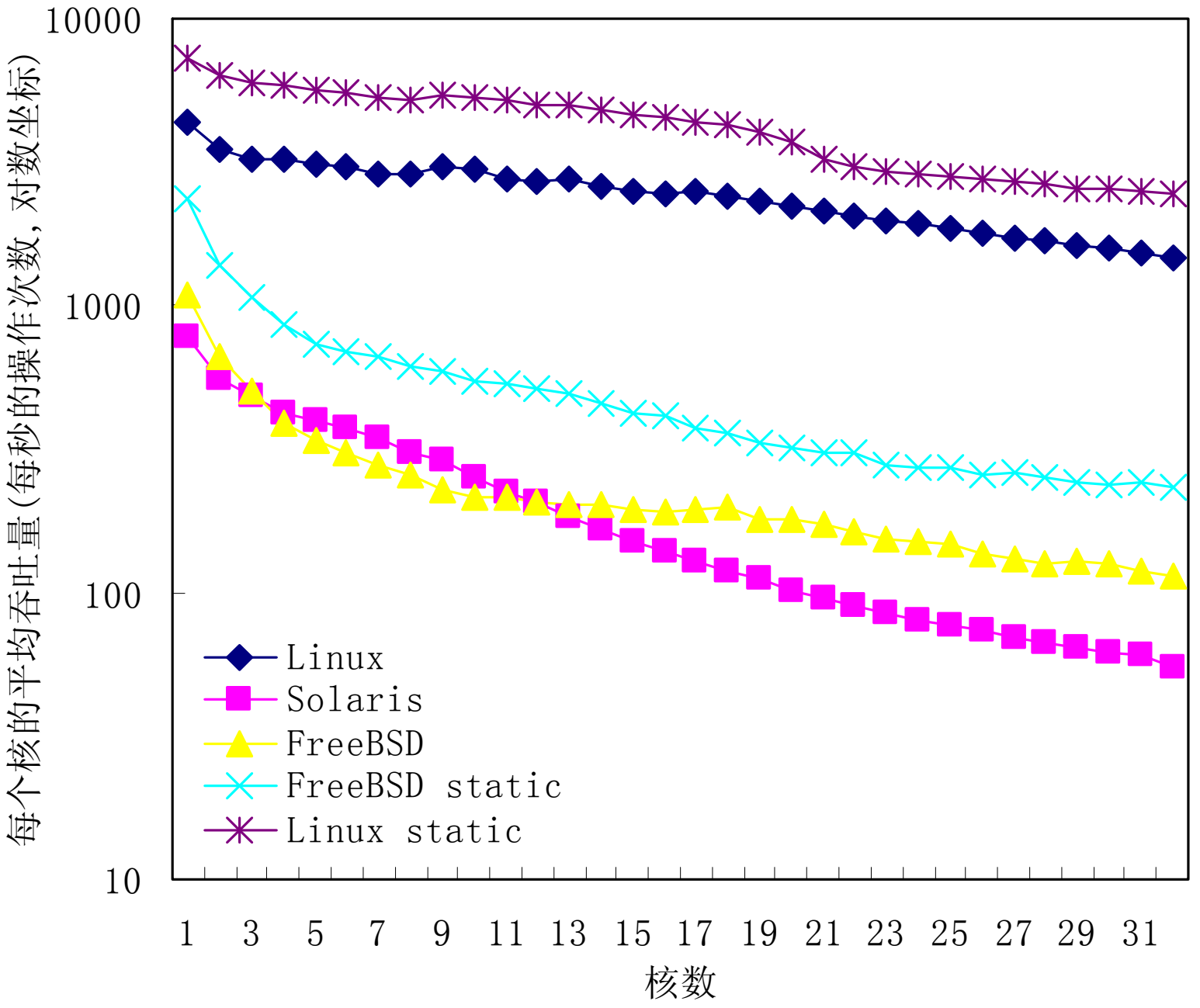

图 2.3 forkbench 随不同核数的平均吞吐量。平均吞吐量被定义成每个核一秒钟内进程创建和删除操作的完成次数。曲线"FreeBSD static"和"Linux static"代表测试程序利用静态链接后的结果。

从图中可见，Linux 比 FreeBSD 和 Solaris 的可扩展性好的多。在单核系统上，Linux 的吞吐量是每秒 4329 次操作，仅是 Solaris 和 FreeBSD 的 5 倍和 3 倍。然而，在 32 核系统上，Linux 的吞吐量分别是 Solaris 和 FreeBSD 的 26 倍和 12 倍。

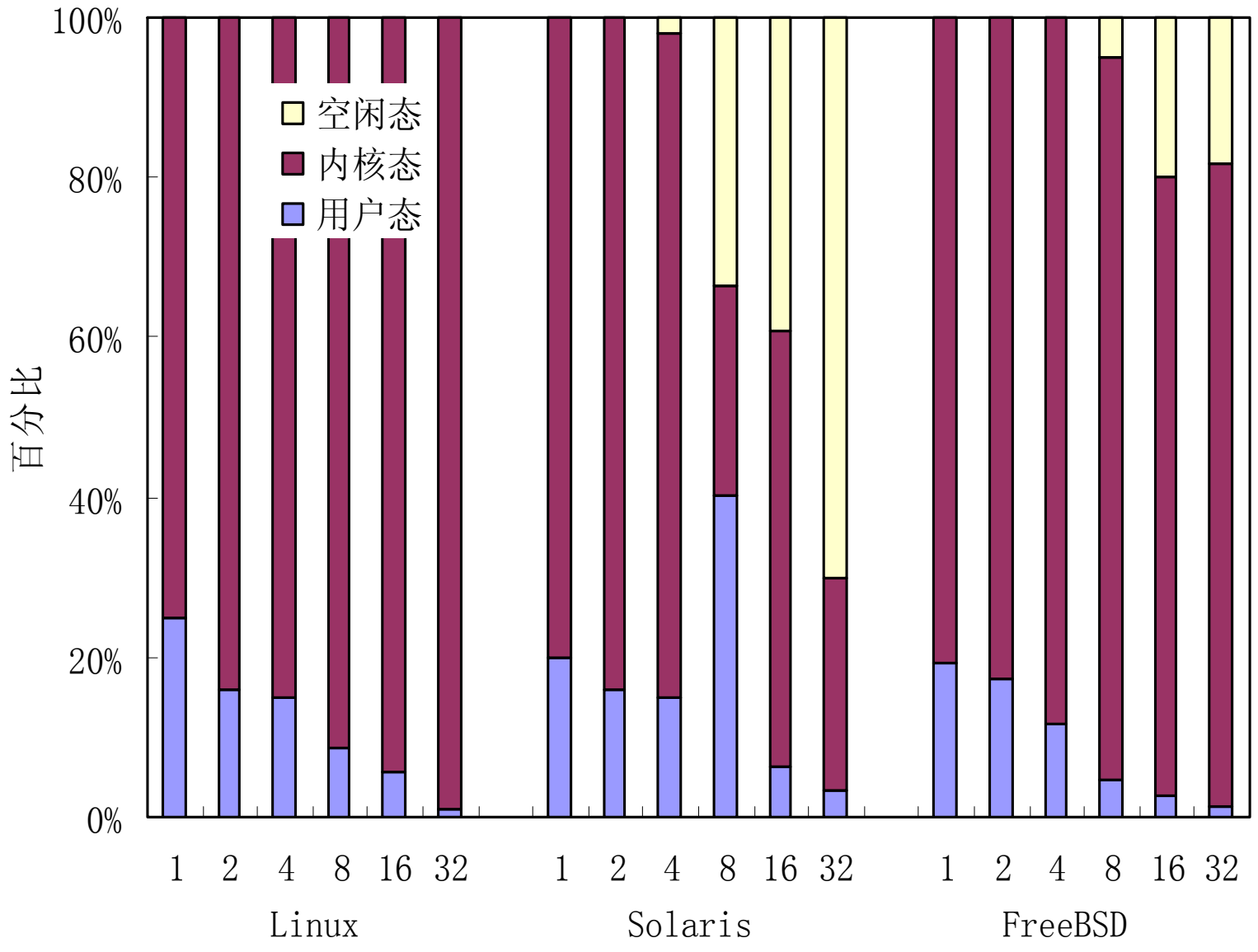

图 2.4 执行时间随着核数的分解

为了得到不同系统的吞吐量下降原因，图 2.4 展示了不同操作系统在不同核数

上总体执行时间的分解图。如图所示，随着核数的增加, Solaris 和 FreeBSD 系统上 CPU 空闲时间百分比随着核数快速地增长。例如,当使用单核时, Solaris 和 FreeBSD 系统上的 CPU 空闲时间百分比是 0%，然而，当使用 32 个核时，这 2 个系统的空闲时间百分比分别增长到了 70.11%和 17.14%。由于用户层的测试程序代码是可扩展的，这一结果清楚的说明了一些系统服务不能充分地利用 CPU。甚至 Linux 系统也未能幸免，因为其系统时间的百分比随着核数在大幅度地增加，表明完成相同次数的操作将会消耗更多的时间。

从测试结果中可以看出 Solaris 系统在 3 个系统中的可扩展性最差，因此我们首先分析 Solaris 的可扩展性瓶颈。当从单核变到多核时，我们按照内核函数增加的执行时间由大到小进行排序后，发现 mach_cpu_idle(), mutex_delay_default()和 mutex_enter()是执行时间增加最快的 3 个函数。这 3 个热点函数和它们占总体时间的比例如表 2.1 所示。这些结果表示系统中的某些锁竞争导致了可扩展性问题。通过使用 lockstat 工具，我们确定出了竞争最为激烈的锁是被 zfs_getpage()调用的一把读写锁。进一步的数据分析表明，函数 zfs_getpage()被频繁地调用是因为在测试程序执行时出现了频繁的小页错 (minor page fault)。例如，当 32 个核同时执行时，每秒发生 120,000 次小页错。在运行时对调用栈采样表明函数 zfs_getpage()是在缺页异常的处理函数中被调用的。为了得知异常产生的地址，我们利用 Dtrace 跟踪每次缺页异常的地址，同时利用 pmap 工具将缺页异常地址对应到进程的地址空间上。可以发现缺页异常是在对 libc.so 和 ld.so 2 个动态链接库的代码段和测试程序的代码和数据段进行访问时引发的。

表 2.1 执行 forkbench 时的 3 个热点函数

| 核数 | 函数 | 百分比 |
|---|---|---|
| | **Solaris** | |
| 32 | mach_cpu_idle | 73% |
| | mutex_delay_default | 8% |
| | mutex_enter | 3% |
| 1 | (user mode) | 34% |
| | mutex_enter | 7% |
| | hment_compare | 6% |
| | **Linux** | |
| 32 | dup_mm | 18% |
| | unlink_file_vma | 18% |
| | page_fault | 9% |
| 1 | page_fault | 23% |
| | handle_mm_fault | 10% |
| | unmap_vmas | 3% |

在操作系统中，当一个新进程通过调用 fork 创建时，库函数文件和程序本身的可执行文件要被映射进它的地址空间中。尽管库函数文件是动态链接到文件中的，

然而，在运行时，所有的进程需要访问同一库函数文件。而且，当工作进程和它们的子进程被创建时，程序的可执行文件是共享的。这样，测试程序开始运行后，如果一个进程需要访问库函数文件或者程序可执行文件的地址，这个地址在进程的页表中还是无效的。因此，从共享的文件中获取对应的页需要触发缺页异常。由于文件通常是缓冲在内存中的，因此缺页异常不需要进行 I/O 操作[28]，这种场景下触发的缺页异常被称作小页错。然而，并发地触发缺页异常处理函数需要获取保护共享文件的锁。而且，当程序结束运行时，解除库函数和可执行文件在进程地址空间的映射关系同样需要获取保护共享文件的锁。因此，锁竞争是进程创建和删除时的可扩展性瓶颈。

尽管 FreeBSD 和 Linux 系统的吞吐量下降程度和 Solaris 系统的下降程度不同，但是在 FreeBSD 和 Linux 上，测试程序的可扩展性瓶颈是类似的。在 FreeBSD 系统上，当使用 32 个核时，每秒钟的缺页次数是 13,000。我们对锁的使用进行分析发现由于 vm_fault()、vm_object_deallocate()和 vm_map_entry_delete()等函数频繁的调用导致了保护虚拟内存对象[29]的互斥锁被激烈地竞争。在 FreeBSD 系统中，除了对动态链接库(libc.so 和 ld.so)的代码段访问会导致缺页异常外，对其数据段和匿名内存的访问同样会导致缺页异常。在 Linux 系统中，保护动态链接库文件和程序的可执行文件的锁在函数 unlink_file_vma()和 dup_mm()中激烈地竞争。2 个函数总共的执行时间占总体函数的执行时间在一个核上仅有 0.5%，而到了 32 个核上却达到了 36% (见表 2.1)。为了验证瓶颈分析的正确性，在编译测试程序时，我们静态地链接系统库。这样，与系统库相关的竞争可以消除。我们对测试程序重新运行，得到的实验结果如图 2.3 所示。在 Solaris 系统中，静态链接后达到的可扩展性无法测量，因为 Solaris 对静态链接已经不再支持[30]。经过了这样的改动后，我们可以看到测试程序的性能在 Linux 和 FreeBSD 上随着核数的增加有所提升。

从该测试中我们可以发现同是锁竞争，但每个操作系统中的影响却不尽相同。其原因是锁实现在每个操作系统中有所不同。Linux 使用排号自旋锁保护共享的内存映射文件，因此，当锁的竞争程度随着核数变大时，由于自旋锁在内核中忙等的特性，CPU 空闲时间不会增长而系统时间却有所增加。相反，Solaris 使用读写锁。当进程要获取竞争激烈的读写锁时，该进程要进入睡眠状态，进而使得 CPU 变成空闲。相似地，FreeBSD 使用的是互斥锁，在竞争激烈时进程会进入睡眠状态[29]。

### 2.3.2 mmapbench

mmapbench 测试程序中，每个工作进程不断的将同一个 500MB 的文件利用 mmap()系统调用映射到自己的进程地址空间中。映射时使用 MAP_SHARED 标记。

然后，每个进程对整个被映射的文件的每一页进行一次读操作，最后将映射删除。这个测试程序用来测量操作系统在内存管理服务上的可扩展性。对于数据库和大规模的网页服务器等应用而言，很多文件首先要被映射到内存中[31]，因此操作系统的内存管理可扩展性至关重要。

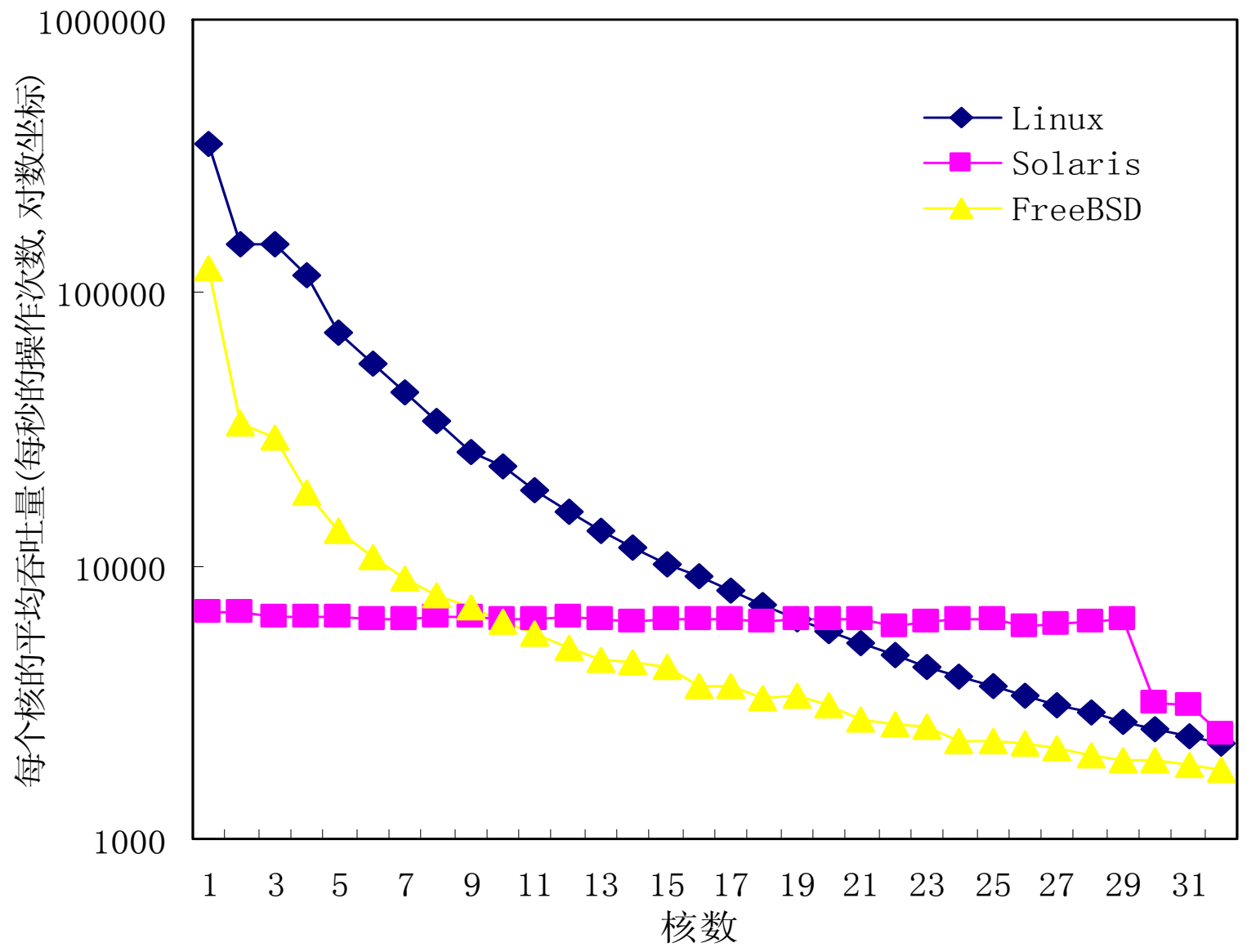

图 2. 5 mmapbench 的平均吞吐量随着核数的变化。平均吞吐量被定义为每个核上单位时间内完成建立映射和解除映射的次数

图 2.5 表示了 mmapbench 的吞吐量在 3 个操作系统上随着核数的变化。从图中可见，尽管在核数较少的情况下，Linux 和 FreeBSD 系统上的吞吐量要高于 Solaris，而当核数到达 29 以前，Solaris 的可扩展性比 Linux 和 FreeBSD 都要好。对 CPU 空闲时间百分比的测量(图 2.6 所示)表示出当使用的核数不超过 29 时，Solaris 系统上的空闲时间百分比不会超过 2%。然而，当增加更多的核数时，CPU 的空闲时间百分比会快速地增加至 50.3%。这个现象表明系统的锁竞争导致了 Solaris 系统上的可扩展性瓶颈。

本文开始时认为竞争产生的原因是访问被映射文件时产生的缺页异常。但是，在 3 个系统上执行这个测试程序时观测到的缺页异常却很少。一个可能的原因是系统中已经采用了类似于预触发异常等优化[32]。我们利用性能分析工具测量内核函数的执行时间。在 Linux 系统中，使用 Oprofile 进行测量。结果发现 2 个被内存映射和解除映射时调用的函数 vma_link()和 unlink_file_vma()的执行时间随着核数的增加增长的最快。例如，它们的执行时间占总时间的比例在单核上分别是 0.4%和 0.3%，但是当使用 32 核时，却高达 46.0%和 50.0% (见表 2.2)。

vma_link()和 unlink_file_vma()在链接和解除链接内存区域(virtual memory area)的时候被调用，而且需要访问每个文件的 address_space 数据结构[33]。该数据结构

由排号自旋锁保护起来，而且在 vma_link()和 unlink_file_vma()函数中要对该数据结构进行操作时，首先要获取该自旋锁。因此，当多个进程并行的对同一文件建立映射和解除映射时，自旋锁的竞争将变得越来越激烈。

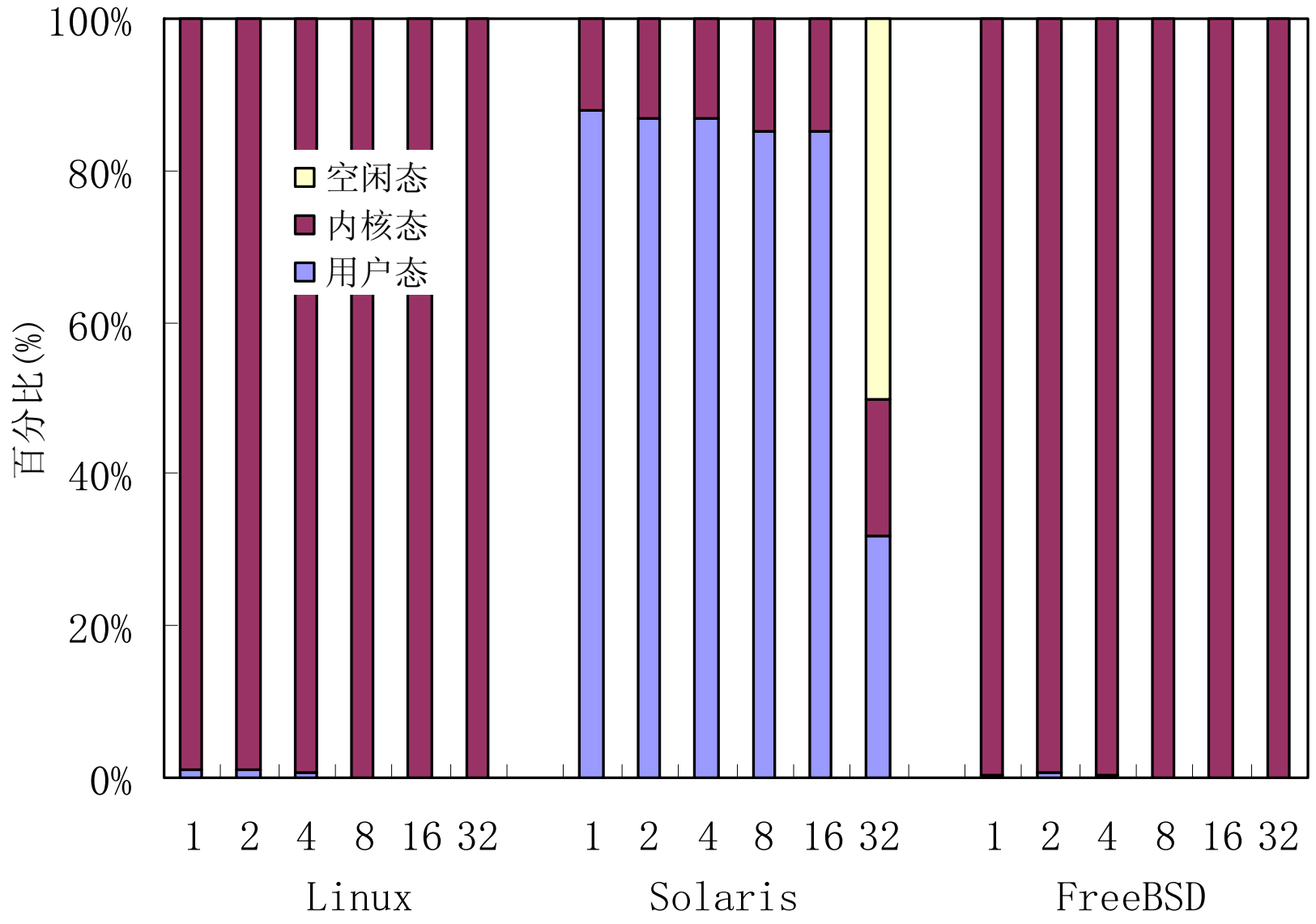

图 2.6 执行时间随着核数的分解

表 2.2 执行 mmapbench 时的 3 个热点函数

| 核数 | 函数 | 百分比 |
|---|---|---|
| | **Solaris** | |
| 32 | mach_cpu_idle | 51% |
| | (user mode) | 34% |
| | mutex_delay_default | 6% |
| 1 | mach_cpu_idle | 97% |
| | (user mode) | 3% |
| | rw_exit | 0% |
| | **Linux** | |
| 32 | unlink_file_vma | 50% |
| | vma_link | 46% |
| | (no symbols) | 1% |
| 1 | __d_lookup | 9% |
| | unmap_vmas | 8% |
| | page_fault | 5% |

FreeBSD 系统上，可扩展性瓶颈产生的原因类似。系统中为每个文件分配的 vnode 数据结构[29]被几个内核函数(vget(), vput(), vrele(), vref(), ufs_markatime()和 ufs_getattr())频繁的访问。当一个文件被映射进地址空间中，或者这种映射被解除时，需要查找或者更新 vnode 数据结构中代表文件属性的域，例如访问文件的时间、引用计数等。因为 vnode 是由互斥锁保护的，所以对于文件映射密集的程序，互斥锁的竞争将会变成可扩展性瓶颈。

FreeBSD 系统上一个有趣的现象是无论使用多少核，执行这个程序时的 CPU 空闲时间总是很少。然而，从原理上分析，当锁竞争比较激烈时，某些进程将会进入睡眠状态[29]，进而增加空闲时间。通过分析 FreeBSD 系统上互斥锁实现机制，作者发现 FreeBSD 的互斥锁针对锁的使用进行了优化。当一个进程 A 请求一把已经由进程 B 持有的锁时，如果进程 B 正在运行，那么 A 进程自旋地等待 B 进程，以期望尽快的释放锁。仅当 B 进程处于睡眠状态时, A 进程才会进入睡眠状态等待 B。当运行 mmapbench 时，工作进程的数目和核数是相同的。因此，持有锁的进程总是处于运行状态。尽管 mmapbench 执行时会竞争互斥锁，但是没有进程会进入睡眠状态。这个场景和 forkbench 的是不同的。在 forkbench 中，每个工作进程不断地创建子进程，导致系统的进程数比核数多。所以，持有锁的进程可能会进入睡眠状态，导致锁的持有时间有所增加，同时也使得其它的进程进入睡眠状态。

对于 Solaris 系统，我们着重研究当核数超过 29 时，可扩展性产生的突然下降。当使用 lockstat 工具对 mmapbench 进行分析后，发现由函数 lgrp_shm_policy_set()请求的一把读写锁随着核数的增加其竞争强度变得异常的激烈。例如，当使用 15 个核时，锁竞争的总次数仅为 192 次(平均每个核 13 次)。然而，当使用 32 个核时，竞争的次数却增加到了 1,571,217 次(平均每个核 49,101 次)。实际上,函数 lgrp_shm_policy_set()负责设置物理内存的分配策略以提升 NUMA 系统的访问局部性[30]。似乎这个函数和文件映射操作没有直接的关系。我们通过研究函数的调用关系和内核代码发现，与 FreeBSD 类似, Solaris 中的每个文件都有一个 vnode 数据结构，而每个 vnode 都有一个域描述内存的分配策略。当文件第一次被某个进程映射进内存中时，进程将会设置 vnode 中的映射关系。由于同一个文件被多个不同的进程不断地映射，保护映射策略的读写锁使得每次写操作都只有一个进程操作以保护数据的一致性。虽然锁竞争在核数较少时不够激烈，产生的性能影响也是可以忽略的，但是当核的数目很多时，锁竞争的概率会大幅度上升，导致吞吐量产生严重的下降。

### 2.3.3 dupbench

dupbench 用来测量操作系统的文件描述符管理功能。在测试程序执行时，每个工作进程不断地复制一个进程私有的文件描述符，然后将复制出的描述符关闭。该测试程序在 3 个操作系统上的吞吐量随核数的变化如图 2.7 所示。正如图中表示的, Linux 系统的可扩展性是完美的，而 Solaris 和 FreeBSD 系统的吞吐量曲线却随着核数的增加一直下降，表示系统的可扩展性存在问题。该测试程序的执行时间分解图(图 2.8)表明内核模式下的某些操作是导致 Solaris 和 FreeBSD 可扩展性问题的原因,

因为内核中的执行时间百分比会随着核数的增加而增加。

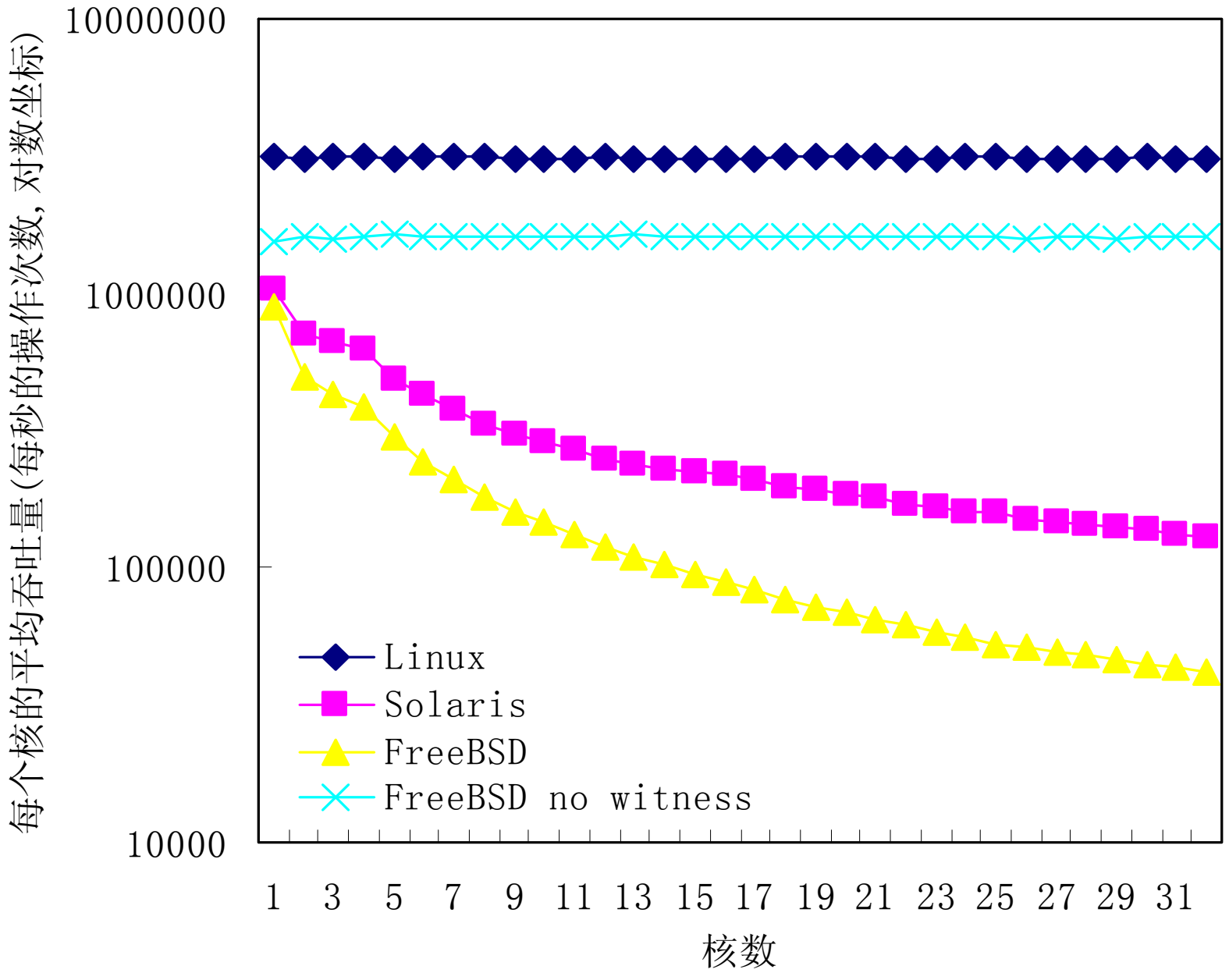

图 2.7 dupbench 的平均吞吐量随着核数而变化的曲线。平均吞吐量的定义是单位时间完成的系统操作次数，即复制和关闭文件描述符的次数。"FreeBSD no witness" 指的是没有 witness 模块的 FreeBSD 内核。

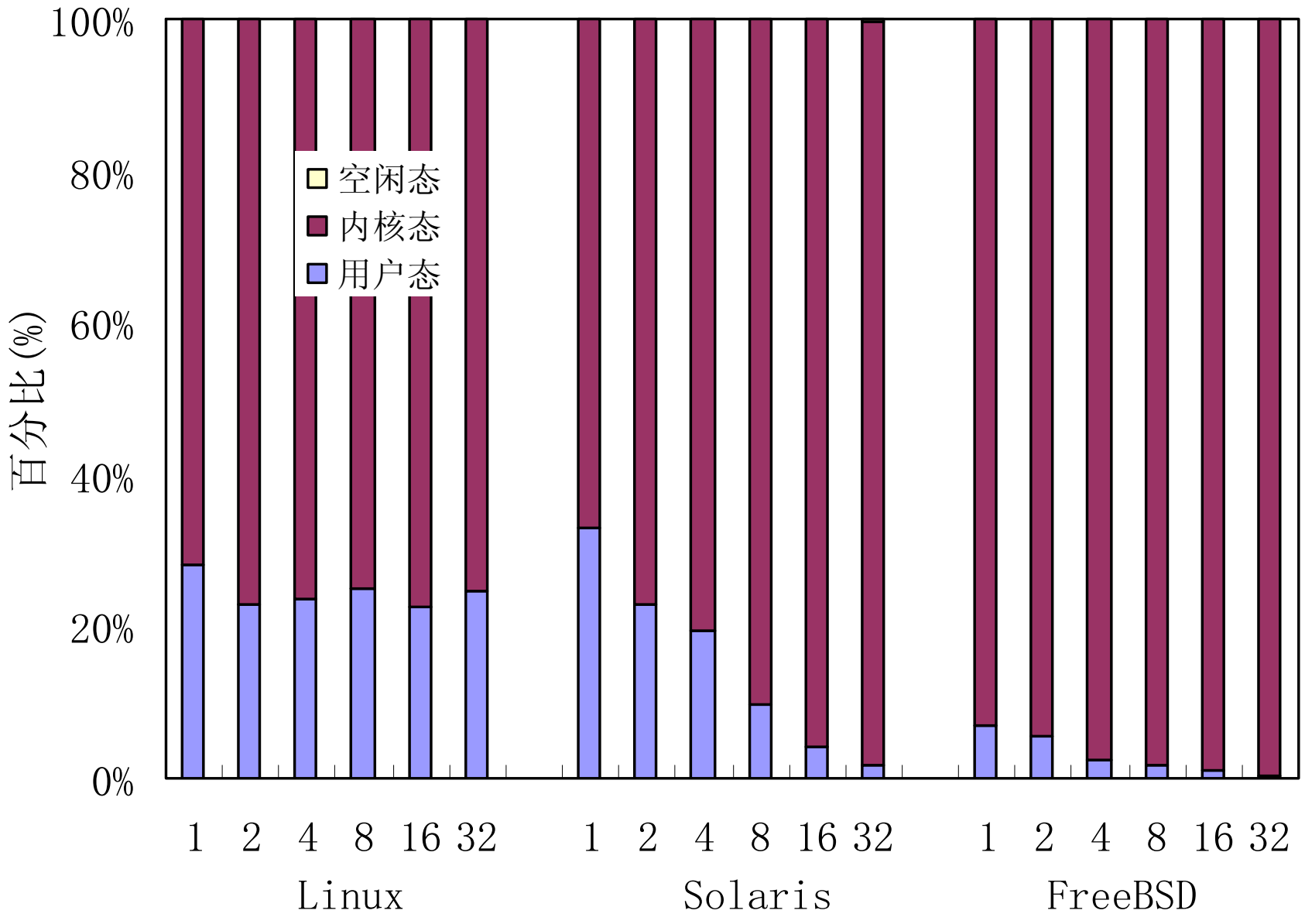

图 2.8 执行时间随着核数的分解

为了找到限制 FreeBSD 系统可扩展性的瓶颈，我们对锁的使用程度进行分析，但是却没有发现某些锁的竞争程度随着核数的增加而变得激烈。而且，在多次的实验中，随着核数的增加，竞争最为激烈的锁在不同的实验中有所不同。对源代码的分析和分析工具的使用发现用于避免内核死锁的 witness 模块是导致 FreeBSD 可扩展性问题的最根本原因。在 FreeBSD 系统中，witness 模块跟踪所有锁的获取和释

放操作。当任务请求一把锁时, witness 模块扫描 2 个链表决定当前的锁请求是否可以授予[29]。当运行测试程序时，尽管没有一把锁被激烈的竞争，但是运行 witness 模块的开销却随着核数的增加变成了可扩展性瓶颈。为了验证上述的分析，我们重新编译生成 FreeBSD 内核，但是没有使用 FreeBSD 8.0-CURRENT 版本中默认的 witness 模块。新内核的吞吐量随核数的变化表在图 2.7 中，其结果表明经过修改的 FreeBSD 系统的可扩展性和 Linux 是一样好的。

Solaris 操作系统中，由函数 flk_get_lock_graph()频繁调用的自适应互斥锁(flock_lock)随着核数的增加其竞争程度变得越来越大。表 2.3 中最热点的 3 个函数也验证了该锁的激烈竞争。在测试程序中，当关闭文件描述符时导致了对该函数的调用。当工作进程在文件描述符上执行 close()操作时, POSIX 语义需要文件中所有的锁都必须被清除。而 Solaris 系统中，这个操作发生在文件的 vnode 数据结构上。在默认情况下，Solaris 系统使用 ZFS 文件系统管理文件。而 ZFS 将系统中所有的 vnode 组织在一个由 flock_lock 保护的全局哈希表中。尽管不同的进程操作不同的 vnode，但是这种组织数据的方式使得每次对一个 vnode 进行操作时都需要获取一个保护所有 vnode 的锁，从而产生了可扩展性瓶颈。

表 2.3 执行 dupbench 时的 3 个热点函数

| 核数 | 函数 | 百分比 |
|---|---|---|
| | **Solaris** | |
| 32 | mutex_delay_default | 51% |
| | default_lock_delay | 9% |
| | mach_cpu_idle | 9% |
| 1 | mach_cpu_idle | 95% |
| | mach_cpu_pause | 2% |
| | mutex_enter | 0% |
| | **Linux** | |
| 32 | system_call | 17% |
| | dupfd | 14% |
| | sysret_check | 11% |
| 1 | system_call | 6% |
| | page_fault | 6% |
| | dupfd | 4% |

### 2.3.4 sembench

sembench 用来测量操作系统中 SystemV IPC 的可扩展性。在测试程序中，每个工作进程和它创建出来的子进程以乒乓模式操作一对 System V 信号量。为了使得该测试程序可以在 FreeBSD 上正确地执行，将内核选项 kern.ipc.semmni 和 kern.ipc.semmns 分别设置成 64 和 100。

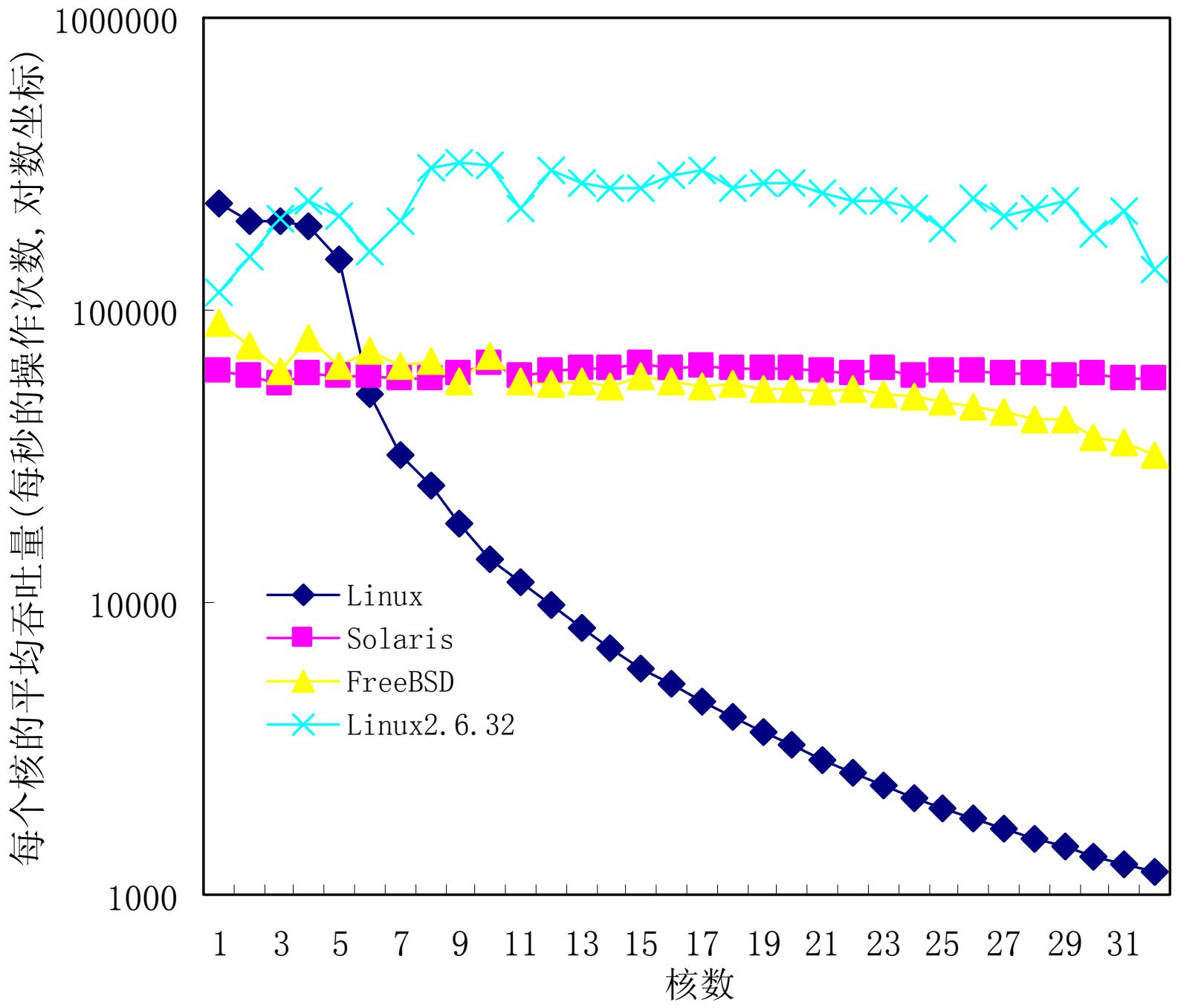

图 2.9 sembench 的平均吞吐量随着核数而变化的曲线。平均吞吐量被定义成每个核每秒完成的 IPC 操作的次数。

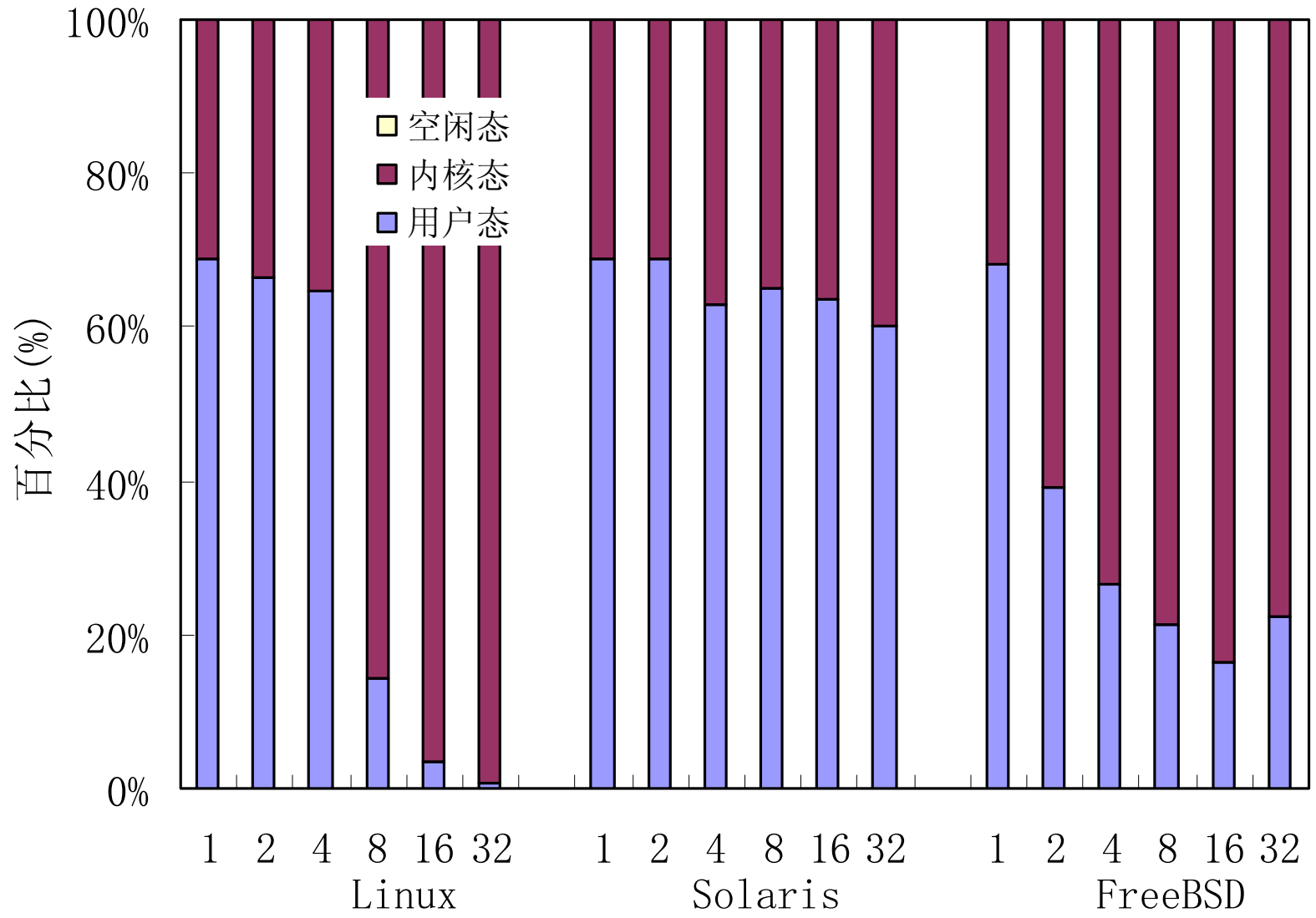

图 2.10 执行时间随核数的分解

吞吐量随着核数的变化和执行时间的分解分别如图 2.9 和图 2.10 所示。从图中可见，Linux 系统在这个操作上的可扩展性是最差的。同时，FreeBSD 系统和 Solaris 系统的可扩展性是完美的。为了找到 Linux 系统的可扩展性瓶颈，我们利用分析工具对函数的执行时间进行分析。函数__down_read()和__up_read()的执行时间随着核数的增加增长最多。例如，从表 2.4 中可以发现当使用 32 核时，大部分的执行时间都消耗在这 2 个函数上(__down_read() 47%, __up_read() 47%)。这 2 个函数用来获取和释放保护每一类 SystemV IPC 资源的全局信号量的读锁。尽管不同

的工作进程操作不同的 SystemV 信号量，但是不同信号量的操作需要获取保护这类资源的全局信号量的锁。因此，对保护信号量锁的竞争是 Linux 的可扩展性瓶颈。

表 2.4 执行 sembench 时的 3 个热点函数

| 核数 | 函数 | 百分比 |
|---|---|---|
| | **Solaris** | |
| 32 | (user mode) | 27% |
| | mach_cpu_idle | 18% |
| | cpu_pause | 11% |
| 1 | mach_cpu_idle | 95% |
| | (user mode) | 1% |
| | mutex_enter | 1% |
| | **Linux** | |
| 32 | __down_read | 47% |
| | __up_read | 47% |
| | (no symbols) | 2% |
| 1 | main | 16% |
| | semop | 8% |
| | __d_lookup | 6% |

### 2.3.5 sockbench

sockbench 用套接字操作的可扩展性来衡量操作系统在网络方面的可扩展性。在测试程序中，每个工作进程不断地调用函数 socket()和 close()。图 2.11 表示了 3 个操作系统在该测试程序上的吞吐量随着核数的变化。如图所示，所有的操作系统都存在可扩展性瓶颈。在 3 个操作系统中，Solaris 比 FreeBSD 扩展的好，而 FreeBSD 比 Linux 扩展的好。图 2.12 表示了每个操作系统的执行时间分解随着核数的变化。如图所示，当执行 sockbench 时，大部分的时间都消耗在了内核态。

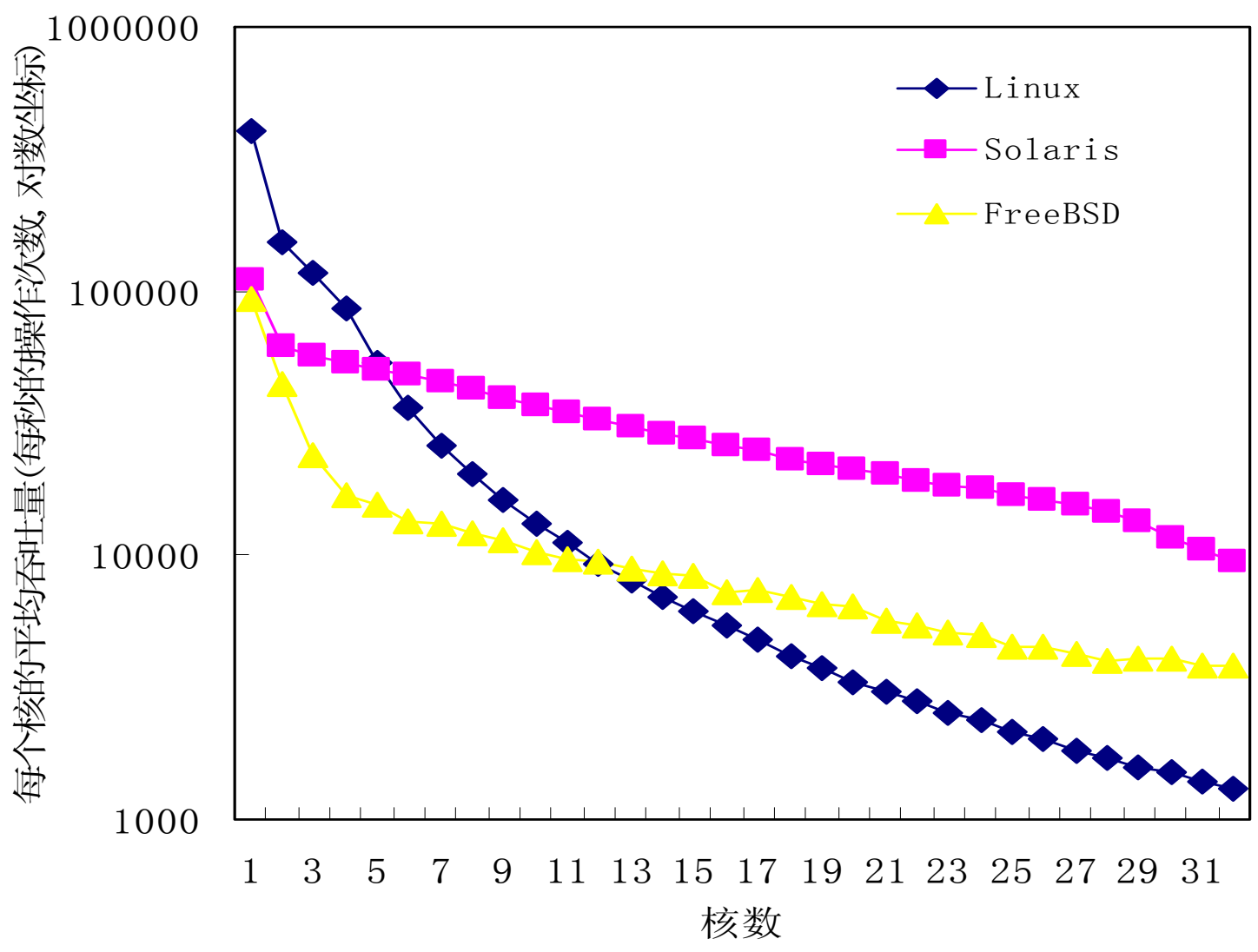

图 2.11 sockbench 的平均吞吐量随着核数变化的曲线。平均吞吐量的定义是每个核在单位时间能够完成的操作次数，即创建或者关闭 socket 的次数

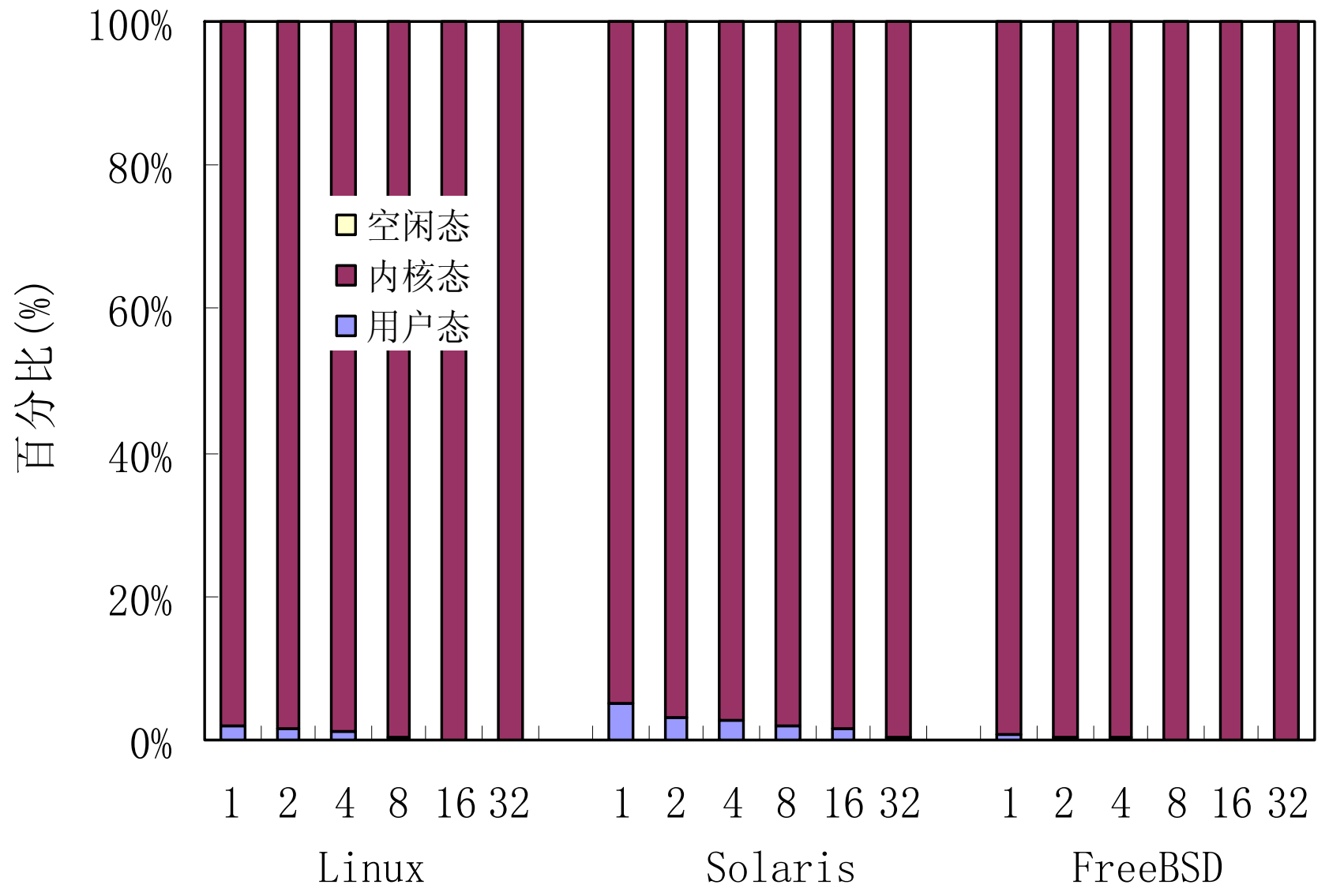

图 2.12 执行时间的分解

表 2.5 执行 sockbench 时的 3 个热点函数

| 核数 | 函数 | 百分比 |
| --- | --- | --- |
| | **Solaris** | |
| 32 | mutex_delay_default | 52% |
| | default_lock_delay | 10% |
| | mutex_enter | 9% |
| 1 | mutex_enter | 26% |
| | kmem_cache_alloc | 6% |
| | tcp_open | 6% |
| | **Linux** | |
| 32 | d_alloc | 33% |
| | d_instantiate | 33% |
| | _atomic_dec_and_lock | 28% |
| 1 | tick_do_update_jiffies64 | 14% |
| | memset_c | 10% |
| | kmem_cache_free | 4% |

Linux 系统中，3 个内核函数(即 d_alloc(), d_instantiate()和_atomic_dec_and_lock())随着核数的增加执行时间的增长是最多的。当使用 32 核时，超过 90%的执行时间消耗在了这些函数上(如表 2.5 所示)。前 2 个函数用来分配和初始化 dentry 实例的，而第 3 个函数在 dentry 或者 inode 实例被删除时调用。在 Linux 系统上，每个套接字和一个 dentry 以及 inode 绑定。当打开或者关闭套接字时，对应的 dentry 和 inode 数据结构分别从全局的 dentry 缓存中和全局的 inode 链表中被分配或者删除。然而，全局的缓存和链表分别是由自旋锁 dcache_lock 和 inode_lock 保护的。这 2 把锁的竞争限制了该测试程序在 Linux 上的可扩展性。

不同于 Linux, Solaris 和 FreeBSD 系统的可扩展性问题是由网络协议栈的操作

而引起的。当套接字被创建或者删除时, Solaris 系统要求 TCP 流需要在应用和网络驱动之间建立或者删除[30]。而用来创建和删除 TCP 流的函数需要更新协议栈数据结构中的引用计数。然而, 这个引用计数是相同协议中的所有流共享的。因此, 当多个核并行地创建和删除套接字时, 将会产生可扩展性瓶颈。FreeBSD 系统上的可扩展性瓶颈类似于 Solaris 上的网络协议栈共享。在 FreeBSD 内核中, 每个网络协议是由 inpcbinfo 数据结构所表示的。而这一数据结构链接了一个协议中的所有协议控制块, 并且由一个锁保护起来。当套接字被创建时, 新的协议控制块需要被插入到链表中,而当套接字被删除时, 需要修改对应的协议控制块以反映当前的状态。然而, 并发操作该链表使得保护链表的读写锁被激烈地竞争。

### 2.3.6 微基准测试总结

表 2.6 总结了 5 个微基准测试程序在 3 个操作系统上的可扩展性并且简要描述了它们的可扩展性瓶颈。尽管由于不同的内核设计和实现导致了相同的测试程序在不同的内核上展示出了不一样的可扩展性, 但是大部分的瓶颈都是由保护内核中共享数据结构的同步互斥操作所导致的。

表 2.6 不同测试程序在 3 个操作系统上的瓶颈描述

| | **Linux** | **Solaris** | **FreeBSD** |
|---|---|---|---|
| **forkbench** | 建立删除 vma 导致了保护内存映射文件的自旋锁竞争 | 小页错在内存映射文件的读写锁上的竞争 | 小页错在内存映射文件的互斥锁上有竞争 |
| **mmapbench** | 建立和删除 vma 导致保护内存映射文件的自旋锁竞争 | 设置内存放置策略导致内存映射文件的读写锁竞争 | 更新和查找 vnode 的统计信息在内存映射文件的互斥锁上有竞争 |
| **dupbench** | 可扩展 | 关闭文件描述符在保护全局哈希表上的自适应互斥锁上有竞争 | Witness 模块的开销随着核数而增加(去掉 witness 后可扩展) |
| **sembench** | 保护全局读写信号量的读锁有竞争 | 可扩展 | 可扩展 |
| **sockbench** | 全局目录缓存的自旋锁竞争和全局 inode 链表的自旋锁竞争 | 建立和删除流导致网络协议的引用计数有竞争 | 保护全局的协议控制块链表的读写锁竞争 |

## 2.4 实际程序的可扩展性预测

从微基准测试程序的评测结果中得到的一个结论是同一测试程序在不同操作系统上的可扩展性会相差很多。例如, Linux 在进程密集和文件描述符密集的操作上比 Solaris 和 FreeBSD 的可扩展性好很多, 而 Solaris 在内存映射文件的操作上比 Linux 和 FreeBSD 好很多。由于每个设计出的微基准测试程序可以反映操作系统中一个最基本的组成部分, 而且通过微基准测试程序的可扩展性评测已经对每个微基准测试程序在不同内核中的可扩展性瓶颈进行了深入的分析, 从而使得多核平

台上的可扩展性预测变得可能。对于一个应用程序，可以将其在内核的执行时间分解成一系列系统调用的时间。而每个系统调用在不同内核上的可扩展性已经通过微基准测试程序有了很好地了解。因此，应用程序的可扩展性可以根据每个系统调用的信息进行预测。在本节中，使用 postmark 的多进程版本来说明这种方法的有效性。

### 2.4.1 可扩展性预测

为了确定 postmark 执行时频繁使用的系统调用，使用 UNIX 和类 UNIX 系统上广泛使用的系统调用分析工具 STRACE。在单核上的统计信息表明程序执行时频繁地使用 4 个系统调用(read(), write(), open()和 close())，而且每个系统调用的使用频率相同。统计信息得到的结论和测试程序中每个事务执行的操作是一致的。

在这 4 个系统调用中, read()和 write()的可扩展性应该是好的，因为每个核在独立的文件集上进行操作。所以，这两个系统调用对可扩展性的影响很小。这样，应用程序的可扩展性与 open()和 close()的可扩展性密切相关。而理解这两个系统调用的可扩展性可以帮助预测应用程序在不同平台上的可扩展性。尽管在微基准测试中没有评价 open()和 close()系统调用的可扩展性，但是可以从测量套接字创建、删除的测试程序 sockbench 的分析中得到一些有价值的结论。在 sockbench 中, Linux 在套接字的创建和删除操作上具有可扩展性问题，因为在分配 dentry 和 inode 实例时，多个进程激烈地竞争全局的 dentry 缓存锁和全局的 inode 链表锁。事实上, Linux 系统中的每个文件都和一个 dentry 和 inode 实例相绑定。因此, open()和 close()在多核平台上会遇到相同的可扩展性瓶颈。不同于 Linux 系统, Solaris 和 FreeBSD 在套接字创建和删除操作中，最大的瓶颈来源于对网络协议栈的共享而非 dentry 缓存和 inode 链表的共享。这意味着 Solaris 和 FreeBSD 在文件创建和删除方面的可扩展性要好于 Linux。根据这些线索，可以预测 Solaris 和 FreeBSD 在 postmark 上的可扩展性要好于 Linux。但是，由于没有对 open()和 close()的可扩展性分析，因此无法确定哪个系统的可扩展性要更好一些。

### 2.4.2 可扩展性分析

图 2.13 测量了 parallel postmark 吞吐量随着核数增加得到的曲线。如图中所示, Solaris 比 Linux 系统的可扩展性好而且在多于 12 核时，Solaris 的吞吐量也要高于 Linux。对于 FreeBSD 系统，尽管其性能比 Linux 差，但是可扩展性却比 Linux 要稍好。总体说来，实验结果表明预测和实际的测量是相吻合的。

为了进一步的验证，对 parallel postmark 在 Linux 上的可扩展性进行分析以确

定可扩展性的瓶颈就是全局 dentry 缓存的锁竞争以及全局 inode 链表的锁竞争。利用 Oprofile 查找函数的执行时间随着核数的增加上升的最快的函数。实验结果表明 d_instantiate(), _atomic_dec_and_lock(), d_delete(), d_alloc()和 d_rehash()是执行时间增加最快的 5 个函数。它们分别占用 26.7%, 15.6%, 15.4%, 13.8%和 13.5%的执行时间。函数的调用关系数据表明这些函数是在分配 dentry 和 inode 实例时被调用的。而且, /proc/lock_stat 的信息也表明 dentry 缓存锁和 inode 链表锁都被激烈地竞争。

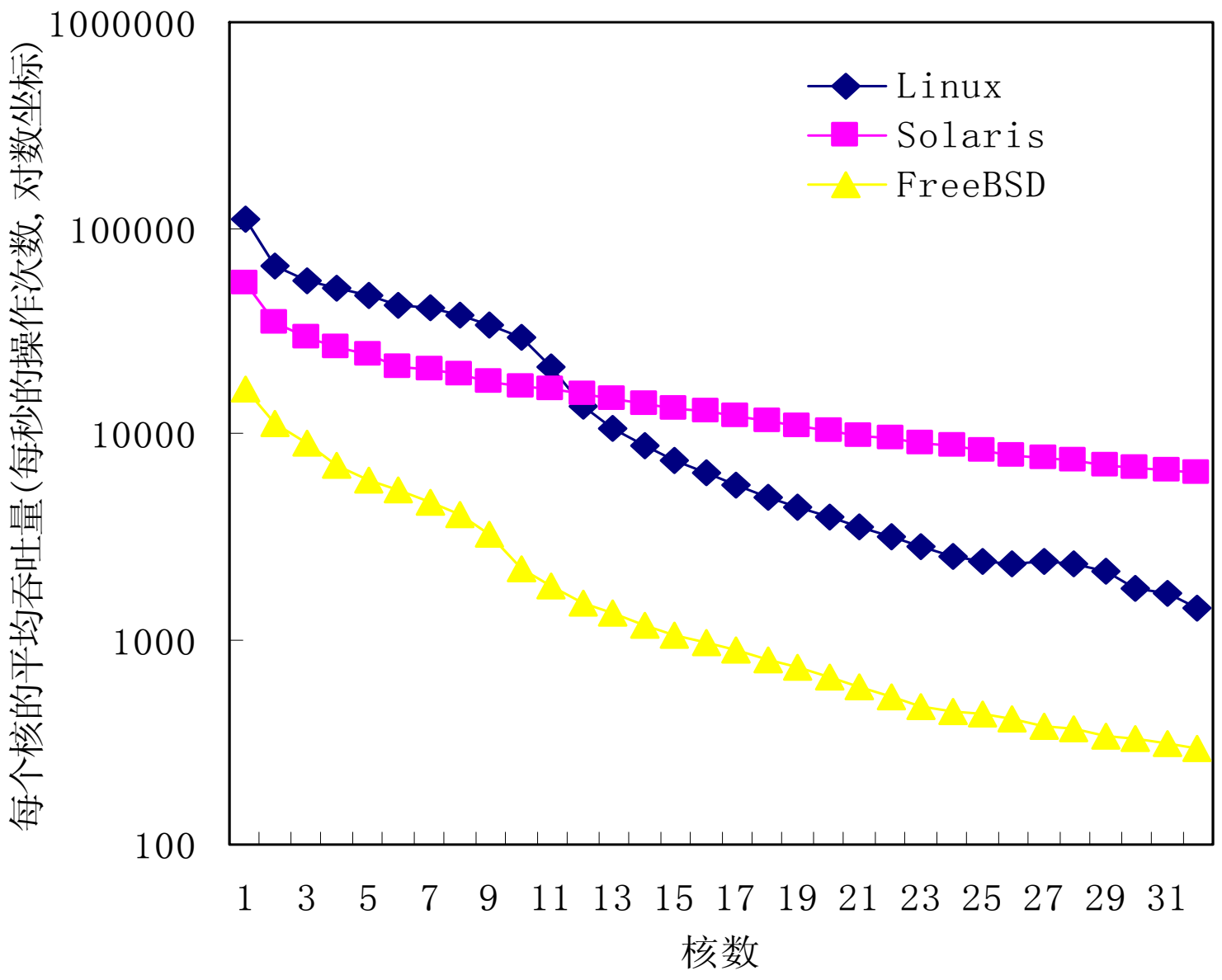

图 2. 13 parallel postmark 平均吞吐量随着核数的变化

我们从测量中发现的另外一个现象是 read()和 write()系统调用对可扩展性的影响比预想中的大。相关的热点内核函数是 shmem_free_inode(), shmem_free_blocks(), shmem_reserve_inode()和 shmem_getpage()。由/proc/lock_stat 报告的统计信息表明，保护内存文件系统中超块元数据的排号自旋锁被激烈的竞争。在 Linux 系统中，每个文件系统具有一个超块。而在超块中的元数据维护着未被分配的 inode 节点和空闲块的数目。当文件被创建或者被追加时，将在内存中被分配新的 inode 或者块。相应地，超块中的元数据会被更新。同时，元数据是被所有执行在相同内存文件系统中的进程所共享的。这样，并行地更新超块中的统计信息会导致保护共享数据的自旋锁竞争。但是，相比于文件创建和删除时导致的锁竞争，这种影响是很小的，因此，预测的结果依然准确。

尽管预测的结果令人激动，但是这仅代表在这个方向上的初步尝试。例如，我们需要很多的微基准测试程序来覆盖操作系统中的每个组成部分。如果具有这样的信息，则可以预测 Solaris 和 FreeBSD 在 parallel postmark 上的可扩展性差别。而且，我们还需要一个更加系统的方法来选择微基准测试程序以发现类似于 Linux 超块的锁竞争问题。在本文的工作中，还没有考虑瓶颈之间的相互影响，后续的工作

中会考虑这种影响。

## 2.5 讨论

### 2.5.1 可扩展性瓶颈

本文总结出多核平台上影响操作系统可扩展性的最根本原因是保护共享数据的同步原语，但是并没有将所有确定的可扩展性瓶颈全部移除来对该结论进行支持，因为这需要对操作系统的子系统进行重新设计。然而，由于两点原因，本章的结论同样使人信服。1) 确定可扩展性瓶颈的方法被充分的验证过。本文查找可扩展性瓶颈的方式是计算从 1 核到 32 核时，函数执行时间的增量。如果函数的时间增量较大，那么说明该函数对总体可扩展性的影响较大，需要首先分析。这个方法在作者的工作[34]中被提出并会在第五章详细地介绍。其中每个函数的执行时间增量被称为可扩展性值(scalability value)。由这个信息做指导，作者已经在多核平台上成功地找到了 2 个在线事务处理应用(OLTP)的可扩展性瓶颈并且提升了可扩展性。2) 有 2 个例子可以支持得出的结论。一个是将 forkbench 的动态链接改为静态链接将导致该程序可以运行的更快，可扩展性更好。forkbench 的可扩展性瓶颈是保护库文件和可执行文件的锁竞争。尽管将链接方式改成静态链接不能移除所有的可扩展性瓶颈，但是确实可以得到实际的可扩展性提升。另外一个例子是更新的 Linux 内核在使用 IPC 资源时避免使用全局的读写信号量。这种改进使得 sembench 在 Linux 上的可扩展性瓶颈被移除。本章利用 Linux 2.6.32 内核来验证这种改进，测量出的结果已经被整合在图 2.9 中。如图所示, sembench 在新版本的内核中可以随着核数很好地扩展。其实验结果也验证了本章瓶颈分析的正确性。

### 2.5.2 选择操作系统的方法

根据微基准测试程序的评测结果(不考虑可扩展性增强后的结果)，没有一个操作系统在所有的方面都比其他的操作系统的可扩展性好。这个实验结果表明应该根据应用程序的特征在多核平台上选择合适的操作系统。具体而言，如果应用在进程创建删除或者文件描述符操作上比较密集，应该为应用选择 Linux 系统,因为它在 forkbench 和 dupbench 上的可扩展性最好。如果应用程序使用 mmap 操作文件，大量的使用 SystemV IPC 或者套接字，那么 Solaris 系统是最好的选择,因为它在这些方面的可扩展性优于 Linux 和 FreeBSD。如果应用大量的使用 SystemV IPC，也可以选择 FreeBSD 系统，因为它在这方面的可扩展性和 Solaris 是一样好的。应用程序的特征搜集可以利用 Dtrace, SystemTap 和 Strace 等工具容易地获取。

### 2.5.3 实验平台

本章使用AMD32核系统作为实验平台比较3个操作系统的系统服务接口的可扩展性。然而，几个最近的研究[8, 35-36]使用 AMD 48 核系统做为实验平台。在 48 核平台上，本文发现的可扩展性瓶颈同样是存在的，而且瓶颈对应用的影响会更大，因为竞争同一把锁的概率会随着核数的增加而变得更大。对于那些在 32 核平台上扩展的比较好的测试程序(例如 dupbench 在 linux 系统上)，新的可扩展性瓶颈可能在 48 核平台上出现。与之类似的一个情景是在 Solaris 系统上运行 mmapbench。当核数少于 29 时，系统没有可扩展性问题，但是当核数多于 29 时，锁竞争会导致吞吐量突然下降。

## 2.6 相关工作

### 2.6.1 操作系统的比较研究

研究者们已经从多方面对主流操作系统进行了分析和比较。在这些工作之中，Lai 和 Baker 等利用微基准测试程序和实际应用在奔腾架构上评测了 3 个主要 UNIX 内核(Linux, Solaris 和 FreeBSD)的性能[37]。他们发现，没有一个操作系统的性能是明显好于其他系统的。在实际应用中，往往是某些非技术性因素(例如易于安装，开源等)是选择某个操作系统的原因。Spinellis 从文件组织，代码结构，代码风格和 C 预编译器等方面对 4 个操作系统(Linux, Solaris, FreeBSD 和 WRK[38])进行了比较。文章表明，尽管不同的操作系统经历了不同的开发过程，但是这些操作系统的代码质量没有明显差别。Bruning 比较了 3 个操作系统(Linux, Solaris 和 FreeBSD)的基本子系统，包括进程调度器，内存管理系统和文件系统[39]。Bruning 发现这些系统在很多方面(如时分复用的调度策略和虚拟文件系统层)都比较相似。也有作者[31]从算法实现角度比较了 5 个内核(Linux2.4, Linux2.6,FreeBSD, NetBSD 和 OpenBSD)的基本网络编程原语(如 socket()和 bind())。本章的工作侧重在不同操作系统的系统服务接口在多核平台上的可扩展性比较。当多核系统逐渐变成了主流的计算机架构，对操作系统的可扩展性分析和比较研究显得更加重要。通过对 3 个主流开源操作系统的可扩展性瓶颈分析，本章的工作提出了另外一个评价系统的方向，该方向和以前的操作系统比较研究工作是相互补充的。

### 2.6.2 操作系统可扩展性研究

Gough 等利用 OLTP 应用测量了 Linux 内核的可扩展性并且发现运行队列上的锁竞争会严重影响系统的可扩展性[41]。Veal 和 Foong 在 Intel 架构的 8 核平台上

运行了网页服务器应用并且发现地址总线的竞争是导致可扩展性的最根本原因[42]。Cui 等利用 2 个 OLTP 应用(TPCC-UVa and Sysbench-OLTP)来研究 Linux 2.6.25 在 Intel 架构的 8 核平台上的可扩展性，发现可扩展性瓶颈存在于应用和内核在 SystemV IPC 上的实现[34]。Boyd-Wickizer 等利用 7 个应用来评测 Linux 在 AMD 架构的 48 核平台上的可扩展性。文中列举了一系列可扩展性问题和每个问题对应的解决方案[8]。许多在该文中发现的问题在本文中和作者前期的工作[43]中都被发现了，而且，本章的工作又研究了其它 2 个操作系统的可扩展性并且找到了可扩展性瓶颈。

为了提升内核的可扩展性，一些研究者提出重新设计操作系统。例如, K42[19]和 Tornado[21]操作系统利用面向对象的设计思路对操作系统进行重新设计,提升了操作系统在对称多处理器系统(SMP)上的局部性和可扩展性。然而，学术界对操作系统在多核系统上的可扩展性仍然了解很少。在最近的研究中[45], Boyd-Wickizer 等发现所有的可扩展性瓶颈都由保护共享数据的锁竞争所引起(这点和本章的观点类似)。基于这个发现，他们为多核系统实现了一个原型操作系统。在这个操作系统中，应用可以控制共享的程度。Baumann 等提出通过搭建不共享任何数据的操作系统以最大化内核的可扩展性[46]。在该系统中，每个核运行独立的操作系统。而所有的操作系统之间通过显式的消息传递而非共享内存进行通信。Wentzlaff 和 Agarwal 提出操作系统中经典的模块(如调度器和内存管理等)应该拆分以更好的支持多核处理器[47]。在本文的工作中，将可扩展性瓶颈产生的根本原因对应到具体的数据结构和内核函数上，可以帮助内核设计者移除操作系统中的可扩展性瓶颈。

## 2.7 本章小结

本章利用一套微基准测试程序集在 AMD32 核平台上测量了 3 个开源操作系统的可扩展性。实验结果表明，没有一个操作系统在所有评价的方面均好于其它操作系统。性能数据和相关的内核代码分析表明保护内核共享数据结构的同步原语是影响可扩展性的主要因素。利用微基准测试程序的结果，本文成功的预测了一个实际应用在不同操作系统上的可扩展性。本章对操作系统的比较工作以及通过分析揭示的不同操作系统的可扩展性瓶颈对应用使用者为应用选择合适的操作系统和系统设计者解决可扩展性问题都是有借鉴意义的。

# 第3章　锁颠簸现象的模拟和避免

## 3.1　本章引言

第 2 章利用一套微基准测试程序集细致地分析和比较了 3 个主流操作系统的重要系统服务接口在多核平台上的可扩展性。实验的结果表明内核中保护共享数据的锁竞争是影响操作系统可扩展性的重要因素。实际上，内核中的锁竞争甚至可以导致应用程序的吞吐量随着核数的增加而下降，这种现象被称作锁颠簸现象(lock thrashing)。图 3.1 表示了用于模拟文件服务器性能的测试程序 parallel postmark 在 AMD32 核系统上总体的吞吐量随着核数的变化曲线。从图中可见，应用程序整体的吞吐量开始时随着核数的增加而增加，但是当核数超过 12 以后，应用程序的吞吐量开始急剧的下降。从性能分析工具中搜集到的数据表明,吞吐量下降是由于 Linux 系统中的两种排号锁的竞争所导致的。

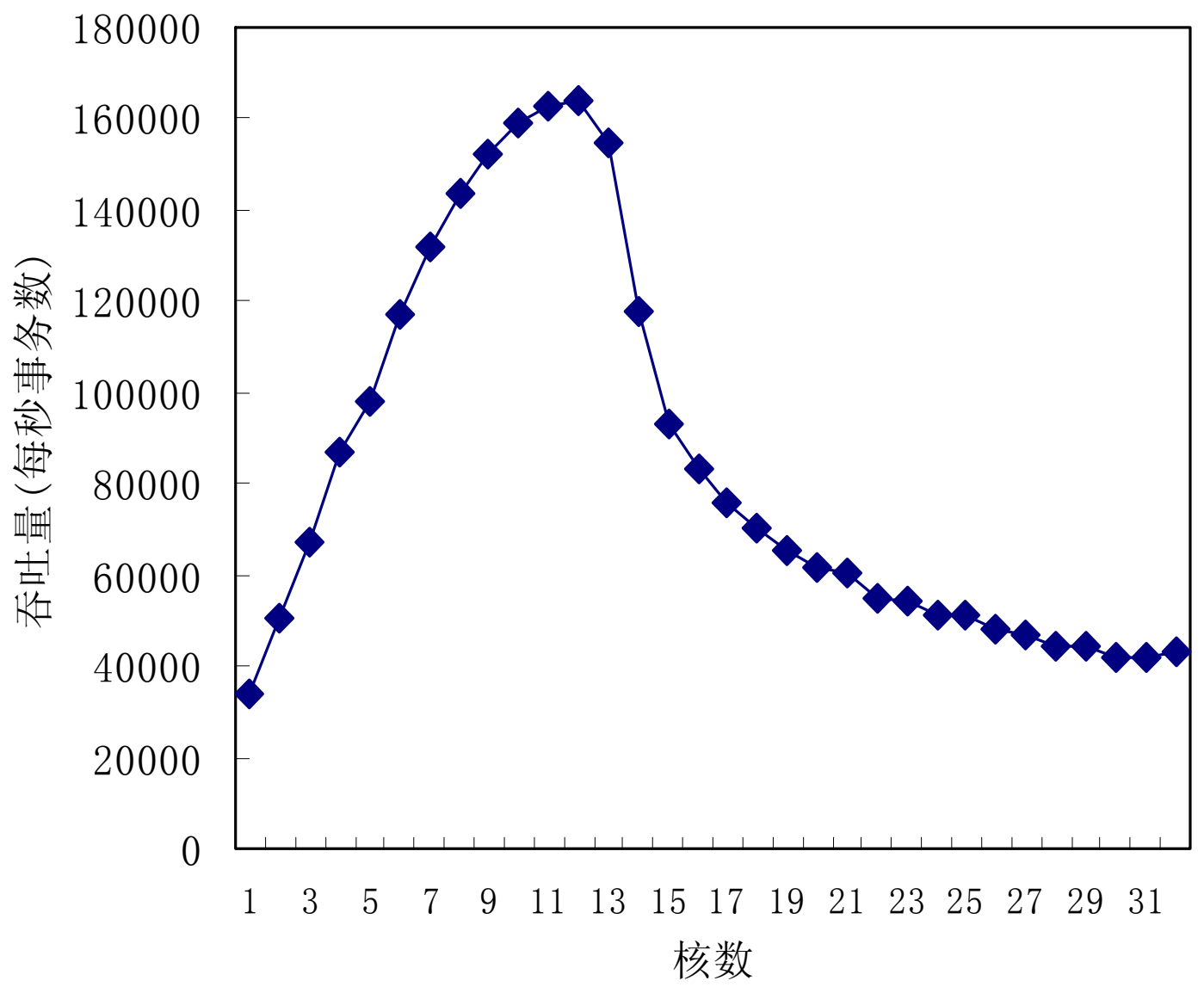

图 3. 1 parallel postmark 在基于 Linux 系统的 AMD32 核上的吞吐量

对于像 parallel postmark 一样内核锁密集的应用程序而言，决定其可扩展性的因素主要有 2 个。一个是由临界区语义导致的串行执行，另外一个是由锁实现导致的开销。当仅有几个任务运行时，等待进入内核代码临界区的任务很少，因此每次锁获取几乎不引入任何缓存失效。在这个时候，尽管增加使用的核数会对整体的吞吐量有提升，但是竞争同一把锁的概率也随之变大，并且每次锁获取和释放操作需要引入额外的缓存交互才得以完成。如果竞争锁的任务数目变的过大以至于某个内核代码的临界区变成了主要的可扩展性瓶颈，每次锁操作都要经历大量的缓

存失效才能够完成。这时，锁颠簸现象就产生了。

锁颠簸现象只有锁被激烈竞争时才会产生。传统的观点认为真实世界的应用(例如网页服务器和在线事务处理等)不太可能由于锁的激烈竞争而导致锁颠簸。然而，随着多核平台被广泛使用，需要更正这个观点。以 parallel postmark 为例，本文的实验结果表明尽管锁竞争的概率在核数较少时是很小的，但是当使用 32 核的系统时，锁竞争的强度会极大地提升。这个事实以及最近的一些研究[8, 47]使作者相信锁颠簸现象有潜力变成多核平台上最有挑战性的问题之一。

正因为该问题的严重性，本章对该问题展开了深入研究。具体而言，研究分为 3 个部分:1 对现象的深刻理解是解决问题的基础。本章首先对锁颠簸现象通过模拟重现。通过分析已有锁模拟器和模型的不足，抽取了可能影响锁密集应用可扩展性的所有因素。其次,利用离散事件仿真技术对所有因素按照应用逻辑进行组织，实验结果可以准确地重现锁颠簸现象。2 设计基于请求者数目的自旋锁。该技术检测自旋锁的等待者数目，如果数目超过一定阈值，所有新来的请求者将进入节能状态，否则仍是以自旋等待方式获取锁。利用实际的内核锁密集应用进行评测表明该项技术可以很好地避免锁颠簸现象，而且比基于阻塞的锁能够提供更好的可扩展性和能耗有效性。3 锁竞争感知的调度策略。该技术检测应用程序中使用内核锁较多的任务，然后通过调度器将这类任务限制在较少的核数上，即通过控制应用的并发程度来避免锁颠簸现象。通过在实际系统上的评测表明，该技术同样可以很好地避免锁颠簸，而且比基于等待者数目的锁具备更好的可扩展性和能耗有效性。

本章的后续章节组织如下:3.2 节介绍了基于离散事件仿真技术的锁颠簸现象模拟器。3.3 节描述了基于请求者数目的锁的设计、实现和评价，而 3.4 节阐述了另外一个用于避免锁颠簸现象的技术，即锁竞争感知的调度策略。

## 3.2 基于离散事件仿真技术的锁颠簸模拟器

### 3.2.1 当前锁模拟器和模型的不足

建模锁竞争最常见的技术就是将其表示成封闭的排队网络(closed queuing network)[48-50]。在模型中，顾客代表核数，排队中心(queuing center)代表临界区而延迟中心(delay center)代表非临界区。当顾客处于某个排队中心时，这意味着它在等待着某个锁；当顾客接受服务时，代表着程序已经执行到了临界区中；当顾客位于延迟中心时，说明程序正在执行非临界区。对于延迟中心而言，它可以同时接纳任意多的顾客，每个顾客将产生指定的延时。

根据文献[48]中的锁模型，延迟中心和服务中心的服务时间都被设置成了常

量。图 3.2 表示了仅当一个临界区存在时利用平均值分析技术(mean value analysis or MVA)对模型求解得到的实验结果。其中, ratio 代表非临界区的执行时间和临界区的执行时间的比例。从图中可见，当 ratio 增大时，应用程序的加速比变得越来越好，这是因为应用程序中串行程序的比例逐渐变小的缘故。然而，对于某个固定的 ratio，加速比随着核数的增加变得稳定而非下降。本文同样验证了代码中有多个临界区时，吞吐量随着核数的变化规律，其实验结果和图 3.2 类似，并不能观察到锁颠簸现象。为了避免由于求解算法的缺陷造成的问题，本文同样利用卷积算法[51]对该模型求解，得到的结果和图 3.2 相同。

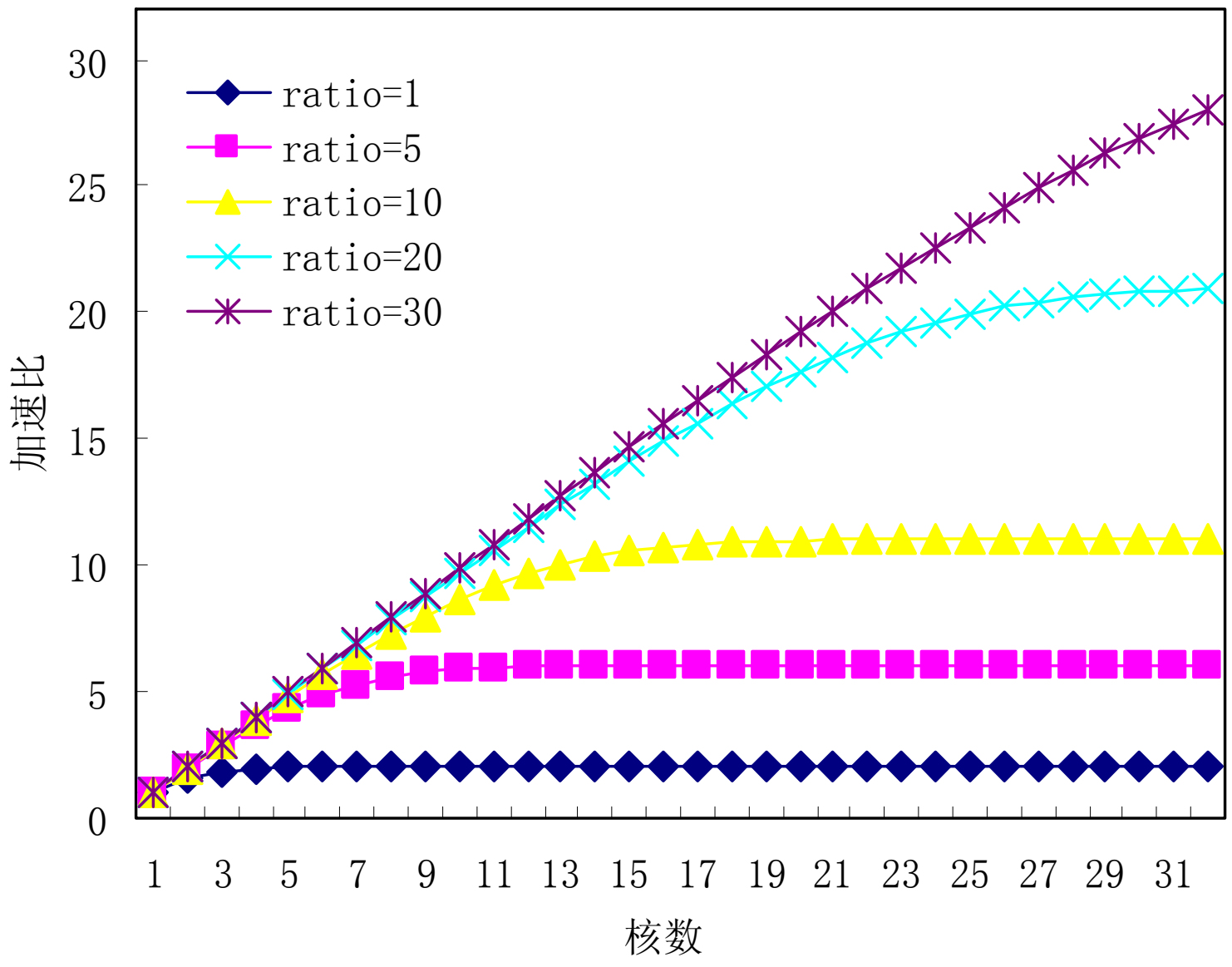

图 3.2 MVA 对锁模型的求解结果, ratio 代表非临界区与临界区的时间比例

文献[52]提出了另一种锁竞争建模的方法，其基本思想是如果工作负载是同构的，那么每个核上执行的代码将展示出循环的特征。根据这个假设,可以计算出应用程序的可扩展性。本文将文中提出的算法进行了实现，实验结果如图 3.3 所示。可以看到其实验结果和图 3.2 几乎相同，只是每个曲线的转折点变得更加突兀。这是因为图 3.2 的求解是利用 MVA 技术，是基于统计平均值的，而图 3.3 的求解方法是对有限次的实验进行分析的结果。然而，可以看到这种方法同样不能重现出锁颠簸现象。

那么，为什么这些锁模型或者模拟器不能重现出实际测量的结果呢？其根本原因是虽然临界区的串行执行语义在模型中被考虑到了，但是却没有考虑到锁本身实现的开销以及临界区和非临界区内的硬件资源竞争。以 Linux 内核中广泛使用的排号自旋锁(ticket spin lock)为例,每个锁被实现成一个 unsigned int 类型的变量。该变量包含两个域，一个称为 owner 域，而另外一个是 next 域。当一个任务要获取一把锁时，需要对 next 域加 1 作为自己排到的号。如果这个号和当前的 owner 域

相同，说明当前的任务可以执行临界区。否则，该任务需要一直自旋直到自己的号和 owner 域相同。图 3.4 展示了排号自旋锁的示意图和代码。而锁竞争带来的开销可以从缓存一致性协议的层面去理解。图 3.5 给出了流程。当在一个锁上排号或者对锁释放时，任务需要对代表锁的变量进行操作。这种操作要在本地的缓存中对缓存变量进行更改。而缓存一致性协议需要保证整个系统中每个缓存对同一数据的副本均相同。因此，一致性协议向系统中其他各核广播缓存更新信息。而当另外的核访问自己缓存中相同的变量时，会发现缓存失效，则又通过一致性协议查找正确的数据副本。如果同时发现缓存失效的核数很多，则每次进入临界区的开销会随着等待进入临界区的核数而线性增长。除了锁实现引入的硬件一致性开销外，临界区和非临界区内的缓存缺失同样会带来额外的一致性开销，其开销产生的原理与图 3.5 类似。

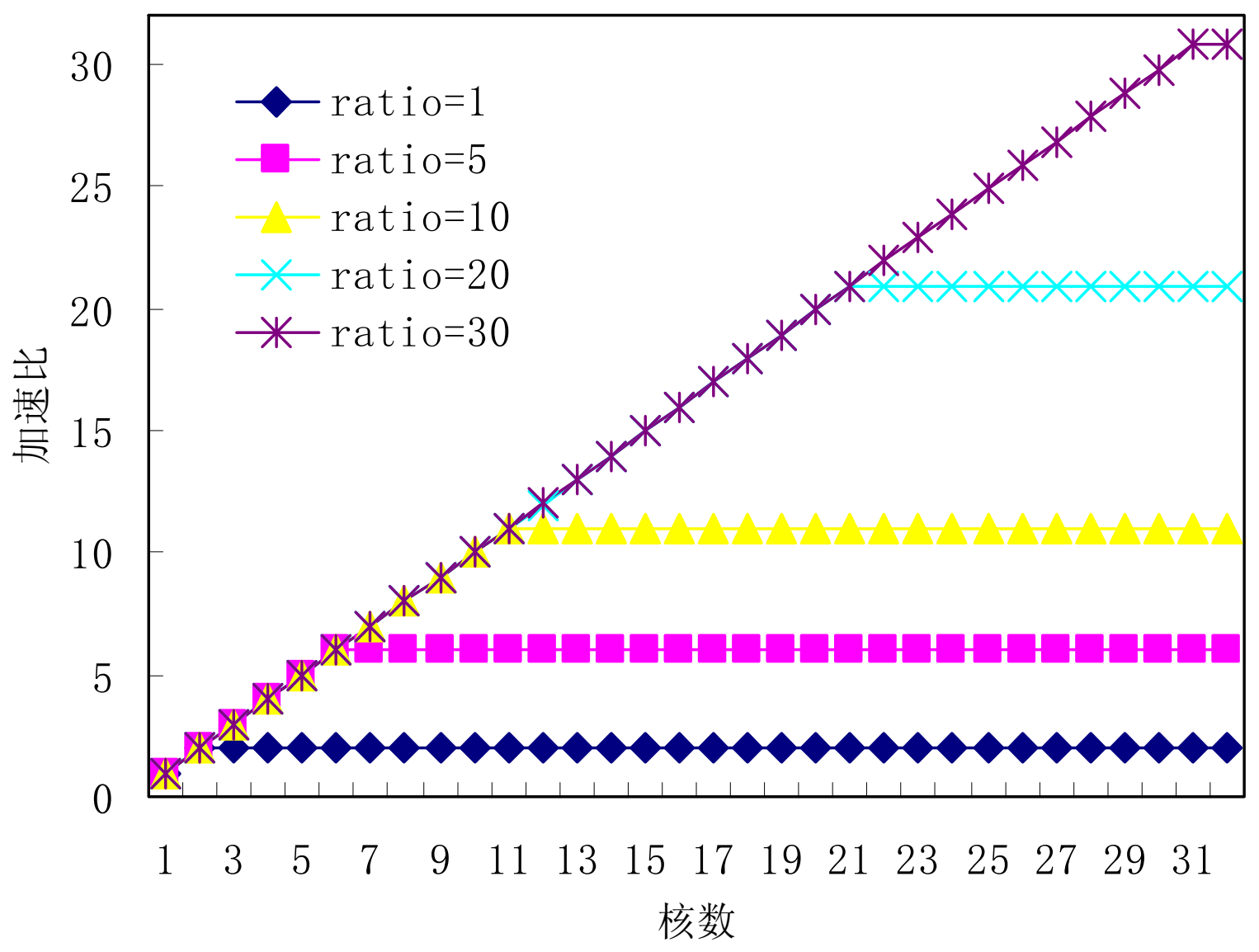

图 3.3 cyclic 算法的实验结果，ratio 代表非临界区对临界区的时间比例

已有工作中，尽管文献[49]同样考虑了锁实现本身带来的影响以及临界区和非临界区内部的缓存缺失开销，但是锁实现开销的计算过于简单，导致根本不能重现锁颠簸现象。D.L.Eager 等[50]利用近似 MVA 技术对锁密集应用提出了分析模型，尽管他们的工作对临界区和非临界区产生的缓存缺失进行了详细的建模，但是并没有考虑到锁实现而引入的开销。Slias 等[53]利用马尔科夫链对锁颠簸现象进行了建模。尽管锁颠簸现象可以被重现，但是该工作没有考虑到临界区和非临界区内的缓存缺失。而且由于马尔科夫链求解困难，因此该方法只适合于小规模系统。

### 3.2.2 模拟器的设计与实现

为了克服当前模型的缺陷，本章提出了一个基于离散事件仿真技术的锁颠簸

现象模拟器，该模拟器不但考虑了锁实现上带来的开销，同样也考虑了临界区和非临界区的缓存缺失对总体加速比带来的影响。

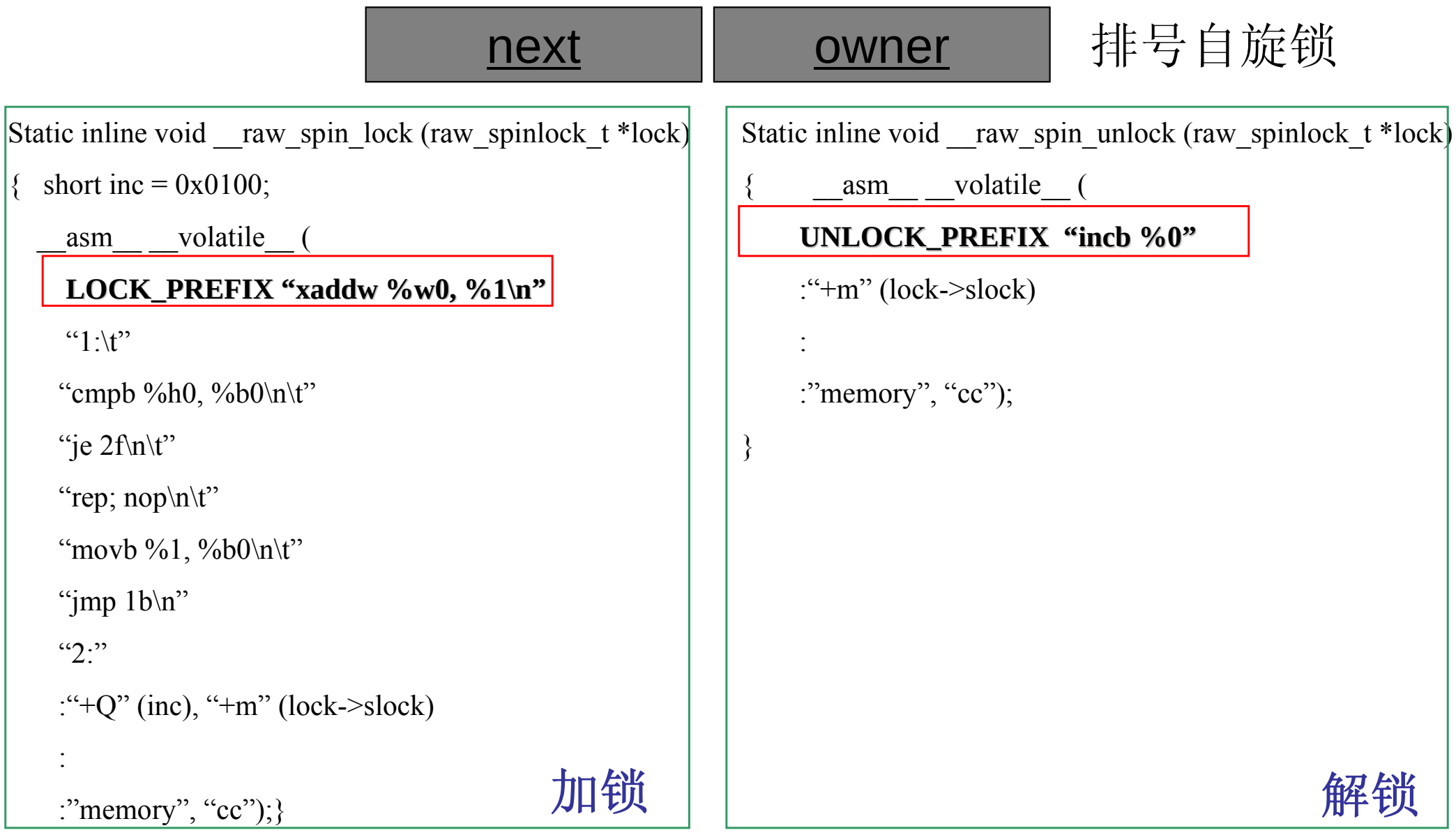

图 3.4 排号自旋锁的示意图和代码

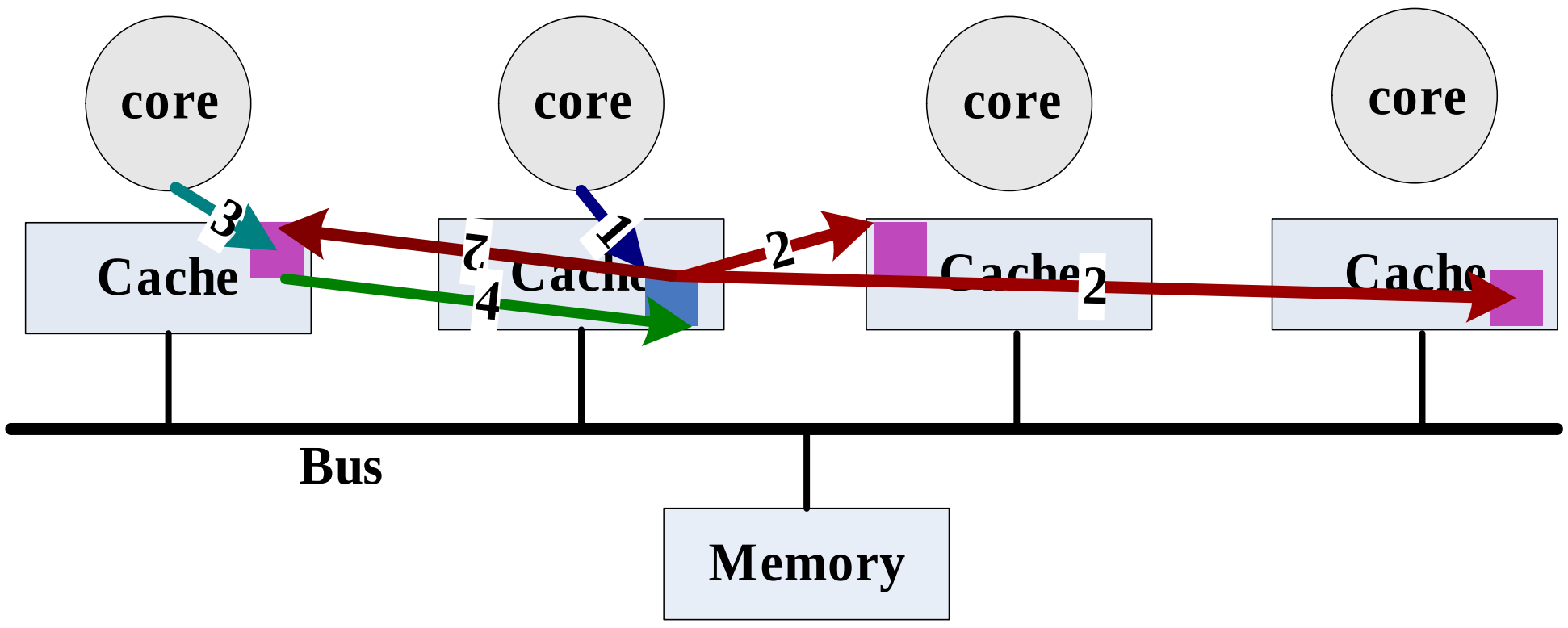

图 3.5 排号锁在缓存一致性上的流程。1.写合法数据, 2.一致性协议广播数据状态，3.读非法数据, 4.一致性协议查找合法数据

#### 3.2.2.1 模型

在该模拟器中，我们使用排队论模型对锁密集应用进行建模。临界区及非临界区的代码和模型的排队中心及延迟中心之间的对应关系如图 3.6 所示。系统中的每个核在临界区和非临界区之间循环。当一个核进入临界区之前，它首先要获取保护该临界区的锁。如果锁被其他的核占用，那么当前的核要自旋等待直到该锁可以被获取为止。假设每个临界区都由内核中广泛使用的排号自旋锁保护。那么，当锁被请求或者被释放时，代表锁的数据结构要在操作锁的核的本地缓存中被修改。而且，缓存一致性协议保证所有等待同一把锁的核产生一次缓存缺失。除了由排号锁引入的缓存缺失外，同样需要考虑临界区和非临界区的缓存缺失。为了保证模型的简

单性，假设临界区中的缓存缺失以及非临界区内的缓存缺失是均匀分布的。尽管如此，每 2 个连续缓存缺失之间的间隔在临界区上和非临界区上可能不同。对于均匀内存访问架构(UMA)，每次缓存缺失需要到内存中查找正确的副本。因此，仅当之前的请求全部完成后才能处理当前的内存请求。可是对于非均匀内存访问架构(NUMA)，每次缓存缺失需要访问的不再是一块每个核共享的内存而是请求数据存在的内存块(memory bank)上。

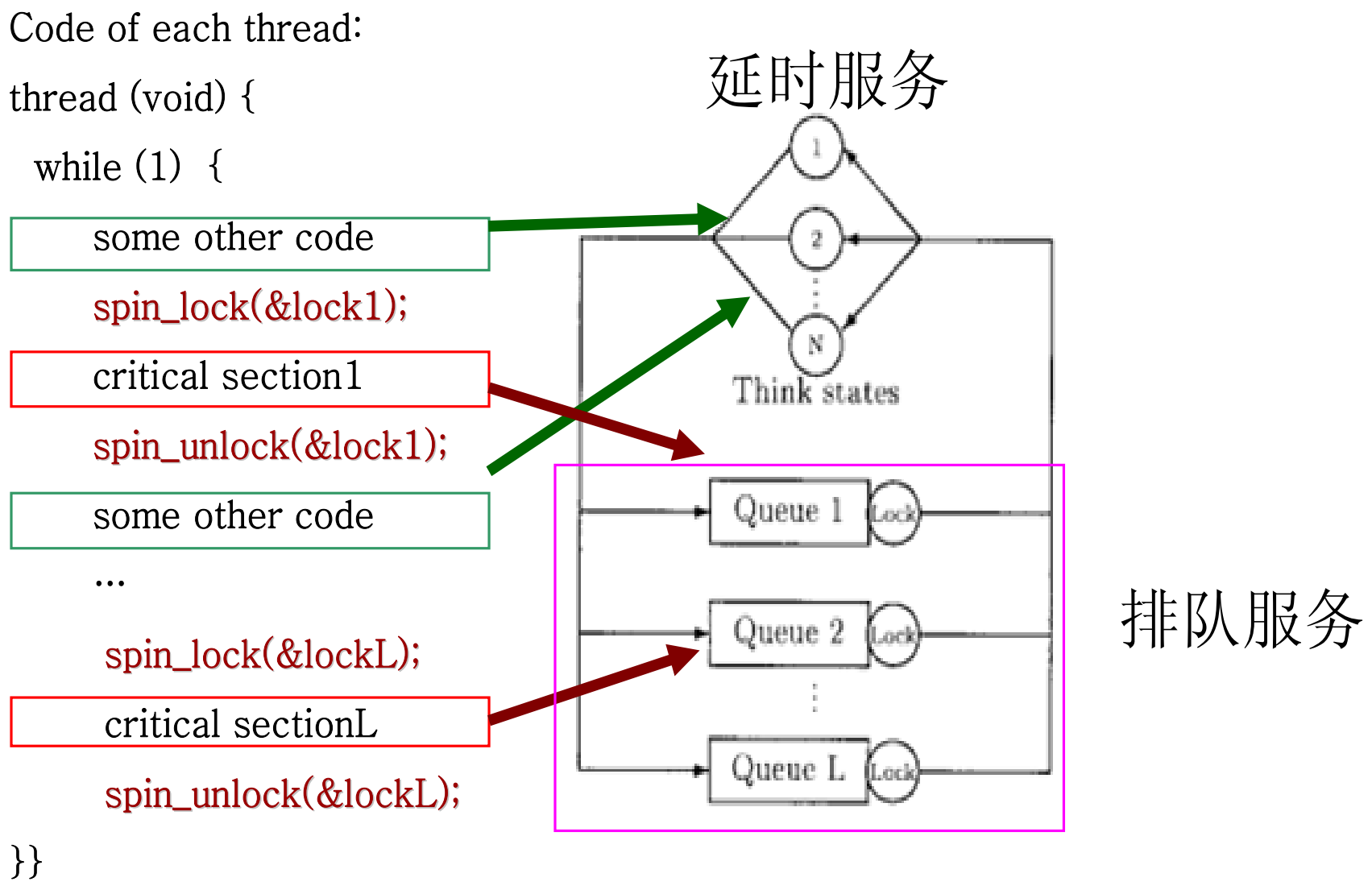

图 3.6 临界区和非临界区和排队论模型的对应关系

表 3.1 模拟器中的事件定义

| 事件 | 类型 | 描述 |
|---|---|---|
| INSTRUCTION | 原子 | 无缓存缺失地执行指令 |
| STORE | 原子 | 对锁数据结构执行写操作 |
| ENTER_C | 复合 | 进入临界区 |
| LOCK_MISS | 原子 | 读锁数据结构时引发缓存缺失 |
| CACHE_MISS | 原子 | 在临界区和非临界区内产生缺失 |
| SPIN | 原子 | 等待排号自旋锁 |
| EXIT_C | 复合 | 离开临界区 |
| ENTER_NC | 复合 | 进入非临界区 |

#### 3.2.2.2 数据结构和事件定义

为了表示整个系统随着时间的演化，模型中的每个核都能够产生一系列事件。模拟器为每个核维护一个事件队列，而每个事件队列都有一个时间戳记录当前事件的开始时间。根据每个事件队列的时间戳，所有的核被组织在一个二叉堆中，而位于堆根部的核具有最小的时间戳。同样地，为每个锁和内存块维护了一个队列用来反映锁的竞争和内存块的竞争。实际上，为每个内存块维护一个队列可以模拟 NUMA 系统上的内存竞争。

为了描述整个系统的行为，模拟器中定义了丰富的事件。表 3.1 列出了整个事件列表和每个事件的描述信息。系统中所有的事件可以被分成两类，一类称为原子事件(atomic event)，而另外一类称为复合事件(composite event)。其中原子事件是不能够进一步分解成其它事件的事件，而复合事件是可以分解成复合事件和原子事件的事件。INSTRUCTION、STORE、LOCK_MISS、CACHE_MISS 和 SPIN 是原子事件，而 ENTER_C、EXIT_C 和 ENTER_NC 是复合事件。对于复合事件 ENTER_C，它可以被分解成如下的一系列事件组合：

ENTER_C=LOCK_MISS+STORE+SPIN+(INSTRUCTION+CACHE_MISS)*+INSTRUCTION+EXIT_C

其中()* 代表括号中的事件可以发生 0 次或者任意多次。在实际使用时，括号内出现的事件次数和临界区的长度直接相关。类似地，EXIT_C 和 ENTER_NC 可以被分别拆解如下：

EXIT_C=LOCK_MISS+STORE+ENTER_NC

ENTER_NC=(INSTRUCTION+CACHE_MISS)*+INSTRUCTION+ENTER_C

EXIT_C 和 ENTER_NC 的最后一个事件分别是 ENTER_NC 和 ENTER_C，这样的事件定义可以反映出临界区和非临界区之间的循环。

#### 3.2.2.3 模拟器引擎

每个核对应的事件队列中的第一个事件被初始化为 ENTER_NC，并且每个事件队列的时间戳均被初始化为 0。在模拟开始之前，根据每个核的事件队列的时间戳将所有的核插入到一个二叉堆中。图 3.7 表示了模拟器的主循环。如图所示，pollEvent()，handleEvent()和 updateTick()是模拟器中 3 个最为核心的函数，下文依次对这 3 个函数进行介绍。

```
/*main loop of the simulator*/
main (void) {
        while (tick < MaxTick) {
            /*remove the core at the heap and maintain the heap*/
            Index index = pollEvent();
            /*handle event according to event type and add the core to the heap again*/
            handleEvent(index);
            /*update the system clock*/
            updateTick();
        }
}
```

图 3.7 模拟器程序的主要循环

函数 pollEvent()的功能是在二叉堆上移除位于根节点的核，然后对剩余的节点维护二叉堆的特性。因为所有的核是根据每个核对应的事件队列的时间戳以二叉堆的形式进行组织，而根节点的核对应的事件队列的时间戳最小，所以根节点对

应的事件队列中的第一个事件是最需要被处理的。事件处理流程由 handleEvent()函数覆盖。具体而言，handleEvent()根据不同的事件类型进行处理，然后更新该队列的时间戳，最后对所有核组成的二叉堆进行维护(利用更新的时间戳重新将核插入二叉堆中)。

如果当前处理的事件是 INSTRUCTION，那么将对应事件队列的时间戳设置为该队列中下一个事件的起始时间。如果事件是 LOCK_MISS、CACHE_MISS 或者 STORE，则需要模拟内存访问的延时。然而，在 NUMA 系统上，每个内存块(memory bank)都要被表示成一个队列，那么，如何决定在哪个队中计算内存延时？如果事件是 LOCK_MISS 或者 STORE，那么说明操作的是代表锁的数据结构，而排在哪个队列的信息已经在模拟器的参数中被提供出来了。如果事件是 CACHE_MISS，那么说明操作的数据并非是代表锁的数据。在这种情况下，随机的选择一个内存队列来计算内存延时。当确定好了计算内存访问延时的队列，则根据如下的规则更新事件队列时间戳。如果当前的内存访问队列为空，那么事件队列的时间戳被更新为当前的时间戳和内存访问延时之和。否则，则更新为上次内存访问的结束时间和内存访问延时之和。

如果当前事件队列中的事件是 STORE，而下一个事件是 SPIN，则说明这个核正进入临界区，那么模拟器需要在对应的锁队列中排序。如果 STORE 的下一个事件不是 SPIN，那么说明这个核正离开临界区，那么，这个核应该从锁队列中被移除。当一个核进入或者离开锁队列时，在锁队列中第一个事件是 SPIN 的核将要导致一次缓存缺失，因为临界区是用排号锁保护起来的。模拟锁数据导致缓存缺失的方法是对每个核的 SPIN 事件之前插入一个 LOCK_MISS 事件。

对于事件 SPIN，模拟器首先需要确定发出该事件的核是否位于锁队列的队首。如果确实位于队首，那么该核对应的事件队列时间戳应该被更新为当前的系统时间，否则事件队列的时间戳被设置成一个非法值以表示该核仍处于自旋状态，并且 SPIN 事件仍然是该事件队列的第一个事件。对于后一种情况，引发 SPIN 事件的核暂时不会被插入到二叉堆中，因为 SPIN 事件还没有被执行完，所以无法根据当前的事件计算出下一个事件的起始时间。

对于复合事件 ENTER_C、EXIT_C 和 ENTER_NC，模拟器需要根据 3.2.2.2 节定义的事件展开规则对其进行展开，并且引发这些事件的核对应的事件队列时间戳不需要被更新。模拟器的参数中已经提供了临界区和非临界区的数目以及特征。当计算单元引发事件 ENTER_C 或者 ENTER_NC 时，则要从模拟器参数中指定的概率分布中挑选某个临界区和非临界区执行。

模拟器中另外一个核心函数 updateTick()负责更新模拟器的时间，更新的值为

位于二叉堆上根节点的核对应的事件队列的时间戳。

#### 3.2.2.4 模型参数

模拟器为用户提供了各种参数。这些参数可以被分成两类, 一类是硬件体系结构相关的参数, 而另外一类是软件同步原语相关的参数。对于硬件体系结构相关的参数, 用户可以指定系统中有多少芯片, 每个芯片的核数，本地内存访问的延时以及内存的架构(UMA 或者 NUMA)。对于软件同步原语相关的参数, 用户可以指定临界区的数目和每个临界区的特征。临界区的特征包括保护临界区的锁的内存位置, 临界区中连续两个缓存缺失的间隔, 临界区中缓存缺失的总数和每个临界区被选择的概率。用户同样可以指定非临界区的数目和每个非临界区的特征。非临界区的特征包括连续两次缓存缺失之间的间隔, 每个非临界区中的缓存缺失总数以及每个非临界区被选择的概率。

### 3.2.3 评测结果

#### 3.2.3.1 锁颠簸现象的重现和特征

硬件体系结构相关的参数被配置成 NUMA 系统。系统中有 8 个芯片, 每个芯片有 4 个核。系统中共有 8 个内存块，每个芯片和一个内存块相连。软件同步相关参数利用了 4 种配置来代表不同的工作负载。表 3.2 详细表示了每种负载的特征。

表 3.2 4 个配置上的软件同步原语相关参数

| 配置 | 非临界区间隔 | 非临界区缺失 | 非临界区概率 | 锁存在的节点 | 临界区间隔 | 临界区缺失 | 临界区概率 |
|---|---|---|---|---|---|---|---|
| C1 | 0 | 0 | 1 | 0 | 1 | 1 | 1 |
| C2 | 34 | 7 | 0.14 | 0 | 20 | 1 | 0.33 |
| | 44 | 8 | 0.18 | 0 | 10 | 1 | 0.33 |
| | 54 | 9 | 0.36 | 0 | 20 | 1 | 0.33 |
| | 44 | 8 | 0.18 | -- | -- | -- | -- |
| | 34 | 7 | 0.14 | -- | -- | -- | -- |
| C3 | 50 | 1 | 0.31 | 0 | 2 | 1 | 1 |
| | 100 | 1 | 0.38 | -- | -- | -- | -- |
| | 50 | 1 | 0.31 | -- | -- | -- | -- |
| C4 | 15 | 1 | 0.16 | 0 | 4 | 1 | 0.25 |
| | 30 | 1 | 0.21 | 0 | 5 | 1 | 0.25 |
| | 125 | 1 | 0.26 | 0 | 3 | 1 | 0.25 |
| | 30 | 1 | 0.21 | 0 | 2 | 1 | 0.25 |
| | 15 | 1 | 0.16 | -- | -- | -- | -- |

由四种配置得到的应用加速比如图 3.8 所示。从图中可以看到, 锁颠簸现象可以在配置 C1、C3 和 C4 中被观察到。而对于展示出锁颠簸现象的配置中, C1 的加速比随着核数的增加而持续下降, C3 和 C4 的加速比分别在使用 17 核和 23 核之后才开始下降。实际上, 一旦硬件架构的参数是固定的, 一个配置是否会产生锁颠簸

现象是由临界区和非临界区的特征所决定的。例如，在一个配置中,如果保护临界区自旋锁的竞争开销相对较大，那么，锁颠簸容易产生。也正是根据这个经验规则，才选择了这样四种负载。值得注意的是，尽管配置 C2 不能在 32 核系统上观察到锁颠簸，但是当使用更多的核时，还是可以观察到该现象的。

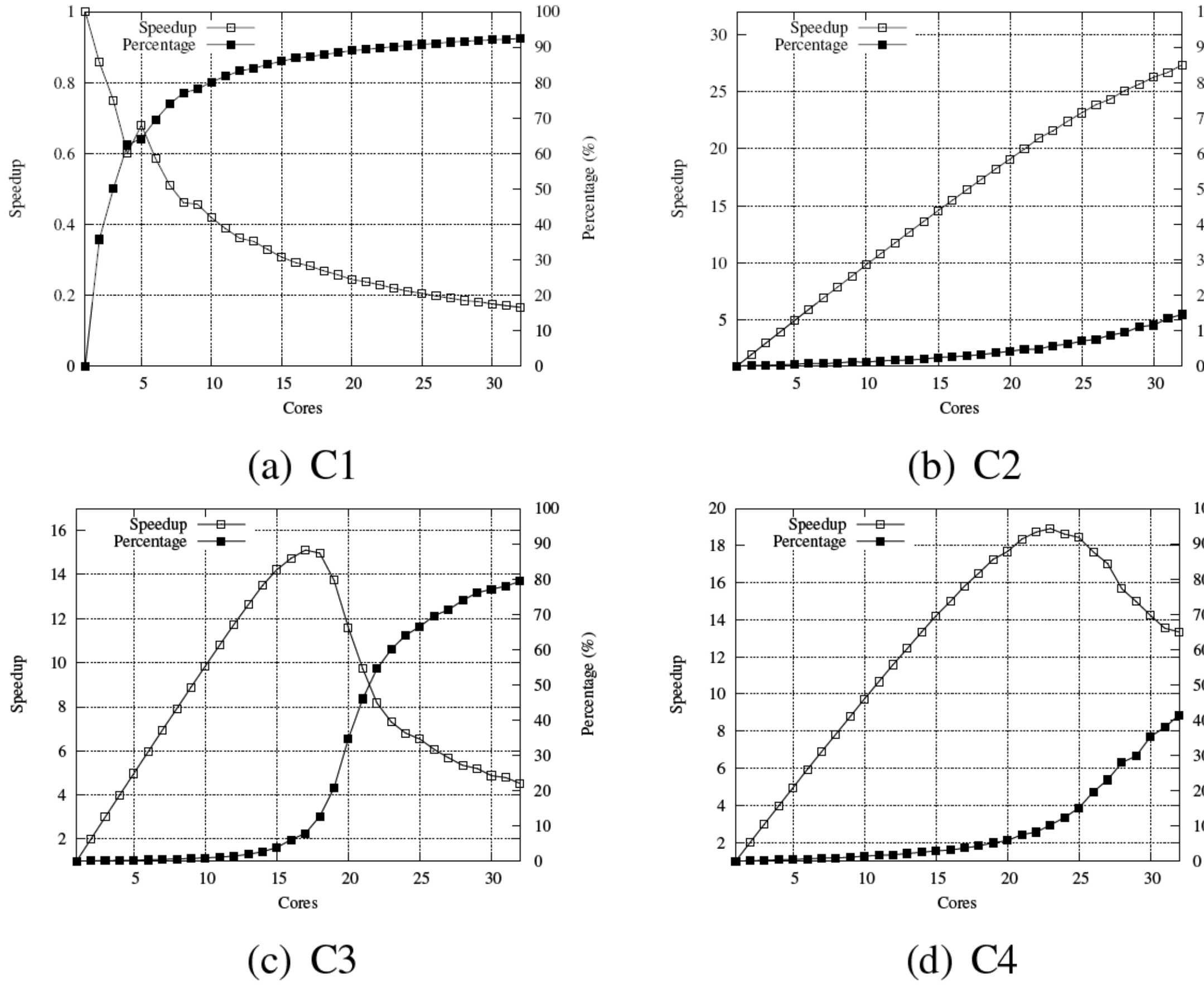

(a) C1　(b) C2

(c) C3　(d) C4

图 3.8 每个配置的加速比以及每个核的平均等锁时间百分比随核数的变化

每个核的平均等锁时间百分比也在图 3.8 中表示以刻画锁颠簸现象发生时相关指标的变化规律。一个有趣的现象是，当锁颠簸现象开始产生时，每个核的平均等锁时间百分比也开始快速的上升。以配置 C3 为例，当系统的核数小于 17 时，该工作负载的加速比随着核数的增加而增加，同时，每个核的平均等锁时间百分比由 0%缓慢的增加至 7.87%。然而，当使用多于 17 个核时，加速比开始突然下降，而等锁时间百分比开始快速增长至 79.31%。这个现象说明，当锁颠簸现象开始后，每个核突然变得忙于等锁以至于有用工作的时间变得很少。

图 3.9 展示了 4 个配置中事件数目的分解随着核数增加的变化。由图中可见，当锁颠簸发生之后，绝大多数事件变成了 SPIN 和 LOCK_MISS，而其它的事件占总体事件的比例开始随着核数的增加而逐渐变少。这个现象产生的原因是一旦锁颠簸现象发生，太多的核将被阻塞在某一个锁上。在模拟器的设计中，SPIN 事件是所有等待某一个锁的核对应的事件队列中第一个将要处理的事件。而当一个核请

求某个锁或者对某个锁释放时, LOCK_MISS 事件被所有等待锁的核所触发。因此, 在锁竞争激烈时模拟器会产生大量的 LOCK_MISS 和 SPIN。

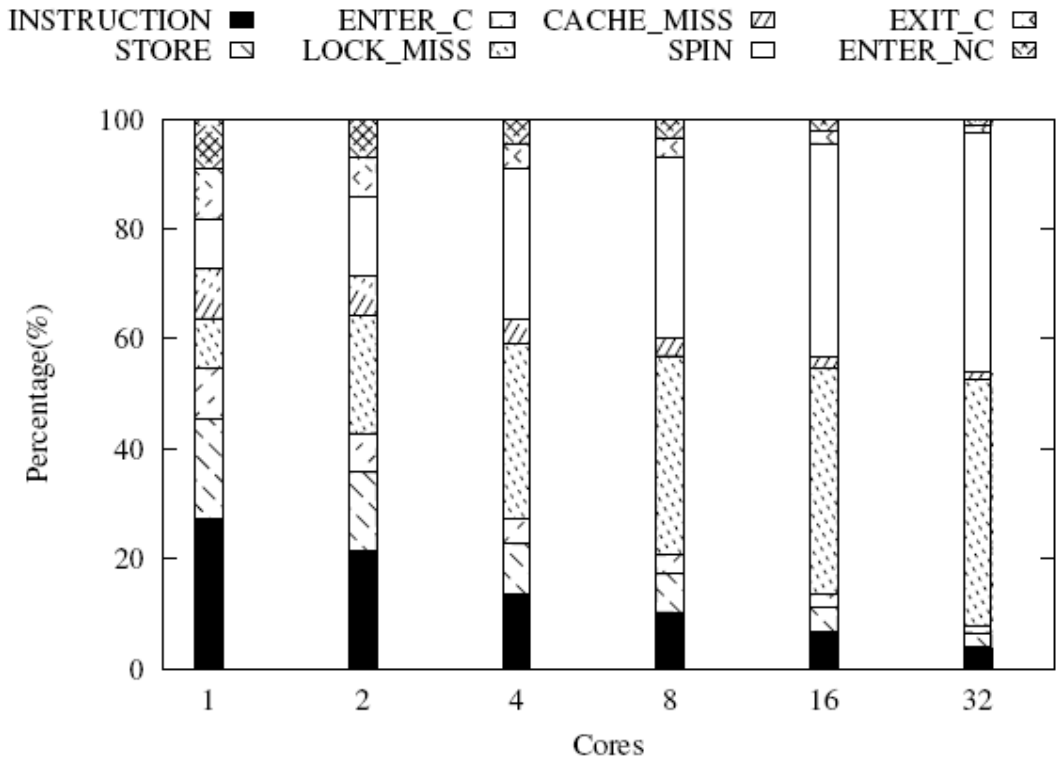

(a) C1

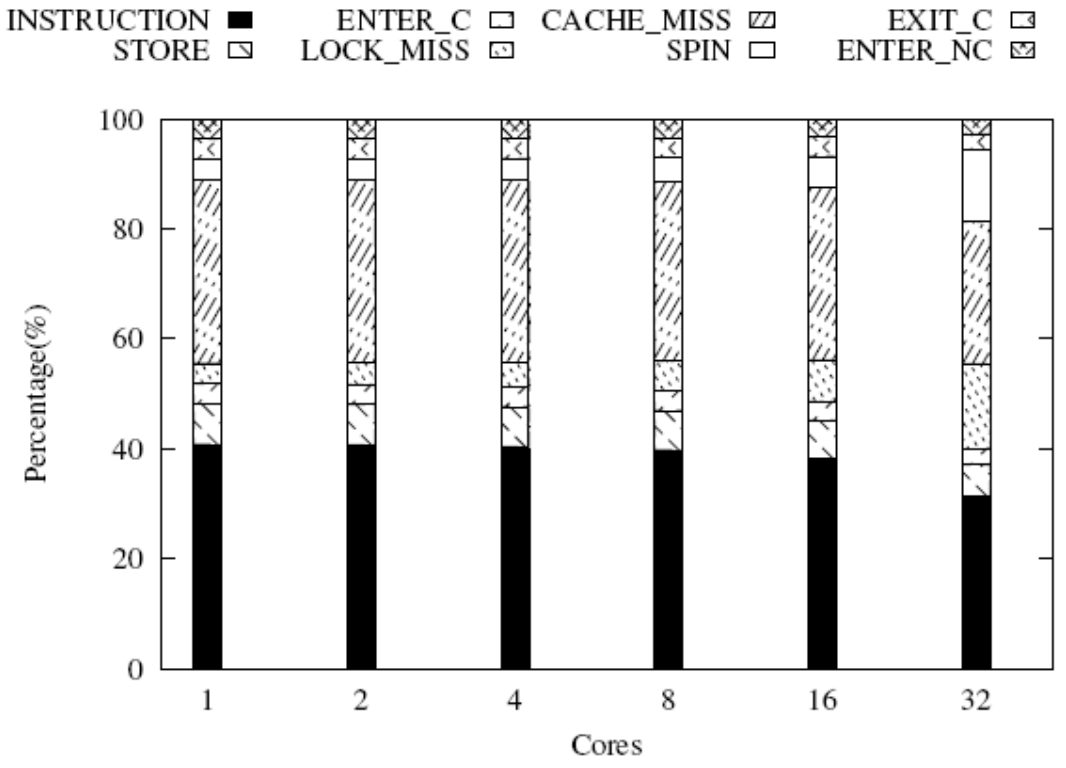

(b) C2

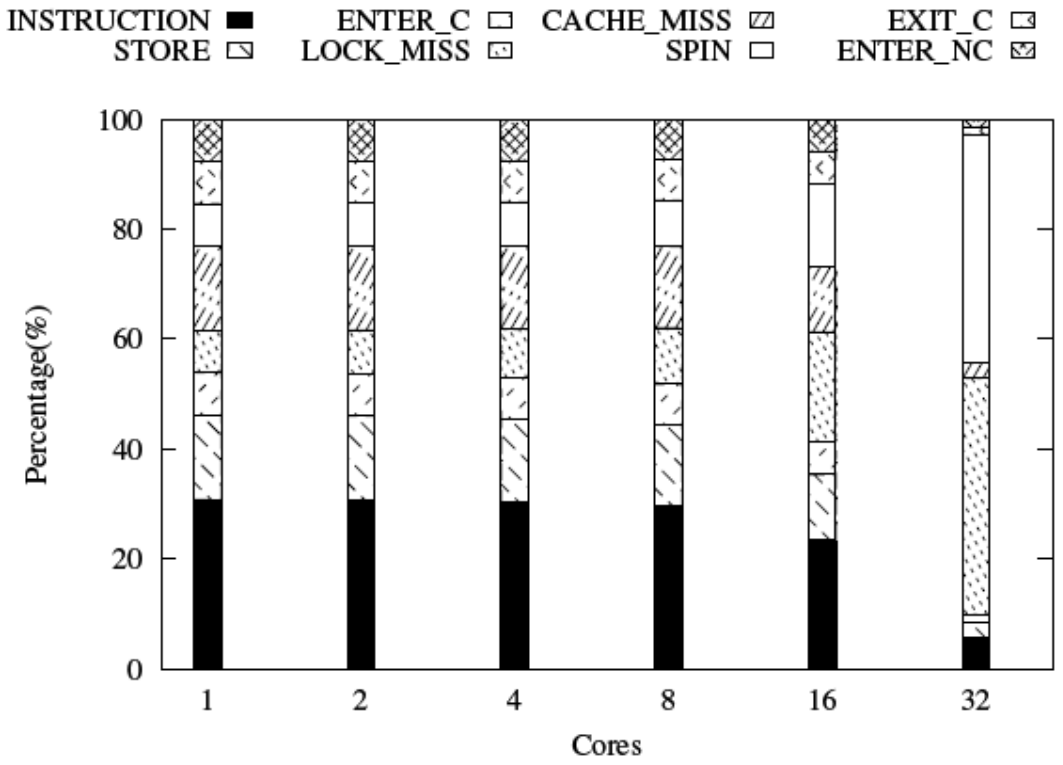

(c) C3

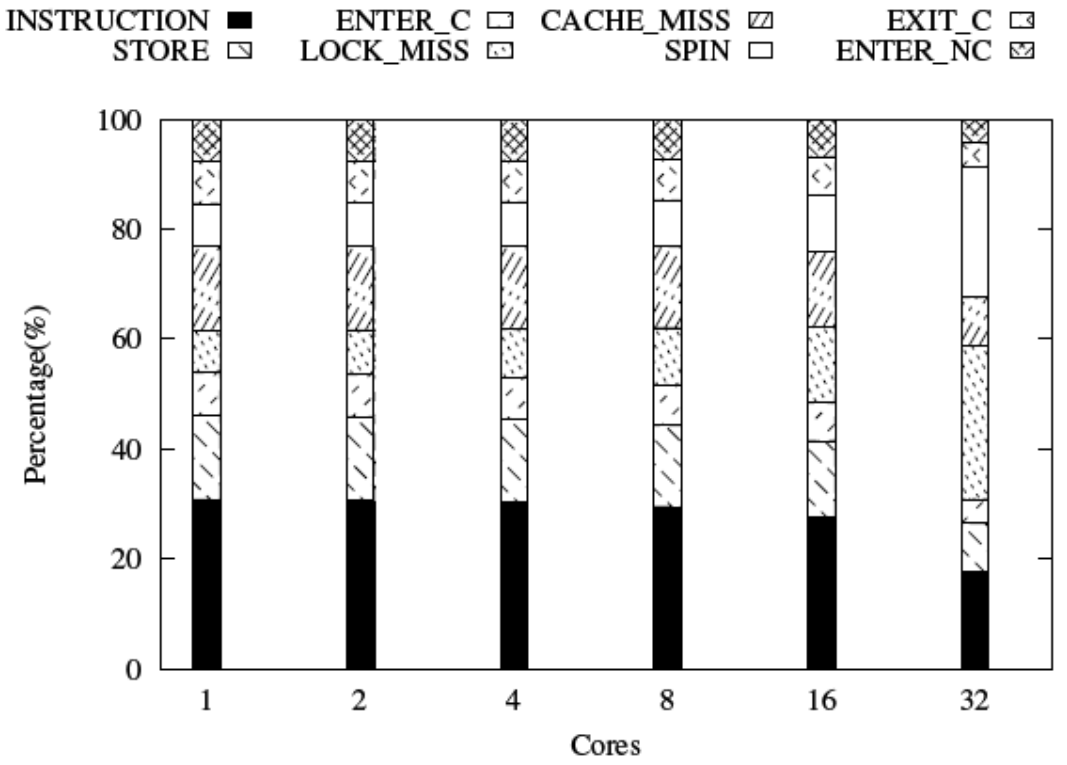

(d) C4

图 3.9 每个配置中事件数目的分解随着核数的变化关系

图 3.10 中表示了每个配置等待每个锁的核数随着核数的变化。可见, 当锁颠簸现象发生后, 等待某个锁的核数会快速的增加。对于加速比随着核数一直增长的配置而言, 所有的锁队列均一直很短。这个结论对于等待内存块的队列依然成立(图 3.11 所示)。即当锁颠簸开始之后, 等待某个内存块队列的对长会大幅度地增加, 而在锁颠簸发生之前, 所有的内存块队列都会很短。由图 3.11 可以看出, 当锁颠簸发生之后, C1, C3 和 C4 配置中内存块 0 上的队列开始变得很长, 这是因为代表锁的数据存在于内存块 0 上(见表 3.2)。

#### 3.2.3.2 硬件架构参数对加速比的影响

为了研究硬件架构参数对应用可扩展性的影响, 除了原来使用的硬件平台(P1)增加 2 个不同的硬件平台。一个平台是 32 个芯片, 每个芯片一个核, 并且每个芯片有一个独立的内存块(P2)。另外一个平台上, 所有的核位于同一个芯片上, 共享一个内存块(P3)。图 3.12 展示了两种配置上每种平台的加速比随着核数的增加的变化(其它配置的结果类似, 故省略)。从图中可以发现, 当锁颠簸现象尚未发生时,

改变芯片的数目和系统中的内存块数不会影响应用程序整体的可扩展性。 然而，当锁颠簸现象发生后，P2 比 P1 的加速比略微好些，而 P1 和 P2 的加速比均远远好于 P3。这些差别的产生是由于系统中内存块数目的不同。对于 P2 而言，系统中的内存块数目最多。当缓存缺失发生时，需要访问的内存块是以均匀分布随机选择的。那么，竞争相同内存块的概率在平台 P2 中是最小的。当锁颠簸现象尚未发生时，不同平台性能差异并不大，因为内存块的竞争不够激烈。

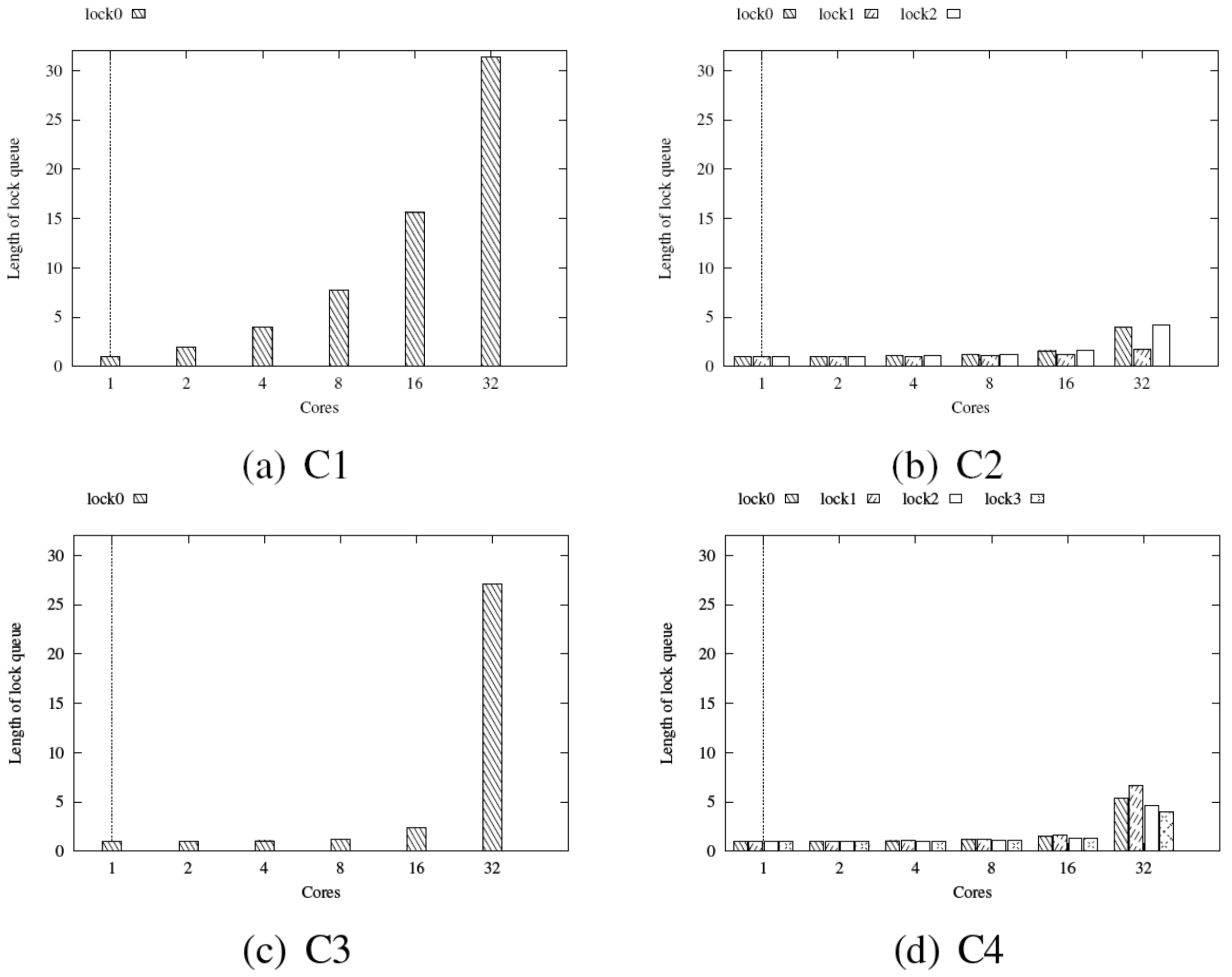

(a) C1 (b) C2 (c) C3 (d) C4

图 3.10 四种配置中等待某个锁的核数随着核数增加的变化

### 3.2.3.3 内存访问延时对加速比的影响

内存访问延时的敏感度是通过使用不同的内存延时值来研究的。在实验中，使用 P1 作为目标实验平台。图 3.13 表示了当内存访问延时分别被设置为 1,5 和 10 时，配置 C3 和 C4 的加速比随着核数增加的变化曲线。通常来说，增加内存访问延时会降低总体的可扩展性，并且锁颠簸现象更容易出现。因为增加内存访问延时同时增大了对某一个锁竞争时引入的开销。以配置 C3 为例，当内存访问的延时被设置成 1 时，整体的加速比在使用 17 核时达到最大。但是，当内存访问的延时分别为 5 和 10 时，最大的加速比分别在 7 和 5 个核时达到。

### 3.2.4 讨论及后续工作

本文提出了基于离散事件仿真技术的锁颠簸现象模拟器。后续的工作由三部分组成：1 模拟器的设计和实现中仅考虑了内存的竞争，然而对其它共享资源的竞争同样会导致应用程序的可扩展性受到影响。在后续的工作中，更多竞争将会在模拟器中被考虑进去。2 用该模拟器预测实际系统的锁颠簸现象。模拟器需要的参数均可以通过离线或者在线的方式测量得到。下一步的工作是将模型得到的加速比曲线和真实的系统中的测量结果相比较。3 该模拟器只适用于每个核的用锁行为类似的工作负载，在后续的工作中，将考虑每个核的锁行为不尽相同的情况。

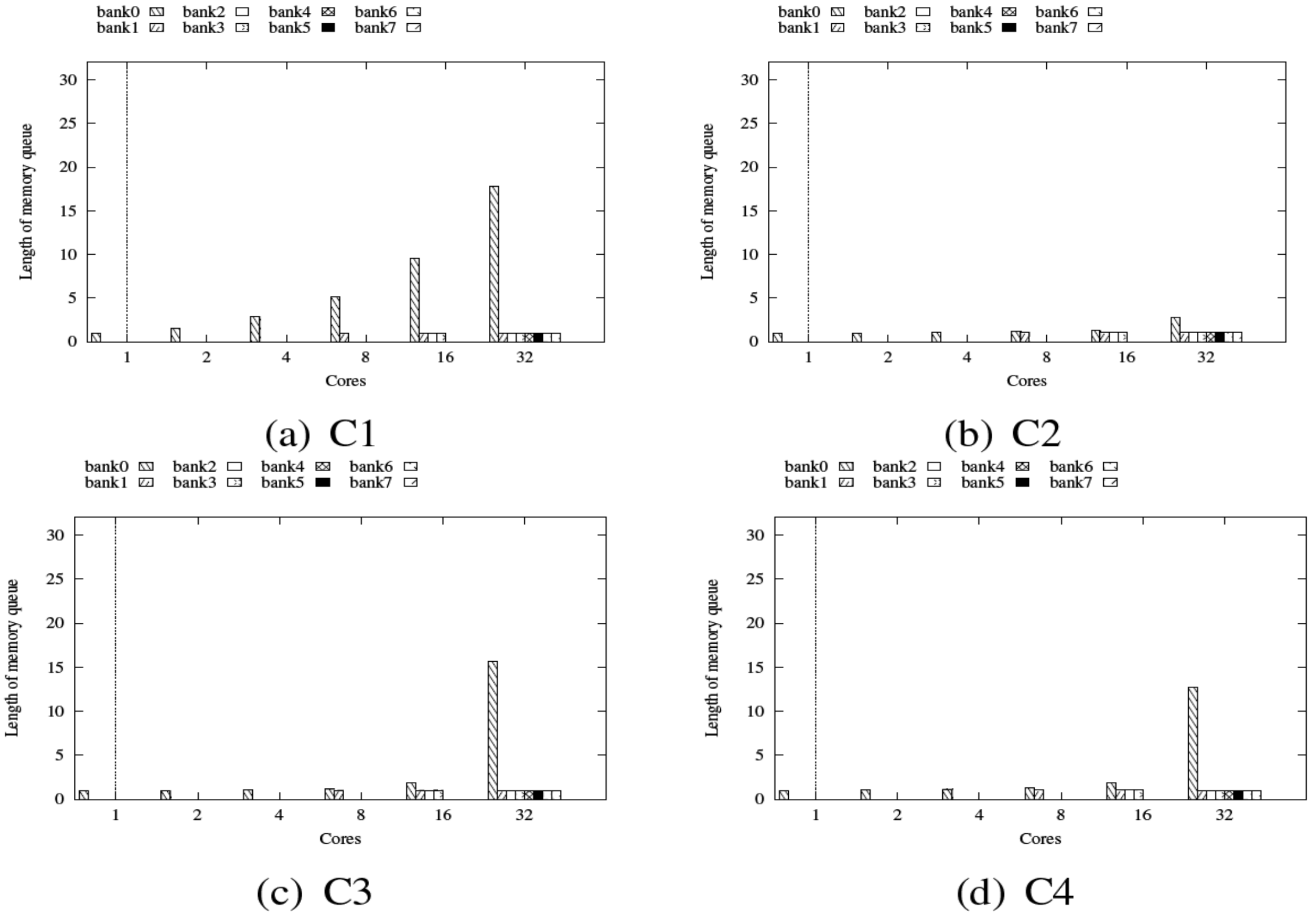

图 3.11 四种配置中等待某个内存块的核数随着核数的变化

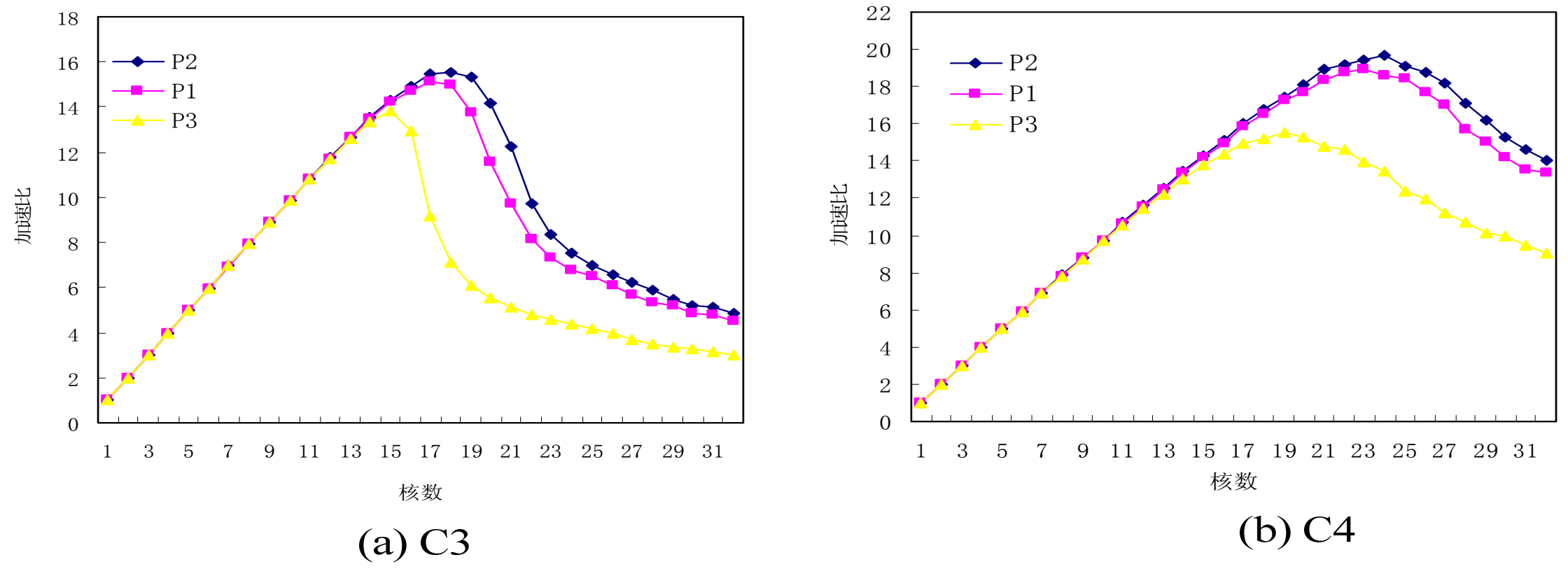

图 3.12 不同架构参数对加速比的影响

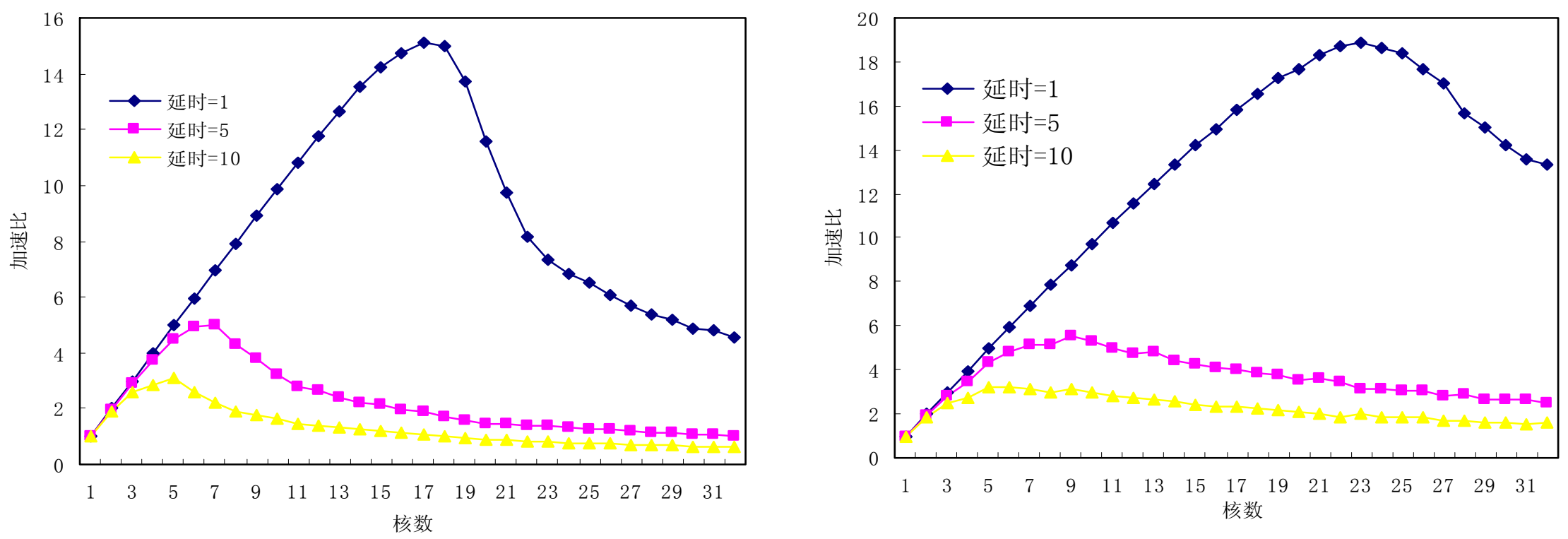

(a) C3　　　　(b) C4

图 3.13 内存访问延时对加速比的影响

## 3.3 基于等待者数目的锁

3.2 节模拟了内核锁竞争导致的锁颠簸现象，并且利用模拟器对该现象进行了深入研究。本节旨在对实际系统进行重新设计以求避免锁颠簸现象。实际上，有许多经典的方法可以被用来避免多核系统上锁竞争导致的锁颠簸。一个方法是任务在等锁时选择睡眠等待而非不断的自旋(例如互斥锁)。然而，这种方法会引入上下文切换的开销，从而影响整体可扩展性。使用自适应锁可以克服互斥锁的主要缺点，因为自适应锁可以在自旋和睡眠之间自适应地进行选择。然而，自旋和睡眠的时机往往是利用启发式规则确定的[29-30, 54]，因此，很难达到自适应锁的最大潜力。例如，决定自旋和睡眠时机的流行做法是根据持锁任务的状态来决定。当一个任务(A)请求一个正由任务 B 持有的锁时，如果任务 B 处于运行状态时，那么任务 A 通过自旋来等待任务 B 以期待锁会马上被释放。仅当任务 B 处于睡眠状态时，A 才通过睡眠来等待获取锁。因此，当应用的任务数小于系统中的核数时，自适应锁和排号自旋锁的行为类似，因为持锁的任务可能一直处于运行状态。因此，当自适应锁被激烈的竞争时，锁颠簸现象同样会出现。使用可扩展锁[55-56]是另外一个避免锁颠簸现象的解决方案。然而，这种锁使得所有的锁等待者自旋在一个局部的变量上，从而导致其它的任务不能利用该核，降低了能耗有效性和资源利用率[57]。为了达到比自适应锁和可扩展锁更好的可扩展性和能耗有效性，本文提出了一种特别的加锁方法。这种方法源于一种对锁颠簸现象的观测，即当锁颠簸发生时，等在某一个锁上的核数开始快速的增加。基于这个观测，本文提出的方法跟踪请求一个锁的核数，如果等待锁的核数大于一个固定的阈值，所有后续的锁请求者将进入节能状态来等待锁，否则，仍然采用自旋等待来获取锁。

本文提出的锁协议称为基于等待者数目的锁，已经实现在了 Linux 2.6.29.4 和 Linux 2.6.32 内核中，并且在 AMD32 核和 Intel40 核平台上进行了评测。利用微基

准测试程序和实际的应用进行评测表明，提出的锁可以在大部分应用中避免锁颠簸现象。对于那些没有受到锁颠簸现象影响的应用，本章提出的实现不会对原来系统的可扩展性有所影响。而且，基于等待者数目的锁比已有的锁颠簸避免方法能够提供更好的可扩展性和能耗有效性。

### 3.3.1 锁颠簸现象分析

本小节将锁颠簸现象的发生和等待某个自旋锁的核数相关联。图 3.14 表示了等待测试程序 parallel postmark 中竞争最为激烈的 2 个自旋锁(即内存文件系统的统计信息锁和文件描述符表锁) 的平均锁请求者数目。图中的数据是 10 次实验的平均值。并且，线程的数目也不断的增加以观测锁等待者的数目随着竞争逐渐激烈而产生的变化。

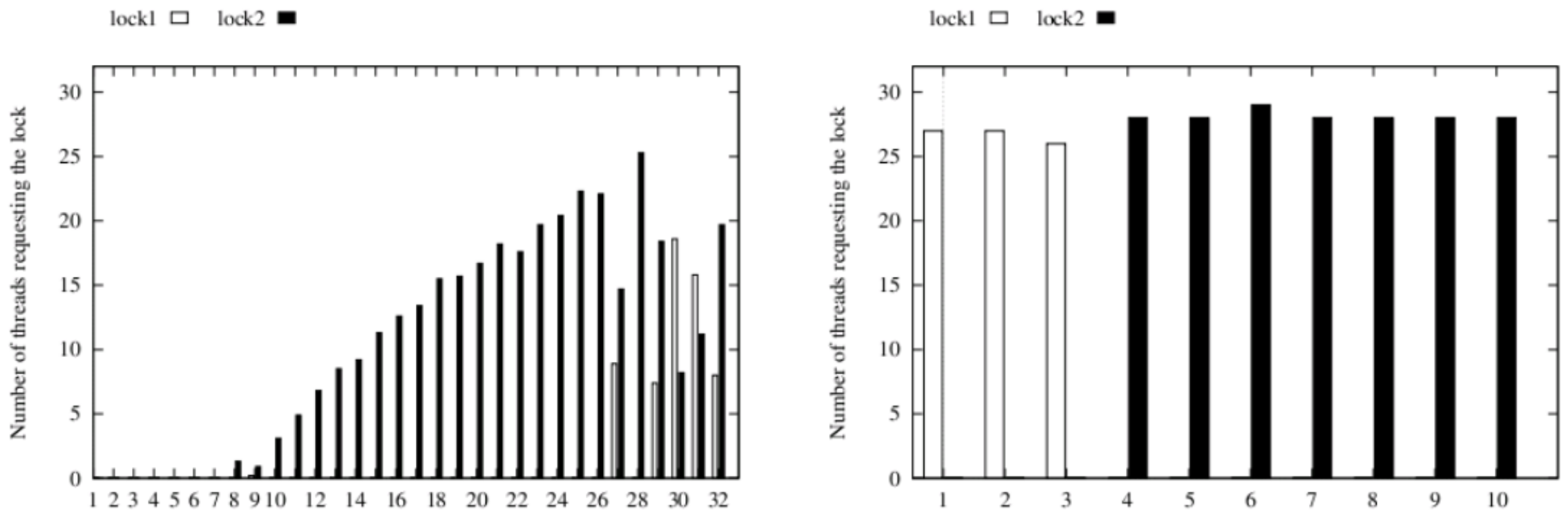

图 3. 14 parallel postmark 在 AMD32 核平台上每个竞争的自旋锁的等待者数目(左)和当使用 32 个核时每个锁的的等待者数目(右)。lock1 代表内存文件系统的统计数据自旋锁而 lock2 代表文件描述符表锁。

如图所示，当线程的数目到达 10 之前，等待文件描述符表锁和内存文件系统的统计信息锁的等待者数目基本为 0。当测试程序的线程数目大于 10 但是小于 25 时，文件描述符表锁变成了最主要的可扩展性瓶颈，因为该锁的平均等待者数目开始随着线程数目的增加而增加。对于内存文件系统的统计信息锁，锁等待者的平均数目同样是 0。当测试程序的线程数目大于 25 时，2 个锁在 10 次实验中轮流的成为可扩展性瓶颈。但是在一次实验中，只有唯一一个锁是瓶颈。10 次实验的平均锁等待者在线程数大于 25 时可能比线程数在 10 至 25 之间的值要小，因为没有一个锁一直是可扩展性瓶颈。

为了进一步地验证 2 个锁在线程数大于 25 时如何轮流成为可扩展性瓶颈，图 3.14 右侧显示了当使用 32 个线程时，每次实验中每个自旋锁的等待者数目。从图中可见，10 次实验中有 3 次可扩展性瓶颈是内存文件系统的统计数据锁，而有 7 次瓶颈是文件描述符表锁。2 个瓶颈之所以可能发生切换是由于进入 2 个临界区的概率基本相同。从实验结果中，可以得到这样一个结论，锁颠簸现象由内核中的锁竞

争所导致，伴随这一现象的是当锁颠簸现象发生之后等待某个锁的等待者数目开始大幅度的增加。实际上，在上一节对锁颠簸现象的模拟中，利用提出的模拟器同样可以得到这一结论。从而也验证了模拟器的正确性。

### 3.3.2 基于等待者数目锁的实现

3.3.1 节对锁颠簸现象的分析产生了设计基于等待者数目锁的动机。基于等待者数目锁的核心方法是检测等待自旋锁上的等待者数目，如果数目大于一个固定阈值，所有新来的锁等待者进入节能状态，否则等待者以自旋方式获取锁。

基于等待者数目的锁对 Linux 排号自旋锁进行了包装以实现更加高级的功能。本文首先介绍 Linux 的排号自旋锁，然后再介绍本文提出的锁。Linux 内核中广泛的使用排号自旋锁。在实现上，每个排号自旋锁由一个 unsigned int 变量代表，该变量被划分为 2 个域，一个称为 next，而另外一个称为 owner[58]。当一个任务请求一把锁时，它利用原子操作增加代表锁变量的 next 域，并且将没有增加之前的锁变量的值存到本地的一个变量中。然后，该任务从本地变量中解析 next 域和 owner 域, next 域被用做该任务排到的号。如果解析出来的 next 域和 owner 域相同，那么说明该任务成功地获取了锁，并且可以进入到锁保护的临界区中。否则，该任务将锁变量的 owner 域拷贝到本地变量的 owner 域，并且在本地比较 next 域和 owner 域。这种拷贝和比较操作会不断的进行直到本地的 2 个域相同。当一个任务释放一个锁时，该任务需要增加 owner 域。通过给每个等待进程一个号，所有的进程以先来先服务(FIFO)的服务策略进行处理。图 3.15，图 3.16 和图 3.17 分别表示了基于等待者数目的锁的定义，获取和释放的代码。基于等待者数目的锁利用与 Linux 的排号锁一样的机制来代表锁的状态。正如在锁的定义中所表示的，有一个 unsigned int (即 slock)变量代表任务是否被允许进入临界区。如果 slock 的 next 和 owner 域不相同，则表明某些任务处于临界区中，并且由锁保护着。这样，新来的锁请求者必须要等待锁的释放。

```
#define CACHE_LINE 64

typedef struct requester_lock{
    unsigned int slock __attribute__((__aligned__(CACHE_LINE)));

    struct {
        spinlock_t wa_lock __attribute__((__aligned__(CACHE_LINE)));
    } wait_lock[NR_CPUS];

    struct {
        struct list_head wa_list __attribue__((__aligned__(CACHE_LINE)));
    } wait_list[NR_CPUS];
} requester_lock_t;
```

图 3.15 基于等待者数目锁定义

```
#define  FIELD_LEN 8
#define  FIELD_MASK ((1<<FIELD_LEN) - 1)
#define  FIELD_SIZE (1<<FIELD_LEN)

void inline requester_lock(requester_lock_t *lock)
{
    struct task_struct *task=current;
    struct lock_waiter waiter;
    int cpu = smp_processor_id(), owner, next, len;
    volatile unsigned int *lockp = &(lock->slock);
    unsigned int lockv = *lockp;

    next = (lockv >> FIELD_LEN) & FIELD_MASK;
    owner = lockv & FIELD_MASK;
    len = (next + FIELD_SIZE - owner)%FIELD_SIZE;

    if (len > waiters) {
            spin_lock(&lock->wait_lock[cpu].wa_lock);
            list_add_tail(&waiter.list, &lock->wait_list[cpu].wa_list);
            waiter.task = task;
            mb();

            lockv = *lockp;
            next = (lockv>>FIELD_LEN) & FIELD_MASK;
            owner = lockv & FIELD_MASK;
            len = (next + FIELD_SIZE - owner)%FIELD_SIZE;

            if (len > 0)
                task->monitor_flag = 0;
            else
                goto done;

            spin_unlock(&lock->wait_lock[cpu].wa_lock);

            while (task->monitor_flag!=1) {
                    monitor((void*) &task->monitor_flag, 0, 0);
                    if (task->monitor_flag!=1)
                            mwait(0, 0);
            }

            spin_lock(&lock->wait_lock[cpu].wa_lock);
    done:
            remove_waiter(&waiter);
            spin_unlock(&lock->wait_lock[cpu].wa_lock);
      }
      requester_spin_lock(lock);
  }
```

图 3.16 获取锁代码

与 Linux 的排号自旋锁相比，本文提出的基于等待者数目锁的不同之处在于当等待的锁被释放时，任务可能会进入节能状态而不是一直地自旋。为了实现这种做法，锁定义中表示出每个 CPU 均有的链表，并且每个链表由一个排号自旋锁保护。

一个锁的所有请求者被放在每个 CPU 的链表中，使得保护链表的锁变得细粒度。这种组织方式比将所有的请求者组织在一个链表中能够达到更好的可扩展性。基于等待者数目的锁的另外一个可扩展性优化是锁中每个域被存于不同的缓存行中以避免伪共享(false sharing)。这个优化对可扩展性的提升很大，因为基于等待者数目的锁的大部分域都是写操作密集的。如果这些域被分配在相同的缓存行中，缓存行会在不同的核之间来回地传递。

当任务请求基于等待者数目的锁时(图 3.16)，任务首先检查 slock 的 next 域和 owner 域之间的差异。如果差值不大于一个预定义的阈值，任务调用 Linux 的排号自旋锁获取代码(requester_spin_lock()函数)。如果一个锁有太多的请求者，基于请求者数目的自旋锁保证后来的请求者进入节能状态。具体而言，请求锁的任务首先获取保护每 CPU 等待链表的自旋锁。该链表用于存储处于节能状态的任务。然后，任务将自己加入到等待链表中。在设置一个用于指示当前的任务是否处于节能状态的变量(monitor_flag)之后，该任务释放每 CPU 的链表锁并且进入节能状态(35-39 行)。计算请求者数目的准确值通常开销比较大，因为在尝试获取锁失败或者锁释放之后需要利用原子操作来更新等待者的数目。但是这种开销在多核系统上是无法忍受的[8]。作为一个优化，基于等待者数目的锁对请求者的数目进行预测而非计算其准确值(13-15 行)。然而，一个重要的问题是这种预测可能是错误的。一个可能的场景是一个任务发现了想要获取的锁已经被其它任务获取，因此决定进入节能状态。但是，当这个任务进入节能状态，该锁已经被释放。这导致一个任务处于节能状态却没有其它的任务将其唤醒。为了避免这种情况，在设置标志一个任务是否处于节能状态时，对临界区内是否有任务进行判断(28 至 31 行)。如果没有任务处于临界区内，表明预测是错误的，那么任务以忙等的方式来获取锁。

当锁被激烈的竞争时，基于等待者数目的锁利用 SIMD (Single Instruction Multiple Data)指令 monitor 和 mwait[59] 来节能。具体而言，monitor 指令可以为硬件建立一个内存线性地址范围来检测，并且将处理器置于检测事件等待状态(monitor event pending state)。当处于检测事件等待状态中，检测硬件检测对指定内存地址范围内的写操作。如果检测到了写操作，处理器则退出检测事件等待状态。mwait 指令利用检测硬件的状态来节能。Linux 的空闲进程[60]和几个研究性的工作[46, 61]已经利用 monitor 和 mwait 指令节能或者提升性能。

图 3.16 展示了如何使用 monitor 和 mwait 指令(35 至 39 行)。monitor_flag 变量被添加在 task_struct 数据结构中作为被检测的变量。在循环中检查被检测变量的值是必须的，因为处于省电状态的处理器可以由任意的检测范围内的写操作、不可屏蔽中断、RESET 以及 monitor 和 mwait 之间的控制转移所唤醒。

当释放基于请求者数目的锁时(图 3.17 所示)，函数 requester_spin_unlock()利用 Linux 自旋锁的实现来改变锁的状态(行 4)。然后，一个非空等待队列的第一个任务被唤醒。为了避免饥饿，在当前的 CPU 之后选择第一个非空队列。换句话说，如果当前的任务正在执行解锁函数并且当前的任务运行在 $CPU_n$ 上，则从 $CPU_{(n+1)\%NR_CPUS}$ 开始查找第一个非空队列。值得说明的是，在检查队列是否是非空时，每个队列状态的获取并没有先获取保护该队列的自旋锁。这个优化极大地减少了在每 CPU 链表锁上的竞争，从而可以提升了可扩展性。然而，现代处理器上的无序执行和编译器的优化可能会引入竞争条件。其正确性的保证是通过在加锁(图 3.16 行 21)和解锁(图 3.17 行 5)时添加额外的内存屏障(memory barrier)来实现的。基于请求者数目的锁用到了一个阈值来决定等待者自旋或者睡眠。该阈值的默认值是 0。在实验和评价一节会讲述如何选择这个阈值。

```
1)  void inline requester_unlock(requester_lock_t *lock)
2)  {
3)      int i, j, found, cpu = smp_processor_id();
4)      requester_spin_unlock(lock);
5)      mb();
6)      found = 0;
7)      for (i=1; i<NR_CPUS + 1; i++) {
8)          j = (cpu + i)%NR_CPUS;
9)          if (!list_empty(&lock->wait_list[j].wa_list)) {
10)             found = 1;
11)             break;
12)         }
13)     }
14)     if (found) {
15)         spin_lock(&lock->wait_lock[j].wa_lock);
16)         if (!list_empty(&lock->wait_list[j].wa_list)) {
17)             struct lock_waiter *waiter = list_entry(lock->wait_list[j].wa_list.next,
18)                                                     struct lock_waiter, list);
19)             waiter->task->monitor_flag = 1;
20)         }
21)         spin_unlock(&lock->wait_lock[j].wa_lock);
22)     }
23) }
```

图 3.17 释放基于请求者数目的锁

### 3.3.3 阈值选择

基于请求者数目的锁的一个关键问题是如何选择锁等待者的阈值。事实上，阈值的选择反映了模式切换的开销和忙等自旋锁的开销之间的折中。当一个核请求正在被另一个任务使用的锁时，它可以选择进入节能状态或者不断自旋来等待锁

释放。如果选择了前者，那么当锁的拥有者释放锁时，唤醒处于节能状态的任务需要付出额外的花销，否则 (选择后者)，锁数据结构将在所有的锁请求者之间来回传递，使得每次锁获取开销都比较大。

对于上面讨论的两种锁等待策略，从节能状态唤醒一个任务的开销是恒定的，而自旋等待锁数据的开销随着自旋锁请求者数目的增加而增加。因此，基于锁请求者数目锁的最优阈值应该使得这两种锁等待方法的开销相同。通过这种方式，可以最小化基于等待者数目锁的开销。最优的阈值在不同的内核锁密集的应用中应该是不变的，因为阈值的选择仅仅依赖于两种锁等待方法的开销，而这些开销和应用是无关的。基于这种观察，选择最优阈值的方式是对一个测试程序的阈值进行敏感性分析，然后根据实验结果选择最佳阈值，最后利用最佳阈值在其它的测试程序上进行验证 (测试程序和运行方式将在 3.3.4.2 节中介绍)。

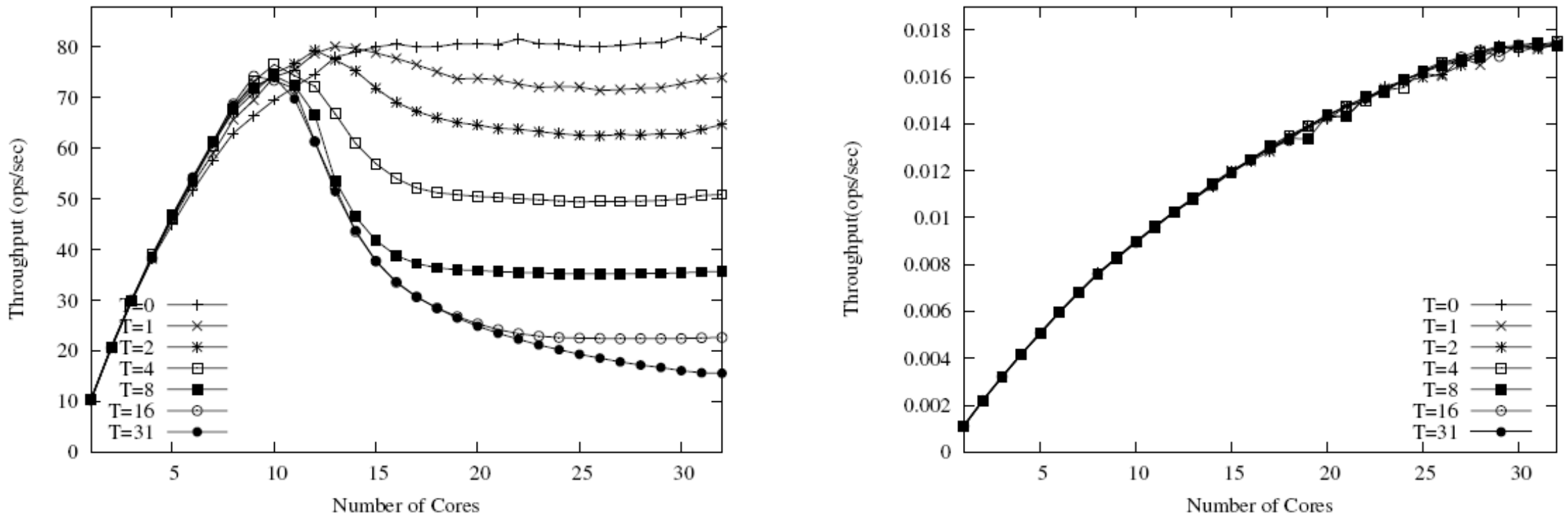

图 3. 18 parallel find(左)和 kernbench(右)中,采用不同阈值(T)时,吞吐量随着核数的变化

测试程序的阈值分别被配置成 0, 1, 2, 4, 8, 16 和 31。对于每个阈值的配置，测量吞吐量随着核数的变化作为可扩展性。图 3.18 表示出了 parallel find 对阈值的敏感程度。对于这个测试程序，阈值选择成 0 时是最好的，因为可以很好地避免锁颠簸现象。然而，随着阈值的增加，锁颠簸现象开始出现并且总体的可扩展性变得越来越差。对于其它的测试程序如 single counter, mmapbench, sockbench 和 parallel postmark，测试程序具有相似的阈值敏感性，并且这些测试程序在阈值设为 0 时，其可扩展性最好。对于 kernbench，尽管阈值由 0 增加到 31，可扩展性却没有受到影响，并且锁颠簸现象不会出现。这是因为 kernbench 中的内核锁竞争非常轻量级。从阈值敏感性分析的结果来看，选择阈值 0 是最好的。实际上，选择阈值为 0 意味着当一个锁被预测为忙时，任务马上进入节能状态。实验结果表明当等待锁时进入节能状态是轻量级的,并且好于自旋等待共享的锁变量。

### 3.3.4 实验和评价

#### 3.3.4.1 多核平台和操作系统

基于请求者数目的锁的有效性在基于 Linux2.6.29.4 的 AMD32 核 NUMA 平台上和基于 Linux2.6.32 的 Intel 40 核 NUMA 平台上被验证。每个系统的硬件参数将会被轮流介绍。对于 AMD32 核系统，共存在 8 个 Opteron 8347HE 芯片，每个芯片有 4 个核。每个核有私有的 L1 数据缓存, L1 指令缓存和 L2 缓存。每个 L1 缓存的大小是 64Kbytes，而 L2 缓存的大小是 512Kbytes。相同芯片上的 4 个核共享相同的 2Mbytes L3。核和芯片之间分别利用交叉互联和 HyperTransport 相连。32G 内存被划分为 8 个内存块，每个内存块与一个芯片相连。

Intel 40 核系统有 4 个 XeonE7 4850 处理器组成，每个处理器有 10 个核。该系统支持超线程，每个核可以虚拟出 2 个核。每个核有 32Kbytes 的指令和数据缓存以及私有的 256Kbytes L2 缓存。同一芯片上的 10 个核共享 24Mbytes 的 L3。独立的芯片利用 QPI(Quick Path Interface)相连。512G 的系统内存被划分为 4 个内存块，每个内存块和一个处理器相连。在测试时，超线程[63]和 Turboboosting[64]功能被关闭以简化可扩展性的分析。后文的实验主要是在 AMD32 核系统上进行的，因此除非有特殊的说明，实验结果都是在该系统中搜集的。

#### 3.3.4.2 测试程序和运行方法

使用微基准测试程序和实际应用验证基于请求者数目的锁的有效性。Single counter[65], mmapbench[66-68], sockbench[66-68] 为微基准测试，而 parallel postmark[66-67], kernbench[69] 和 parallel find[70]是实际的应用程序。本文使用这些测试程序，因为它们最大的可扩展性瓶颈是锁竞争，并且这些测试程序的竞争强度是变化多样的。所有的测试程序均是同构的，意味着每个任务都执行着相似的操作。

所有的微基准测试程序都被实现成多进程程序并且利用相同的框架同步。Single counter 测试程序中，每个进程不断增加一个利用自旋锁保护的计数器；mmapbench 测试中，每个进程不断将相同文件的 500Mbytes 数据以 MAP_SHARED 方式映射进自己的地址空间，对映射进的每一页读取第一个字节，然后将该映射解除；sockbench 中，每个进程不断的创建一个套接字然后将其关闭。Parallel postmark 是多线程的测试程序，用来模拟提供邮件和网络信息服务的文件服务器。Parallel postmark 中的每个线程在一组独立的文件(大小在 0.5K 至 10K 之间)集合上执行事务(transaction)。每个事务由两步组成, (1)创建或者删除一个文件(2)读或者追加一个文件。需要操作的文件，文件的 I/O 操作(创建或者读数据)和文件的大小都是按照均匀分布来选择。为了测量 Linux 内核的可扩展性而不是 I/O，所有的文件均被存储在内存文件系统中。同时，文件操作关闭了 I/O 缓存的功能以避免 glibc

对可扩展性的影响。测试程序的性能以总体事务的吞吐量来衡量。Kernbench 通过启用多个并行执行的任务编译软件来测量某个系统的性能。在实验中，使用 kernbench0.42 编译 Linux 内核 2.6.29.4。对于每次测试，仅仅改变启动的任务数目，同时开启快速运行选项，关闭半负载运行和最优负载运行选项。和 Parallel postmark 类似, kernbench 使用的内核被置于内存文件系统中以消除 I/O 的影响。Parallel find 创建多个进程，测试程序中的每个进程不断的调用 GNU 的 find 命令在一个私有的 Linux 内核代码目录下查找一个不存在的文件名。为了避免 I/O 的影响，在实际测试之前对硬件缓存进行预热。

#### 3.3.4.3 热点内核锁

使用 Linux 中的/proc/lock_stat[23]工具来确定测试程序运行时哪些内核锁是被激烈竞争的。每个测试程序的热点锁在表 3.3 中被列出。表中省略了 single counter 的结果因为这个测试程序的锁是作者自己定义的。

表 3.3 每个测试程序的热点内核锁

| 测试程序 | 热点内核锁 |
|---|---|
| mmapbench | mmapings lock (i_mmap_lock) |
| sockbench | dentry cache lock (dcache_lock)<br>inode lock (inode_lock) |
| parallel postmark | file descriptor table lock<br>tmpfs statistics lock |
| parallel find | files lock (files_lock)<br>dentry cache lock (dcache_lock) |
| kernbench | files lock (files_lock)<br>inode lock (inode_lock)<br>zone lru lock (lru_lock) |

#### 3.3.4.4 测量方法

为了测量提出的基于请求者数目的锁的可扩展性，将/proc/lock_stat 报告出的每个热点锁的定义和全部调用点替换成基于请求者数目的锁。值得说明的是，尽管没有将 Linux 内核所有锁的定义和调用替换成基于请求者数目的锁，替换所有热点锁已经足够验证提出的想法，因为系统性能受热点锁竞争的影响是最大的。而且，实验结果表明替换所有的热点锁已经可以很大程度的提升多核的可扩展性，这一事实已经充分说明提出的锁在更大规模系统上的潜力。为了和提出的方法进行对比，相同的测量方式也被应用在其它的锁实现上。

#### 3.3.4.5 可扩展性提升

图 3.19 表示了当阈值设置为 0 时，基于等待者数目的锁达到的吞吐量随着核数的变化(曲线 T=0)。默认系统(使用排号自旋锁)可以达到的可扩展性同样在图中表示出来以便于比较(曲线 spin lock)。从图中可见，对于每个展示出锁颠簸现象的

应用而言(single counter, mmapbench, sockbench, parallel postmark 和 parallel find)，使用基于请求者数目的锁可以大幅度减少锁颠簸带来的影响。除了 single counter 以外，锁颠簸现象基本上被很好地避免。而且，对于 mmapbench, parallel postmark 和 parallel find，提出的锁可以在默认系统出现锁颠簸时维持系统的峰值。对于唯一一个没有受到锁颠簸影响的程序 kernbench 而言，提出的锁引入的开销基本可以忽略不计。下文将结合具体测试程序仔细分析基于等待者数目锁的可扩展性。

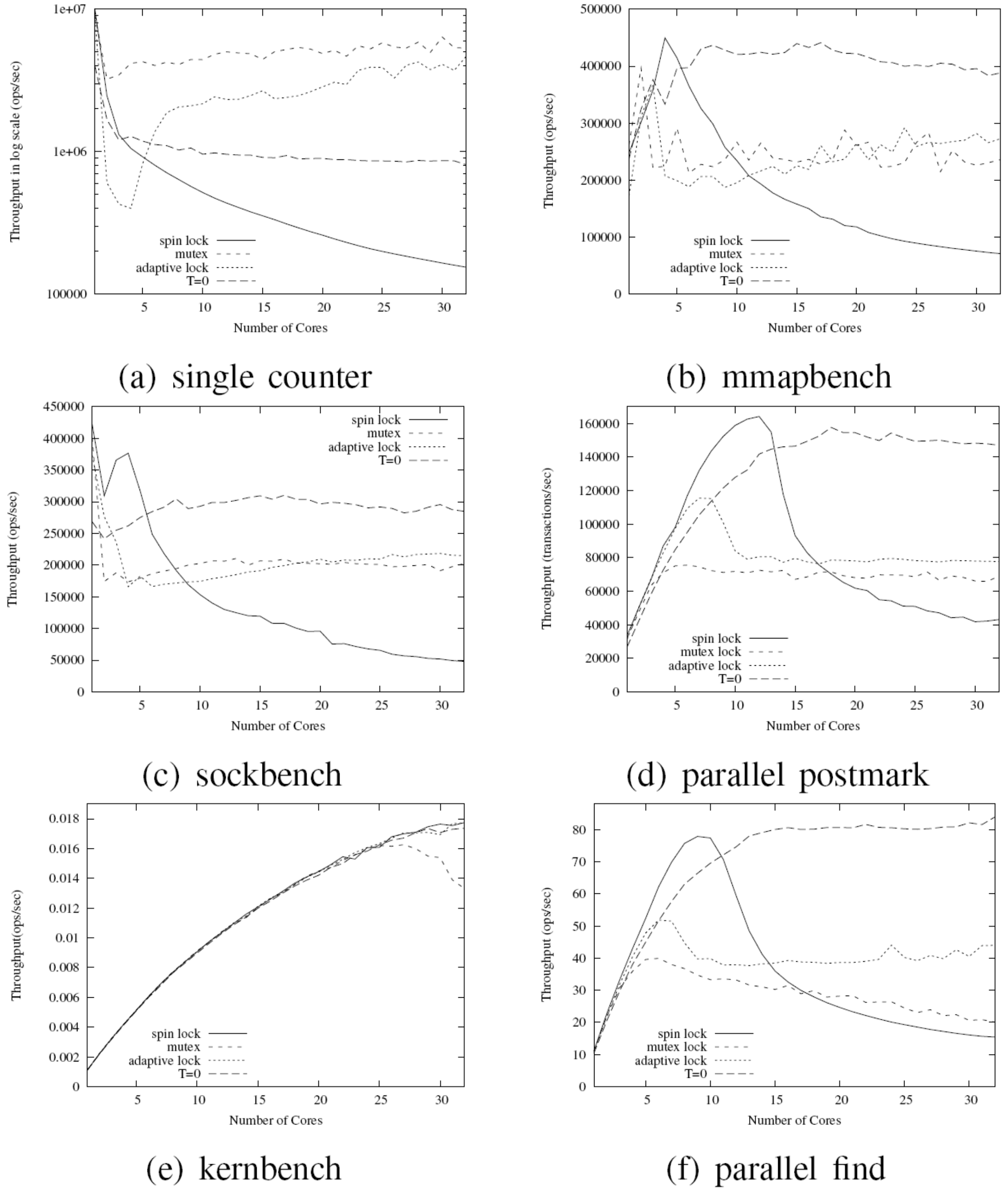

图 3.19 不同加锁方式的吞吐量随着核数的变化

对于 single counter，当使用的核数多于 3 时，利用基于等待者数目的锁使得吞

吐量变得稳定，因为所有竞争进程会进入节能状态等待锁的释放而非不断的自旋。然而，使用单核的达到的吞吐量任然高于使用 3 核的吞吐量，因为当使用多于 1 核时，将开始出现竞争。此时，进入节能状态和预测锁请求者数目将引入一定的开销。对于测试程序 mmapbench, sockbench, parallel postmark 和 parallel find, 锁颠簸现象可以被很好地避免。这是因为提出的基于等待者数目的锁利用了很多可扩展性优化，如轻量级的锁等待机制和每 CPU 的等待队列。值得指出的是，基于等待者数目的锁会比排号自旋锁的性能在某些核数的时候稍差(例如，parallel find 在核数小于 10 时)。这是因为提出的锁在加锁和解锁过程中引入的开销。具体而言，在加锁时，需要预测锁请求者的数目，如果数目大于一定阈值，需要获取等待队列的排号自旋锁。尽管这些操作是轻量级的，但是和 Linux 的排号锁相比，毕竟会引入一定的开销。在解锁代码中，需要查找第一个非空等待队列，当找到后，需要获取该队列的排号自旋锁，然后对队列进行操作。这样，队列的搜索过程需要访问其他核的变量，从而导致额外的延时。除此之外，队列的加锁开销同样会被累积到最终的性能中。

与其他应用程序不同，kernbench 没有受到锁颠簸现象的影响，因此可以被用来测试当竞争不够激烈时的可扩展性。图 3.19(e)表明基于等待者数目的锁和排号自旋锁的可扩展性几乎一样，从而表明提出的锁不会降低总体可扩展性。

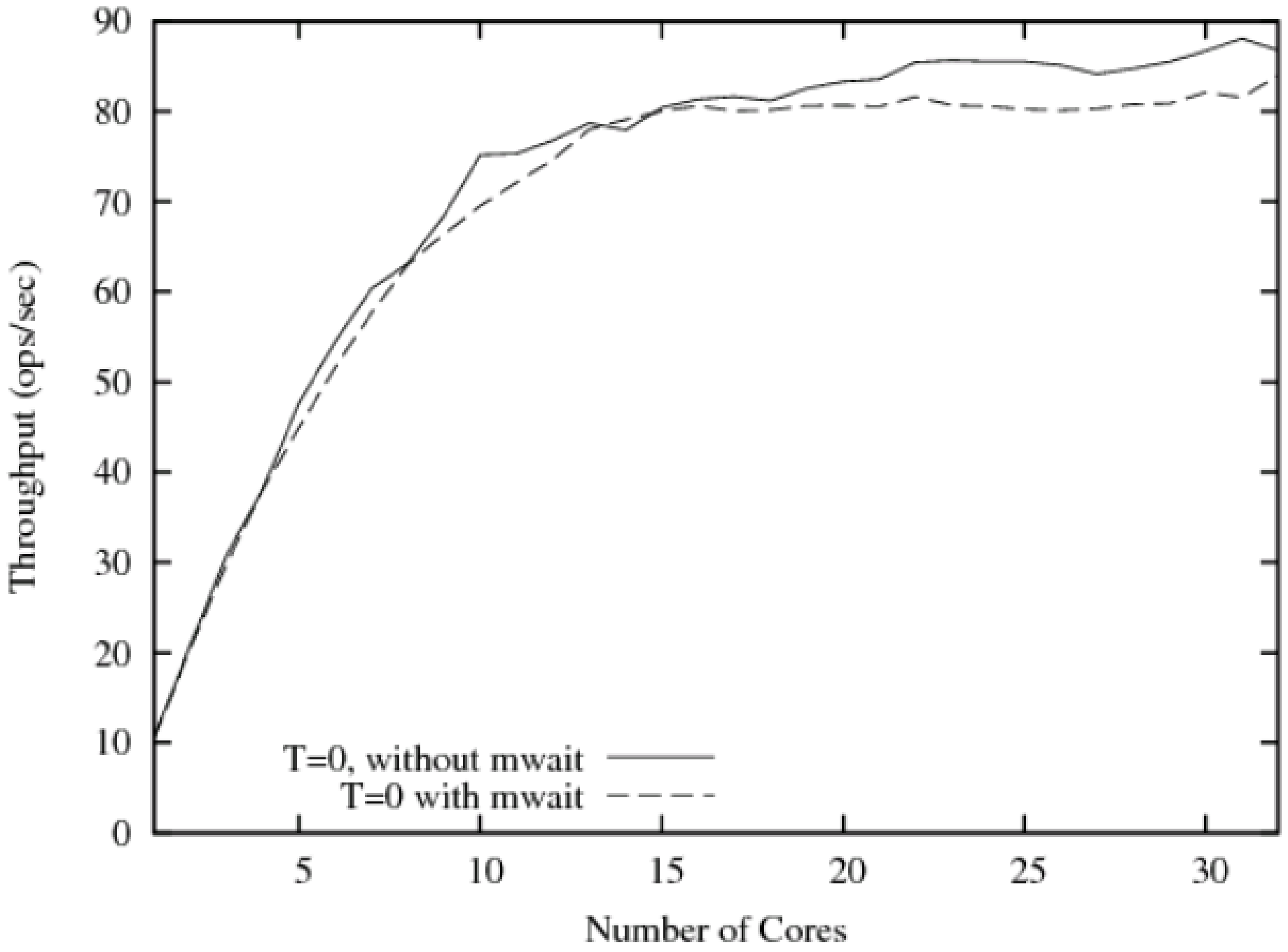

图 3.20 不带 mwait 和带有 mwait 的 parallel find 吞吐量随核数的变化

### 3.3.4.6 维护节能状态的开销

本节定量分析利用 monitor 和 mwait 指令可以引入的开销有多大。开销的测量

是通过删除进入节能状态的代码(图 3.16, 36 至 38 行)。通过这种方式，当锁被释放时，将等待任务从节能状态唤醒的延时可以被避免。改动后的实现相对于基于请求者数目的锁可以获得的性能提升被视作维护节能状态的开销。图 3.20 表明了具备节能功能和不具备节能功能的实现产生的吞吐量随着核数的变化。使用的测试程序是 parallel find。由图中可见，维护节能状态的开销不超过 8%。仅仅是 parallel find 的实验结果被表示因为维护节能状态的开销在这个例子中是最大的。对于 single counter, sockbench, mmapbench 和 kernbench，具备节能功能和不具备节能功能的曲线几乎相同，从而表明引入的开销可以忽略。对于 parallel postmark，开销小于 7%。

#### 3.3.4.7 锁请求者的数目

锁颠簸现象的发生和等待某一个锁的请求者数目具有很强的相关性。正是基于这种观测，本文设计并且实现了基于请求者数目的锁。提出的锁可以使得锁请求者进入节能状态直到锁被释放为止。提出的锁可以减少对相同锁的竞争并且节省电能。图 3.21 表示使用基于请求者数目的锁后，parallel postmark 中每个竞争锁中处于自旋状态的平均等待者数目。如图中所示，与默认的 Linux 系统相比(图 3.14)，每个锁上处于自旋状态的锁等待者已经大大减少。右侧的子图表示当使用 32 核时，每次实验中的每个锁的自旋等待者数目。结果表明，其数值依然很小。图 3.21 的实验结果符合作者的预期，因为大部分的锁等待者处于节能状态而非自旋等待。除了 kernbench，其它测试程序的实验结果和 parallel postmark 类似。对于 kernbench，由于锁竞争不够激烈，因此无论是否使用基于等待者数目的锁，锁等待者都为 0。

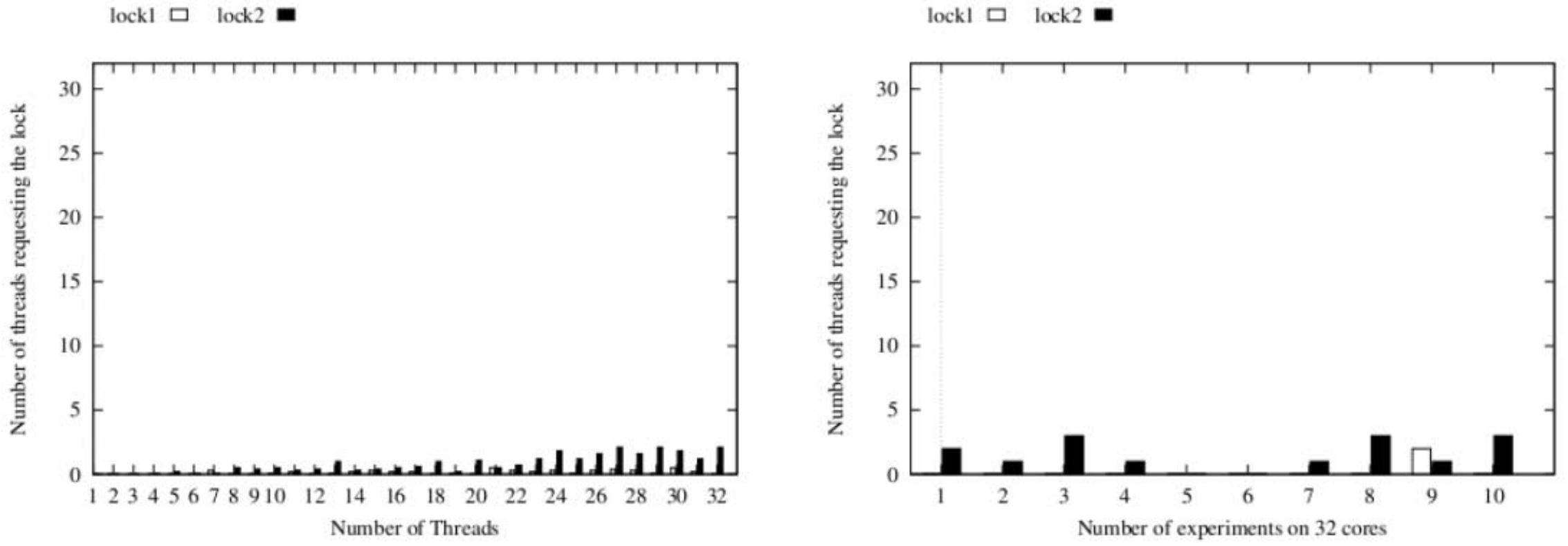

图 3. 21 parallel postmark 在 AMD32 核平台上每个竞争的自旋锁的等待者数目(左)和当使用 32 个核时每个锁的的等待者数目(右)。lock1 代表内存文件系统的统计数据自旋锁而 lock2 代表文件描述符表锁。

#### 3.3.4.8 硬件架构及内核版本的影响

对于基于请求者数目的锁的一方面考虑是这种方法是否依然适用于其它的系统架构和内核版本。在多核架构方面，使用本文提出方法的最大障碍是利用特殊指

令(monitor 和 mwait)进入节能状态。幸运的是，很多流行的架构已经在指令集中提供了节能指令。例如，X86 平台提供了 monitor 和 mwait 指令进入节能状态，而在 ARM 平台上引入的 wfi 和 wfe 指令[71]同样可以提供类似的功能。对于内核版本，相信操作系统会继续利用锁来维护一致性，因此，提出的技术可以用在将来的多核平台上以提升可扩展性和能耗有效性。

本节将展示基于等待者数目的锁在基于 Linux2.6.32 的 Intel40 核平台上的可扩展性提升以研究架构及内核版本带来的影响。图 3.22 表示了 3 个实际应用的实验结果。由于实际应用可以代表真实世界的行为，因此本文对这类程序更加感兴趣，所以在此省略微基准测试的结果。在 Intel40 核系统上的性能和 AMD32 核的结果基本类似:当采用基于请求者数目的锁后, parallel postmark 和 parallel find 中的锁颠簸现象可以被避免。对于 kernbench，基于请求者数目的锁和默认系统结果非常相近，从而说明该测试程序的锁竞争是轻量级的。

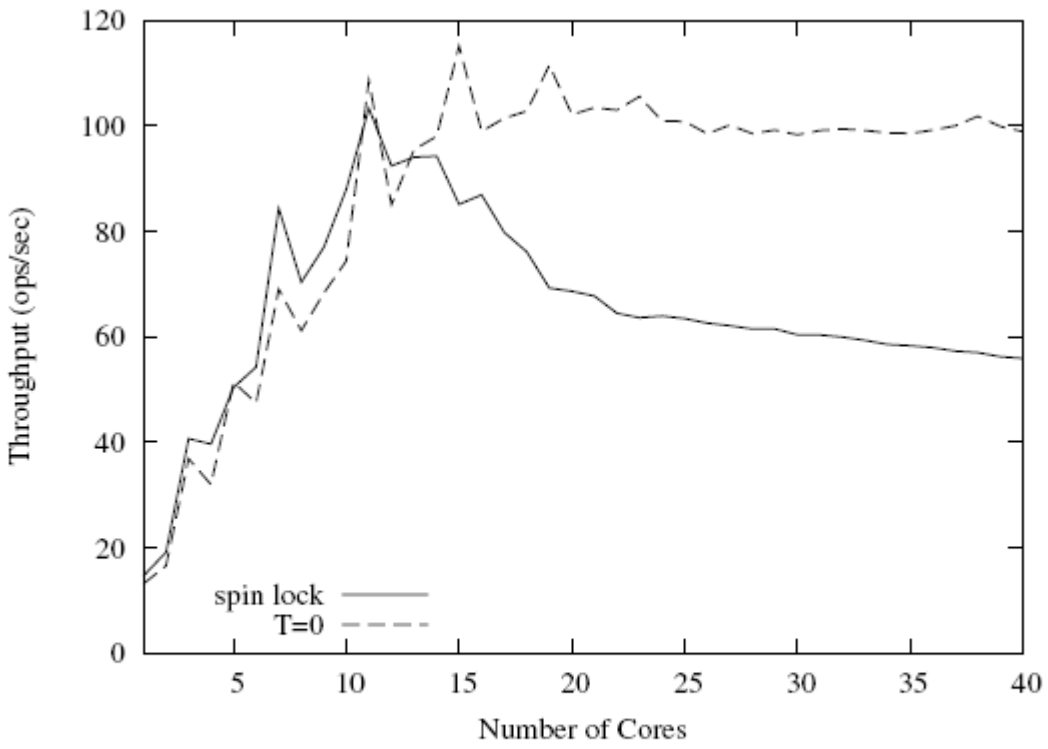

(a) parallel find

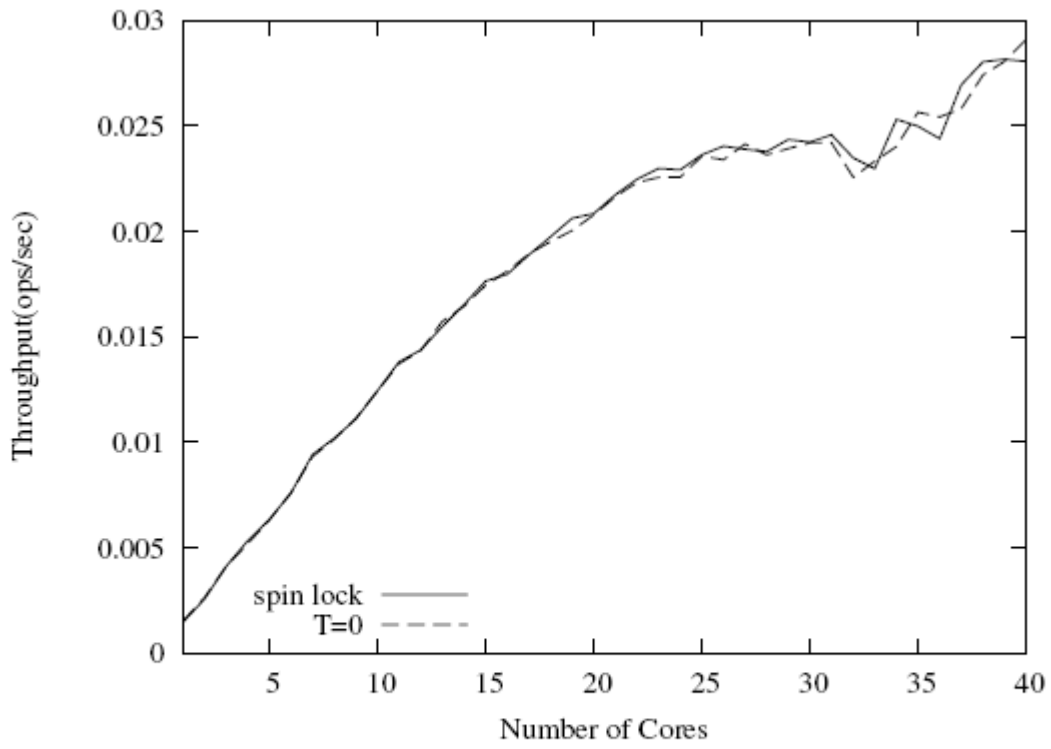

(b) kernbench

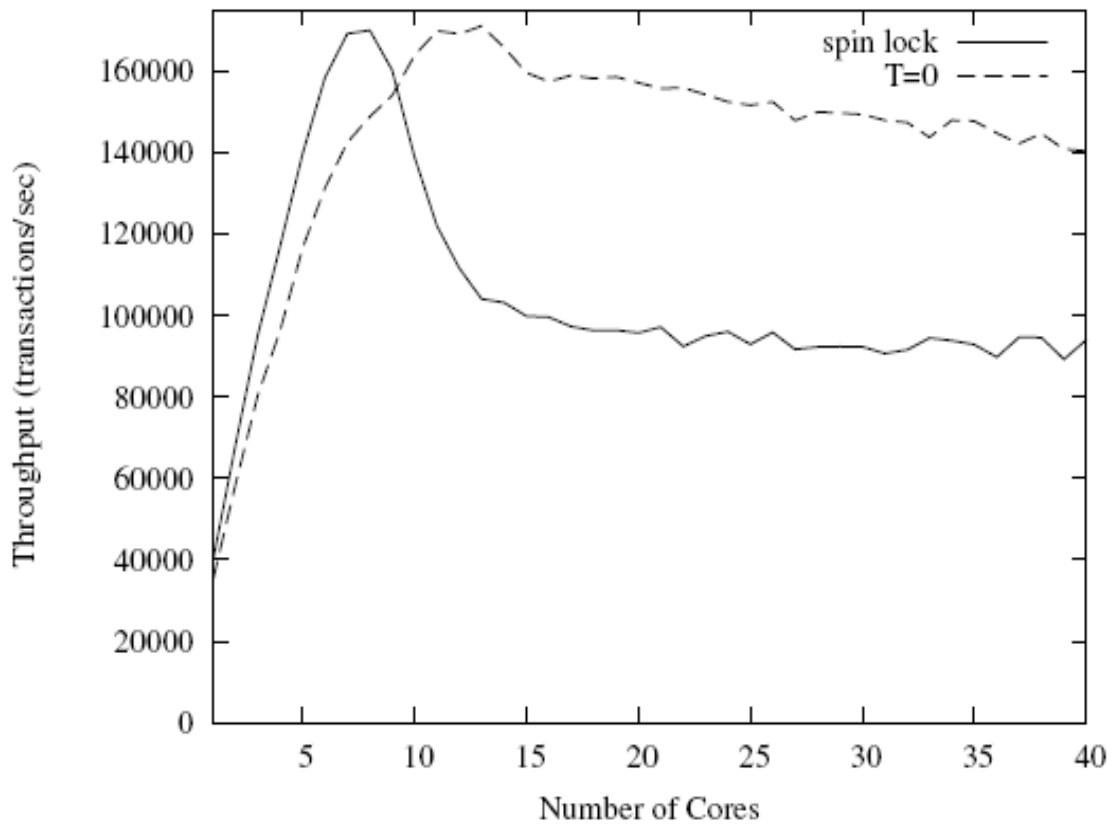

(c) parallel postmark

图 3.22 基于等待者数目的锁(T=0)和自旋锁(spin lock)在 Intel 40 核系统上的实验结果

#### 3.3.4.9 能耗有效性的提升

基于等待者数目的锁使得锁请求者进入节能状态，从而具有较好的能耗有效

性。在本节中，定量验证能耗有效性的提升。能耗有效性以每焦耳完成的工作来衡量，其公式是 efficiency = throughput×time/power×time = throughput/power。为了测量程序执行时的功耗，本文采用 380801 能耗分析仪[72]。测量的原理图如图 3.23 所示。在执行测试程序时，由分析仪报告的结果相对稳定。对每个测试程序，功耗的结果被读取 3 次，利用 3 次的平均值作为最终的功耗。本文测量的是系统的功耗而并非 CPU 的功耗，因为提出的锁同样减少了芯片互联和内存的操作，而这些都不能被 CPU 功耗所覆盖。

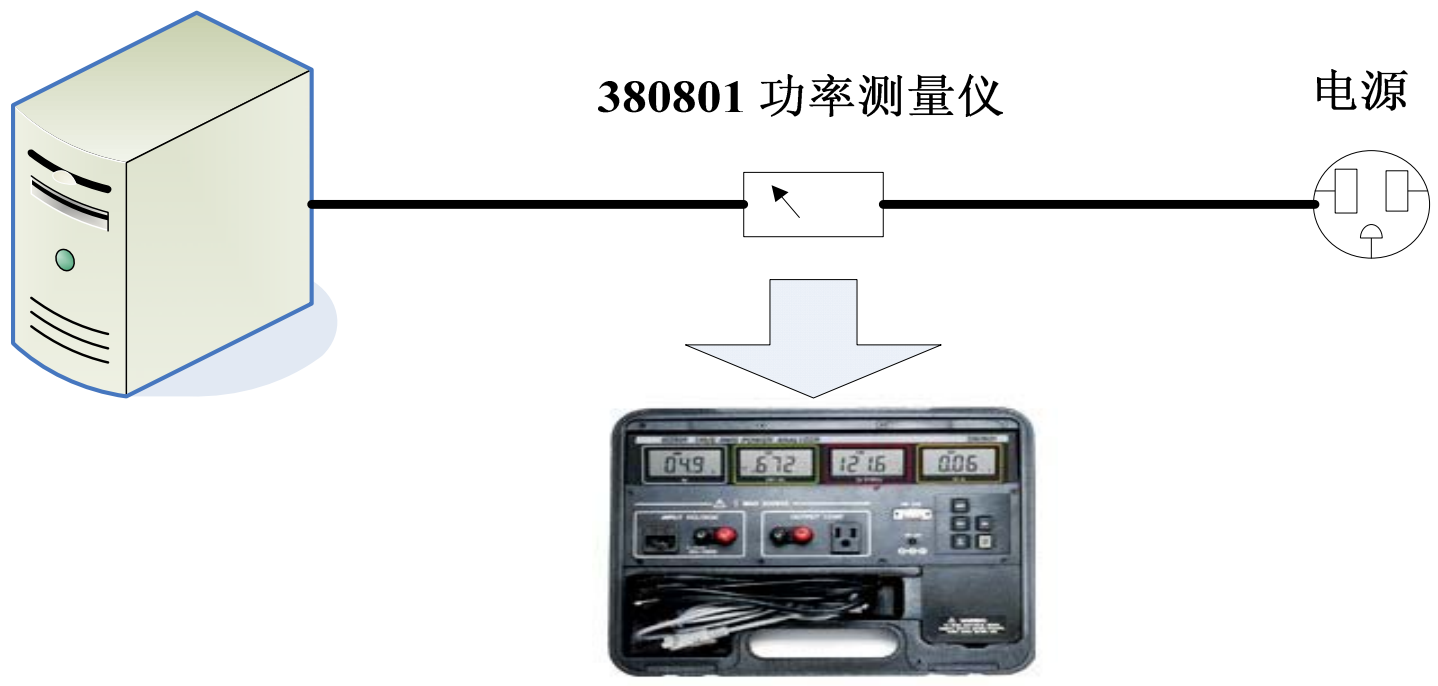

图 3. 23 功耗测量架构示意图

图 3.24 展示所有测试程序在能耗有效性方面的提升。和预期的结果相同，基于等待者数目的锁的能耗有效性在锁颠簸发生时开始超过默认系统。这是因为当锁颠簸现象发生后，某个锁的等待者数目开始大幅度增加，并且大部分请求者开始进入节能状态。如果锁颠簸现象没有发生，采用提出的方法也不会带来能耗有效性的提升。

### 3. 3. 4. 10　与其它加锁方式的对比

为了和基于等待者数目的锁进行对比，本文利用互斥锁(mutex)和自适应锁(adaptive lock)[60]来避免锁颠簸现象。互斥锁使得每个锁等待者在锁不能被立即获取时进入睡眠状态。对于自适应锁，锁等待者根据锁持有者的状态自适应地自旋或者睡眠。

1、可扩展性比较 图 3.19 表明了不同加锁方式产生的吞吐量随着核数的变化。除了 single counter，基于等待者数目的锁比排号自旋锁，互斥锁和自适应锁展示出了更好的可扩展性。下面将针对每个例子具体分析。对于 single counter，实验结果表明基于等待者数目的锁表现最差，而互斥锁要好于自适应锁。互斥锁好于自适应锁，因为在睡眠之前的自旋对内存总线造成了较大压力。基于等待者数目的锁差于基于睡眠的同步原语(互斥锁和自适应锁)的原因有 2 个。1) 基于等待者数目的锁在该测试用例中展示出的开销较大。开销大小可以通过单核的吞吐量被验证出来。

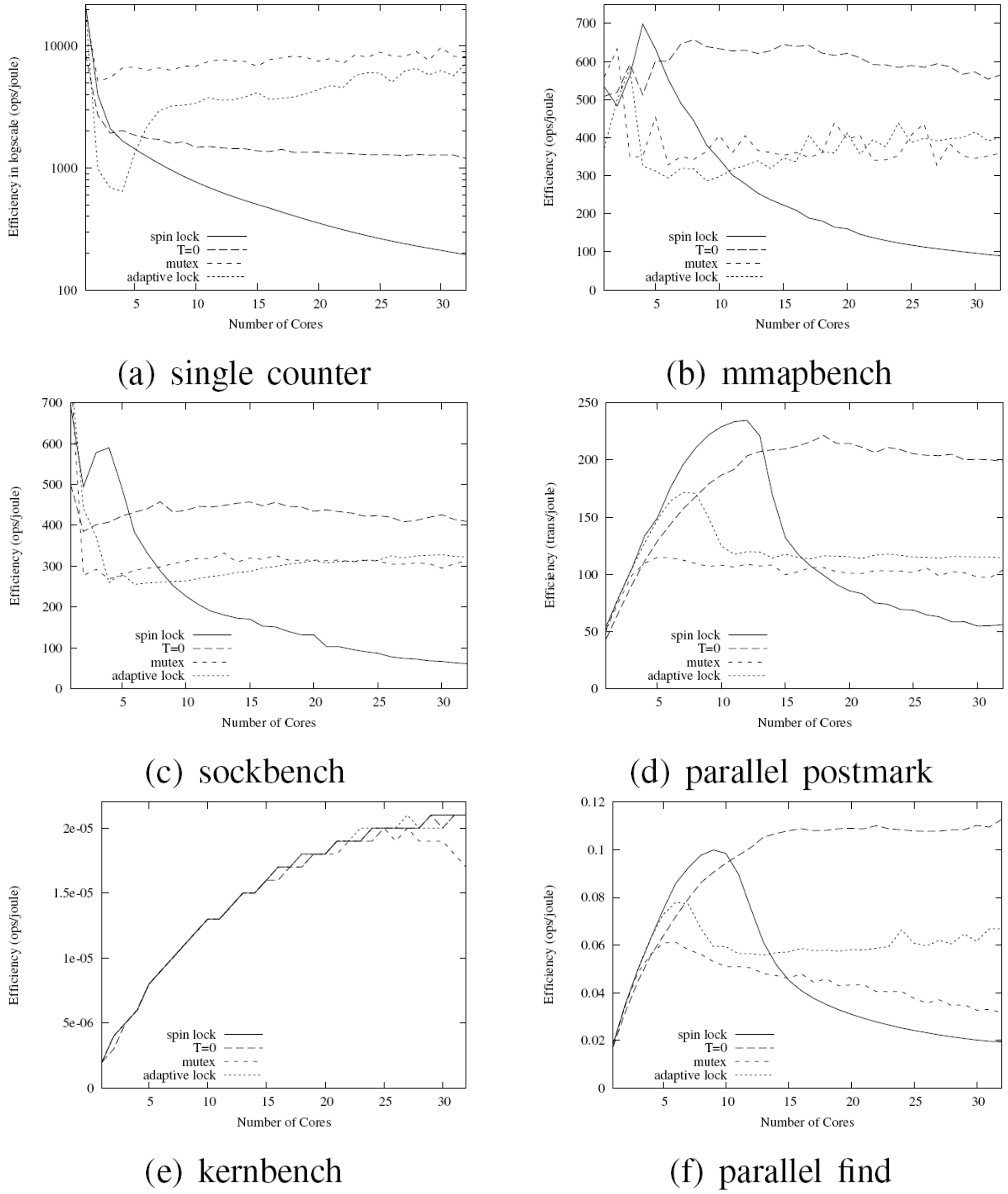

图 3.24 不同加锁方式的能耗有效性随核数的变化。single counter 的有效性是对数坐标。

如图 3.19 所示，基于请求者数目锁的吞吐量是每秒 $4.00\times10^6$ 次操作，而对于互斥锁和自适应锁却是每秒 $9.09\times10^6$ 次操作。2)当一个任务由节能状态被唤醒时，缓存一致性协议将引入额外的开销。这种开销可以从图 3.16 中被分析出来。处于节能状态的核要检测 monitor_flag 的值，如果该值被其它的进程设置成 1，处于节能状态的核将被唤醒。然而，monitor_flag 是由 2 个进程共享的变量。该变量的副本由缓存一致性协议自动更新。在实验平台上，每次更新的开销是 380 个时钟周期。这种开销的大小任然远远小于 1 次上下文切换的开销(11624 时钟周期)。然而，single counter 中的同步特征导致了基于等待者数目的锁表现差。在 single counter 实现中仅有一个临界区，刚刚释放锁的任务在通知一个睡眠任务去获取锁之后马上又变成了锁等待者。在 Linux 的互斥锁和自适应锁中，一个锁请求者要在睡眠之前尝试获取锁。实际上，锁经常是在这个过程中被获取到的，而不是等着正在被唤醒的任务来获取锁。通过这种方式，互斥锁中上下文切换的开销不会被计算在整体的时间内。然而，对于基于等待者数目的锁，缓存一致性的开销会被加到整体的时间内。基于等待者数目的锁只有遇到类似 single couner 一样特殊的程序时才会展现出较大的开销。对于 mmapbench, sockbench, parallel postmark 和 parallel find 测试程序，基于等待者数目的锁具有最好的可扩展性，并且其性能分别在使用多于 4 核, 4 核，12 核和 10 核时开始高于其它的实现。这是因为基于等待者数目的锁采用了很多可扩展性优化如可扩展的请求者数目的预测方式，轻量级的锁等待机制，可扩展的等待链表检测和每 CPU 数据结构组织。在 kernbench 中，默认 Linux 的锁，自适应锁和基于等待者数目的锁产生的吞吐量几乎相同。而一个有趣的现象是尽管默认的系统没有出现锁颠簸现象，但是当使用互斥锁时锁颠簸现象却出现了。这是因为互斥锁引入的上下文切换开销过大。

2、能耗有效性比较 图 3.24 比较了不同加锁方式的能耗有效性。实验的结果和图 3.19 类似，表明不同方法的功耗不会差的太大。总体而言，基于等待者数目的锁展示出了合理的能耗有效性，并且当使用多于 4 核, 4 核, 12 核和 10 核时，提升了 mmapbench, sockbench, parallel postmark 和 parallel find 的能耗有效性。对于 kernbench，基于等待者数目的锁的能耗有效性和默认系统及自适应锁类似，所有的这些实现均比互斥锁在多于 25 核时展示出了更好的有效性。

## 3.4 锁竞争感知的调度策略

3.3 节提出了基于等待者数目的锁以避免锁颠簸现象。从实验的结果来看，该方法不但可以避免锁颠簸现象，而且可以利用特殊的指令节能。然而，在提升应用程序的可扩展性上，该方法却存在着可以改进的空间。例如，尽管基于等待者数目的锁可以避免锁颠簸现象的发生，但是，当竞争过于激烈时，该方法不会维持默认系统的峰值性能。该问题可以通过测试的结果体现出来(如图 3.25 所示)。从图 3.25 可见，尽管基于等待者数目锁在 sockbench 上表现优于互斥锁和自适应互斥锁,并且可以避免锁颠簸现象，但是该方法会降低默认系统的峰值性能。该问题产生的原因是由于本文的改进是侧重在锁上的。为了避免锁颠簸现象，需要将锁的实现变得更加复杂。然而，这一方法的副作用是当竞争较大时，引入的开销也比较大。本节从另外一个角度(任务调度)入手以避免锁颠簸现象。该方法被提出的动机是锁颠簸

现象出现时，每个用锁密集任务的等锁时间百分比会大幅度的增加。基于这种相关性，本文提出了锁竞争感知的调度算法。具体而言，每个任务持续的检测用锁时间百分比。如果该百分比超过一个预定义的阈值，那么，该任务被分类成用锁密集类任务，并且被迁移到一组特殊的核上(称为 Special Set of Cores，简称 SSC)。通过这种方式，并行运行的锁密集类任务的个数是受限的，因此，锁竞争的程度也被控制住了。使用该方法最为核心的挑战是如何决定 SSC 中核的数目。在提出的调度算法中，最优的核数是通过模型驱动的搜索确定的。

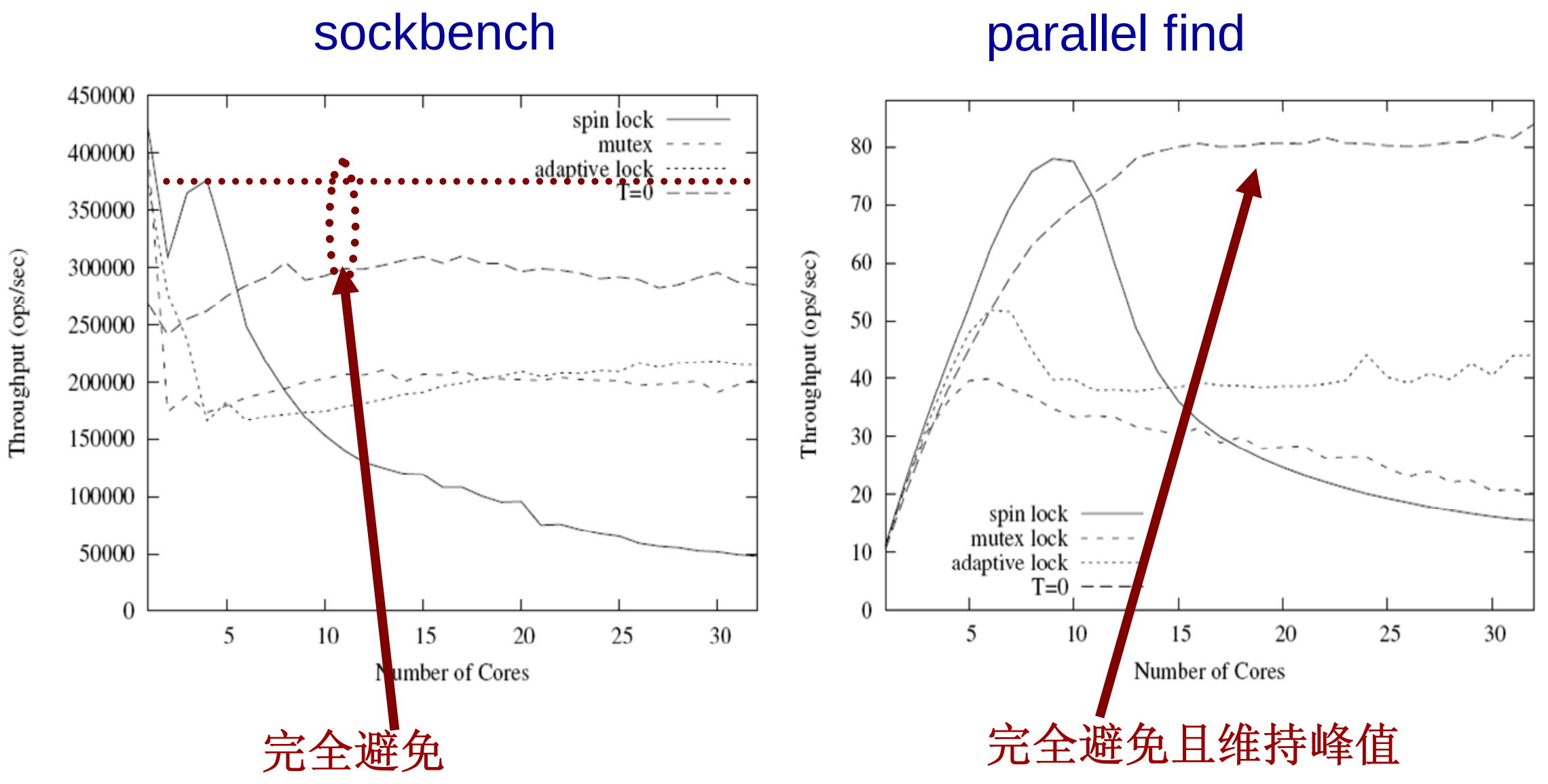

图 3. 25 基于等待者数目锁，互斥锁，自适应互斥锁及默认系统在 sockbench 和 parallel find 上的可扩展性

提出的调度算法在 Linux 内核中进行实现，并且在 Intel 32 核 NUMA 系统和 AMD32 核 NUMA 系统上利用微基准测试程序和实际应用进行了细致的评测。实验结果表明提出的方法不但可以很好地避免锁颠簸现象，而且对于大部分应用可以维持默认系统的峰值性能。而且，任务的等锁时间百分比可以降低达 84%。与基于等待者数目锁，互斥锁和自适应互斥锁相比，该方法能够表现出更好的可扩展性，功耗消耗和能耗有效性。

### 3. 4. 1 锁颠簸现象分析

本节将展示锁颠簸现象的产生和任务等锁时间百分比之间的强相关性。图 3.26 显示了在测试程序 parallel postmark 中每个任务等待 2 个激烈竞争的排号自旋锁(内存文件系统的统计数据锁和文件描述符表锁)的时间总和占总体执行时间的百分比。如图 3.26 所示，当锁颠簸现象发生之前，等锁时间百分比只是缓慢的从 3.81% 增长至 21.77%。然而，当核的数目超过 16 时，锁颠簸现象开始出现，同时等锁时间百分比迅速地增长至 92.24%。从实验的结果可以得到的结论是当锁颠簸现象出

现时，每个任务的等锁时间百分比会大幅度增加以至于用来处理有用工作的时间却所剩无几。

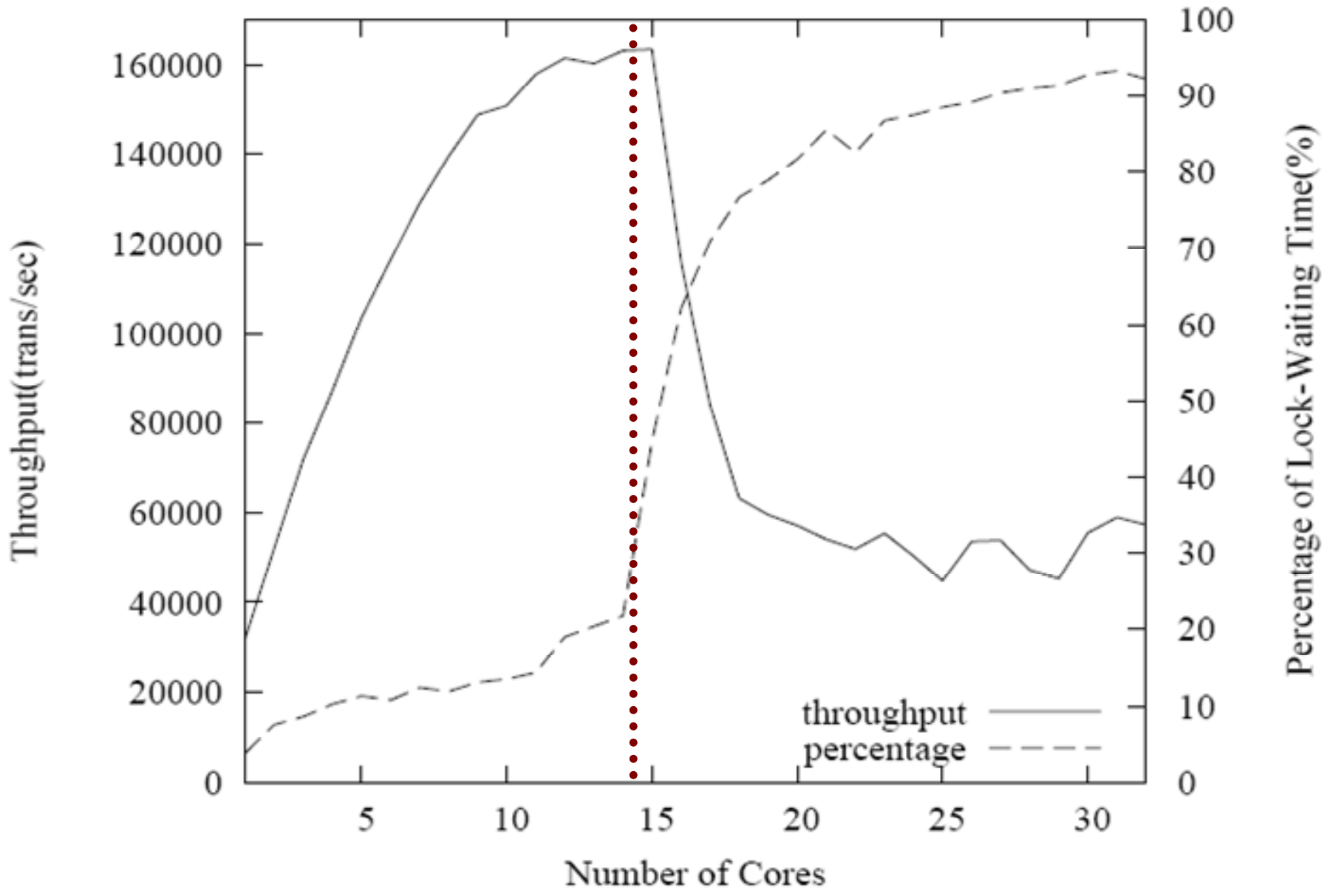

图 3.26 parallel postmark 的吞吐量和等锁时间百分比随着核数的变化

## 3.4.2 锁竞争感知的调度策略

### 3.4.2.1 方法概要

3.4.1 节发现的任务等锁时间百分比和锁颠簸现象的强相关性是设计锁竞争感知调度策略的根本动机。总体而言，该调度策略分为 4 个部分。

1、锁颠簸现象检测 调度器可以管理的任务在创建之后持续地检测自身的等锁时间百分比。如果该百分比大于一个预定义的阈值，该任务被认作是用锁密集类任务，并且需要运行在一组特殊的核上(简称 SSC)。

2、 SSC 中的处理器分配 所有用锁密集类任务都运行在 SSC 上。该部分决定 SSC 中要被分配多少核以达到用锁密集类任务的最大吞吐量。

3、分离式负载均衡 为了克服默认调度器的全局负载均衡，提出的调度策略对用锁密集类任务和普通任务(非用锁密集)进行分离式的负载均衡。

4、锁行为变化的自适应 锁行为的变化是通过对调度器中关键参数(例如，在 SSC 中的核数等)的变化和定时器来检测的。如果程序的用锁行为发生了变化，SSC 中的核数会被重新确定。

### 3.4.2.2 锁颠簸现象检测

当一个任务花费了大量的时间用于等锁，锁竞争感知调度器将该任务划分成用锁密集类任务并且将其迁移至 SSC 上。为了确定一个任务是否是用锁密集类任

务，在每个调度时间片内(即相邻的两次上下文切换的时间间隔)，为该任务计算其等锁时间百分比。该百分比可以形式化为$\sum_j W_{i,j}/D_i$，其中，$W_{i,j}$代表在时间片 i 内对第 j 把锁的等待时间，而$D_i$代表时间片 i 的长度。为了计算公式中的$\sum_j W_{i,j}$，利用函数 sched_lock()代替内核中获取排号自旋锁的函数 spin_lock()。sched_lock()的伪代码如图 3.27 所示。可以看到, sched_lock()是 spin_lock()的一个重新包装。除此之外，sched_lock()添加了任务迁移逻辑(migrate_to_special _cores())和利用时间戳计数器(TSC)对锁获取时间的细粒度测量。对于$D_i$，通过在每个时间片前后读取 TSC 计数器进行测量。尽管测量锁获取时间和时间片长度时需要对 TSC 计数器进行频繁的读取，但是这项功能是通过指令 rdtsc 完成的，因此引入的开销非常小。在评价一节，可以发现其开销几乎可以忽略不计。

```
/*current task is migrated based on the accumulated lock-waiting time*/
sched_lock(spinlock_t *lock)
{
        migrate_to_special_cores(lock);
        curr->start = read_tsc();
        spin_lock(lock);
        curr->acc_lock_time+=read_tsc()-curr->start;
}
```

图 3. 27 sched_lock()伪代码

在每个时间片的末尾，调度器将会在 Linux 内核中用于负责任务调度的函数 schedule()中检查当前的任务是否已经被分类成用锁密集类任务(图 3.29)。如果没有，则调度器计算该任务的等锁时间百分比。如果计算出的百分比大于一个预定义的阈值 T，则代表每个任务的 task_struct 数据结构中的域 is_mig 会被设置成 QUALIFY_TO_MIGRATE 以表明该任务已经被认作用锁密集类任务。为了确定预定义的阈值 T，使用经验值 10%。实验和评价一节，本文验证了该经验值的有效性。

```
void migrate_to_special_cores()
{
      local_core_bound = special_core_bound;
      if (curr->is_mig==QUALIFY_TO_MIGRATE) or (curr->is_mig==HAVE_MIGRATED
and curr->special_core_bound!=local_core_bound)
      {
            if (curr->is_mig==QUALIFY_TO_MIGRATE)
            {
                  inc_lock_intensive_tasks();
                  curr->is_mig = HAVE_MIGRATED;
            }
            /*get a core whose id is chosen from 0 to local_core_bound*/
            target_core = get_target_core(local_core_bound);
            /*migrate curr task to target_core*/
            if (this_core!=target_core)
                sched_migrate_task(curr, target_core);
            curr->special_core_bound = local_core_bound;
      }
}
```

图 3. 28 migrate_to_special_cores()伪代码

锁密集类任务迁移至 SSC 的时机是当一个任务打算获取一把排号自旋锁时。图 3.28 表示了在 sched_lock()中调用的函数 migrate_to_special_cores()的伪代码。可

以看到，如果当前任务的is_mig域被设置成为QUALIFY_TO_MIGRATE，那么，它的值将会被更新成 HAVE_MIGRATED，然后任务通过调用函数sched_migrate_task()被迁移。迁移的目标使得 SSC 中处理器的负载尽量均衡。当前的任务在 is_mig 为 HAVE_MIGRATED 时同样可以被迁移。这种迁移是必要的，因为 SSC 中的最优处理器数目是通过模型驱动搜索被确定出来的。下一节将阐述搜索的过程。

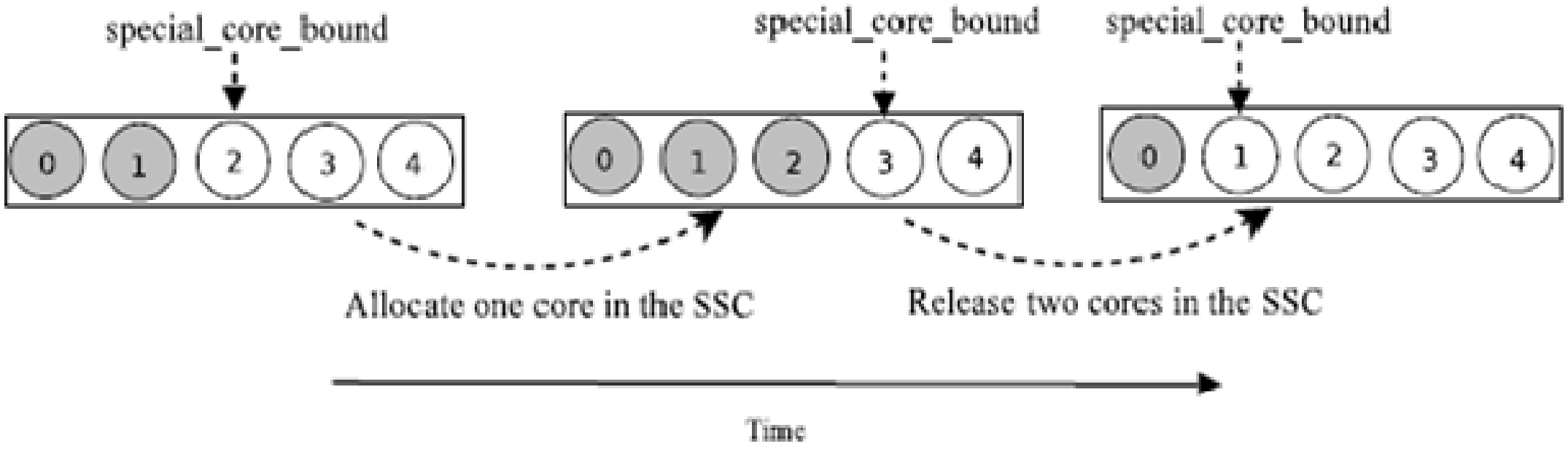

图 3.29 SSC 中的计算单元分配和释放

#### 3.4.2.3 SSC中的处理器分配

所有被分类成用锁密集类的任务都只能在 SSC 上执行。在本文提出的锁竞争感知的调度算法中，具有连续 ID 的计算单元会被分配。图 3.29 展示了 SSC 中的核是如何从初始状态(双核)被分配和释放的。一个全局变量 special_core_bound 用来区分 SSC 中的核和剩余的核。提出的核分配方法充分的考虑到了本文使用的硬件平台的特征(具有相邻 ID 的核共享最后级缓存)。例如，在实验的 AMD32 核平台上，ID 在[4×i , 4×(i+1)), 0≤i≤7 区间的核共享最后级缓存。因此，这种计算单元分配方法可以确保在 NUMA 系统上使用最少的节点，进而减少了节点之间的缓存行传递。除此之外，这种方法同时保证了最后级缓存的共享程度是最大的，这也使得 SSC 内部的核通讯变得更加迅速。然而,有些多核系统中 ID 相邻，但是核却位于不同的芯片上。对于这样的系统，本文提出方法的优势可以在软件层上增加 ID 映射关系来获得。

在锁竞争感知的调度策略中，最为关键的问题是在 SSC 中需要分配多少计算单元用于处理用锁密集类任务。如果核数分配的过少，锁竞争的可能会变得很小同时共享数据也不会频繁的传递。然而，应用程序的并行程度却受到了限制。相反地，如果核数分配的过多，任务可以并行的执行，但是锁颠簸现象可能同样存在。事实上, SSC 中的最优核数(即 $n^*$)可以通过求解如下的最优问题得到:

$$n^* = \arg\max_{1 \le n \le minimize\{M,N\}} T(n) \qquad 3\text{-}1$$

其中 T(n)代表当 SSC 中被分配 n 个核时，所有用锁密集类任务可以达到的归

一化吞吐量, M 代表用锁密集类任务的数目,而 N 代表系统中所有的核数。正如公式(1)所表示的, 优化的目标是最大化所有用锁密集类任务的吞吐量。

假定 $p_i(n)$代表 SSC 中被分配 n 个核时,第 i 个核的等锁时间百分比, 而 $\overline{p(n)}$ 代表 SSC 中所有核的平均等锁时间百分比。那么, T(n)可以由如下的等式表示:

$$T(n) = \sum_{i=1}^{n}(1 - p_i(n)) = n\times(1-\sum_{i=1}^{n} p_i(n)/n) = n\times(1-\overline{p(n)}) \quad 3\text{-}2$$

其中 $1-p_i(n)$代表核 i 上的有用工作百分比。在等式(2)中, 只有锁竞争瓶颈被考虑进去, 因为本文的目标应用是用锁密集类程序。尽管该模型比较简单, 却可以有效的用于找到 SSC 中的最佳核数。

求解最优化问题(1)的一种可能方式是通过穷举找到最优的 n。初始时 n 需要被设置成 1, 每次计算 T(n)完毕后 n 增加 1。一旦 n 被更新, $\overline{p(n)}$ 和 T(n)需要被重新计算。这个过程一直持续，直到最大的 T(n)被找到。这种穷举法的时间复杂度是 O(N)，其中 N 代表系统中的核数。但是, 根据摩尔定律, 未来计算机系统的核数将呈指数规模增长, 因此, 这种复杂度在实际系统中显得过高。

为了减少这种开销，本文提出了一种启发式规则以加速搜索的过程。如果 T(n)大于上次的计算值, 那么, n 值被更新为原来的两倍, 并且搜索过程继续。否则, 搜索过程停止, n 值被重置为上一轮的值。加入了启发式规则的搜索复杂度是 O(log(N))。尽管减少了时间复杂度, 但是却会对最优值的精度有所影响。这种影响需要通过实际应用的评测来确定, 在实验和评价一节将会看到, 这种影响对于大部分应用程序都是可以接受的。

为了计算 T(n), $\overline{p(n)}$ 是一个关键的参数。锁竞争感知的调度器使用投票策略来计算 $\overline{p(n)}$。具体而言, 投票子系统的状态由 voting_locking 和 voting_slice2 个全局变量表示。第一个变量记录总共的等锁时间而第二个变量累计所有投票任务的时间片总和。使用这两个变量, $\overline{p(n)}$ 通过 voting_locking/voting_slice 来计算。

SSC 中计算单元分配和释放的伪代码如图 3.30 所示。在每个时间片结束的时候, 每个用锁密集的任务有机会开始投票(图 3.30 12 至 23 行)。每个打算投票的任务尝试去获取保护投票变量的锁。如果锁可以被成功的获取, voting_locking 和 voting_slice 分别累加等待锁的总时间(即 acc_lock_time 的值)和当前任务的时间片长度(即 slice_len 的值)。保护投票变量的锁不会变成可扩展性瓶颈, 因为这个锁是通过 spin_trylock()来获取的, 其语义保证了当一把锁不能被立即获取时, 锁不会被等待(事实上, 该函数立即返回)。为了通过实验来验证这一结论, 本文使用称为 single counter 的微基准测试程序。在该测试程序中, 每个进程不断的增加由全局排号自旋锁保护的计数器。使用两种方法来获取这把锁。一个是 spin_trylock()而另外

一个是 spin_lock()。对这两种获取锁的方法进行测量发现，使用 spin_trylock()获取锁的开销仅是使用 spin_lock()开销的 1/15。

```
 /*runs at the end of each time slice*/
curr->slice_end = read_tsc();
curr->slice_len = curr->slice_end － curr->slice_begin;
if (curr->is_mig==0)
{
     if (curr->acc_lock_time>T.curr->slice_len)
     {
          curr->is_mig = QUALIFY_TO_MIGRATE;
          curr->special_core_bound=special_core_bound;
      }
 } else {
       if (curr->is_mig==QUALIFY_TO_MIGRATE or curr->is_mig==HAVE_MIGRATED)
       {
            /*the second condition indicates special_core_bound has been modified by another task*/
            can_vote=(this_cpu<special_core_bound and curr->record_bound=special_core_bound);
            if (can_vote)
            {
                if (spin_trylock(voting_lock))
                {
                     if (curr->record_bound==special_core_bound)
                         VOTE(1, n_lock_intensive_tasks, curr->acc_lock_time, curr->slice_len);
                     spin_unlock(voting_lock);
                }
                is_sufficient=voting_slice>special_core_bound.256000;
                if (is_sufficient and spin_trylock(voting_lock))
                {
                     determine_new_bound();
                     spin_unlock(voting_lock);
                }
            }
       }
 }
 /*time out logic*/
 if (timer_fires() and spin_trylock(voting_lock))
 {
       if (timer_fires())
       {
            special_core_bound = 1;
            CLEAR_VOTING();
             climbing=1;
             remig_next=jiffies + remig_timeout_interval;
       }
       spin_unlock(voting_lock);
  }
  curr->acc_lock_time = 0;

   /*runs at the begin of each time slice*/
   curr->slice_begin = read_tsc();
   curr->record_bound = special_core_bound;
```

图 3. 30 schedule()函数中在上下文切换之前和之后增加的代码

```
void determine_new_bound()
{
      moving = 0;
      useful_perc = 1-voting_locking/voting_slice;
      work_cores = n_lock_intensive_tasks>special_core_bound?special_core_bound:n_lock_intensive_tasks;
      thro = useful_perc.work_cores; //calculate the normalized throughput
      //climbing is initialized to be 1;
      if (climbing)
      {
          if (thro > thro_last_step)
          {
              if (climbing < 0)
                 climbing = 1;
              else {
                 if (climbing==2) {
                     new_bound = 2.work_cores;
                     if (new_bound > max_special_cores)
                        new_bound = max_special_cores;
                     if (new_bound > n_lock_intensive_tasks)
                        new_ bound = n_lock_intensive_tasks;
                     if (new_bound > work_cores) {
                        special_core_bound = new_bound; //bound update;
                        RECORD_LAST_VOTING(n_lock_intensive_tasks, work_cores, thro);
                        CLEAR_VOTING();
                        moving=1;
                      }
                      climbing = 1;
                 } else
                      climbing++;
              }
          } else {
              if (climbing > 0)
                 climbing = -1;
              else {
                  if (climbing==-2) {
                      special_core_bound = work_cores_last_step;
                      CLEAR_VOTING();
                      moving = 1;
                      climbing = 0;
                  } else
                      climbing--;
              }
          }
      } else
           if (work_cores!=work_cores_last or throughput>a.throughput_last_step or throughput <b.throughput_last_step)
              remig_next = jiffies;
      if (moving==0)
          SHRINK_VOTING(1,2);
      else
          remig_next = jiffies + remig_timeout_interval;
```

图 3. 31 determine_new_bound()伪代码

在投票之后，当前的任务通过计算每个在 SSC 中核的总共投票时间片长度来判断是否足够多的核数已经投票完毕。如果该值充分大，那么当前的任务再次尝试去获取保护投票变量的排号自旋锁。如果该锁可以被成功的获取，那么，当前的任

务要负责驱动 SSC 中最优核数的搜索。本文的方法不会使得多个任务同时去做不同的决策，因为在任意时刻只有一个核可以获取锁。而且，当一个核驱动完搜索的过程，投票变量将要被清除或者缩减，这种操作也避免了搜索过于频繁的进行。

所有的搜索逻辑均在函数 determine_new_bound()中被处理。该函数在图 3.30 的伪代码中被调用，而其自身的实现逻辑如图 3.31 所示。从图 3.31 中可以看到，需要首先计算归一化的吞吐量(图 3.31 中 行 4-6)。然后，将计算出来的归一化吞吐量和上一步的计算值相比较来决定是否将当前分配在 SSC 中的核数加倍或者恢复上一步的值。然而，测量误差会使得吞吐量的值不够准确。为了克服这个问题，仅当当前计算的吞吐量比上一步的计算值连续大或者小两次时，才决定改变 SSC 中核的数目。这种逻辑是通过一个有限状态自动机进行描述的 (图 3.31 中行 7 至 44)。一旦 SSC 中核的数目被成功更新，所有的投票变量被清零以标志着新一轮的投票开始，否则，所有的投票变量都要除以 2 以反映时间对投票系统的影响(行 47 至 49)。如果调度器决定将 SSC 中核的数目减半，搜索过程将要停止直到程序的用锁行为发生变化。

一个自然的问题是用锁密集类任务如何感知到 SSC 中更新后的核数。在锁竞争感知调度器中，即使一个用锁密集类任务已经被迁移过一次它仍然可以继续迁移 (图 3.28)。而这种迁移发生在当前 SSC 中的核数与任务记录的值不同时。通过使用这种机制，每个用锁密集类任务在获取排号自旋锁时得知 SSC 中最新的核数，这样，所有用锁密集类任务将会执行在核 ID 处于范围 0 至 special_core_ bound-1 的核上。

#### 3.4.2.4 分离式负载均衡

锁竞争感知调度策略将所有的用锁密集类任务管理在 SSC 上。然而，负载可能并非全局均衡。这样，本文提出的调度策略在 SSC 的内部和外部进行分离式的负载均衡以避免默认系统的全局均衡带来的影响。具体而言，在分离式负载均衡中，执行负载均衡函数(find_busiest_group(),find_busiest_queue()和 run_rebalance_ domains())的任务首先检查 special_core_bound 变量。如果当前任务运行的 CPU ID 小于该值，所有的均衡在核 ID 0 至 special_core_bound-1 内进行，否则，负载均衡在剩余的核上进行。尽管负载均衡是分离式的，在每个集合中的均衡算法与默认的系统相同。

#### 3.4.2.5 用锁行为自适应

应用程序的用锁行为可能会发生变化，因为系统中的任务是动态的创建和删除或者用锁密集类任务频繁使用的排号自旋锁可能会发生变化。一旦用锁行为发

生变化，锁竞争感知的调度器需要重新为用锁密集类任务决定SSC中的最优核数。本文提出的调度器利用 SSC 中核数的变化以及模型计算出的归一化吞吐量来决定程序的用锁行为是否发生变化。这部分逻辑如图 3.31 中所示(行 45 至 46)。从图中可以看到，如果 SSC 中的核数和上一轮的值不相同或者当前计算出的归一化吞吐量比上一轮的计算值过大或者过小，那么用于激活 SSC 中核数重新分配的计时器会被激活。超时事件的处理函数如图 3.30 所示(行 33 至 44)。在该处理函数中，special_core_bound 和投票系统的状态被重新初始化。另外一种触发定时器的方式是 SSC 中的处理器个数在 remig_timeout _interval 时间间隔内保持不变。在系统中，remig_timeout_interval 间隔凭经验设置成 30 秒，用户同样可以根据自己的需求进行设置。

#### 3.4.2.6 讨论

1 公平性 锁竞争感知的调度策略将用锁密集类任务限制在SSC上执行。然而，从任务等待队列中选择下一个要执行的任务的方法和 SSC 内进行负载均衡的策略与默认系统一样。这样，对于用锁密集类任务，尽管系统中的 CPU 并未全部使用，但是其公平性没有任何降低。

2 目标工作负载 锁竞争感知调度策略的目标应用是同构的用锁密集类应用。在这样的应用中,每个任务执行类似的操作并且相同的一组排号自旋锁被所有的任务所竞争。实际上很多应用都属于同构的用锁密集类应用，例如文件服务器和邮件服务器等。

### 3.4.3 实验和评价

#### 3.4.3.1 多核平台和操作系统

本文提出的调度策略的有效性在基于 Linux2.6.29.4 的 AMD32 核 NUMA 平台上和 Intel32 核 NUMA 平台上被验证。每个实验平台将分别介绍。对于 AMD32 核平台，系统中有 8 个 Opteron8347HE 芯片，每个芯片有 4 个核。每个核拥有私有的 L1 指令缓存, L1 数据缓存和 L2 缓存。每个芯片上的 4 个核共享最后级 L3 缓存。芯片内部的核和芯片之间分别通过内部互联和 HyperTransport 相连。总共 32G 的内存被划分为 8 块，每块和一个芯片相连。本文利用微基准测试程序测量了在该平台上的任务迁移开销。在相同芯片上的 2 个核之间进行迁移的开销是 9.3 微秒，在距离一跳的 2 个核之间迁移任务的开销是 9.7 微秒，两跳是 10.2 微秒而三跳则是 11.6 微秒。

本文使用的另外一个平台是 Intel32 核 NUMA 系统。系统中共有 4 个 Xeon

X7560 处理器，每个处理器有 8 个核。该系统支持超线程，系统中的每个核都可以虚拟出 2 个逻辑核。每个核均有私有的 L1 级指令数据缓存和 L2 缓存。芯片上的 8 个核共享 L3 缓存。系统中所有的内存划分成 4 个内存块，每个内存块和一个芯片相连。在测试时，关闭超线程和睿频加速(turboboosting)功能以简化可扩展性的分析。后文的实验主要使用的是 AMD32 核系统。因此，除非特别声明，实验结果都是在本系统上被搜集出来的。

#### 3.4.3.2 测试程序和运行方法

本文使用微基准测试程序和实际应用来评测锁竞争感知调度策略的有效性。使用的微基准测试程序有 single counter, mmapbench, sockbench，而使用的实际应用包括 parallel postmark, kernbench, parallel find 和 parallel grep。其中大部分测试程序均在评价基于等待者数目锁时被使用过，在此不再赘述。在这里主要介绍 2 个测试程序,一个是 parallel postmark 另外一个是 parallel grep。对于 parallel postmark，尽管前文使用过该测试程序用于评测基于等待者数目的锁，但是在调度算法的评测上，本文添加了额外的程序配置。具体而言，每个线程的初始文件个数使用两种配置，一个是 500 个文件，另外一个是 10000 个文件。对于 parallel grep，程序由多个进程组成，每个进程不断的调用 GNU 的 grep 命令在一组小文件里面搜索一个找不到的字符串。文件集合的总体大小是 4G 字节。所有的小文件在相同的目录下，目录深度是六。在 parallel grep 中有 2 个可以调节的参数。一个用于控制启动的进程数，另外一个控制每个进程搜索的次数。当运行这个程序时，所有的文件存在 tmpfs 文件系统中。

在使用的 7 个测试程序中, parallel postmark, kernbench 和 parallel grep 使用内存文件系统 tmpfs 存储工作目录和文件。本文使用 tmpfs 而非真正的文件系统有两点原因。1 本文尝试使用过实际文件系统，然而，最大的可扩展性瓶颈变成了 I/O 而非锁竞争，但是本文的关注点是解决锁瓶颈。2 在工业界, I/O 并不一定是文件操作密集应用的最大瓶颈，因为 RAID 会被用来提升系统的 IOPS (每秒钟的 IO 操作次数)。在本文的评价中，使用内存文件系统来模拟 RAID，因为本文的实验系统并没有 RAID 功能。

由于在测试程序运行时，本文保持使用的进程或者线程数目和测试用的核数相同，因此每个测试程序在多核系统上的可扩展性可以用吞吐量随核的变化来衡量。在可扩展的系统中，吞吐量的增长速度和核的增长速度是相同的。

#### 3.4.3.3 热点内核锁和其调用函数

使用 Linux 内核中的锁检测工具/proc/lock_stat 来确定测试程序运行时哪些内

核锁被激烈的竞争。每个测试程序频繁使用的内核锁和每个锁的热点调用函数如表 3.4 所示。

表 3.4 热点内核锁和调用该内核锁的热点调用函数。parallel postmark 中 2 个配置的热点锁和热点函数均相同。

| 测试程序 | 热点内核锁 | 调用函数 |
|---|---|---|
| mmapbench | mmapings lock(i_mmap_lock) | unlink_file_vma() vma_link() |
| sockbench | dentry cache lock (dcache_lock)<br><br>inode lock (inode_lock) | d_alloc() _atomic_dec_and_lock()<br>d_instantiate() d_rehash()<br>new_inode() __mark_inode_dirty()<br>generic_sync_sb_inodes() |
| parallel postmark | file descriptor table lock<br><br>files lock (files_lock) | sys_close() fd_install()<br>get_unused_fd_flags()<br>file_kill() file_move() |
| kernbench | files lock (files_lock)<br>dentry cache lock (dcache_lock) | file_kill() file_move()<br>d_alloc() _atomic_dec_and_lock()<br>d_path() dcache_readdir() |
| parallel find | files lock (files_lock)<br>dentry cache lock (dcache_lock) | file_kill() file_move()<br>_atomic_dec_and_lock() |
| parallel grep | files lock (files_lock)<br>per dentry lock (d_lock) | file_kill() file_move()<br>__d_lookup() dnotify_parent() dput()<br>inotify_dentry_parent_queue_event() |

#### 3.4.3.4 测量方法

为了测量本文提出的锁竞争感知的调度策略，将/proc/lock_stat 报告的每个热点函数对应的热点调用点利用 sched_lock()进行替换。对于某一热点函数，尽管并不是所有的调用点都会被替换，但是这种方式对于验证本文提出的观点已经足够了，因为程序的锁竞争是由热点锁的热点调用路径所主导的。而且，当替换掉热点锁的热点路径后，系统的可扩展性有很大的提升。这一事实已经充分的验证了本文提出的调度算法在大规模系统上的潜力。

在实验和评价一节，锁竞争感知的调度策略将会和其它的避免锁颠簸的方法进行比较，其中包括互斥锁，自适应互斥锁以及前文提出的基于等待者数目的锁。这些方法的测量方式是替换所有热点函数的全部调用点。不同于锁竞争感知的调度策略，对于这些方法，仅仅替换热点锁的热点调用路径并不可行，因为不同的同步原语不能被同时使用。

#### 3.4.3.5 可扩展性提升

图 3.32 表示了锁竞争感知的调度策略(曲线 lock-contention-aware)在 7 个负载上的吞吐量。默认 Linux 系统的实验结果(曲线 default)也被整合在相同的图中以方便比较。如图中所示，对于受到锁颠簸现象影响的 6 个测试程序中的 4 个(即 mmapbench, sockbench, parallel postmark, 和 parallel grep)，提出的锁竞争感知调度策略可以准确的感知锁颠簸现象发生的时机，并且维持默认系统的最大加速比。对于不受锁颠簸现象影响的测试程序 kernbench，锁竞争感知的调度策略和默认系统

的表现几乎相同。在下一节将会看到，在 kernbench 的执行过程中，基本没有任务被划分为用锁密集类任务。因此，在该测试用例中并不会有任务迁移的开销。实际上，锁竞争感知调度策略的唯一开销来源于持续地检测每个任务的等锁时间百分比。根据 kernbench 的实验结果，读取 TSC 的开销可以忽略不计。

然而，2 个测试程序表现的较为异常。对于 single counter，尽管锁竞争感知的调度算法在使用多于 1 核时可以很好地避免锁颠簸现象，但是，默认系统的最大加速比并不能被维持。对于 parallel find，锁竞争感知调度器的吞吐量在使用多于 15 核时，开始出现略微的下降。尽管吞吐量不像默认调度器下降的那样严重，但是这种略微的下降表明在锁竞争感知的调度器上运行 parallel find 时，存在着某些不可扩展的因素。这些异常现象将在介绍每个测试程序确定的用锁密集类任务数目和分配的核数时进行阐述。

#### 3.4.3.6 用锁密集类任务数目以及SSC中分配的核数

图 3.33 表示了锁竞争感知调度器中运行每个测试程序时，分配的核数和确定的用锁密集类任务随着程序启动的任务数目的变化。当固定了程序启动的任务数目，分配的核数和用锁密集类任务信息被搜集三次，所有的数据均被整合在一张图中。在提出的调度算法中，SSC 中的最优核数通过模型驱动的搜索来确定。而且，当程序的用锁行为发生变化时，这种搜索将被重新触发。这样，SSC 中分配的核数和确定的用锁密集类任务将会随着时间而变化。图 3.33 是当搜索稳定时搜集到的。

从 mmapbench, sockbench, parallel postmark 和 parallel grep 的实验结果中可以看到，锁竞争感知的调度算法将所有的任务均分类成用锁密集类任务。这个结论是合理的，因为本文使用的测试程序是同构的，意味着所有的任务均有类似的用锁行为。当与默认系统的实验结果(图 3.32)相结合时，可以看到 SSC 中分配的核数对这些测试程序也是最优的。而且，SSC 中分配的核数和确定的用锁密集类进程在每次运行时得到的结果基本类似，表明提出的调度策略可以持续的找到 SSC 中的最优核数。对于 kernbench，几乎没有进程被确定为用锁密集类进程，因为该测试程序中的锁竞争相当轻量级。这样，SSC 中分配的核数和确定的用锁密集类任务仍然维持着初始值。从图 3.33 中可以发现 single counter 和 parallel find 在锁竞争感知调度器上运行时的无效性。对于 single counter，锁竞争感知的调度策略在 SSC 中分配了 2 个核以处理用锁密集类任务。因此，最大的吞吐量不能被维持。SSC 中的核数并不是最优的，因为核分配算法运行时会受到测量误差的影响。一个形成误差可能的原因是一旦用锁密集类任务被确定，投票过程立即开始。然而，投票过程可能在所有的用锁密集类任务被确定出来并被迁移到 SSC 上之前。如果这种情况发生，

voting_locking 变量可能会比理想值大很多，这样 SSC 中分配的核数就不会是最优值。本文提出的锁竞争感知调度器已经在实际系统中采用了很多方法(如当票数充分大时才对 SSC 中的核数进行更新)来屏蔽这种影响，但是对于 single counter 这种测试程序，测量的误差同样会影响 SSC 中的最优核数,因为所有的任务激烈地竞争同一把排号自旋锁，而临界区和非临界区内几乎没有任何代码。

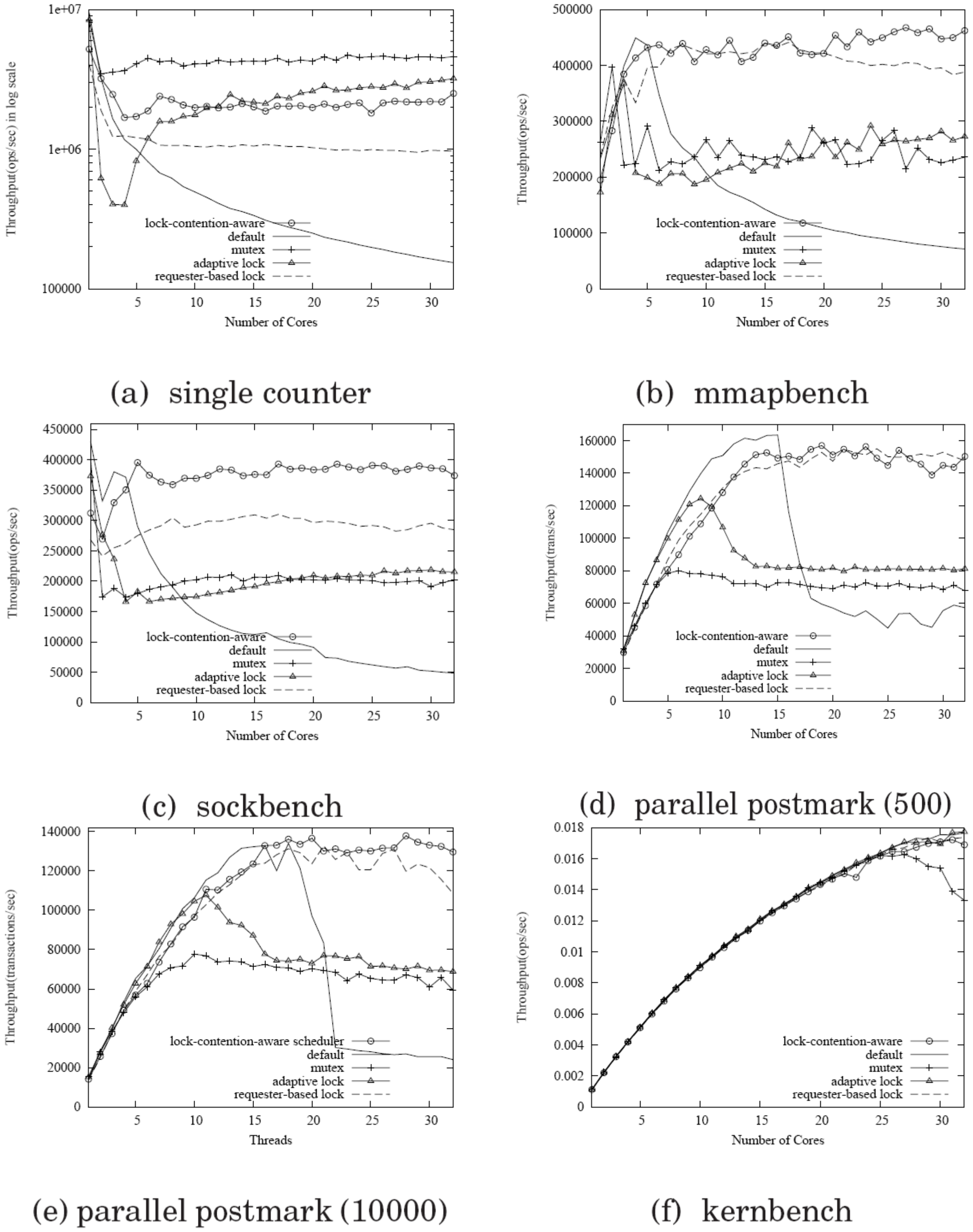

图 3.32 不同锁抖动避免方法的吞吐量随着核数的变化。对于 parallel postmark，每个线程的初始文件数目也被显示出来。而 single counter 的吞吐量是以对数规模被展示出来的。

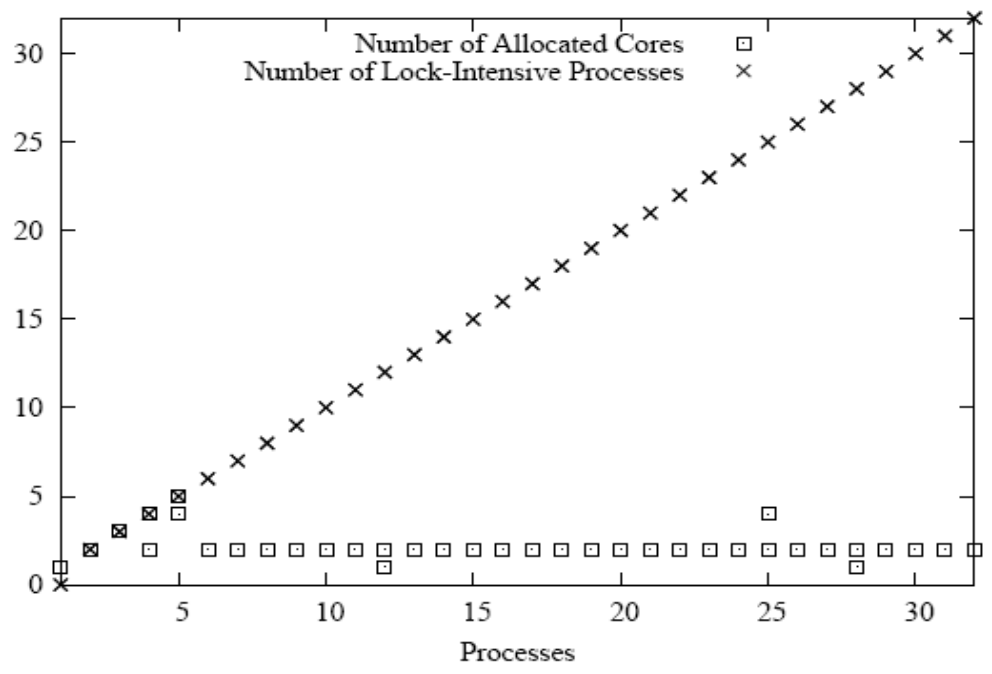

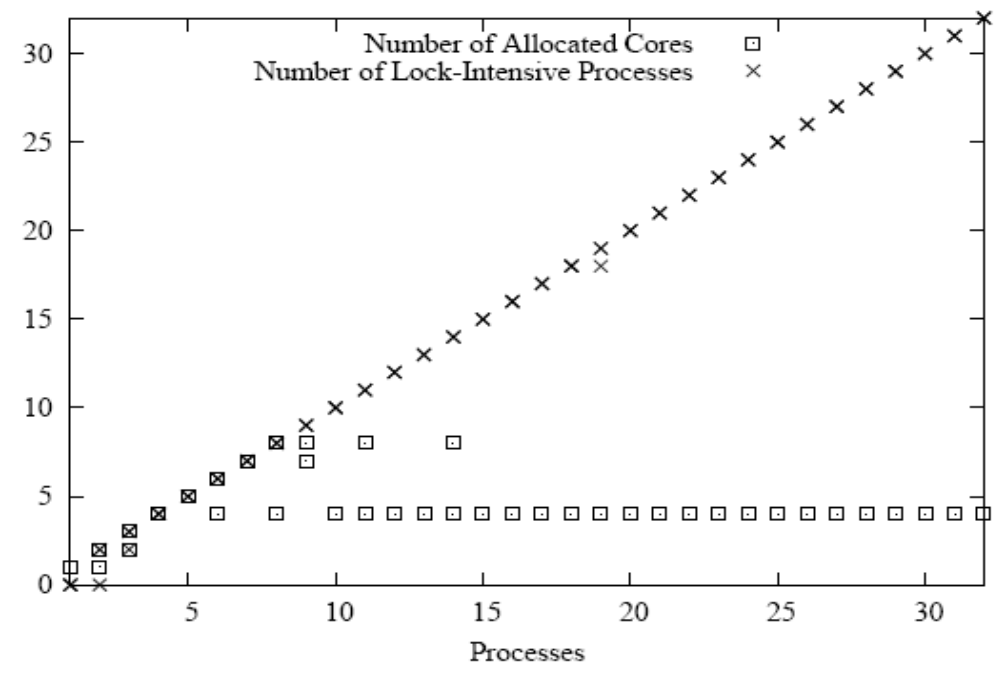

(a) single counter　　(b) mmapbench

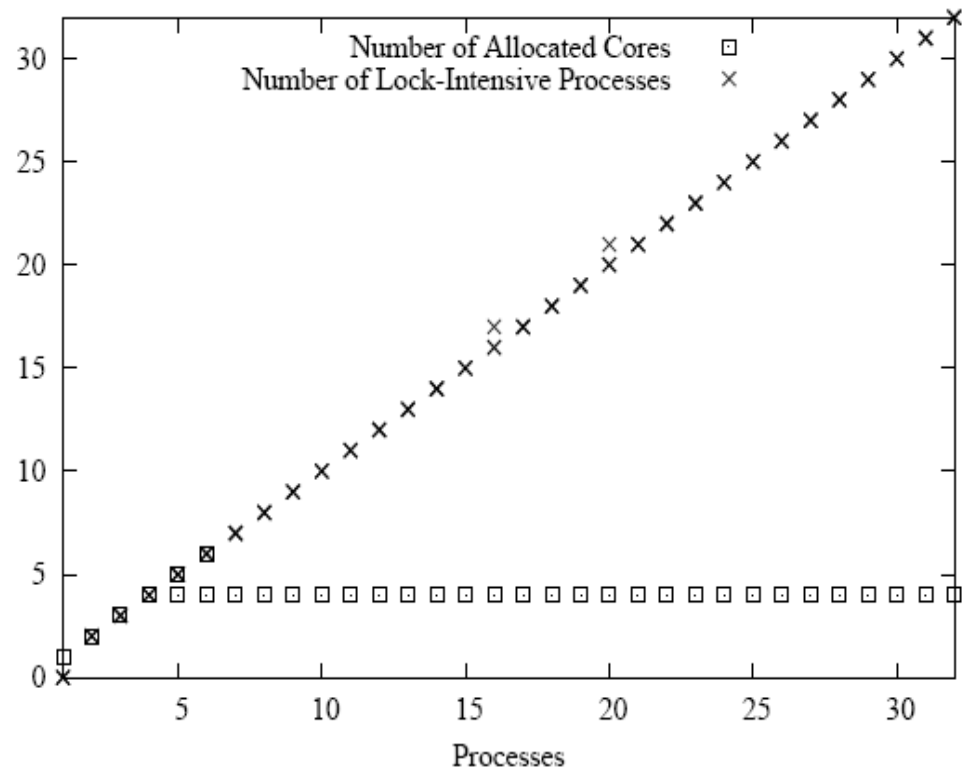

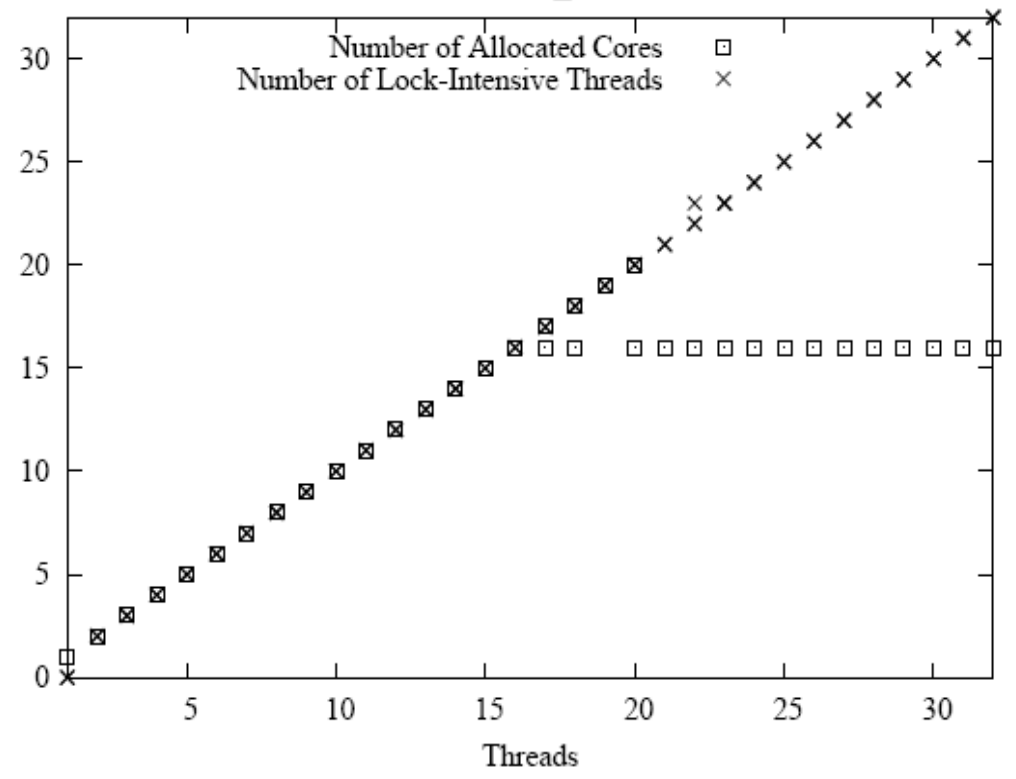

(c) sockbench　　(d) parallel postmark (500)

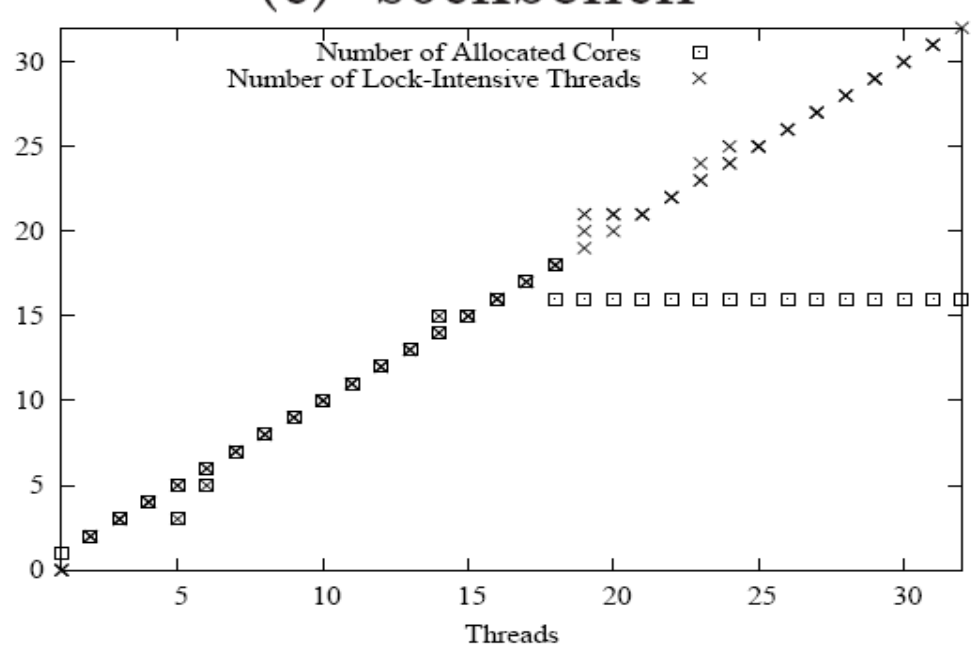

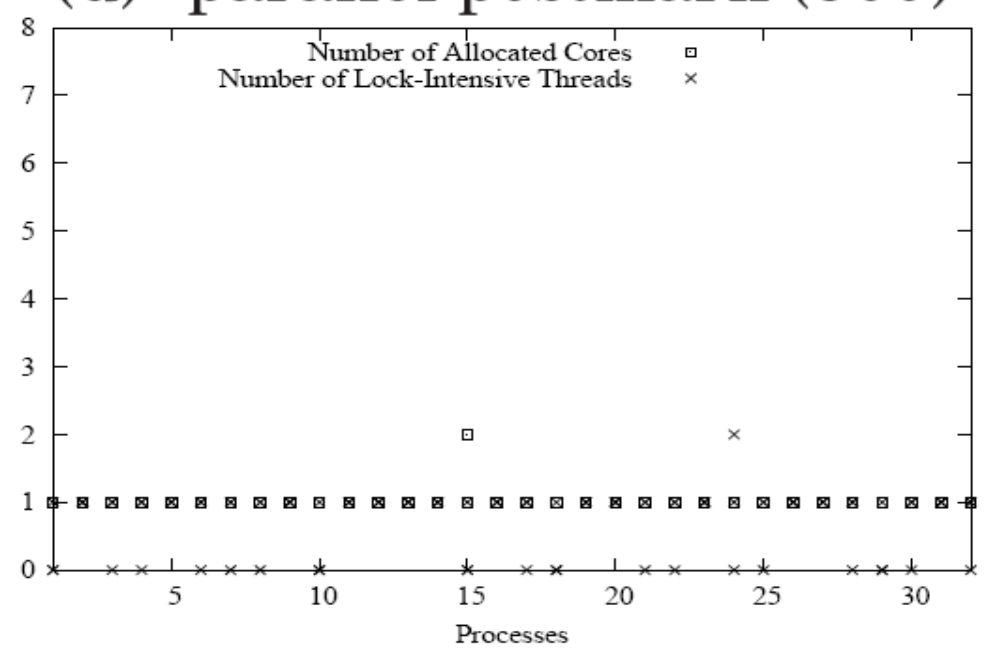

(e) parallel postmark (10000)　　(f) kernbench

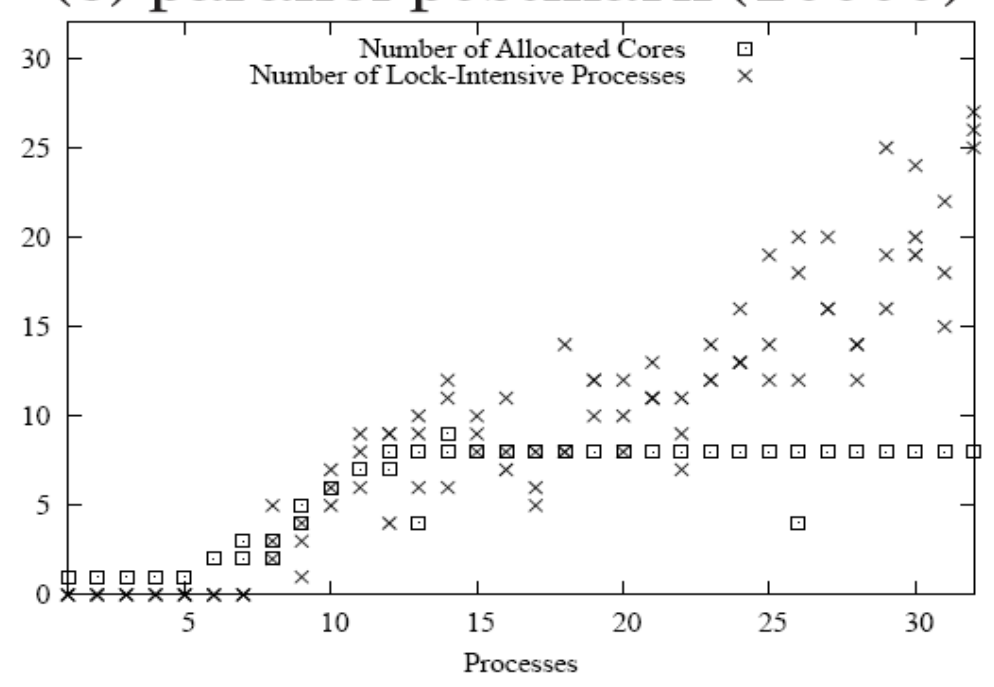

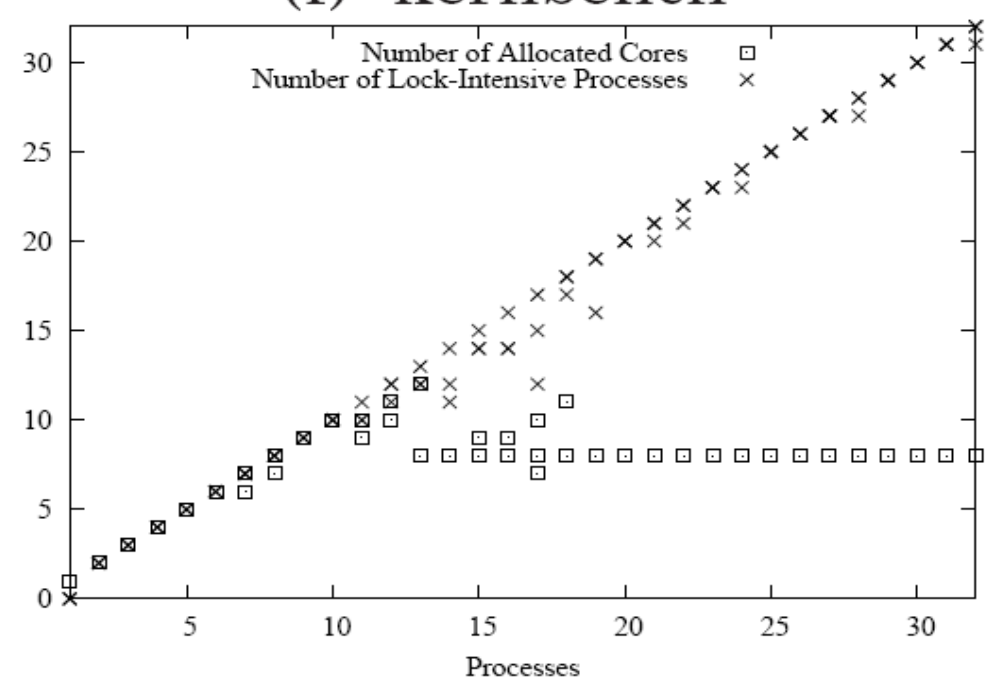

(g) parallel find　　(h) parallel grep

图 3.33 竞争感知调度器中运行每个测试程序时分配的核数和用锁密集类进程的数目

对于 parallel find，当启动的进程数目超过 7 个时，确定的用锁密集类进程在 3 次执行中波动很大。这是因为 parallel find 中，每个创建出的进程的执行时间过于短暂。因为每个创建出来的进程很快就执行完毕，内核则会频繁的创建和终结进程。然而,一个任务的 task_struct 数据结构的 is_mig 域被设置成 QUALIFY_TO_MIGRATE 之后，锁竞争感知的调度策略才会将其记为用锁密集类任务。如果任务执行的过于短暂，在其被标记为用锁密集之前该任务就可能已经退出了。这样，在每次运行中确定出的用锁密集类任务可能会发生变化。锁竞争感知的调度策略在检测到程序用锁行为的变化后会重新决定SSC中的最优核数。在parallel find中，进程的频繁创建和终结使得程序的用锁行为不断的发生变化。而且，这种变化的频率随着启动的工作进程的数目而增加。因此，程序的吞吐量会随着工作进程的增加而渐渐下降。当 paralell find 开始出现锁颠簸现象时, SSC 中分配的核数是 8 而不是最优值 10。这是因为本文提出的调度策略在每轮投票结束时将当前的核数乘以 2。

图 3.33 中也显示出了锁竞争感知调度的另外一种无效性。当锁颠簸现象刚刚出现时，本文提出的调度策略倾向于比最优值分配更多的核数（例如，parallel postmark 中当使用 17 或者 18 个工作进程)。这表明锁竞争感知的调度策略的负载均衡算法同样存在着改进的空间。幸运的是，这种无效性带来的影响并不大。

#### 3.4.3.7 CPU利用率和锁竞争的降低程度

图 3.34 显示了每个测试程序运行时，所有处于运行状态的核在默认调度策略和锁竞争感知策略中平均执行时间的分解图。从图中可以发现，在默认系统中和锁竞争感知调度策略中的执行时间分解基本类似。single counter, mmapbench, sockbench, parallel postmark 和 parallel grep 的 CPU 利用率均为 100%并且大部分时间执行在内核态。对于 kernbench，大部分时间处于用户态，并且当使用超过 4 个核时，每个核开始出现空闲时间。对于 parallel find，每个核的空闲时间是最大的，因为每个进程的执行时间过于短暂并且很多核都处于工作状态。

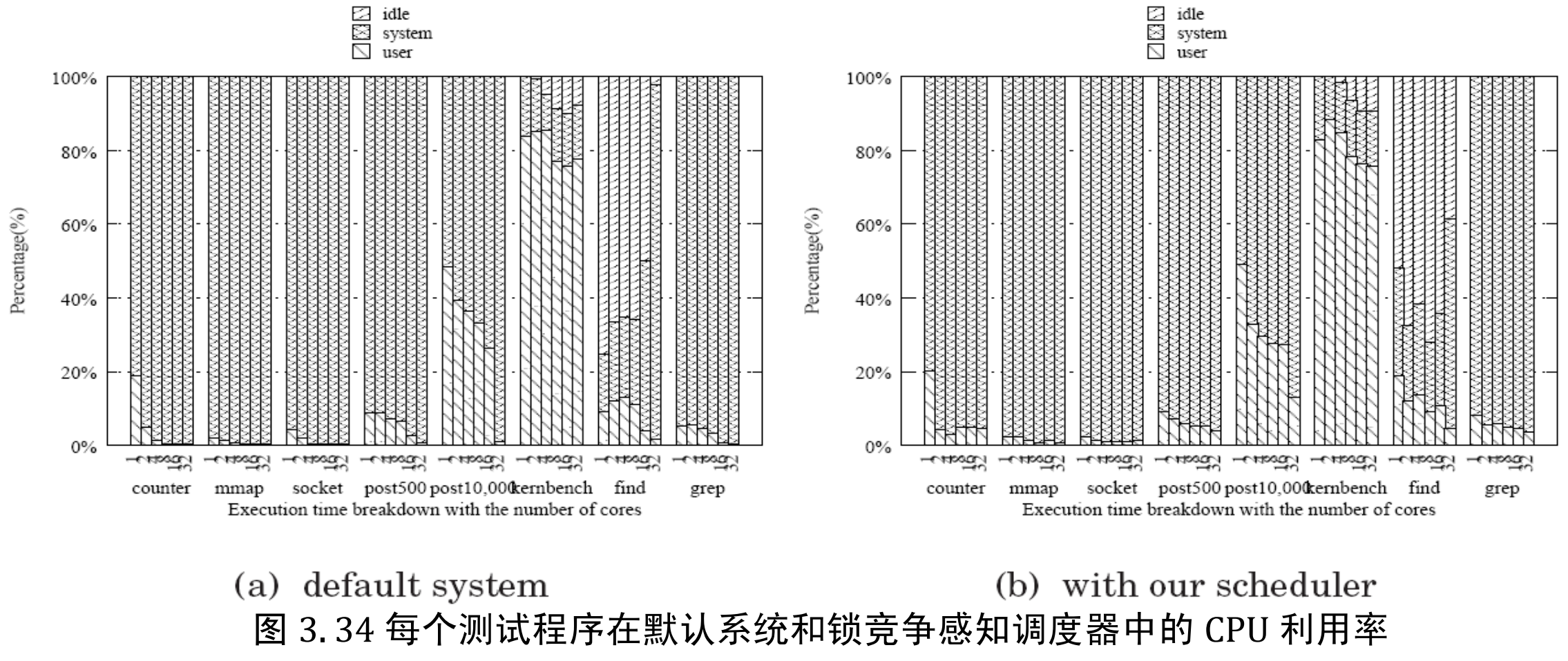

(a) default system (b) with our scheduler

图 3.34 每个测试程序在默认系统和锁竞争感知调度器中的 CPU 利用率

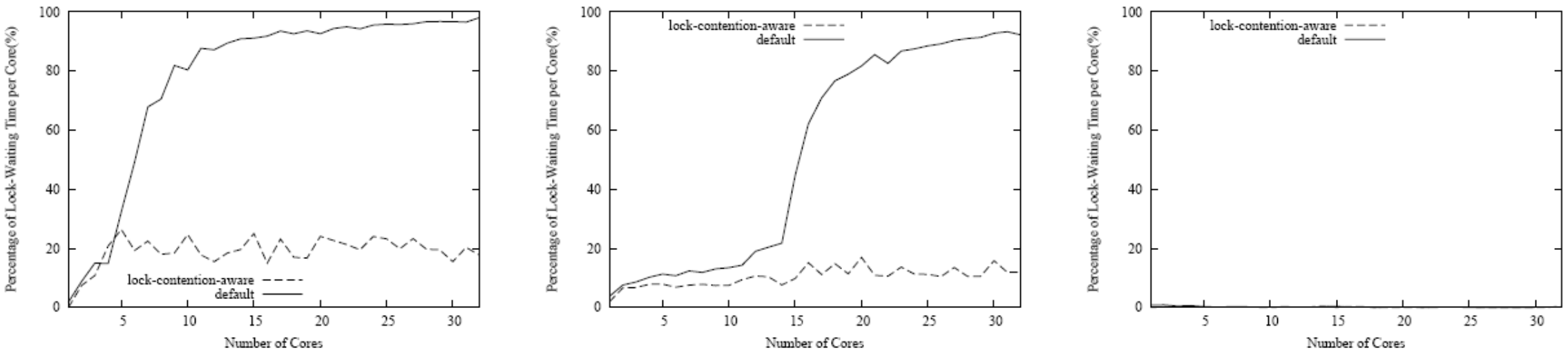

(a) mmapbench　(b) parallel postmark (500)　(c) kernbench

图 3.35 默认系统和锁竞争感知调度策略中平均等锁时间百分比随核数的变化

锁竞争的降低程度通过每个任务等锁时间百分比的降低程度来衡量。图 3.35 展示了一个微基准测试程序(mmapbench)和两个实际应用(parallel postmark 和 kernbench)的实验结果，其他测试程序的实验结果在此省略。与预想的一样，一旦锁颠簸现象发生，锁竞争感知调度策略开始大幅度地减少任务的等锁时间百分比，因为所有的用锁密集类任务被限制在 SSC 上执行。而且，降低的程度随着工作进程数目的增加而变得越来越大。当应用程序使用 32 个工作进程时，等锁时间百分比可以被降低至 84%。

#### 3.4.3.8 计算机架构影响

锁竞争感知调度策略的一个考虑是在其它的多核架构上该方法是否依然有效。事实上，许多架构参数例如缓存大小，缓存一致性协议，内存控制器带宽，互联拓扑等均可能影响锁竞争感知算法的有效性。本节通过在基于 Linux2.6.29.4 的 Intel 32 核平台上表示锁竞争感知调度策略的有效性以研究可能的架构影响。

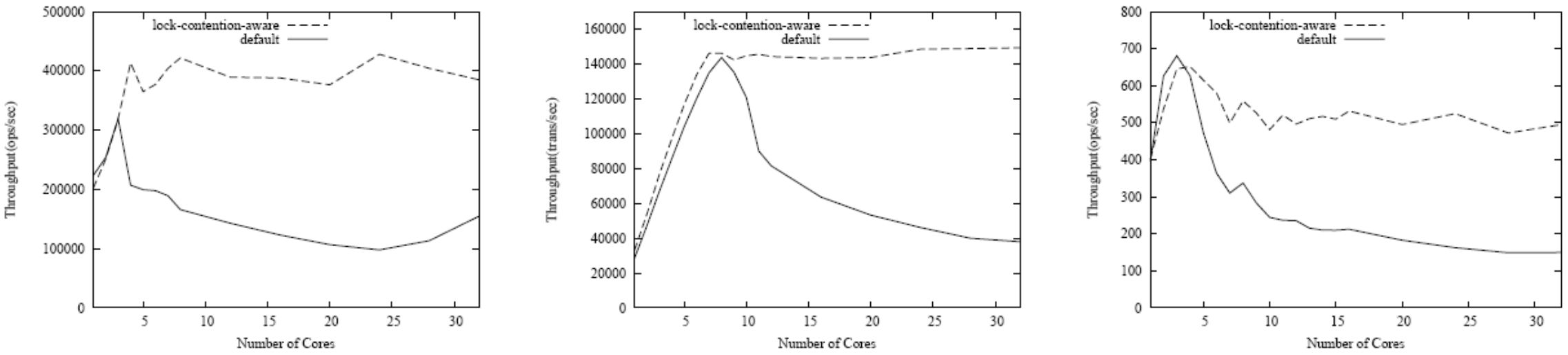

(a) sockbench　(b) parallel postmark　(c) parallel find

图 3.36 锁竞争感知调度器和默认调度器在 Intel32 核系统上的吞吐量随核数的变化

图 3.36 展示了 3 个测试程序(sockbench, parallel postmark 和 parallel find)的实验结果，其它测试程序的结果均被省略。从图 3.36 中可以发现，Intel32 核系统上的实验结果和 AMD32 核系统的实验结果类似。测试程序 sockbench 和 parallel postmark 中出现的锁颠簸现象在采用锁竞争感知的调度器后可以被很好地避免。而且，默认系统的最大吞吐量可以被维持住。对于 parallel find，可扩展性在本文提出的调度算法中被大幅度的提升。然而，吞吐量会随着工作进程数目的增加开始逐

渐下降，因为 parallel find 中进程的执行时间相当短暂。这样，在 SSC 中搜索最优核数的开销随着工作进程数目的增加而逐渐变大。

Intel 32 核系统上的实验结果和 AMD 系统上的结果基本类似。这一结果表明锁竞争感知的调度器对架构参数的改变并不敏感。这个结论符合作者的预期,因为本文的调度器并不依赖于任何具体的硬件参数，尽管参数的变化会影响应用程序的性能。

#### 3.4.3.9 与其它锁颠簸现象避免方法的比较

1、比较目标 本文的目标是为内核锁密集类应用提供可扩展的性能和较好的能耗有效性。为了达到这一目标，一般有两种方式。1 自旋锁仍然被用于同步，而其它的策略与自旋锁共同使用以避免锁颠簸现象或者提供较好的能耗有效性。例如，本文介绍了如何将调度策略和排号自旋锁相结合，而基于等待者数目的锁[73]是将排号自旋锁和节能指令进行结合。在这个研究方向中，与自旋锁共同使用的策略与各种自旋锁的实现是正交的关系。2 系统中放弃使用自旋锁，转而去使用基于阻塞的同步原语。在这个研究方向中，节能策略往往是隐含在锁协议中的。例如，如果一把互斥锁不能被立即获取，锁等待者将会进入睡眠状态。这样，锁等待者不会继续占用处理器资源，也不会耗费电能。

为了与锁竞争感知的调度策略进行比较，本文从第一个方向上选择基于等待者数目的锁(requester-based lock)，从第二个方向上选择互斥锁(mutex)和自适应互斥锁(adaptive lock)。

本文没有直接和经典的可扩展自旋锁(例如，MCS 和 CLH 锁)相比较因为本文的调度策略和可扩展的自旋锁是正交的关系，两者应该相结合以提供一个完整的解决方案。本文同样没有和带有节能策略的可扩展锁(如文献[57])相比较，因为当前的锁竞争感知调度策略是基于排号自旋锁的。这样，自旋锁实现的差异会使得实验结果变得无法解读。在后面的小节中，本文提出的调度策略与可扩展锁的结合将会被详细介绍。

在选择的 3 个锁颠簸现象避免的解决方案中，基于等待者数目锁根据当前锁的等待者数目决定锁等待的方式(自旋或者进入节能状态)。如果当前的锁等待者数目大于 0，新来的锁等待者通过使用 monitor 和 mwait 指令进入节能状态，否则，该请求者通过自旋来获取锁。对于互斥锁，如果不能被马上获取，锁等待者将进入睡眠状态。对于自适应互斥锁，锁请求者会根据当前持锁进程的状态选择自旋或者睡眠。

2、可扩展性比较 图 3.32 表示了不同的锁颠簸避免方法获取的吞吐量随着核

数的变化。在微基准测试 mmapbench 和 sockbench 中，锁竞争感知的调度方法和基于等待者数目的锁比互斥锁和自适应互斥锁的可扩展性好很多。实验的结果符合作者的预期，因为基于睡眠的同步原语在避免锁颠簸现象时引入的上下文切换的开销非常大。当与基于等待者数目的锁相比较时，锁竞争感知的调度策略在这 2 个测试实例中好于基于等待者数目的锁，尽管基于等待者数目的锁同样可以很好地避免锁颠簸现象。

对于 single counter，锁竞争感知的调度算法好于基于等待者数目的锁，然而，这两种方法比基于阻塞的同步原语的可扩展性要差。这是因为在该测试程序中，上下文切换的开销不会影响基于阻塞的同步原语的性能。具体而言，互斥锁和自适应锁尝试在睡眠之前再次获取锁。而在 single counter 中，只有一个极短的临界区。那么，刚刚释放锁的任务将尝试再次获取锁。这种机制保证相同的锁会被同一任务不断的获取，而不会受到上下文切换开销的影响。实际上，基于睡眠的同步原语表现出的较好的可扩展性是通过牺牲进程的公平性换来的。

在实际应用的测试中，锁竞争感知的调度策略和基于等待者数目的锁在 parallel postmark, parallel find 和 parallel grep 上的可扩展性比基于睡眠的同步原语的可扩展性好。而将这两种方法进行比较时，锁竞争感知的调度算法在 parallel postmark 和 parallel grep 上不会比基于等待者数目的锁表现的差(尤其是 paralell postmark 的每个工作进程的初始文件数目在 10000 时)，但是在 parallel find 测试实例中，当开启的进程数目多于 16 时，锁竞争感知的调度策略的可扩展性稍差。parallel find 中较差的可扩展性是由于频繁的进程创建和删除从而导致较大的调度开销。对于 kernbench，锁竞争感知调度，基于等待者数目锁和自适应锁达到的可扩展性和默认系统的几乎相同，因为该测试程序中锁竞争的量级相当轻。一个有趣的现象是尽管锁颠簸现象没有在默认系统中被观察到，然而在使用互斥锁之后，锁颠簸现象反而出现了，这是因为互斥锁引入的上下文切换开销过大而导致的。

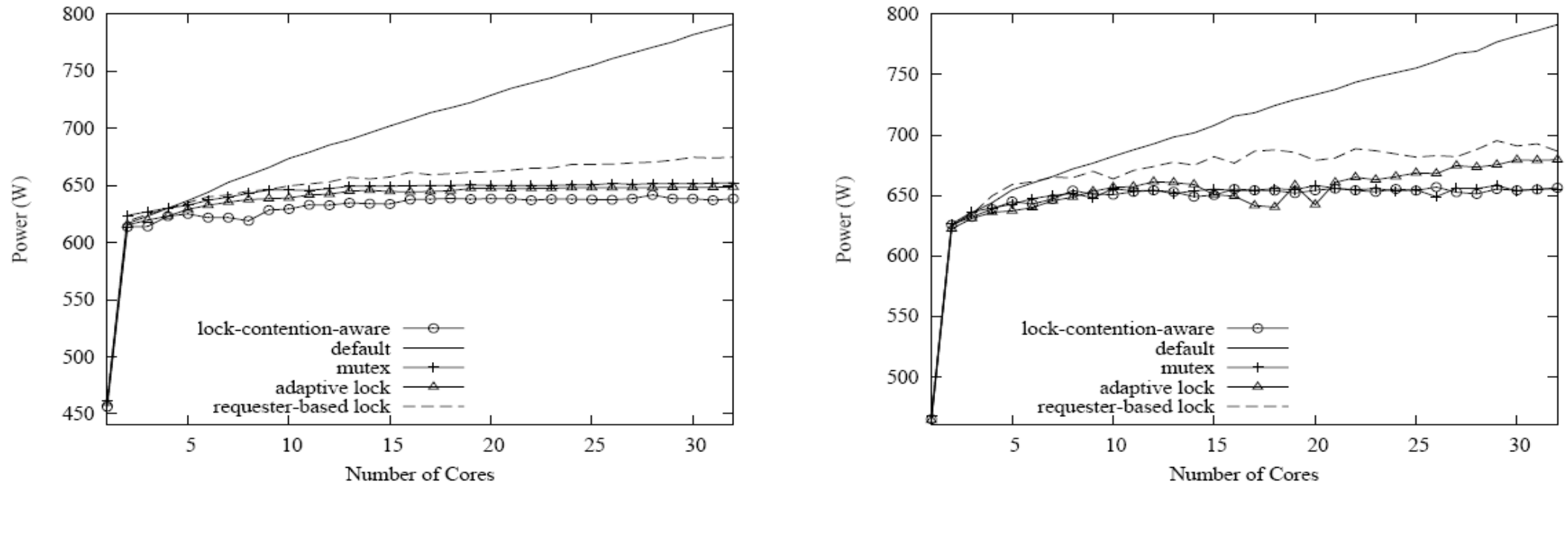

(a) single counter　　(b) mmapbench

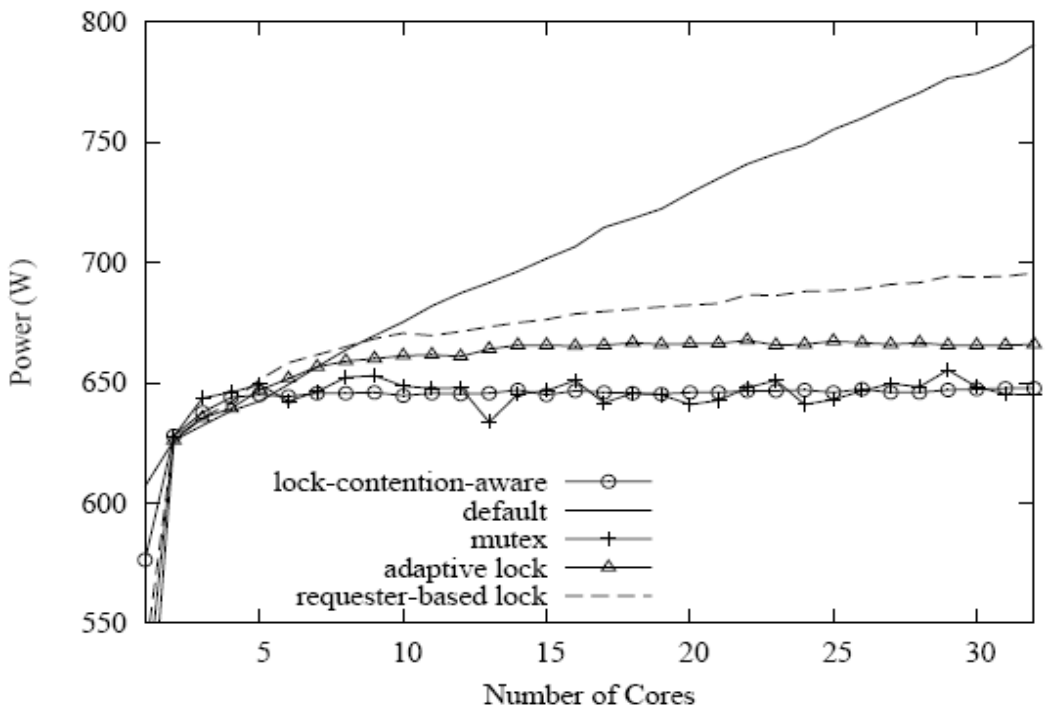

(c) sockbench

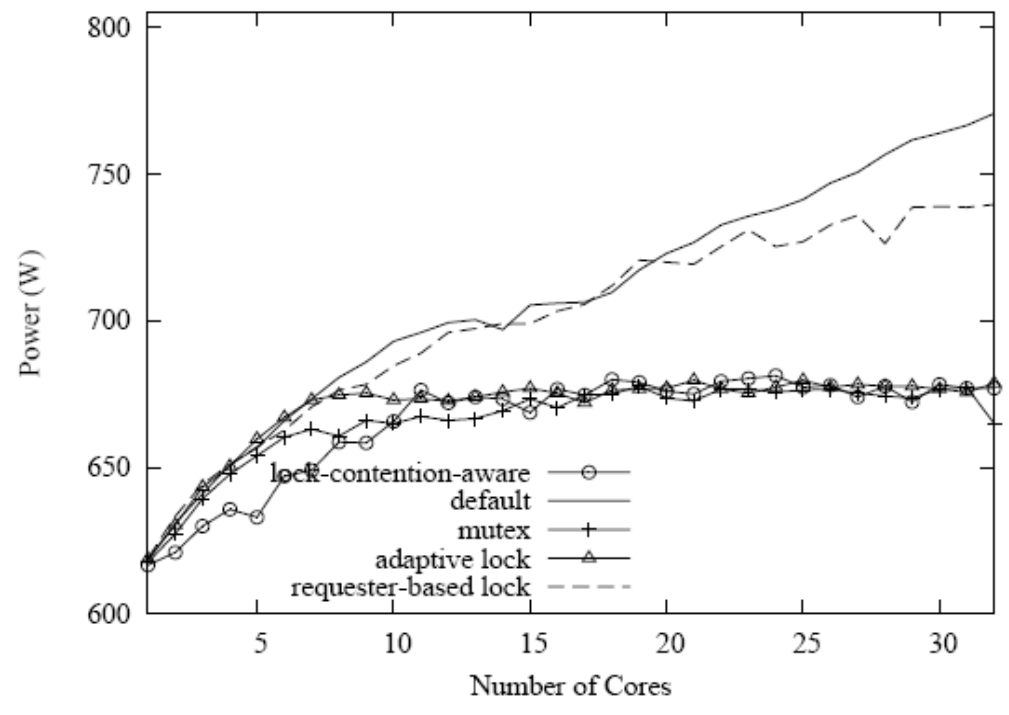

(d) parallel postmark(500)

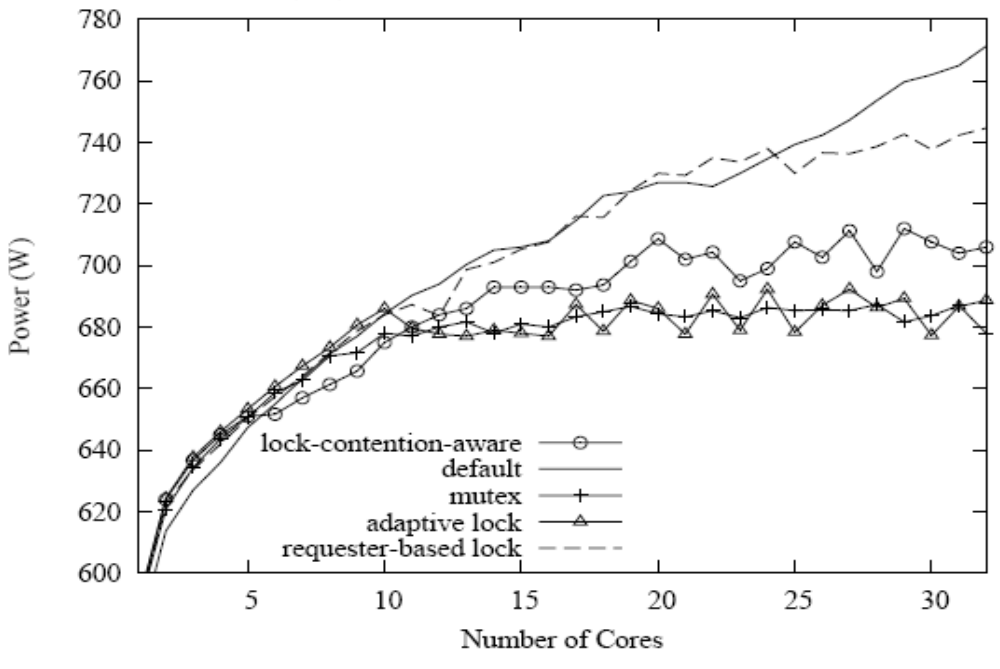

(e) parallel postmark(10000)

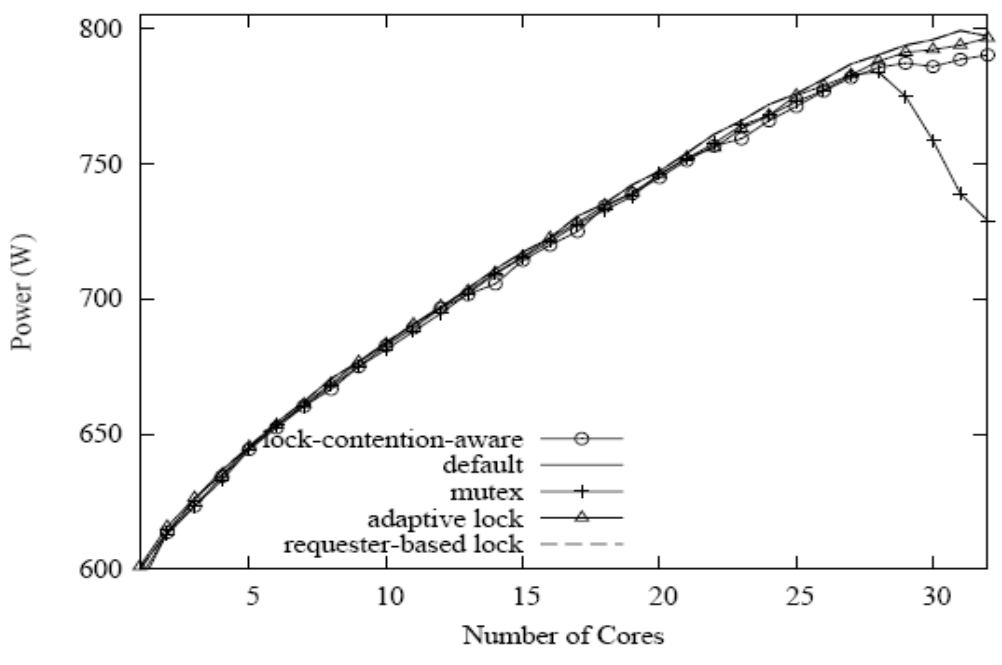

(f) kernbench

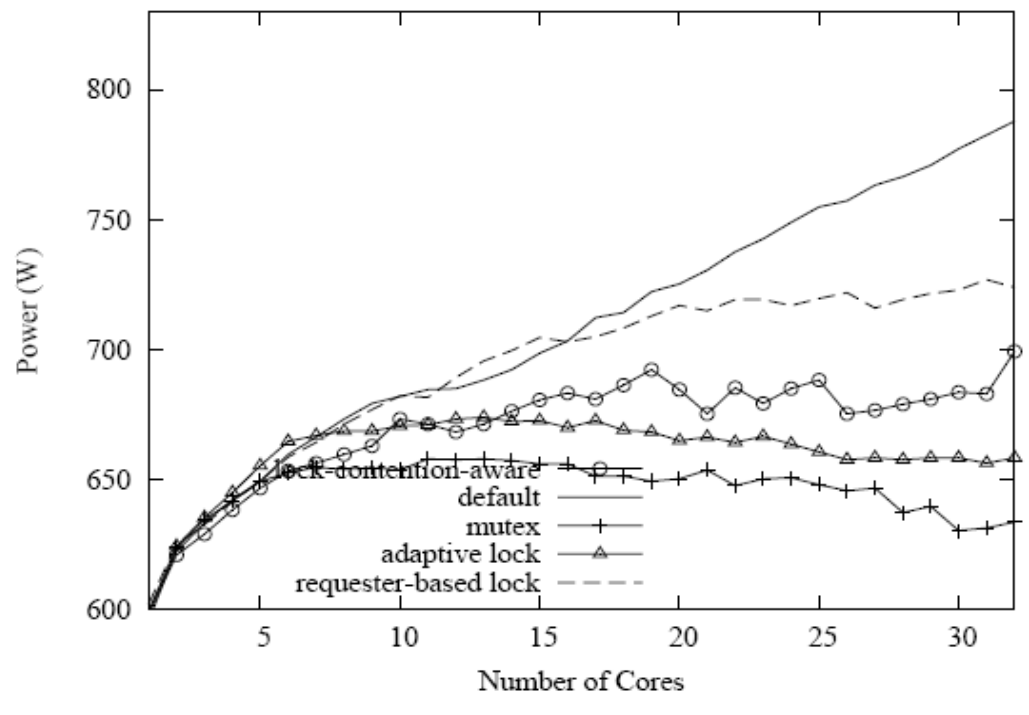

(g) parallel find

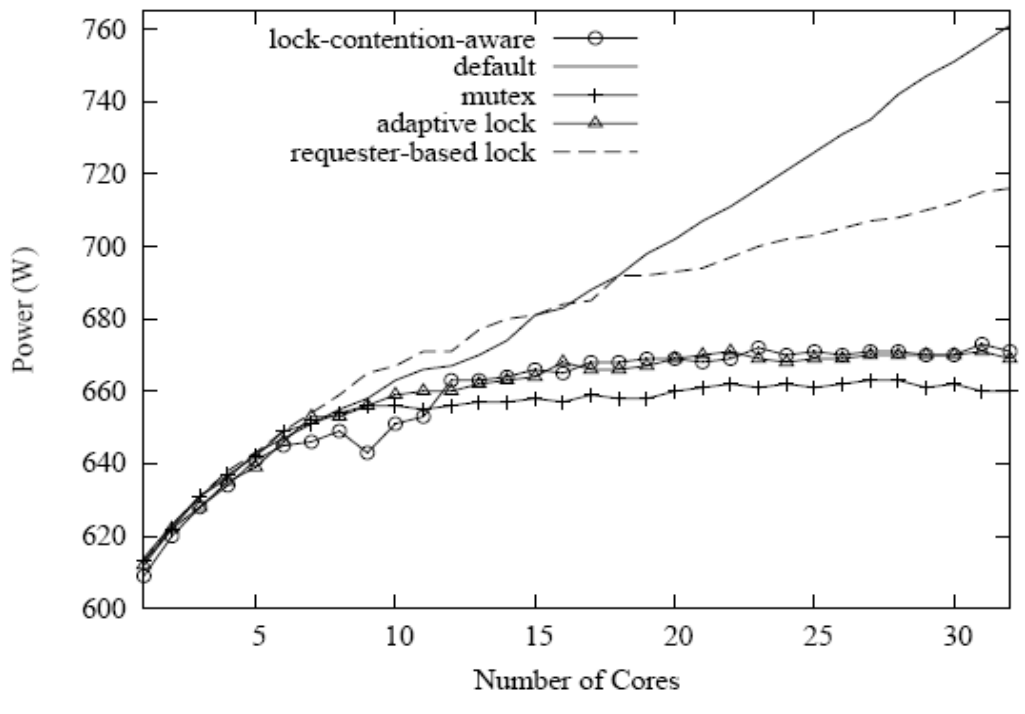

(h) parallel grep

图 3.37 不同的锁颠簸避免算法的功率随着核数的变化

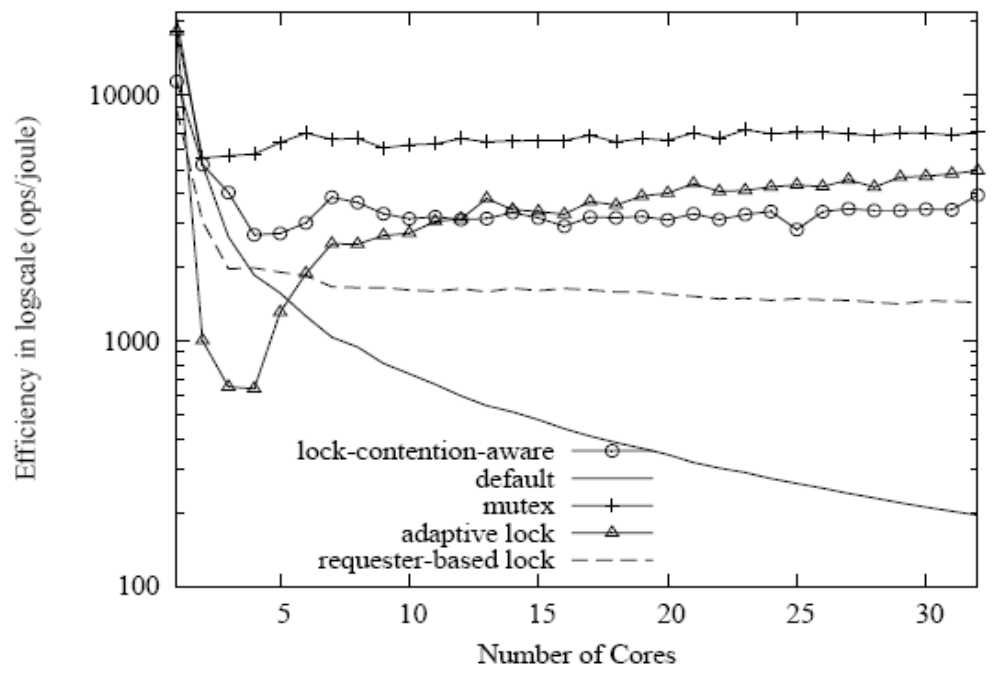

(a) single counter

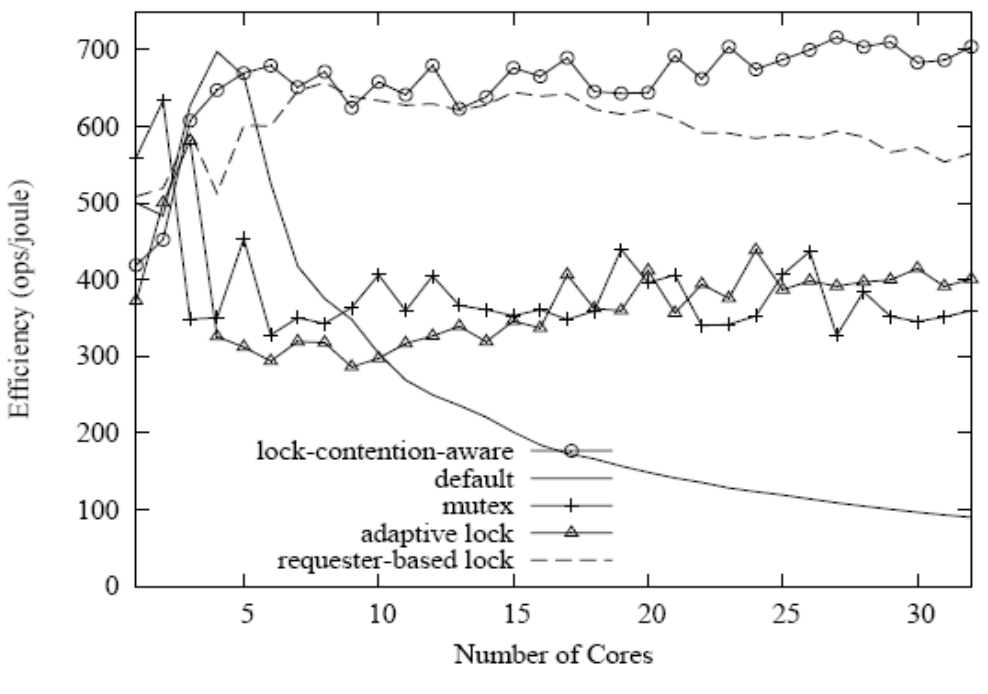

(b) mmapbench

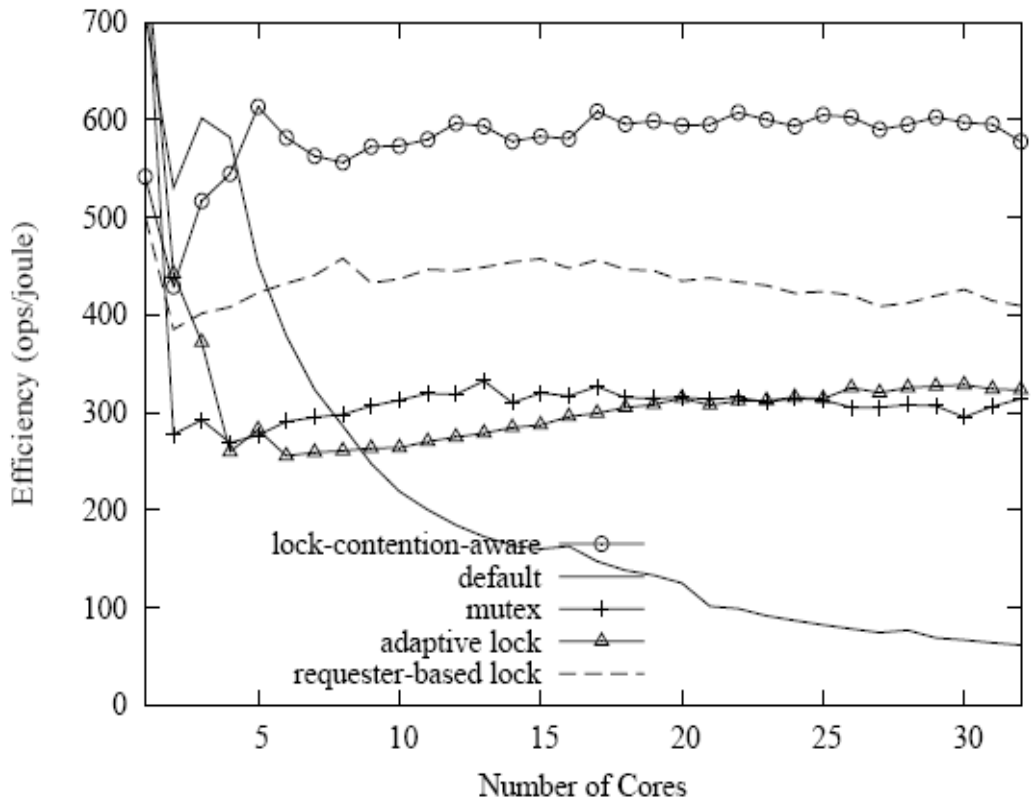

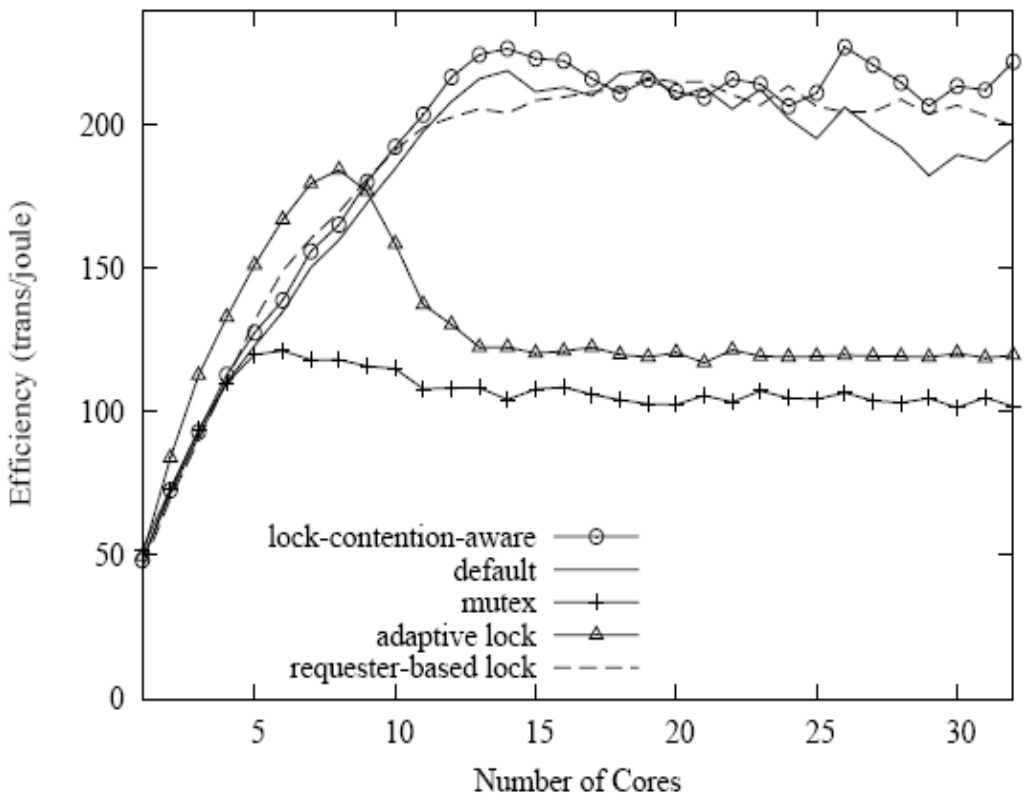

(c) sockbench　　(d) parallel postmark (500)

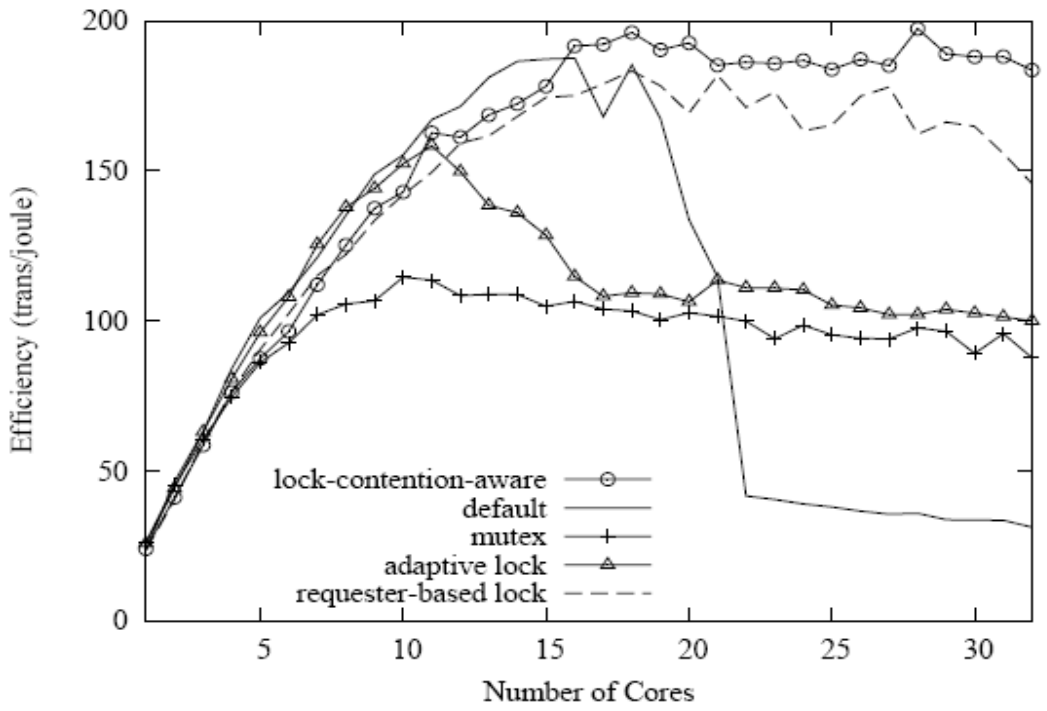

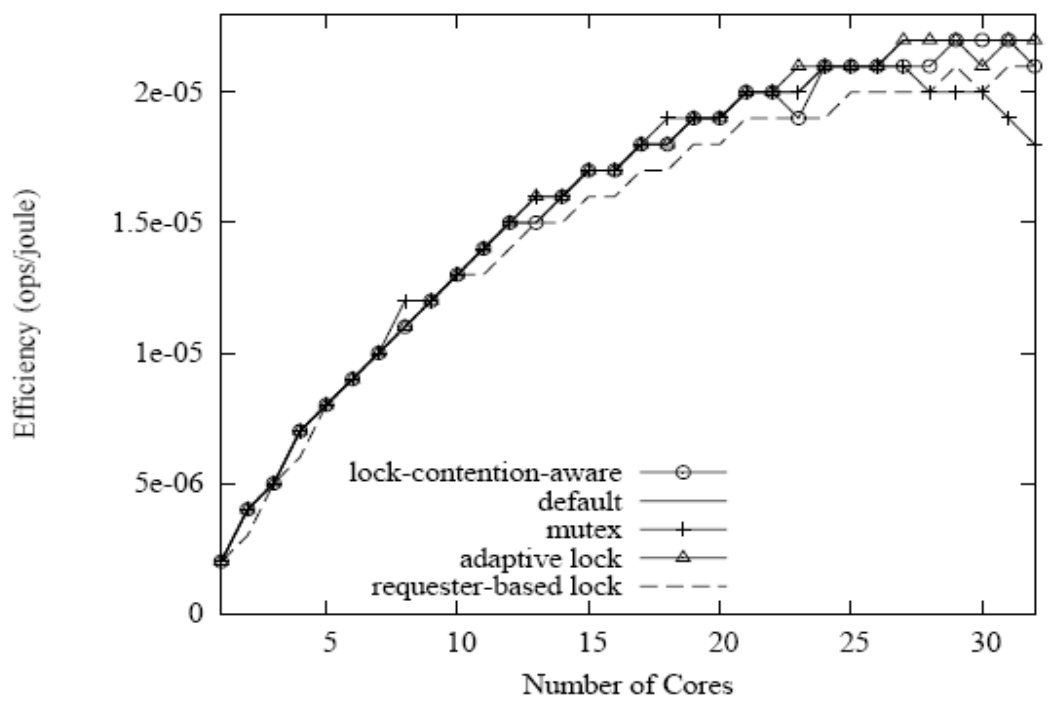

(e) parallel postmark (10000)　　(f) kernbench

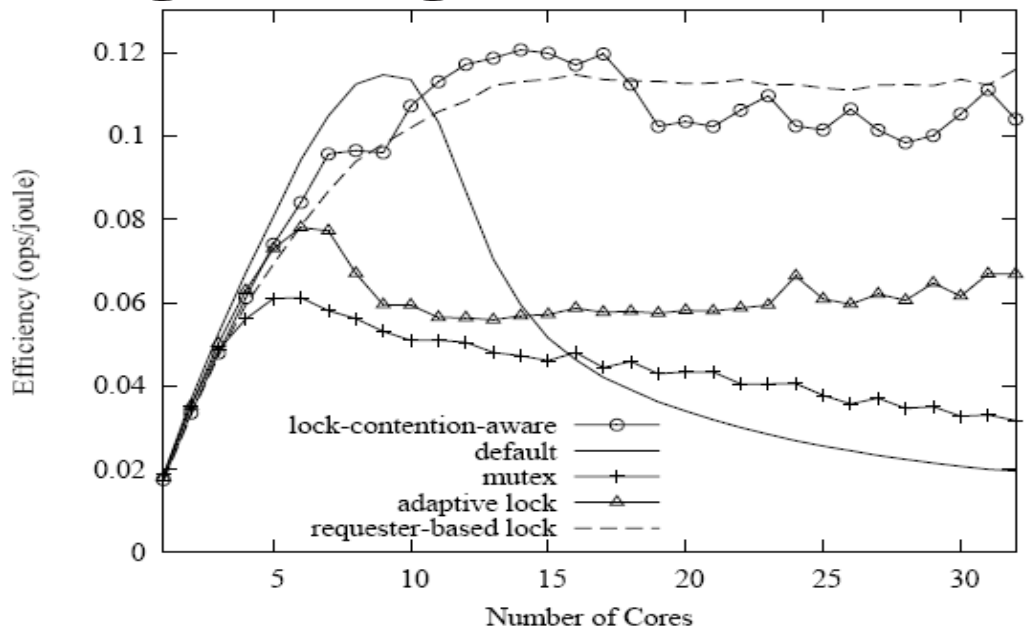

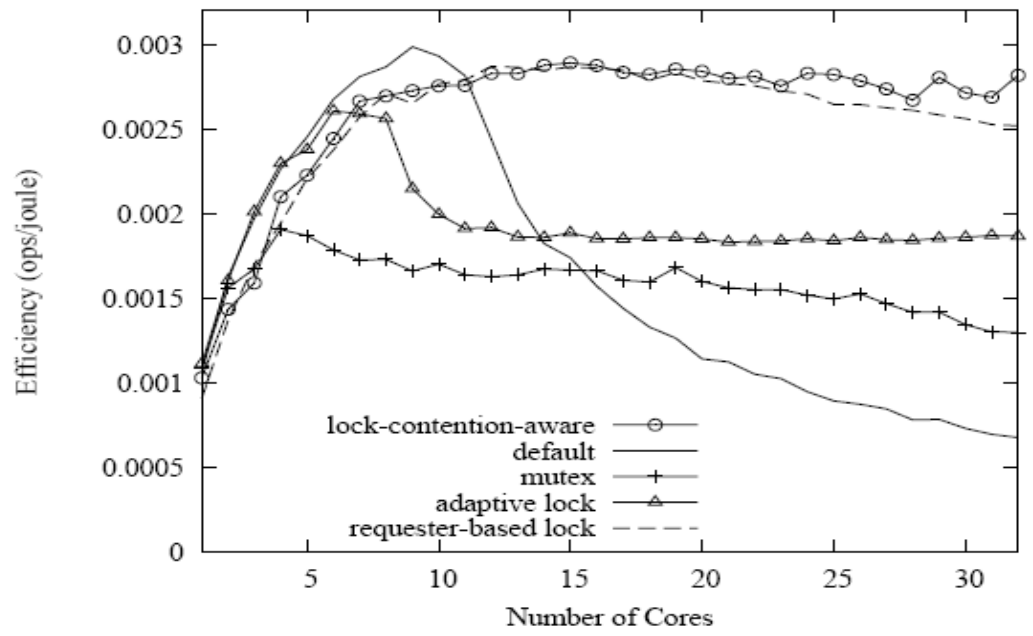

(g) parallel find　　(h) parallel grep

**图 3.38 不同锁颠簸现象避免方法的有效性随核数的变化**

3、 功率消耗比较 锁竞争感知的调度策略使用 SSC 来处理用锁密集类任务。剩余的核处于空闲状态。因此消耗的功耗不会像 SSC 中的核那样多。这一节比较不同锁颠簸现象避免方法的功率消耗。图 3.37 给出了实验结果。对于 single counter, mmapbench, sockbench, parallel postmark 和 parallel grep, 实验结果都是类似的。默认系统(使用排号自旋锁)的功率在所有的方法中随着核的增长速度是最快的, 因为每个锁等待者通过自旋等待的方式来等待锁的释放。对于基于等待者数目的锁, 其功率同样随着核数的增加而增加, 但是增加的速度会慢很多。这是因为大部分的锁

等待者在等待的过程中会利用 monitor 和 mwait 指令进行节能。锁竞争感知的调度策略，互斥锁和自适应锁是消耗功率最小的三种方法。从图中可以看出这些方法的功率消耗随着核数的增加而变得稳定。这里的实验结果也表明将处理器处于空闲状态是一种有效的节能机制。

对于 kernbench，锁竞争感知的调度策略，基于等待者数目的锁，互斥锁和自适应锁的功率消耗基本相同，因为该测试程序的锁竞争不够激烈。唯一的异常是互斥锁。当使用多于 29 个核时，其功率消耗开始下降。实际上，当使用多于 29 个核时，互斥锁会导致锁颠簸现象。一旦这种现象发生，某一把锁会被激烈的竞争，因此，所有的锁等待者将会进入睡眠状态，从而达到节能的效果。

4 、能耗有效性比较 图 3.38 表示了不同锁颠簸现象避免方法的有效性随核数的变化。总体而言，锁竞争感知调度展示出了合理的能耗有效性的提升。具体而言，对于微基准测试程序 mmapbench 和 sockbench，本文的调度算法比基于等待者数目的锁和基于阻塞的同步原语的可扩展性要好。对于 single counter，基于调度的方法比基于等待者数目的锁表现要好，但是两者均差于基于睡眠的同步原语，这是因为基于睡眠的同步原语牺牲了任务的公平性。对于实际应用，锁竞争感知的调度策略和基于等待者数目的加锁方法在 parallel postmark, parallel find 和 parallel grep 上比互斥锁以及自适应互斥锁的能耗有效性要好。当与基于等待者数目的锁相比较时，基于调度的方法在 parallel grep 和 parallel postmark 中均有优势。在 parallel find 中，由于频繁的任务迁移，基于调度的方法达到的有效性在多于 18 核时要差于基于等待者数目的锁。

#### 3.4.3.10 其它自旋锁上的可应用性

本文的锁竞争感知调度算法实现假定系统使用的是排号自旋锁。然而，提出的调度算法并不依赖于自旋锁的具体实现机制。因此，提出的算法可以和任意一种自旋锁同时使用。为了验证这个结论，本节将锁竞争感知调度策略和可扩展自旋锁(MCS 锁)结合使用来提升可扩展性和能耗有效性。

MCS 锁[55]是一种著名的用于避免锁颠簸现象的可扩展锁实现。当等待一把 MCS 锁时，每个锁等待者在局部的而非全局的变量上自旋以最小化核间的缓存行传递。当一把锁被释放时，仅有等待时间最长的等待者被通知以获取锁。图 3.39 中(a)和(d)表示了将默认系统的排号自旋锁替换成 MCS 锁后(曲线为 MCS)，一个微基准测试程序 (mmapbench)和一个实际应用 (parallel find)获得的可扩展性提升。从图中可以看到，使用 MCS 锁可以在内核锁竞争不够激烈时很好地避免锁颠簸现象(例如 parallel find)，但是当锁竞争比较激烈时，锁颠簸现象仍然存在(例如

mmapbench)。使用了 MCS 锁的 mmapbench 达到的吞吐量会随着核数略微下降，因为 MCS 锁在实现时使用了不可扩展的复杂原子指令。在微基准测试中，原子指令的不可扩展性会充分的体现出来。对于 parallel find，尽管锁颠簸现象被很好地避免，但是其消耗的功耗会随着核数快速的增加(子图(e))。

将锁竞争感知的调度策略和 MCS 锁相结合得到的优势是很清晰的。对于使用 MCS 就可以很好避免锁颠簸现象的应用，这种结合可以换来更少的功耗消耗和更好的能耗有效性。对于使用 MCS 锁不能很好地避免锁颠簸现象的应用，这种结合还可以提供更好的可扩展性。图 3.39 表示了将锁竞争感知的调度策略和 MCS 锁相结合后在 mmapbench 和 parallel find 上的吞吐量，功耗和能耗有效性(曲线为 lock-contention-aware scheduler)。默认系统和带有 MCS 锁的系统的实验结果也被表示出来以便于比较。从图中可以看到，对于 mmapbench，将 MCS 锁和本文提出的调度策略相结合比只使用 MCS 锁能够达到更好的可扩展性，因为 MCS 锁使用了复杂的原子指令而导致在锁激烈竞争时可扩展性不佳。对于 parallel find，将 MCS 锁和本文的调度算法相结合得到的可扩展性和单独使用 MCS 锁时基本相同。然而，结合之后消耗的功率却大幅度降低，因为本文的调度器将存在竞争关系的进程限制在 SSC 上，从而使得其它的处理器是处于空闲。因为与 MCS 锁结合之后的调度算法获得的吞吐量不会比只使用 MCS 锁得到的吞吐量低，则结合之后系统具有更好的能耗有效性。

## 3.5 相关工作

本章首先提出了基于离散事件仿真技术的模拟器以模拟锁颠簸现象，然后介绍了两种锁颠簸现象避免的方法，即基于等待者数目的锁和锁竞争感知的调度策略。相关工作分为两部分介绍，一部分是锁颠簸现象的模拟和建模，另外一部分是锁颠簸现象的避免方法。

### 3.5.1 锁颠簸现象的模拟和建模

自旋锁竞争对系统可扩展性的影响在很多文献中都被研究过。然而，绝大多数的模型或者模拟器均不能重现在锁激烈竞争时导致的锁颠簸现象。其根本原因是排号自旋锁的实现开销或者缓存缺失导致的硬件资源竞争没有被抽象和建模。下面对相关文献进行逐一分析。D.C.Gilbert 等建立了基于排队论的模型来刻画自旋锁的竞争对系统吞吐量的影响[48]。Bjorkman 等提出了循环算法来建模自旋等待的影响[52]。然而，这两种方法均不能重现出锁颠簸现象，因为模拟器或者模型中仅仅考虑了临界区的串行语义，但是并没有考虑排号自旋锁和缓存缺失带来的开销。

在后续的工作中，Bjorkman 等又提出了基于排队论的模型来建模自旋锁对系统可扩展性的影响[49]，并且在实际系统上对两种网络协议(TCP 和 UDP)操作密集的应用进行了验证。尽管该方法比循环算法考虑了更多的细节，比如缓存缺失带来的内存竞争等，但是作者将排号自旋锁抽象成可扩展自旋锁，因此，实际测量发现的锁颠簸现象同样无法通过该模型来重现。与这个工作有相同问题的是文献[74]以及文献[50]。以后者为例，作者详细的建模了缓存缺失导致的性能影响，然而锁实现的细节却未被考虑到。

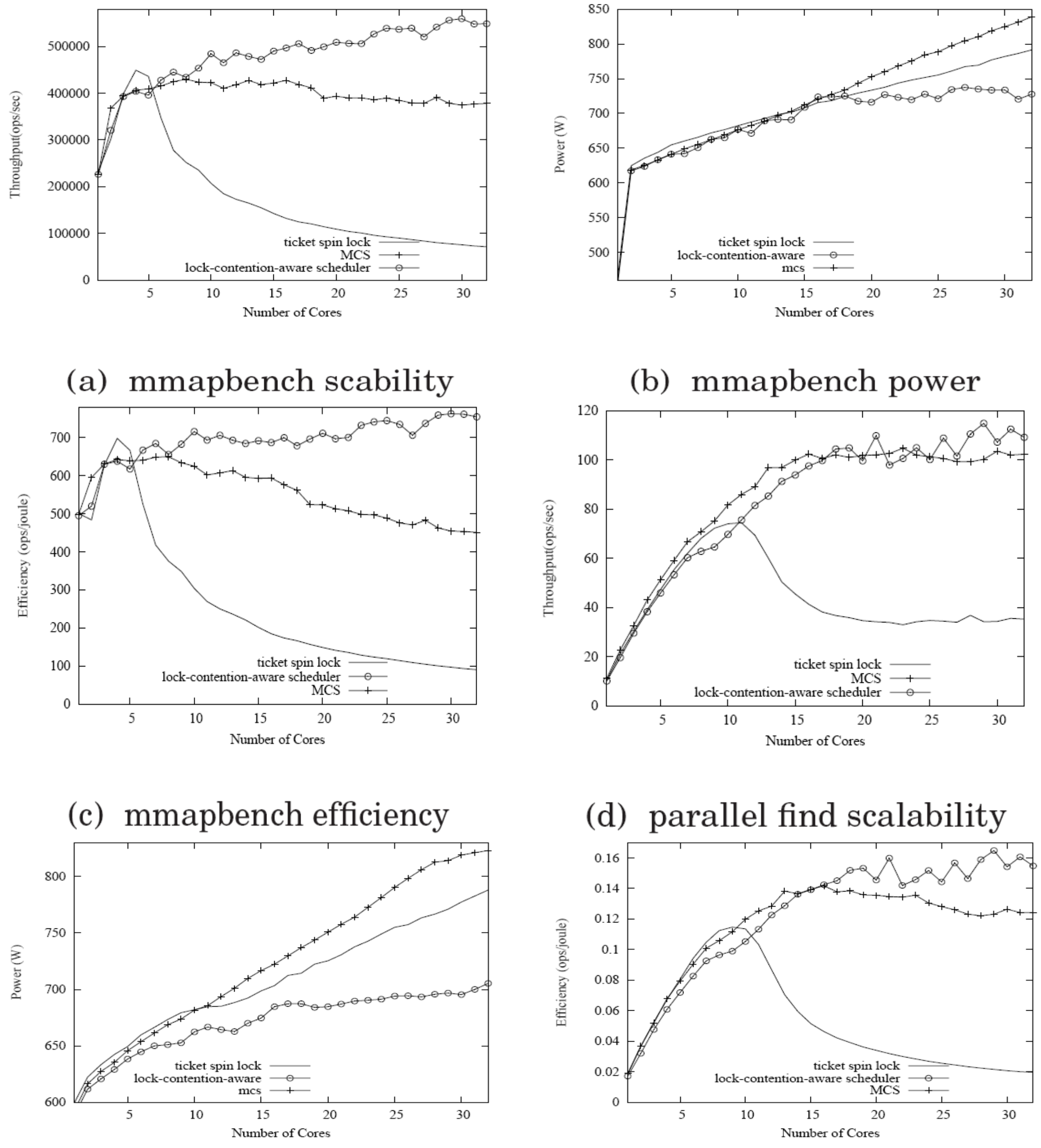

(a) mmapbench scability (b) mmapbench power

(c) mmapbench efficiency (d) parallel find scalability

(e) parallel find power (f) parallel find efficiency

图 3.39 mmapbench 和 parallel find 在默认系统，带有 MCS 的系统和带有锁竞争感知调度策略系统上的吞吐量，功耗和能耗有效性

在 OLS 2012 上的一个工作[53]考虑了排号自旋锁的实现细节，并且提出了基于马尔科夫链的模型来对锁颠簸现象进行重现。该工作还在实际的系统上验证了该模型的准确性并且取得了较好的结果。然而，基于马尔科夫链的方法只适合于建模锁的数目较少的情况，否则其计算复杂度将是不可忍受的。另外，临界区内部和非临界区内部的缓存缺失带来的硬件资源竞争并没有在模型中被考虑到。与这些工作相比，本文提出的基于离散事件仿真技术的模拟器不但考虑了临界区的串行执行，同时考虑到了锁实现和缓存缺失带来的资源竞争问题。因此，本文的方法更加真实的反映出了实际系统的行为。通过对模拟器的实验和评价，锁颠簸现象可以准确地被重现。

### 3.5.2 锁颠簸现象的避免方法

为了避免锁颠簸现象，学术界已经提出了很多方案。互斥锁使得每个锁等待者进入睡眠状态而非不断的自旋。当锁被激烈的竞争时，使用互斥锁会得到比排号自旋锁更好的可扩展性。然而，基于睡眠的同步原语的缺点是每次锁获取和释放时，都需要进行上下文切换，但是这种开销是很大的。自适应锁可以在自旋和睡眠之间取得折中。在大部分的现代操作系统内部都已经采用了自适应锁[29-30, 54]。自适应锁中，自旋和睡眠的时机往往是经验决定的，因此，很难达到其最佳性能潜力。自适应锁也在实时系统中被用于降低延时[75]。其使用的等待策略是在睡眠之前自旋固定的次数，而次数的多少是可以配置的。Gupta 等也提出了一种自旋锁，其思想是在睡眠之前自旋固定的时间[76]。与这些自适应锁相比较，基于等待者数目的锁使用等待者的数目来决定是否自旋或者进入节能状态。Boguslavsky 等利用排队论分析了自旋和睡眠的开销并且提出了自旋和睡眠切换的最佳策略[77]。与该文献不同，本文在实际的系统上探索了自旋和睡眠(将进入节能状态看成是轻量级睡眠)的最佳折中。

可扩展自旋锁[55-56]同样可以被用于避免锁颠簸现象。然而，可扩展自旋锁使得所有的锁等待者以自旋方式等待，因此其资源使用率和能耗有效性均不高[57]。

另外，可扩展自旋锁在实现中使用了复杂的原子指令，因此,当锁被竞争激烈时，锁颠簸现象同样会出现。而且，基于等待者数目锁中的轻量级睡眠机制以及锁竞争感知的调度策略均可以和可扩展自旋锁共同使用。

使用细粒度锁是在多处理器系统和多核系统上降低锁竞争的一种有效方式。不幸的是，随着核数的增加，这种方法显得更加困难和耗时[47]。本文提出的两种锁颠簸现象避免算法和细粒度锁是正交的关系。

事务内存避免直接地使用锁，相关的工作[78]已经证明了它在性能和可编程性

上的有效性。然而，当临界区被激烈竞争时，锁颠簸现象同样会出现(事务内存的临界区和传统的不同,因为代码中并没有显示的锁操作，但竞争同样存在)。

利用调度策略解决锁颠簸现象已经被 Xian 等使用过[79]。在这个工作中，作者提出利用锁竞争感知调度器降低多线程 Java 程序的锁竞争。本文提出的调度算法和文献[79]最大的不同是目标应用。对于他们的调度器，其目标应用是异构的 Java 应用，而核心的挑战是如何根据使用锁的信息对线程进行聚类。然而，对于本文的锁竞争感知调度器，其目标应用是使用内核锁较多的同构负载。其最大的挑战是如何确定 SSC 中最优的核数。与锁竞争感知调度相关的另外一个工作是文献[80]。作者根据任务之间共享数据的关系将系统中所有的线程分类以减少芯片之间的访问延时。尽管锁数据的共享也会导致芯片之间的访问延时，这种方法并不能提高用锁密集类同构应用的可扩展性，因为该方法仅针对异构负载有效。早期的锁竞争感知调度策略受到了加速临界区(ACS)[82]和 Corey 操作系统中[45]kenrel core 抽象的影响。ACS 方法通过在高性能的处理器上执行临界区提升系统可扩展性。当遇到临界区时，计算在低性能的核和高性能的核之间来回迁移。在 Corey 操作系统中，kernel core 抽象允许应用指定核来管理系统调用和其它核的硬件驱动请求。计算的迁移是通过共享内存的 IPC 来实现的。然而，这些方法并不适用于本文的情况，因为频繁迁移的开销几乎无法忍受。本文中的竞争感知调度器根据吞吐量进行迁移，这种迁移比在每个临界区前后进行迁移的开销要小的多。实际上，本文的锁竞争感知调度策略属于非预留调度，这种调度策略可以将某些 CPU 资源空闲，尽管仍然有任务等待处理[82-83]。除了自旋锁造成的吞吐量下降现象可以由非预留调度解决，这种方法同样可以适用于其它原因造成的吞吐量下降。例如，Fedorova 等利用非预留调度算法避免了 SMT 上由于缓存竞争造成的吞吐量下降[84]。

## 3.6 本章小结

本章工作对锁颠簸现象进行了分析和解决，主要分为 2 个部分。第一，提出了基于离散事件仿真技术的锁模拟器。与已有工作相比，该模拟器不但考虑了临界区的串行执行语义，同样考虑了排号自旋锁实现的开销和缓存缺失带来的硬件资源竞争。通过对该模拟器的评价发现锁颠簸现象可以被准确地重现。第二，提出了基于等待者数目的锁和锁竞争感知的调度策略以避免锁颠簸现象。基于等待者数目的锁根据等待者的数目以决定等待者自旋还是进入节能状态，而锁竞争感知的调度策略将用锁密集类进程限制在部分处理器上执行。两种策略均在 Linux 中实现，并且在多核系统上进行了评测。对于大部分遭受锁颠簸现象影响的应用，两种方法均比现有工作能够达到更好的可扩展性和能耗有效性。

# 第4章 共享硬件资源竞争避免

## 4.1 本章引言

与单核处理器和传统的对称处理器(SMP)不同，多核系统上的最后级缓存(LLC)是共享的。这种共享意味着除了最后级缓存，许多微架构硬件资源(如内存控制器，最后级缓存的预取硬件)等都是共享的。然而，共享这些性能关键的硬件资源是一把“双刃剑”。一方面，共享提升了硬件资源的利用率，进而提高计算机系统的总体性能[85]；另一方面，这些硬件资源可能同时被多个核并行访问而产生额外开销，导致计算机系统的总体可扩展性及独立程序的性能均会受到影响。这种资源竞争问题在强调性能隔离和要求服务质量(QoS)的计算环境下[86-87]将变得尤为严重。本文通过同时执行 NAS-SER[88]测试程序集中的 2 个测试程序，可以说明这种资源竞争问题的严重性。这 2 个测试程序分别运行在共享最后级缓存的 2 个核上，硬件架构的详细信息可以在后文的实验和评价一节中找到。为了避免由于测试程序不同执行时间带来的不确定性影响，本文采用 Tuck 等的方法[89]，将每个测试程序执行连续多次。仅当目标测试程序的全部执行时间均被背景测试程序(与目标应用共同运行，并且共享最后级缓存)的执行时间覆盖时，才搜集目标程序的时间。

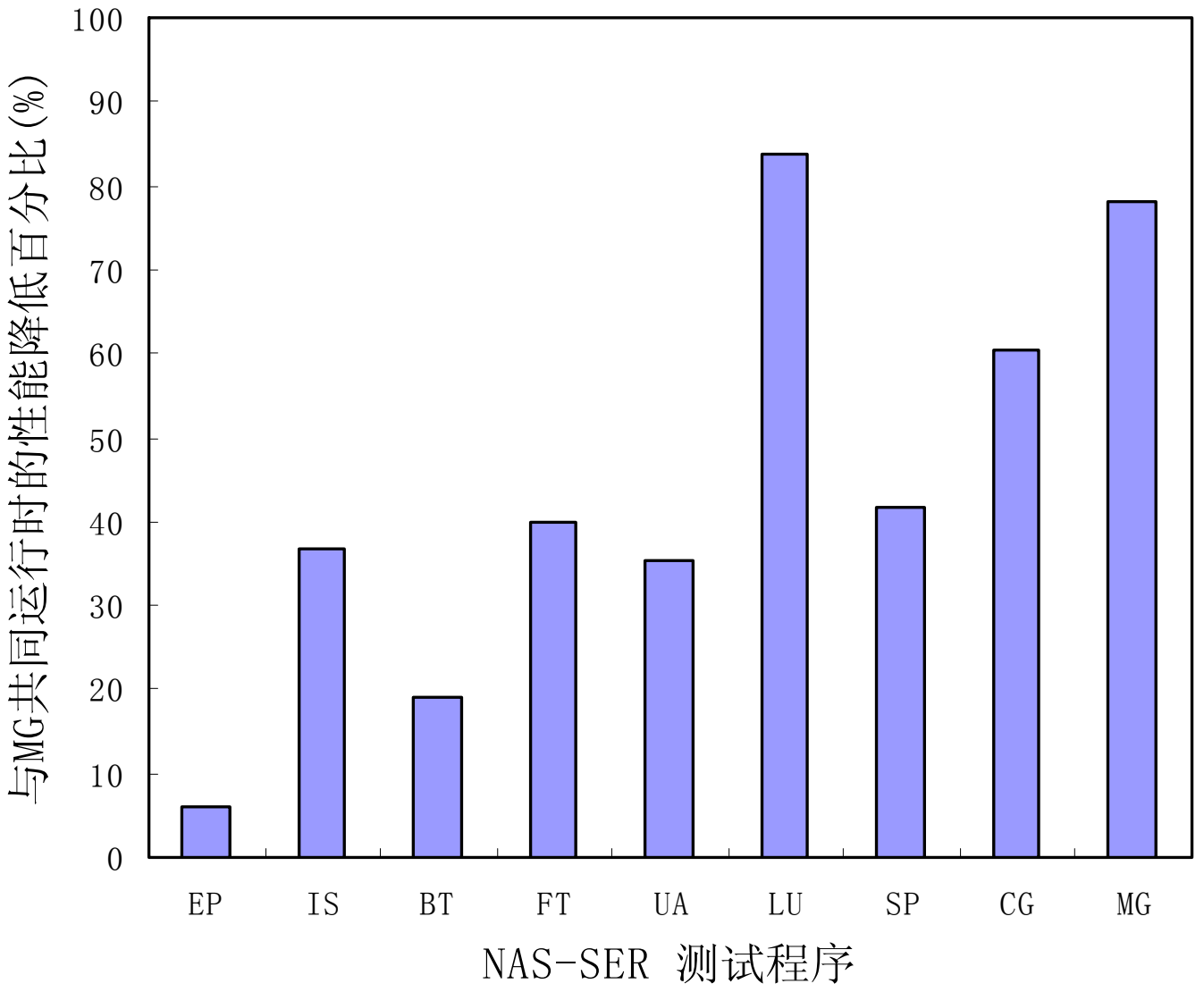

**图 4.1 与测试程序 MG 共享缓存造成的性能降低百分比。性能降低是由执行时间增加的百分比来衡量的。所有 NAS-SER 程序都使用规模 A**

图 4.1 表示出了 9 个 NAS-SER 程序(目标测试程序)与程序 MG(背景测试程序)共同执行时，每个 NAS-SER 程序执行时间增加的百分比(即性能降低程度)。从图中可以观察到 2 个现象:1、当与 MG 共同运行时，每个测试程序都遭受了性能影响。

2、这种性能降级程度会大幅度的变化，最小的有 6%，而最大的达 84%。目标测试程序遭受的性能降级是由共享的硬件资源竞争所导致的。其中共享的硬件资源不仅包括最后级缓存，也包括最后级缓存中的预取硬件，前端总线(FSB)和内存控制器等。每种硬件资源的竞争均会反映在性能降低的数据中。

性能降低程度的变化是因为不同的测试程序具有不同的资源需求。例如，MG 是一个内存密集类程序。如果它与另外一个内存密集类程序(例如，LU)同时运行，那么，二者将激烈地竞争共享的硬件资源。然而，如果选择一个计算密集类程序(例如 EP)和 MG 同时运行，这 2 个程序的总体硬件需求则更容易被满足。

为了解决当多个程序共享最后级缓存而引入的共享资源竞争问题，本章提出了资源竞争感知的调度策略。具体而言，本文首先使用现代处理器上广泛存在的性能计数器(PMU)获取一个应用程序的资源需求信息。基于这种信息，操作系统通过调整调度顺序和迁移核间任务，使得具有互补资源需求的任务可同时执行。尽管将互补任务同时执行的调度策略在很多文献中[11-13,15,90]已经被提出，本文的工作和以前的工作相比具有 4 个显著的不同。

1、本文提出了一种选择启发式性能指标的方法。该启发式指标利用性能计数器在线获取，其值可以代表应用程序对共享硬件资源的需求程度。本文的启发式指标的选择过程不仅考虑到了指标描述程序资源需求的准确性，而且也考虑到了在线获取该指标的稳定性。尽管已有的几个相关研究工作也利用性能计数器来获取程序的资源需求[11-14]，但是其选择的过程是基于经验获得的。文献[15]中描述了一个选择启发式指标的系统方法。在这个工作中，作者发现最后级缓存缺失率是比最后级缓存访问率和每时钟周期内的指令数更好的启发式指标。然而，该工作仅考虑了用启发式指标描述程序资源需求的准确性，没有考虑稳定性因素。由于指标是在线获取的，对共享资源的竞争会使得实验的结果比单独执行某一程序时大很多。本文在科学计算应用(NAS-SER)上的实验结果表明，每周期的总线事务数是代表程序共享资源需求的最准确指标，然而，当加入了指标稳定性的考虑之后，最后级缓存的访问率是本文最终的选择。

2、很多现有的工作通常仅考虑通过调整任务到核的重新映射来减少资源竞争[11,13,15-18]。然而，当系统中的进程数多于核数时，不仅需要考虑核间负载均衡的方法，而且需要考虑在每个核上选择哪个任务运行。与以前的工作不同，本文将这两部分相结合以提升系统总体的可扩展性和性能。因此，本文的资源竞争感知调度策略具有更广泛的适应性。除此之外，将任务映射到核时，很多工作依赖于中心化的排序策略[14-17,90]。然而，在实际的多核系统中，这种方法是不可扩展的，尤其是当任务的数目较大时。相反地，本文实现的在 Linux 内核中的调度算法是分布

式的。

3、本文提出的资源竞争感知调度算法实现在 3 个最主流的操作系统调度器(即 CFS, RSDL 和 O(1))中,并进行了详细的评测和比较。现有工作中, 大部分只在一个调度器中进行了验证[12, 15, 91-92]。因为每个调度器均被广泛的使用, 在这些调度器中加入资源竞争感知的功能具有重要的意义。而且, 调度器的差异会导致资源竞争感知功能在实现上的不同。一个有趣的例子是饥饿问题在 CFS 调度器中需要被显式地避免, 而在 RSDL 和 O(1)调度器中, 该问题可以被默认的调度机制避免。因此, 本文的工作在如何将资源竞争感知功能融合进主流调度器方面做得更加全面。

4、本文对资源感知调度器在实际多核系统上的可扩展性和性能进行了深入的分析。通过分析,本文提出的竞争感知调度策略可以有效的避免资源竞争。具体而言, 独立任务的执行时间可以被降低至 21%, 而系统的可扩展性可以提升达 13%。同时, 对调度器的深入分析也发现了一些有趣的现象。例如, 加入了资源竞争感知功能的 O(1)调度器会周期性地将一个核上的一些任务迁移到另外一个核上, 尽管 2 个核的负载是均衡的。而且, 本文发现缩短连续两次上下文切换的间隔(即时间片长度)有助于提升带有资源感知功能的 O(1)调度器和 RSDL 调度器的总体性能。

本文的剩余部分进行如下组织。4.2 节对强度和敏感度进行了量化定义, 并从强度和敏感度的角度来分析科学计算应用。4.3 节介绍选择启发式指标的方法, 指标的获取方式和利用该指标指导调度的方式。4.4 节介绍用于评测的测试程序, 测试程序的运行方式以及调度开销, 系统可扩展性的分析。4.5 节介绍相关工作, 而 4.6 节对本章工作进行总结。

## 4.2 共享资源竞争分析

本节借助强度和敏感度的概念对科学计算应用进行分析。

### 4.2.1 定义

使用性能降低的程度(即执行时间增加百分比)可以构造出一个性能降级矩阵 D。矩阵元素 $d_{i,j}$ 代表与任务 $i$ 共同运行时, 任务 $j$ 的性能降级,

$$d_{i,j} = (T_{i,j} - T_j)/T_j \quad 4\text{-}1$$

其中 $T_{i,j}$ 和 $T_j$ 分别代表应用 j 和应用 i 共同执行以及应用 j 独立执行的时间。对于构造出的矩阵 D, 第 i 行代表每个应用与应用 i 一起运行时的降级程度。如果第 i 行的数据较大, 那么, 应用 i 在资源竞争上对其它进程形成的影响也较大。矩阵的第 j 列代表应用 j 受所有其它应用的影响程度。如果第 j 列的元素相对较大, 那么, 应用 j 对资源竞争表现的较为敏感。实际上, 矩阵的每一行和列均可以被看做是一

个向量。而向量的模被用于量化强度和敏感度的大小。等式 4-2 中的$I_i$和$S_j$表明如何从矩阵 D 中计算应用 i 的强度和应用 j 的敏感度。

$$I_i = \sqrt{\sum_j {d_{i,j}}^2} \quad S_j = \sqrt{\sum_i {d_{i,j}}^2} \qquad 4\text{-}2$$

### 4.2.2 分类

应用可以利用测量得到的强度和敏感度信息进行分类。具体而言，如果一个应用的强度和敏感度相对较大，该应用倾向于具有较大的工作集并且在运行时会占用较多的共享资源。相反地，如果一个应用的强度和敏感度均较小，那么，该应用使用的工作集较小，而且消耗的共享资源也较少。如果一个应用的强度较大,但是敏感度较小，那么，这样的应用被称为“流式应用”，对其分配更多的共享资源不会带来总体性能的提升。

表 4.1 每个测试程序在默认配置下以及和其它程序共同运行时的执行时间。第一列代表所有的背景应用，而第一行代表所有的目标应用。执行时间的单位是秒。例如，第二行的数值 111.01 代表与 EP 共同运行时 UA 的执行时间是 111.01 秒

| | EP | IS | BT | FT | UA | LU | SP | CG | MG |
|---|---|---|---|---|---|---|---|---|---|
| **EP** | 41.32 | 1.97 | 152.15 | 9.60 | 111.01 | 151.51 | 109.68 | 4.12 | 4.73 |
| **IS** | 41.22 | 2.14 | 157.37 | 10.01 | 114.61 | 165.61 | 117.20 | 4.58 | 5.07 |
| **BT** | 41.58 | 2.10 | 163.53 | 10.35 | 116.56 | 168.78 | 118.70 | 4.61 | 5.19 |
| **FT** | 43.00 | 2.21 | 167.85 | 11.66 | 125.09 | 197.59 | 127.62 | 4.94 | 5.63 |
| **UA** | 42.05 | 2.36 | 168.14 | 11.45 | 130.37 | 204.95 | 133.43 | 5.42 | 6.47 |
| **LU** | 41.78 | 2.30 | 164.82 | 11.19 | 132.30 | 217.39 | 133.10 | 5.43 | 6.51 |
| **SP** | 42.57 | 2.49 | 177.77 | 12.26 | 138.27 | 232.29 | 142.18 | 5.74 | 7.04 |
| **CG** | 43.27 | 2.79 | 182.58 | 13.12 | 133.77 | 222.88 | 147.25 | 6.23 | 7.19 |
| **MG** | 43.76 | 2.67 | 179.13 | 13.19 | 144.95 | 258.14 | 149.99 | 6.29 | 7.96 |
| -- | 41.28 | 1.95 | 150.60 | 9.43 | 107.20 | 140.44 | 105.94 | 3.92 | 4.47 |

本文的应用分类方法被用在 9 个 NAS 串行程序中(NPB3.3-SER)。所有的测试程序均使用问题规模 A，因为这种问题规模在使用的多核系统上展示出了各种各样的资源需求。为了得到 4.2.1 中的矩阵，每个测试程序不但单独运行，而且和所有的测试程序搭配运行。当将测试程序 i 和 j 共同运行以计算$d_{i,j}$时，测试程序 j 是目标应用而测试程序 i 是它的背景应用。这 2 个测试程序分别被绑定到共享最后级缓存的 2 个核上。为了避免不同应用的不同执行时间带来的影响，背景应用被执行很多次使得当目标应用从头执行到尾时，背景应用一直处于运行状态。每个测试程序在默认配置下以及和其它程序共同运行时的执行时间如表 4.1 所示。图 4.2 表示了 NAS 串行程序的强度和敏感度信息。从图中可见，强度和敏感度具有很强的相关性。另外，在 NAS-SER 程序集合中不存在流式应用。这样，应用程序的资源需求可以由强度或者敏感度来刻画。本文选择使用应用程序的强度信息来度量。

本文使用所有任务的平均资源需求作为一个应用程序是否是内存操作较为频

繁的应用的阈值。图 4.2 中，实心点代表具有较高资源使用需求的程序，而空心点代表具有较低资源使用需求的程序。分类的有效性可以通过图 4.1 中的实验结果进行验证。从图 4.1 中得到的一个结论是被分类成低资源需求的任务在与 MG 共同运行时受到的影响比其它的进程小。

基于强度和敏感度的任务分类方法仅是决定一个应用在所有目标应用中的相对资源使用情况。如果增加一个应用的问题规模，其共享资源需求也会随之提升。一个应用的强度和敏感度信息从性能降级矩阵中计算得到。虽然本文使用的性能降级矩阵是利用具有问题规模 A 的科学计算应用计算出的，但是强度和敏感度具有强相关性这一结论在其它测试程序集中也是成立的[101]。这一事实增强了作者对这一结论正确性的信心，并且相信在其它问题规模上，该结论也同样成立。在 4.3.1 节中，将提出最有效的指标来代替强度和敏感度，因为从性能降级矩阵计算强度和敏感度极其耗时，不利于在线获取。一旦找到最有效的指标，应用程序的资源需求可以方便地在线获取。

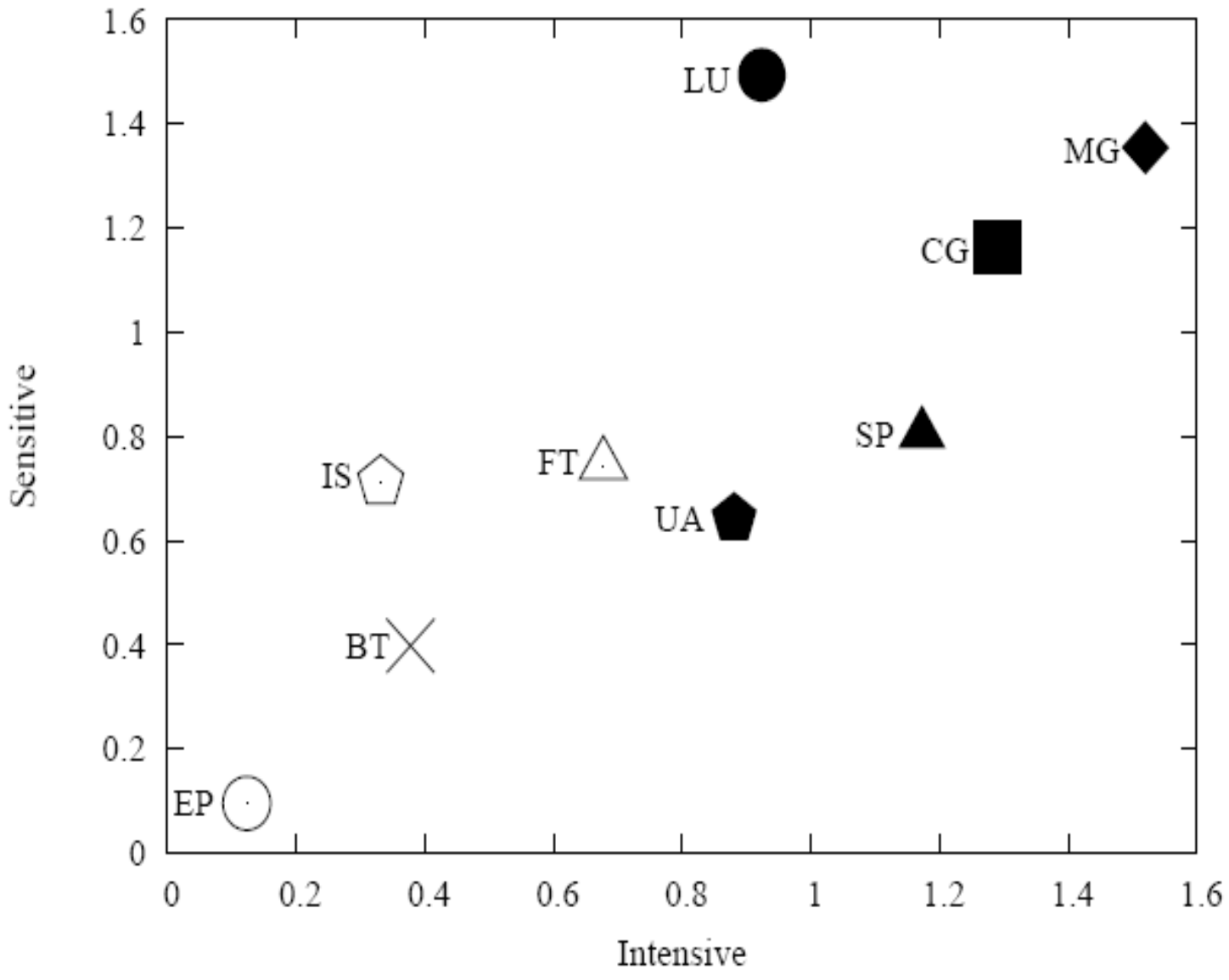

图 4.2 NAS 串行程序的强度和敏感度

### 4.2.3 资源竞争的相关性

利用强度或者敏感度衡量应用程序使用资源程度的一个考虑是指标的粒度可能过粗，因为很多共享资源同时都被程序竞争。然而，对于每种资源的竞争都存在着相关性。例如，激烈竞争最后级缓存的 2 个任务同样倾向于竞争前端总线和地址总线。这是因为最后级缓存的竞争将导致大量的缓存缺失，这些缺失需要通过前端总线和地址总线从其它的核上和或者内存中取指令或者数据来解决。这种由硬件逻辑建立的竞争相关性使得强度或者敏感度可以代表应用程序的资源需求。

## 4.3 竞争感知的调度策略

### 4.3.1 获取应用的资源需求

在 4.2 节中，本文提出利用强度信息来代表应用程序的资源需求。然而，获取强度的过程极为耗时，不能在线使用。因此，本文的竞争感知调度策略利用性能计数器近似应用程序的强度信息。而实现这一目标的最核心的问题是需要检测哪些硬件事件来近似强度信息。本文考虑以前工作中[12-15]最有效的 5 个性能指标。具体而言，这些指标分别是最后级缓存的缺失率，每周期的最后级缓存缺失率，最后级缓存的访问率，每时钟周期的停等周期，和每时钟周期的总线事务数。每个指标都在相关的工作中被考虑成度量程序使用共享资源多少的最有效指标。计算每个指标的公式如表 4.2 所示。接下来，本文将提出一种方法从 2 个候选指标中选择出最合适的指标。

实验系统的性能计数器被配置成同时检测 2 个事件，因为每个计数器只能检测一个事件[93]，而每个候选指标同时需要 2 个事件(见表 4.2)。对于一个候选性能指标 $k$，本文构造向量 $V_k$，向量的第 i 个元素是测试程序 i 在该指标上的取值。回顾一下，前文得到的结论是程序的共享资源需求可以通过强度信息来表示。但是强度信息不能在线获取，因此需要找到合适的启发式指标。使用每个程序的强度信息也可以得到一个向量 $I$。实际上，本文要寻找的指标 k 是使得对应的向量 $V_k$ 和 I 具有较大的相关系数的那个指标。

表 4.2 不同性能指标的相关系数，竞争稳定性和前两者的比率。从第 2 至第 6 行分别代表最后级缓存缺失率,每时钟周期的最后级缓存缺失率，每时钟周期的停等周期,每时钟周期的总线事务数和最后级缓存访问率。

| 指标 | 相关系数 | 稳定性 | 比率 |
|---|---|---|---|
| $\#LLC cachemisses/\#\text{instructions}$ | 0.69 | 227.20 | $2.49\times10^{-3}$ |
| $\#LLC cachemisses/\#cycles$ | 0.81 | 549.60 | $1.47\times10^{-3}$ |
| $\#stall cycles/\#cycles$ | 0.81 | 0.54 | 1.5 |
| $\#bus transactions/\#cycles$ | **0.99** | 293.48 | $3.37\times10^{-3}$ |
| $\#LLC accesses/\#instructions$ | 0.82 | **0.0099** | **82.83** |

除了指标的准确性，指标的稳定性同样需要考虑。这是因为选择的指标是在线测量的，因此资源竞争会大幅度地影响指标的测量值。如果选择的指标稳定性不好，那么，测量出的程序资源需求是不准确的。在本文中，指标的稳定性是指当应用由独立运行切换到与其它任务共同运行时，维持原测量值的能力。为了进一步说明指标稳定性的概念，本文这里举个例子。当使用 2 个性能指标($m_1$和 $m_2$)分别来衡量程序 A 在独立运行时的资源需求时，可以得到测量值 $v_1$和 $v_2$。当 A 与另外一个程序 B 共同运行时，同样使用 $m_1$和 $m_2$来衡量 A 的资源需求。这时，可以得到测量

值 $v_1^*$ 和 $v_2^*$。如果 $v_1^*$ 更加靠近 $v_1$，那么 $m_1$ 比 $m_2$ 的稳定性好。为了测量指标 k 的稳定性，通过使用指标 k 的取值增加的百分比来构造降级矩阵 $MD_k$。矩阵元素 $md_{i,j}^k$ 代表与任务 i 一起运行时，任务 j 在指标 k 上的增量。事实上，矩阵 $MD_k$ 中的元素与性能降级矩阵 D 中的元素计算方式相同。2 个矩阵唯一的不同是 D 中的元素是利用执行时间的增量百分比计算出来的，而 $MD_k$ 是利用指标 k 的增量百分比计算出来的。任务 j 对资源竞争的稳定性是利用矩阵第 j 列形成的向量 $A_k$ 的模来衡量的。而性能指标 k 对资源竞争的稳定性通过所有应用稳定性的平均值来衡量。为了确定最佳启发式指标，本文计算相关系数和竞争稳定性的比率。指标选择过程

$$K^* = \arg\max_k \frac{E_k}{S_k} \tag{4-3}$$

其中

$$E_k = \frac{I \times V_k}{\|I\| \times \|V_k\|} \qquad S_k = \frac{\sum_j \|A_k\|}{n} \tag{4-4}$$

表 4.2 表示了 5 个指标的相关系数，竞争稳定性以及前两者的比率。对于相关系数和比率而言，其值越大越好。相反地，对于稳定性而言，其值越小越好。如表所示，每时钟周期的总线事务数具有最大的相关系数，但是这个指标的稳定性不佳。在 5 个候选指标中，最后级缓存的访问率是最终的选择，因为它具有最大的比率，意味着该指标在准确性和稳定性上取得了较好的折中。

### 4.3.2 调度方法

调度器基于应用的资源使用需求进行调度。本文先考虑系统中只有 2 个处理器的情况。将其中一个作为主核，而另外一个作为从核。主核主要执行资源使用较多的任务，而从核主要执行资源使用较少的任务。如果系统中存在更多的核，所有的核两两一组被分成很多对。对于每一组，都存在一个主核和一个从核。

在任务的task_struct数据结构中增加一个域access_rate来代表任务的最后级缓存访问率。这个域的信息在每个时间片末尾被更新。本文之前的分析表明一个程序从开始运行到执行完毕之间的最后级缓存访问率被选作代表程序资源需求的启发式指标。然而，现代操作系统上的任务是以时间片为执行单位的。因此，这里的关键问题是如何利用程序的部分执行结果来近似整体的最后级缓存访问率。为了解决这个问题，本文尽量多地使用历史信息。一个任务在时间片 i 的资源需求

$$R_i = \sum_{j}^{i-1} S_j \Big/ \sum_{j}^{i-1} I_j \tag{4-5}$$

其中 $S_j$ 和 $I_j$ 分别代表时间片 j 内的最后级缓存访问次数和执行的指令。这样，

为了计算一个任务的 access_rate，在任务的 task_struct 数据结构中增加 2 个域 accesses 和 instructions 以分别记录任务的总体最后级缓存访问次数和执行的指令。如何更新任务的 access_rate 的伪代码如图 4.3 所示。

```
1)  /* p is scheduled on a core, ReadPMU() reads a
    performance monitor unit */
2)  a0 = ReadPMU(pmu0);
3)  i0 = ReadPMU(pmu1);
4)  /* p runs out of the time slice */
5)  p → accesses += ReadPMU(pmu0) - a0;
6)  p → instructions += ReadPMU(pmu1) - i0;
7)  P → access_rate = p→accesses/p→instructions ;
```

**图 4.3 更新任务的 access_rate 代码**

基于任务 access_rate，需要修改操作系统内核以支持资源竞争感知调度策略。本文的改动体现在 2 个方面。1 当在从核上发生上下文切换时，需要调整从核上任务的调度顺序。具体而言，在从核上搜索一个任务使得该任务和主核上当前执行任务的 access_rate 总和接近一个中间值。从核上的搜索过程要在前 maxcnt 个任务上进行。为了计算每个运行队列的最后级缓存访问率的平均值，每个核上的运行队列需要维护 2 个域(即 total_access_rates 和 total_counts)。这 2 个域的值均在上下文切换时被更新。total_access_rate 变量需要累加当前执行任务的 access_rate 而 total_counts 增加 1。使用这 2 个值，可以计算出每个队列上的平均最后级缓存访问率( $rq\text{-}>total_access_rates/rq\text{-}>total_counters$ )。而调度器中需要的平均值被设置为两个队列上的平均最后级缓存访问率的总和。调整默认调度顺序的伪代码如图 4.4 所示。2 在负载均衡结束时，在主核上重新进行任务到核的映射。这种重新映射是通过对 Linux 内核函数 run_rebalance_domains()进行改写而完成的。具体而言，如果一个任务不在 CPU 上执行并且其 CPU 掩码集合包含目标 CPU， 那么，这个任务则被定义成”允许迁移”任务。对于每次负载均衡，主核寻找并且交换 2 个运行队列上的一对”允许迁移”任务。这种交换的触发条件是在主核上存在的一个”允许迁移”任务的 access_rate 小于中间值，而同时存在从核上的一个”允许迁移”任务的 access_rate 大于中间值。在 2 个核互换任务时，2 个核上的运行队列是保持均衡的。其伪代码如图 4.5 所示。

当使用多于 2 个核时，将有多对主核和从核。一个核是否是主核或者从核是由共享最后级缓存的核数来决定的。例如，8 个核共享最后级缓存，那么，只需确保 4 个核是主核而另外 4 个是从核。如果芯片的数目多于一个，在每个芯片内部分配主核和从核的方式是相同的。例如在本文使用的八核系统上(图 4.7) 共有四对主核和从核 (即 C0 和 C4, C1 和 C5, C2 和 C6, C3 和 C7)。这样，算法 4.3, 4.4, 4.5 会在每

对主核和从核内执行。

```
m = rq0→total_access_rates/rq0→total_counts +
        rq1→total_access_rates/rq1→total_counts ;
a = GetCurrTask(core_m)→ access_rate;
min = +∞;
for i = 1 to maxcnt do
    p = GetNextTask(core_s);
    d = CalculateDelta(p → access_rate + a, m);
    if d < min then
        min = d; minp = p;
    end if
end for
return minp;
```

图 4.4 从核上调整默认调度顺序

```
m = rq0→total_access_rates/rq0→total_counts +
rq1→total_access_rates/rq1→total_counts;
p = GetCurrTask(core_m);
while p != NULL do
    q = GetCurrTask(core_s);
    while q != NULL do
        If p → access_rate < m/2 and AllowToMigrate(p, core_s)
                and AllowToMigrate(q, core_m) and q→access_rate > m/2  then
            ExchangeTask(p, q);
            Break;
        end if
        q = GetNextTask(core_s);
    end while
    p = GetNextTask(core_m);
end while
```

图 4.5 主核上重新进行任务到核的映射

除此之外，本文需要一种机制在每对主核和从核之间进行任务迁移以达到竞争感知调度的全部潜力。一种简单的方法是将所有的任务根据资源使用需求降序排序。在每轮的算法迭代中，将当前的未分配任务链表中前 n 个和后 n 个任务置于一对核上。n 的选择使得所有核上的负载都是均衡的。这个方法的缺点是它依赖于中心化的排序，从而可扩展性不好。本文的解决方案是在负载均衡结束时对连续两对核的主核之间交换任务，同时在从核之间也执行这种操作。通过控制任务交换的方式，可以保证最后的结果和依靠中心化排序得到的结果是相同的。在 2 个主核间交换任务的伪代码如图 4.6 所示，在此省略 2 个从核间交换任务的伪代码。

本文的资源竞争感知调度在 CFS，RSDL 和 O(1)3 个调度器中进行实现。涉及的 Linux 内核版本有 Linux2.6.32(CFS)和 Linux2.6.21.7(RSDL 和 O(1))。默认的

Linux2.6.21.7 内核使用 O(1)调度器。为了在 RSDL 调度器中实现资源竞争感知调度，本文使用 RSDL 调度器的补丁[94]。

```
/* core_m1 should run tasks with larger resource
   requirements than core_m2 */
p = GetMinAccessRateTask(core_m1);
q = GetMaxAccessRateTask(core_m2);
if p != NULL and q != NULL and p → access_rate < q → access_rate then
    ExchangeTask(p, q);
end if
```

图 4.6 2个主核间交换任务

尽管将资源竞争感知功能整合进调度器中的设计原则是相同的，不同的调度器会使得竞争感知功能的实现有所区别。一个问题是如何在不同的调度器中避免饥饿。RSDL 和 O(1)的默认调度机制保证了增加资源竞争感知功能之后的调度器不会出现饥饿现象。具体而言，这 2 个调度器将所有的任务组织在 2 个独立的队列中，一个队列中的任务是等待处理的任务(活跃数组)，而另外一个队列中的任务是已经处理了的任务(超期数组)[33,95]。仅当活跃数组中没有任务执行时，超期数组中的任务才可能重新被调度。尽管 RSDL 和 O(1)对 2 个数组的切换时机有所不同，但是这种切换机制保证了饥饿不会存在。而对于竞争感知调度功能，调度顺序的调整仅仅是针对超期数组。因此，改进后的调度算法同样不存在饥饿问题。

然而，CFS 调度器中，所有的任务以虚拟执行时间被组织在一棵红黑树中[96]。尽管在默认的 CFS 调度器中并不存在饥饿问题，但是加入了资源竞争感知功能的调度器可能会有饥饿问题，因为任务是基于资源使用情况进行调度而非虚拟时间。为了克服这个问题，在 CFS 中增加一个阈值以判断具有最小虚拟执行时间的任务是否已经等待了足够长的时间。如果是这样的，则立即运行该任务，否则资源使用情况是最高优先级的指标，任务的调度要使得共享资源均衡使用。

### 4.3.3 讨论

本文提出的调度器解决的是多个串行程序导致的硬件资源竞争问题。然而，程序可能利用多线程或者多进程来提升性能。因此，当多个并行程序同时运行时，一个自然的考虑是提出的调度策略是否依然有效。事实上，该问题很难一般性地回答，因为在该场景中涉及到了复杂的数据共享模式。具体而言，应用内部可能存在数据共享，这样在共享最后级缓存的核上运行共享数据的线程会提升线程之间的通信速度。除此之外，应用内部和应用之间也存在着资源竞争。实际上，本文的调度策略是否在新的场景下依然有效取决于数据共享和资源竞争的程度。然而，在绝大多

数情况下，资源竞争是影响系统总体可扩展性的最主要因素[14,90]。因此，本文的调度方法在新的场景下依然有效。

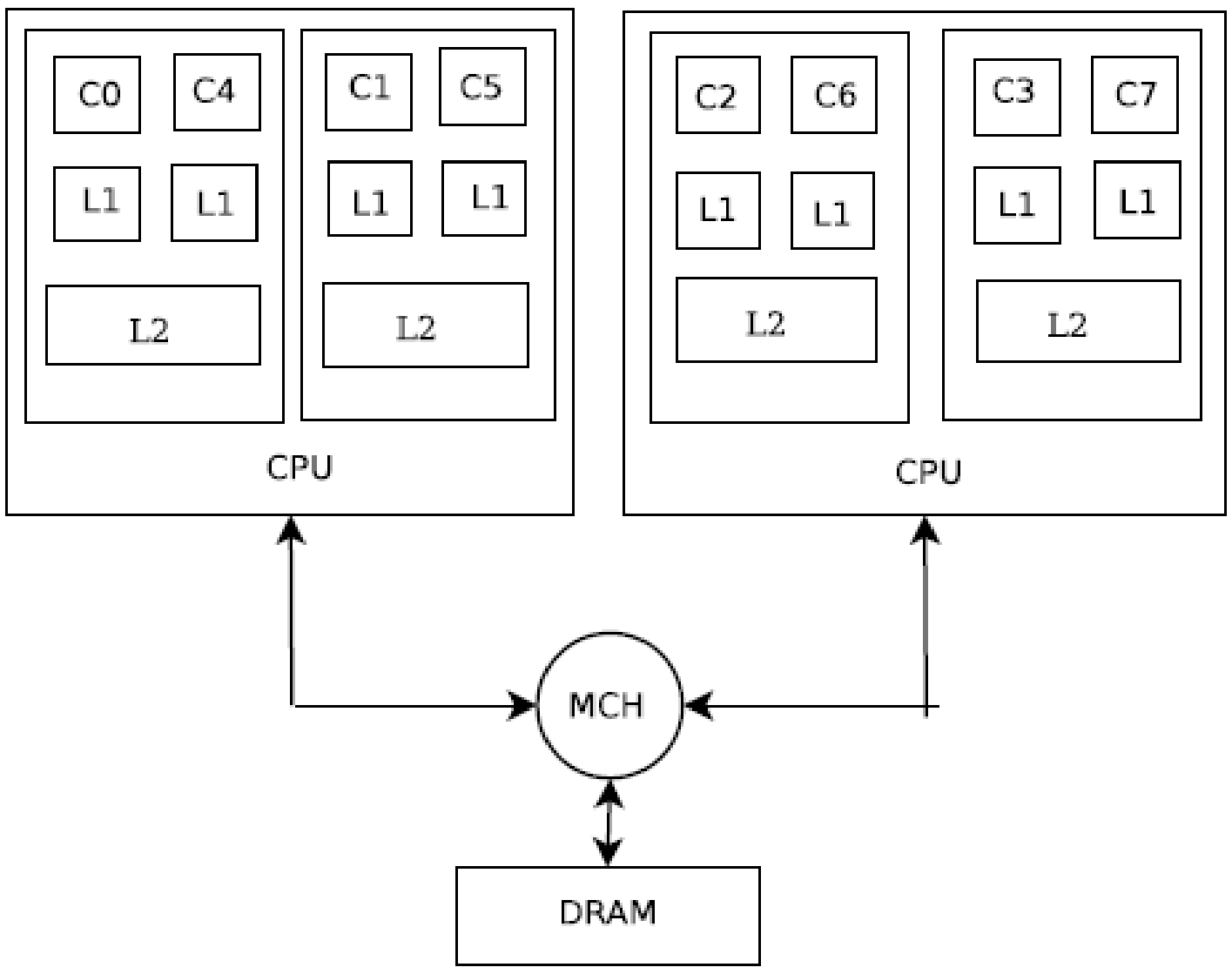

图 4.7 实验中使用的 Intel 八核系统架构图，带有 C 标记的方块代表处理器，带有 L1 标记的方块是 L1 指令和数据缓存，带有 L2 标记的是 L2 缓存，带有 DRAM 的方块是内存，带有 MCH 标记的圆代表内存控制器

## 4.4 实验和评价

### 4.4.1 实验平台

所有的实验均是在 Intel 8 核平台上进行的。这个架构的平台如图 4.7 所示。系统中有 2 个 Xeon E5310 芯片，每个芯片上有 4 个核。每个核拥有私有的 L1 指令和数据缓存，其大小均为 32K 字节。每个芯片上有两个 L2 缓存，每 2 个核共享一个 L2 缓存。L2 缓存的大小是 4M 字节。每个芯片上的 4 个核与 4G 内存通过前段总线相连。处理器和总线的频率分别是 1.6GHZ 和 1.033GHZ。正如从图 4.7 看到的，每个最后级缓存由 2 个核共享。而本文的竞争感知调度算法对核数的最小要求是 2(主核和从核)。如果核的数目多于 2，那么，需要构造多个主核从核对。作为一个例子，如果系统中有 4 个核(C0, C4, C1 和 C3), C0 和 C1 可以作为主核，而剩余的核是从核。这样，系统中存在两对主核从核对。由于 2 个核共享最后级缓存的情况是多个核共享多个缓存的基本组成部分，本文的评测方法是先在 2 个核的情况下研究竞争感知调度策略的有效性，其次，使用更多的核进一步验证。因此，除非特殊指明，后文的实验结果是在双核上收集的。

对本文使用的多核平台的一个考虑是很多处理器已经开始使用共享的 L3 缓存(如 AMD Opteron 处理器)，并且 L3 缓存的大小比本文硬件平台的 L2 缓存大很多。

然而，共享最后级缓存的核数也有增加，并且最后级缓存容量的增长速度赶不上核数的增长速度[122]。因此，每个核分配到的最后级缓存容量实际上正在逐渐缩小。由于这个原因，资源竞争问题在具有较大 L3 缓存的处理器上反而会更加明显。因此，本文提出的调度策略同样适用于这种处理器。

### 4.4.2 测试程序

NAS 串行程序集 NPB3.3-SER 被用于验证资源竞争感知调度的正确性。在 NPB3.3-SER 中有 10 个串行程序。所有的串行程序均是计算密集类程序，而一个异常是测试程序 DC。在该程序运行时，可以观察到大量的 I/O 操作。这样，本文的研究不将 DC 考虑在内，因为关注的是计算密集类应用。每个程序有 5 个问题规模(S,W,A,B,C)。本文使用规模 A 作为输入，因为使用该规模的测试程序在图 4.7 的平台上展示出了各种各样的资源使用需求。使用较大的问题规模将使程序的内存操作大大增加。在本文的实验平台上，作者已经尝试运行由更大问题规模的测试程序组成的负载。但是，大部分负载在运行时会导致系统无任何响应，这是因为内存子系统的压力过大而导致的。读者可以参考文献[97]来了解每个测试程序的详细描述。

本文没有选择著名的 SPECCPU 测试程序集，因为本文关注的是科学计算应用导致的硬件资源竞争问题。本文研究的一个可应用的场景是云计算环境。在该场景中，多个应用运行于同一个系统上。而且,最近的研究表明，科学计算可以在云上实现，因为云计算环境使得获取强有力的计算资源更加容易和方便[123]。因此，提出的调度器可以在该环境下解决资源竞争问题。

本文的目标是在多道程序(multiprogramming)执行环境下避免共享资源的竞争问题。其中多道编程环境是指很多串行程序同时运行在一个多核系统上。在现代的计算机系统上，核数越来越多。然而，软件只有有限的并行性。为了利用多核系统的计算资源，一个直接的方法是将多个串行程序同时运行在一个系统上。本文选择多道程序环境作为目标，因为本文展示出的资源竞争问题可以容易的在实际场景中被重现 (例如，当浏览网页时听音乐)。

### 4.4.3 调度参数和运行方法

从核上搜索的任务数目(即 maxcnt)在 3 个调度器中均被设置成 10。对于带有资源竞争感知功能的 CFS 调度器，用于避免饥饿的阈值被设置成 50,000,000。当执行一个负载时，每个测试程序需要不断执行以避免不同测试程序的不同执行时间。具有最长执行时间的任务被连续的执行 5 次后停止测量。

### 4.4.4 调度开销

调度开销通过同时运行多个相同程序的多个实例来测量。通过这种方式，本文的调度算法不会好于默认的调度算法，因为所有测试程序的资源需求均相同。对于资源竞争感知调度器而言，在每次上下文切换时，需要在从核上搜索 maxcnt 个任务和主核上的任务匹配。每次负载均衡时，需要对任务到核的映射进行重新调整。默认调度器和竞争调度器的时间差异即为本章提出的调度策略的开销。

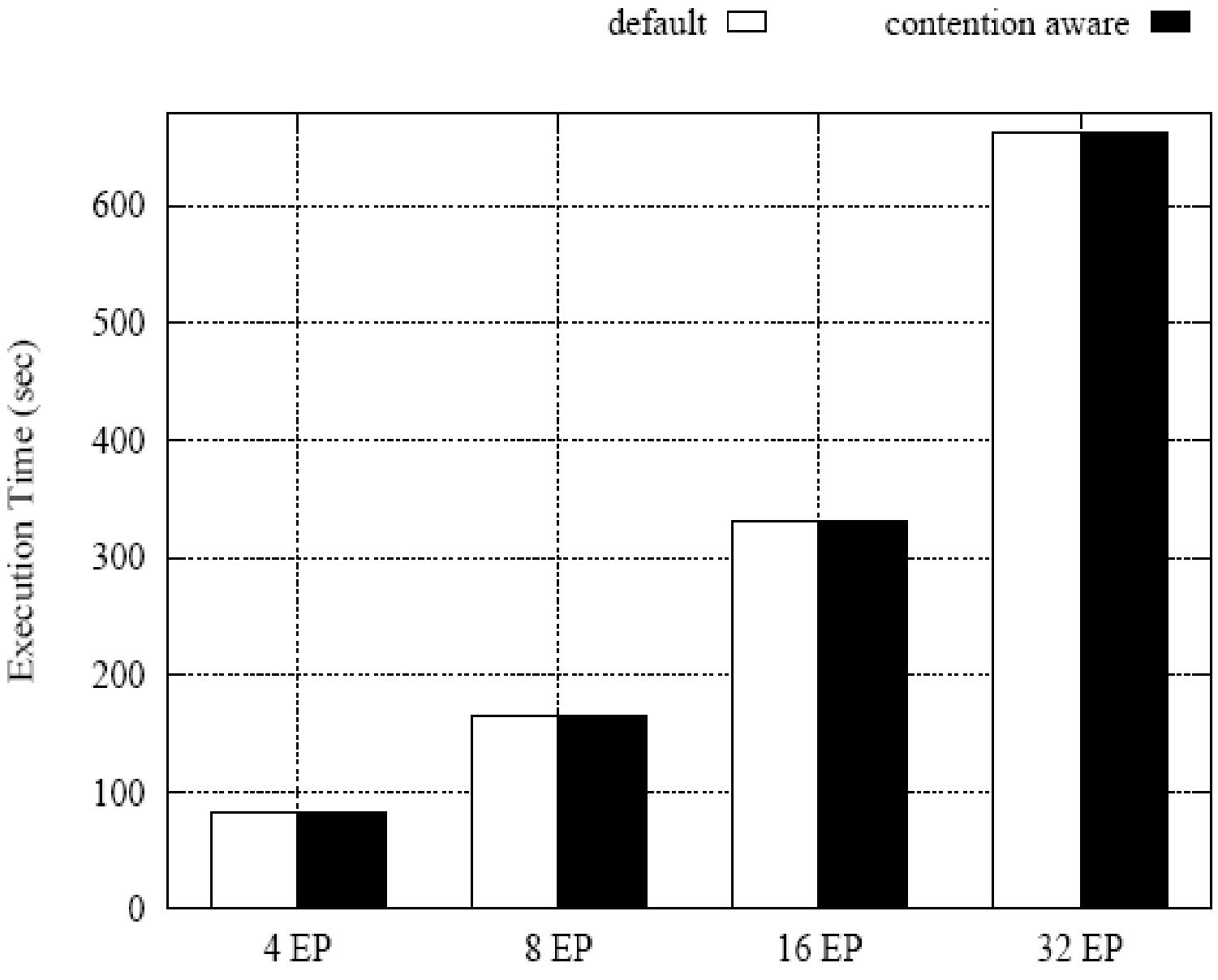

图 4.8 运行多个 EP 时，默认 O(1)调度器与竞争感知调度器的执行时间

本文分别运行 4, 8, 16 和 32 个 EP 测试程序。对于每个配置，最终的结果是所有测试程序的平均执行时间。图 4.8 表示了默认调度器和资源竞争感知调度器的执行时间随测试程序数目的变化。该图只表示了 O(1)调度器的实验结果，其它调度器上的结果类似。从图中可以看出，随着测试程序数目的增加，两种调度算法的执行时间基本类似，进而表明其开销可以被忽略不计。

事实上，不管 EP 程序的数目如何变化，使用资源竞争感知调度的程序执行时间要稍微少一些。这是因为任务的资源需求在不同的时间片内可能不同。这样，共享资源的竞争仍然存在，尽管同时执行的是多个相同的测试程序。例如，如果 2 个程序在当前时间片的资源需求都较大，那么，共享资源的压力也会相对较大。但是，如果一个程序在当前时间片的资源需求较大，而另外一个程序在当前时间片的资源需求较小，那么，同时运行这 2 个程序产生的竞争则会相对较小。本文的资源竞争感知算法是在时间片粒度上调整共同运行的程序。因此，该算法具有较好的性能。多个 MPI 进程在相同集群节点上的行为可以由这个测试用例进行模拟。在集群计算中，每个应用的代码由多个 MPI 进程执行[98]。为了降低进程间通信的延时，属于同一应用的 MPI 进程倾向于在相同的节点上执行。那么，本文提出的资源竞

争感知算法在这种场景下同样有效。

### 4.4.5 执行时间降低

为了衡量资源竞争感知调度器的有效性，本文使用 9 个 NPB3.3-SER 测试程序来构造 8 个工作负载(如表 4.3 所示)。每个负载均由相同数目的内存密集型程序和计算密集型程序构成。除了负载 4，其它的负载由 16 个测试程序组成。对于负载 4，一共有 12 个测试程序。例如，负载 2 中包含 8 个 MG 和 EP。

表 4.3 8 个构造的负载，每个负载既包含内存密集型程序也包括计算密集类程序

| 工作负载 | 内存密集应用 | **CPU** 密集应用 |
| --- | --- | --- |
| 1 | LU | EP |
| 2 | MG | EP |
| 3 | SP | EP |
| 4 | MG | IS |
| 5 | MG | BT |
| 6 | CG | FT |
| 7 | SP UA | EP |
| 8 | CG SP | BT EP |

本文在 3 个调度器上对两种调度策略进行评测。一个调度策略是资源竞争感知调度，而另外一个是将内存密集类应用和 CPU 密集类应用分别绑定到共享最后级缓存的 2 个核上产生的调度。对于每个负载，最终的实验结果通过每种测试程序在所有执行中的平均执行时间来度量。例如，负载中的第 k 种测试程序的最终结果

$$\mathrm{C}_k = \sum_{j}^{m}\sum_{i}^{n} E_{k,i}^{j} \Big/ (m \times n) \qquad 4\text{-}6$$

其中 $E_{k,i}^{j}$ 代表第 k 种测试程序中，第 i 个程序在第 j 次执行中的执行时间。

图 4.9 表示了每个负载中的每个测试程序在两种调度下相对默认系统的执行时间降低百分比。从图中可见，几乎所有负载的内存密集型任务在使用了竞争感知调度策略之后均有性能提升。对于 CFS, RSDL 和 O(1)调度器而言，单独任务的执行时间降低百分比分别可达 18.66%, 14.37%和 21.34%。而对于大部分 CPU 密集型应用，采用竞争感知调度后，在 CFS 和 RSDL 调度器中的性能是下降的。而在 O(1)调度器中，大部分 CPU 密集型应用的性能却有略微的提升。内存密集型应用的性能提升和 CPU 密集型应用的性能下降符合作者的预期，因为本文的调度器会将具有互补资源需求的任务同时调度。这样，默认调度器中 2 个内存密集型应用相互竞争的情况将会被避免。而对于 CPU 密集型应用，与其共享最后级缓存的另外一个任务对它的影响在资源竞争感知调度器中比在默认调度器中要大，因为另外一个任务有很大概率是内存密集型应用。O(1)调度器中的 CPU 密集类应用可以获得一些性能提升，因为 O(1)中的负载均衡使 CPU 密集类应用执行的更加频繁。通过对内核日志的分析，发现增加竞争感知功能之后的 O(1)调度器会周期性地将一些任

务(个数记为 n)从从核迁移至主核, 尽管 2 个核上的任务数目已经相同。经历了几次负载均衡, n 个任务又从主核迁移至从核上。

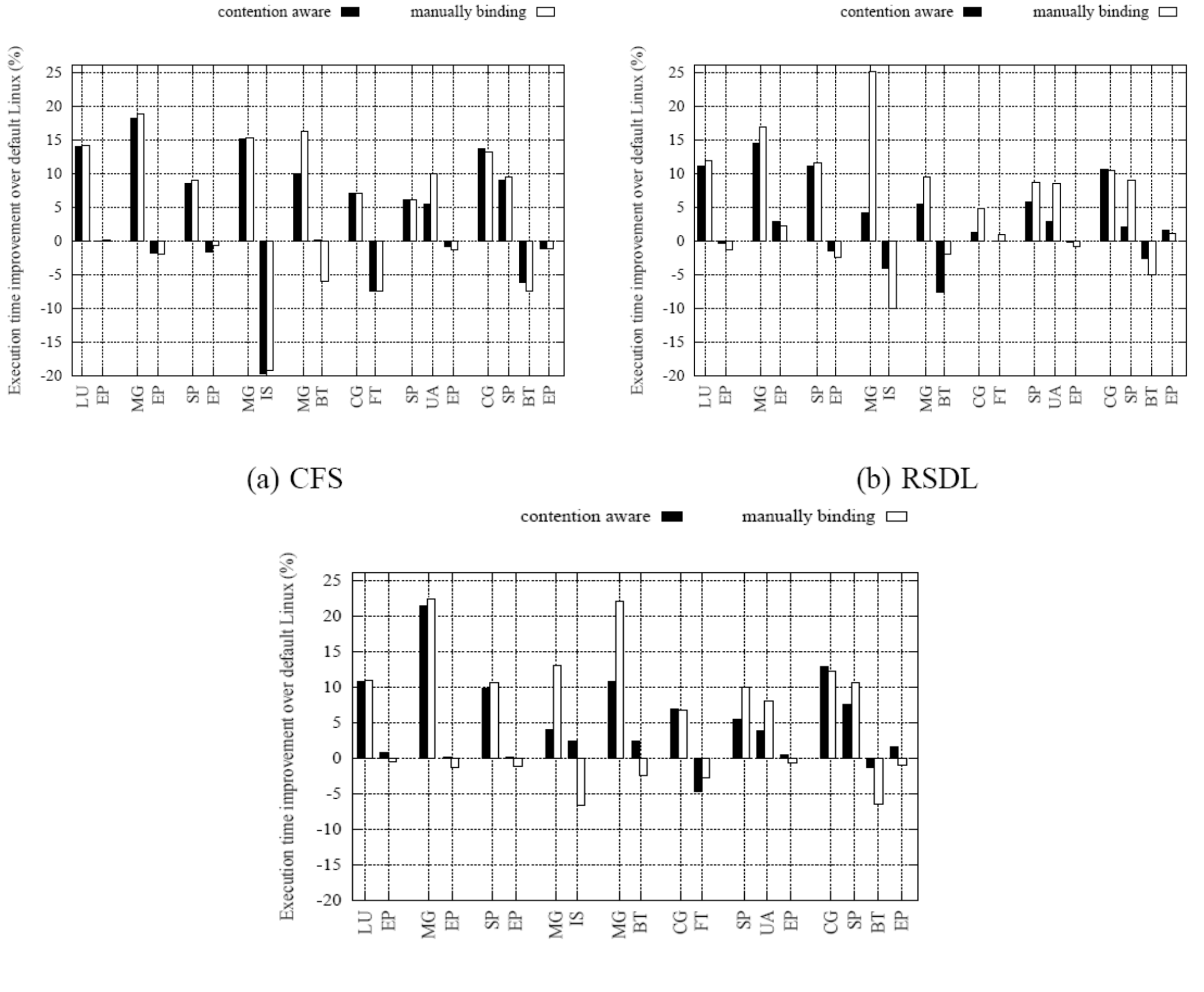

(a) CFS (b) RSDL

(c) O(1)

图 4.9 每个负载中的每个测试程序相对默认系统的执行时间降低百分比

图 4.10 表示了在执行负载 1 时, O(1)调度器在 200 次负载均衡过程中主核和从核上的任务数目。从图中可以看出, 2 个核上的任务数目并不是总是相同。相反, 其变化相当剧烈。这种负载均衡行为对内存密集类任务具有不利的影响, 而对 CPU 密集类任务却有好处。这是因为负载均衡导致从核上的任务执行的更快一些。而在本文的调度算法中, 从核上的所有任务几乎均是 CPU 密集类任务。这种负载均衡行为是由于竞争感知调度策略对任务调度顺序的调整。一些在从核上的任务不能被立即调度, 但是这些任务在负载均衡过程中具有较大的权重。在 O(1)调度器中,events 内核线程是造成这种负载不均衡的原因。在 O(1)中, 观察到的负载不均衡现象在使用 RSDL 调度器的补丁之后消失, 因为 RSDL 调度器中的 events 线程的权重和其它普通任务相同。这使得 events 线程的权重没有达到足以导致系统负载不均衡的程度。

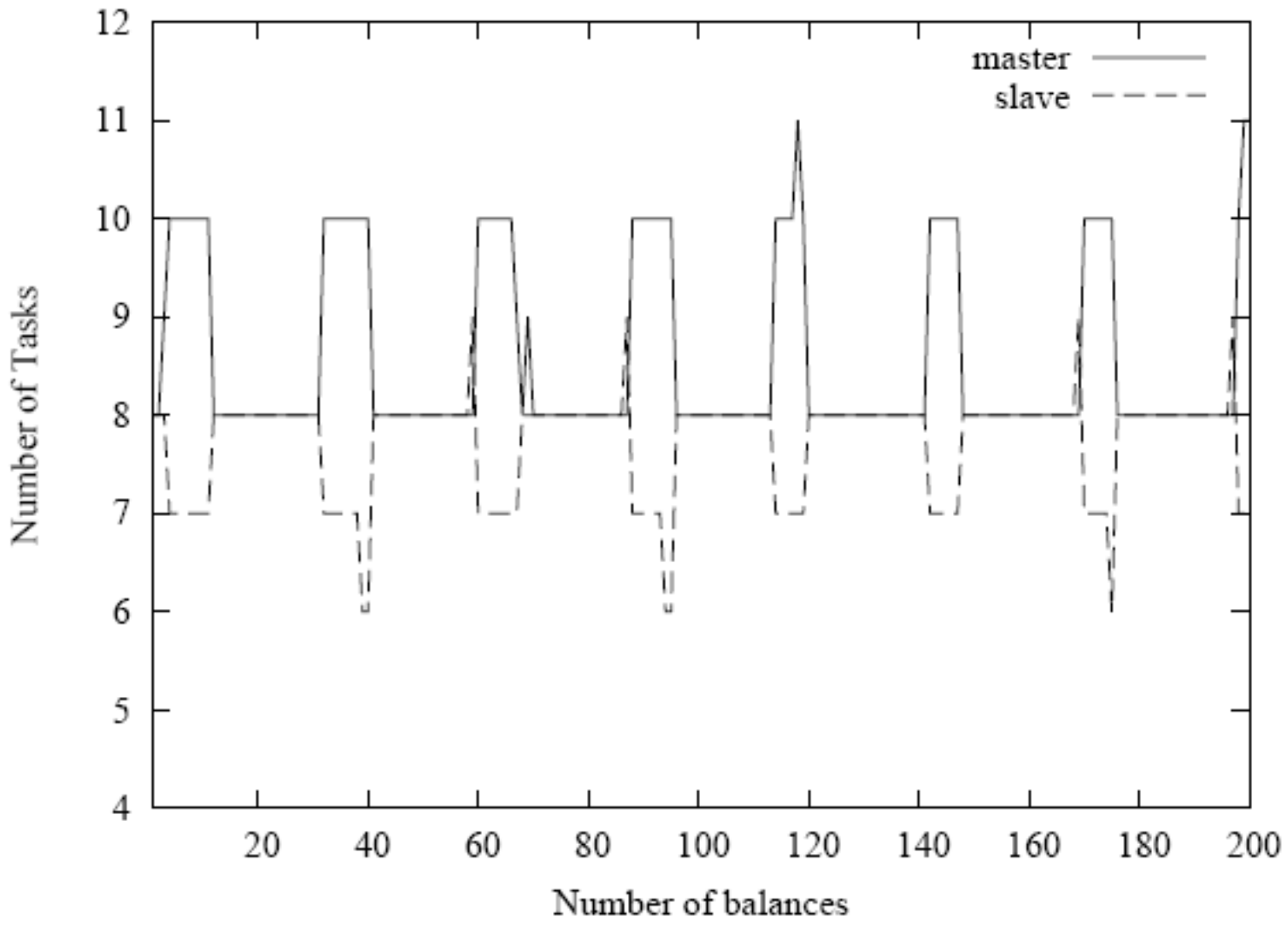

图 4.10 200 次负载均衡过程中，主核和从核的任务数目

与手工绑定的调度相比，采用竞争感知调度器会使得内存密集类任务的性能稍微差一些，而 CPU 密集类任务的性能要好一些。对于内存密集类任务，竞争感知调度策略表现出的稍差性能是由于 3 个原因。1、负载均衡时任务到核的重新映射对于某些任务是无效的。一个典型的场景是一个任务需要被迁移到另外一个核上以均衡硬件资源的使用，但是由于该任务正在处理器上运行，所以这种迁移不能发生。这种场景可能会导致 2 个内存密集型任务被同时调度。2、本文的竞争感知调度需要在内核中学习一个中间值。基于这个值，具有互补资源使用需求的任务被同时调度。但是，2 个关键的因素可能影响中间值的大小。一个因素是维护的平均值需要一些时间才能达到统计平均，另外一个因素是该中间值会受到应用使用资源特征的影响。这样，不正确的调度可能会被产生。3、在 O(1)调度器中，特殊的负载均衡行为会影响内存密集应用的性能。竞争感知调度器中的 CPU 密集类应用的性能比手工绑定调度中的 CPU 密集类应用要好，因为与该类任务共享最后级缓存的任务在手工绑定调度中更可能是内存密集类应用。而且，O(1)调度器中的特殊负载均衡会提升这类任务的性能。

本文发现了几个负载中展示出的有趣现象。第一个有趣的例子是 MG 和 IS 的组合。在带有竞争感知功能的 RSDL 和 O(1)调度器中，MG 相对于默认调度器的提升仅小于 5%，而手工绑定达到的提升分别是 25%和 12.98%。第二个是 MG 和 BT 的组合。MG 在带有资源竞争感知功能的 CFS 和 O(1)调度器中得到的提升远远小于手工绑定获得的提升。而 BT 在具有资源竞争感知功能的 RSDL 调度器中获取的性能甚至差于手工绑定的结果。第三个是 CG 和 FT 的组合。在 RSDL 中 CG 仅仅表现出了有限的提升。在 O(1)调度器中 FT 的性能要差于手工优化的结果。所有的这些现象均由复杂的最后级缓存访问模式引起，该问题在后面会进行详细阐述。

### 4.4.6 匹配率

在本文的资源竞争感知调度策略中，从核在本地的运行队列上寻找合适的任务和主核上当前的任务共同运行。如果从核发现一个和主核上当前运行任务具有互补资源需求的任务时，则认为从核和主核相匹配。在从核上，匹配次数占总体调度次数的百分比被称作匹配率。实际上，匹配率可以反映资源竞争感知调度器产生均衡调度的能力。

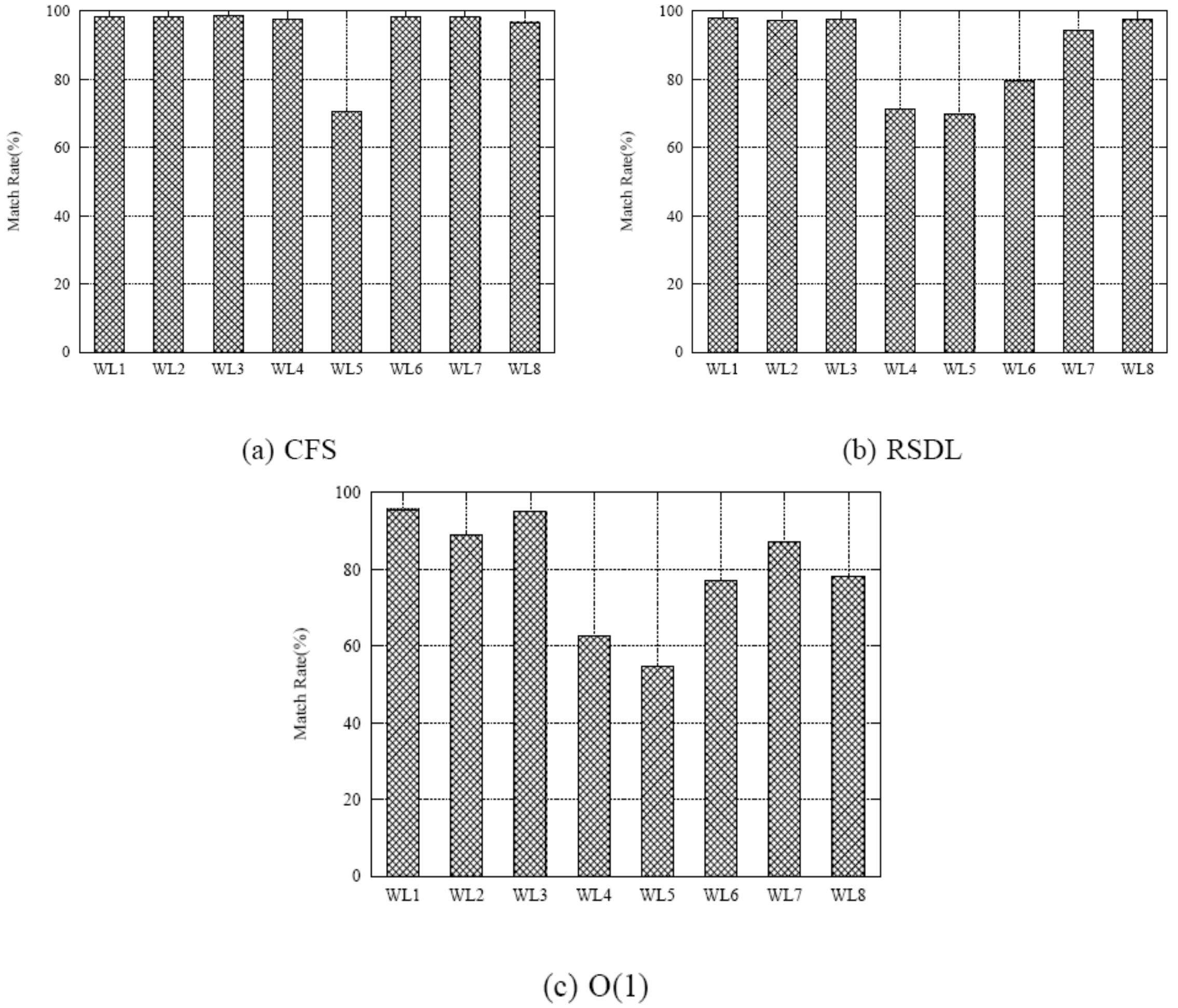

(a) CFS (b) RSDL

(c) O(1)

**图 4.11 不同调度器上的匹配率**

图 4.11 表示 3 个调度器上每个负载的匹配率。如图所示，负载 1，2，3，7 和 8 在 CFS 和 RSDL 调度器上的匹配率接近 100%，而在 O(1)调度器上的匹配率要差一些。O(1)调度器上的匹配率较差,因为负载均衡函数周期性地将任务由从核迁移到主核上，尽管 2 个核的任务数目已经相同。这种行为使得从核大部分时间执行的是 CPU 密集类任务，而主核要执行内存密集类应用和 CPU 密集类应用。这样，当主核调度一个 CPU 密集的任务，从核找不到合适的任务和其一起运行。对于负载 4 和 6，RSDL 和 O(1)调度器上的匹配率较低，而 CFS 调度器上的匹配率却接近 100%。负载 4 和 6 在 RSDL 和 O(1)上的匹配率较低，因为某些测试程序的最后级

缓存访问率在每个时间片内波动太大。具体而言，负载 4 中的测试程序 IS 和负载 6 中的程序 CG 和 FT 展示出了不稳定的最后级缓存访问率。图 4.12 表示了 O(1)调度器中运行负载 4 和 6 时，每个测试程序在每个时间片内的最后级缓存访问率。

图 4.12 表明负载 4 和 6 中至少有一个程序的最后级缓存访问率会大幅度地波动。资源竞争感知调度利用任务的历史信息预测任务的资源需求。当程序执行时，这种方法实际上降低了最后级缓存的访问率抖动。然而，测试程序 IS, CG 和 FT 的执行时间相对短暂(分别是 1.95 秒, 3.92 秒和 4.47 秒)。这样，这些测试程序的最后级缓存访问率在测试程序执行完毕之前不能达到一个相对稳定的状态。本文提出的调度器中需要的中间值在执行这些负载时是不准确的，得到的匹配率也比较低。

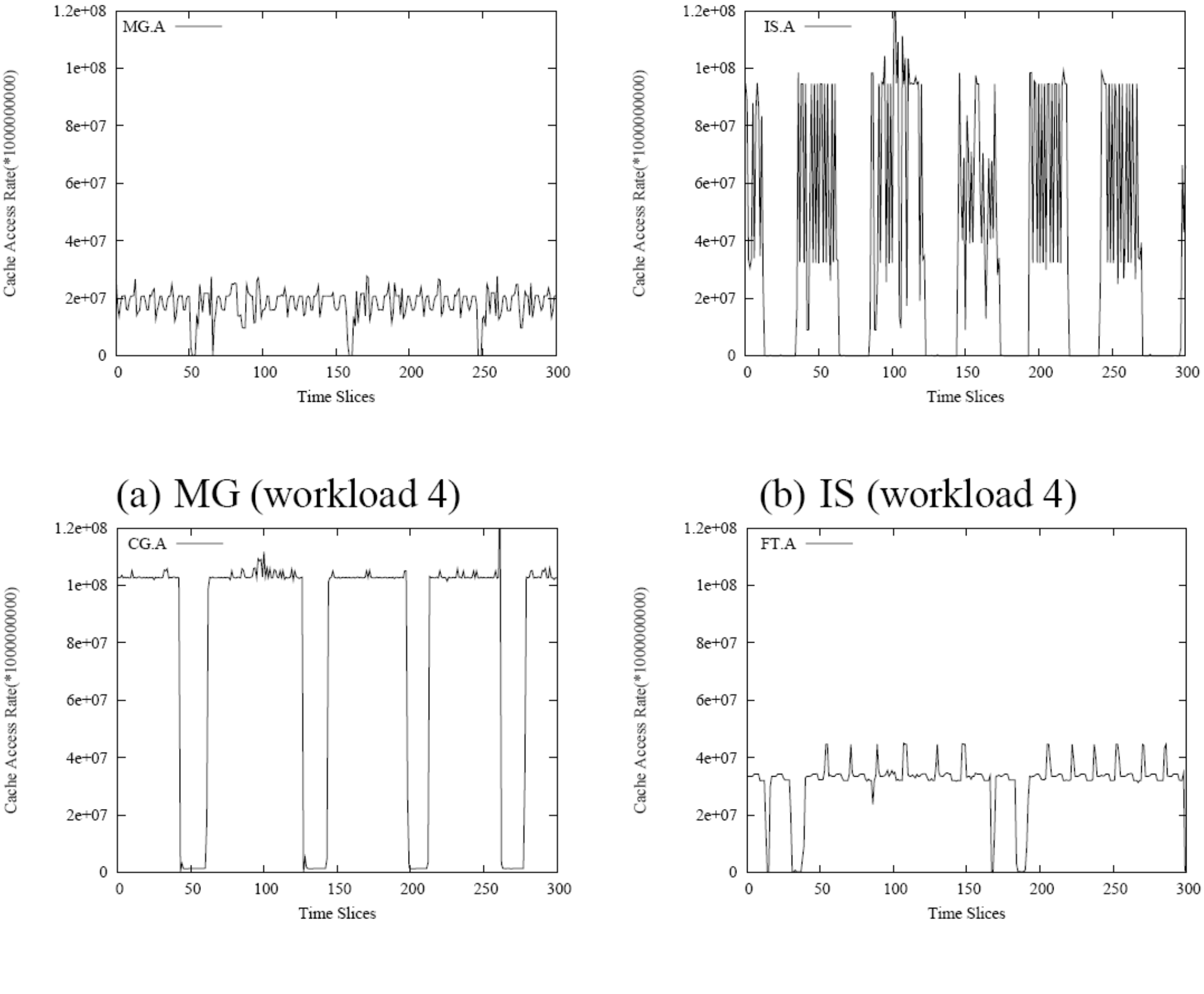

(a) MG (workload 4) (b) IS (workload 4)

(c) CG (workload 6) (d) FT (workload 6)

图 4.12 负载 4 和 6 中测试程序的最后级缓存访问率

对于 RSDL 调度器而言，负载 4 和 6 的匹配率较低不仅是因为最后级缓存访问率的抖动，而且因为活跃数组和超期数组之间较长的切换间隔。在 RSDL 调度器中，当一个任务使用完了当前的时间片，该任务的优先级降低而不是进入超期数组。仅当任务被降低到最低的优先级并且使用完了分配的时间片，该任务才进入超期数组。实际上，本文提出的竞争感知调度策略可以提升一个任务的优先级以平衡硬件

资源的使用。因此，可能一个任务被连续的调度直到它进入超期数组中。当竞争感知调度器中的中间值由于最后级缓存访问率的波动而变得不准确时，活跃数组和超期数组之间较长的切换间隔使得不准确的中间值难以矫正。相反地，CFS 调度器对最后级缓存访问率波动的表现不够敏感，这是因为每个任务被执行一次的总体时间和每个任务的时间片均较短(小于 8 毫秒)的缘故。与之相比，O(1)调度器中，普通优先级任务的时间片长达 100 毫秒[33]。对于 RSDL 调度器而言，尽管时间片的长度也相对较短(15 毫秒)，但是一个普通任务由活跃数组进入超期数组的时间间隔却达到了 300 毫秒。CFS 调度器的调度特征使得竞争感知调度器中需要的中间值很准确，进而可以准确地捕获任务的最后级缓存访问率波动。对于负载 5，3 个调度器得到的匹配率均很低，因为该负载中的 2 个测试程序(MG 和 BT)具有相似最后级缓存访问率。MG 和 BT 的最后级缓存访问率分别是 0.0185 和 0.0171。因为本文提出的调度策略利用程序的最后级缓存访问率来代表资源需求，因此调度器无法仅根据这一指标来区分 MG 和 BT。

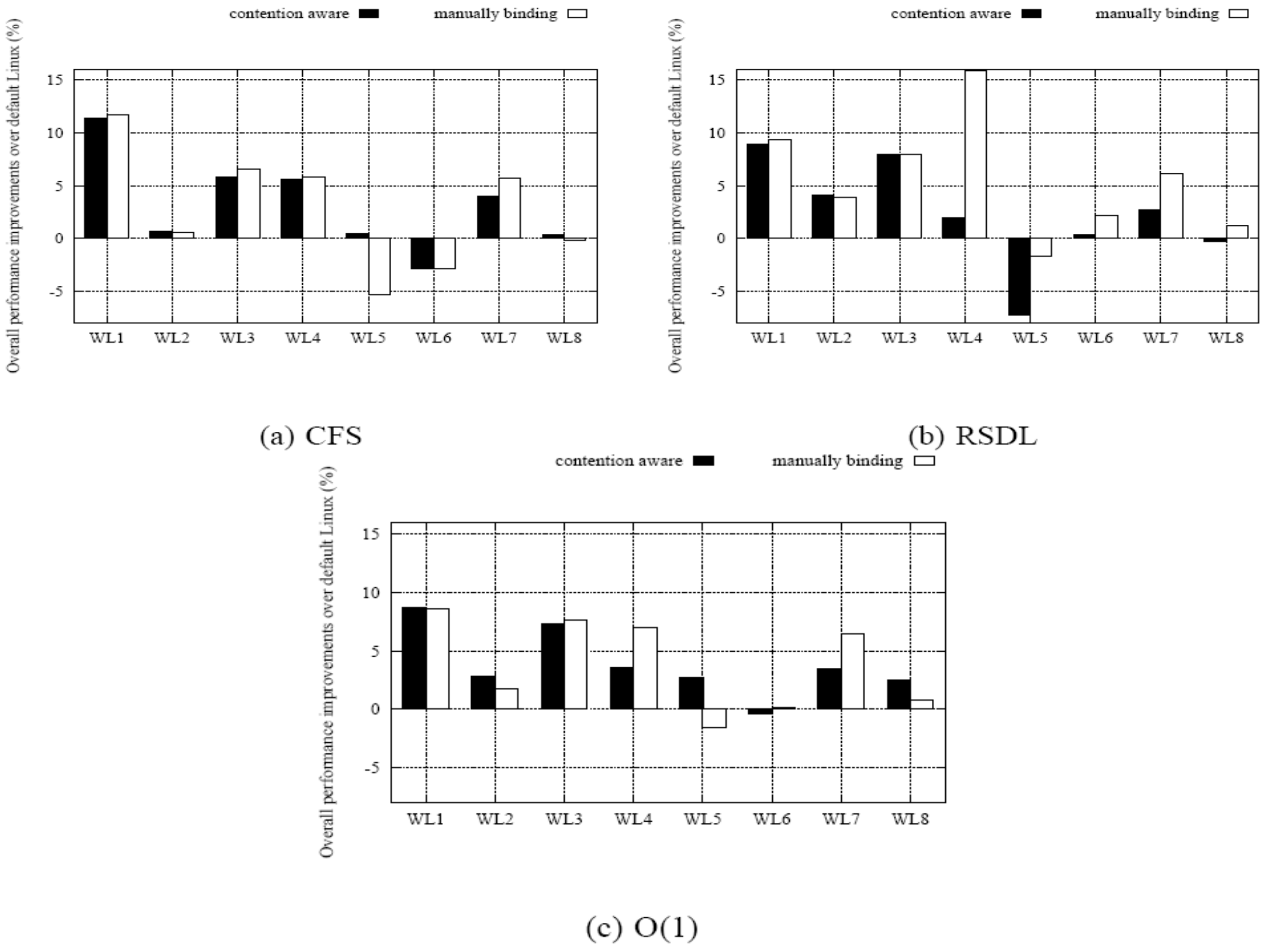

图 4.13 相对默认 Linux 系统的总体性能提升

### 4.4.7 总体性能和可扩展性提升

为了测量系统的总体性能，我们需要计算负载的总体执行时间。图 4.13 表示了每个负载相对默认 Linux 系统的总体性能提升。手工调度优化的结果也在该图中

表示以方便比较。如图中所示，大部分负载在采用资源竞争感知调度策略后均会提升总体性能。最大的提升有 11.39%，转化成系统可扩展性是 12.85%。与手工调度优化的结果相比，有很多负载表现出了类似或者更好的性能。然而，RSDL 调度器中的负载 4，5 和 6 和 O(1)调度器中的负载 4 均还有改进的空间。这些负载性能不佳的原因是复杂的最后级缓存访问模式。负载 4 和 6 在 RSDL 和 O(1)调度器上的性能可以通过缩短时间片的长度和每个任务均执行一次的时间来达到。

### 4.4.8 执行时间的稳定性

本文提出的竞争感知调度策略的一个有趣特征是应用程序的执行时间变得更加稳定，因为具有互补资源需求的任务被同时调度运行。对于每个程序而言，和其共同运行的任务对它的影响更加稳定。为了测量执行时间的稳定性，使用负载中每种测试程序的执行时间方差来衡量

$$D_k = \frac{\sum_{j}^{m} \left( \frac{\sum_{i}^{n} E_{k,i}^{j}}{n} - C_k \right)^2}{m} \qquad 4\text{-}7$$

其中 $D_k$ 代表第 k 种测试程序的执行时间方差。$E_{k,i}^{j}$ 代表第 k 种测试程序中第 i 个测试程序在第 j 次执行的执行时间。$C_k$ 是第 k 种测试程序在所有执行中的平均执行时间。

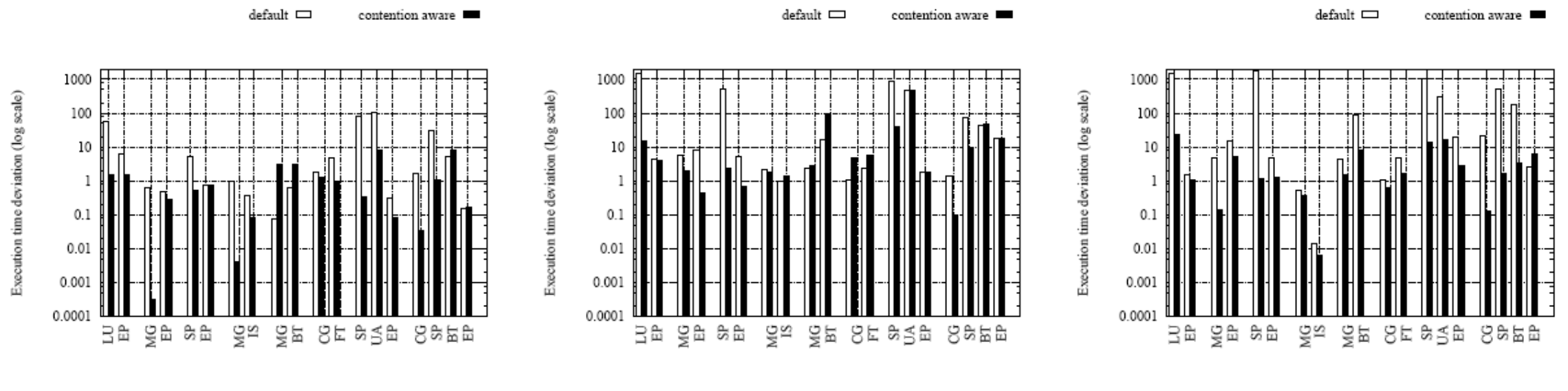

(a) CFS (b) RSDL (c) O(1)

图 4.14 每个负载中每个测试程序的执行时间方差

图 4.14 表示了所有负载在 3 个调度器中的实验结果。默认调度器中每个负载的程序执行时间方差也被整合进去以方便对比。和预期的相同，当采用了本文提出的竞争感知调度策略后，内存密集型应用和 CPU 密集型应用的执行时间方差被大幅度地降低。从图中可见，几个负载展示出了异常的执行时间方差。例如，CFS 和 RSDL 调度器中的 MG-BT 组合，以及 RSDL 调度器中的 MG-IS 和 CG-FT 组合。根据我们的分析，这些异常是由复杂的最后级缓存访问模式所导致的。

### 4.4.9 Intel 8核系统的实验结果

本文同样在 Intel8 核平台验证了带有资源竞争感知功能的 CFS 调度器的有效性。图 4.15 表示了每个工作负载的总体执行时间提升。可以看出 8 个负载中的 4 个展示出了性能提升，而其余的负载则有一些性能下降。本文发现在 8 核平台上，采用了资源竞争感知功能之后得到性能提升的负载数目比在双核平台上要少。这是因为 CPU 密集类任务在 8 核平台上遭受到了更大的性能影响 (图 4.16)。然而，通过内核日志分析，发现大部分负载均可以同时调度具有互补资源需求的任务，而且每个负载的匹配率很高。这表明本文的调度算法可以在 8 核平台上正确地工作。

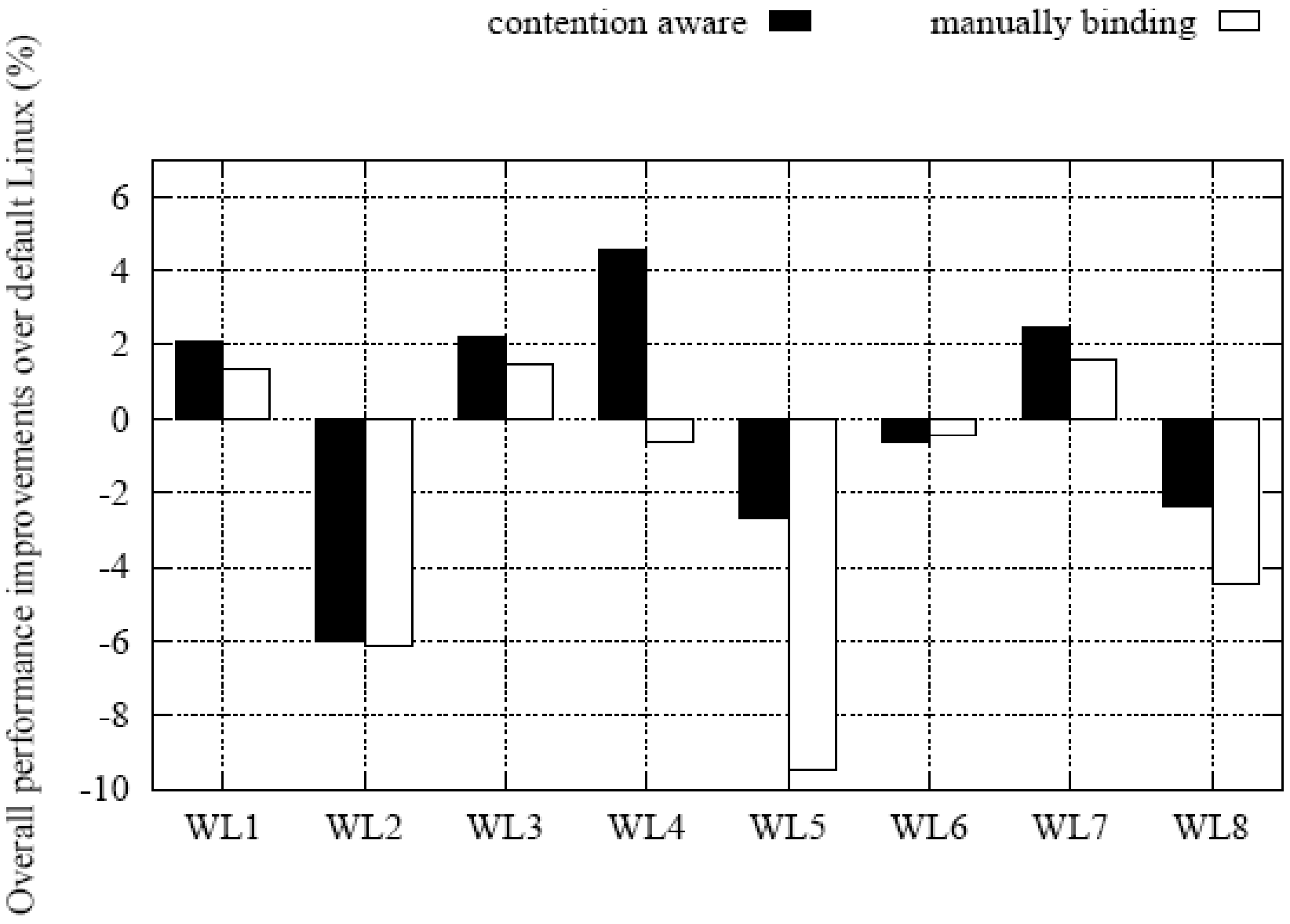

图 4.15 总体执行时间提升

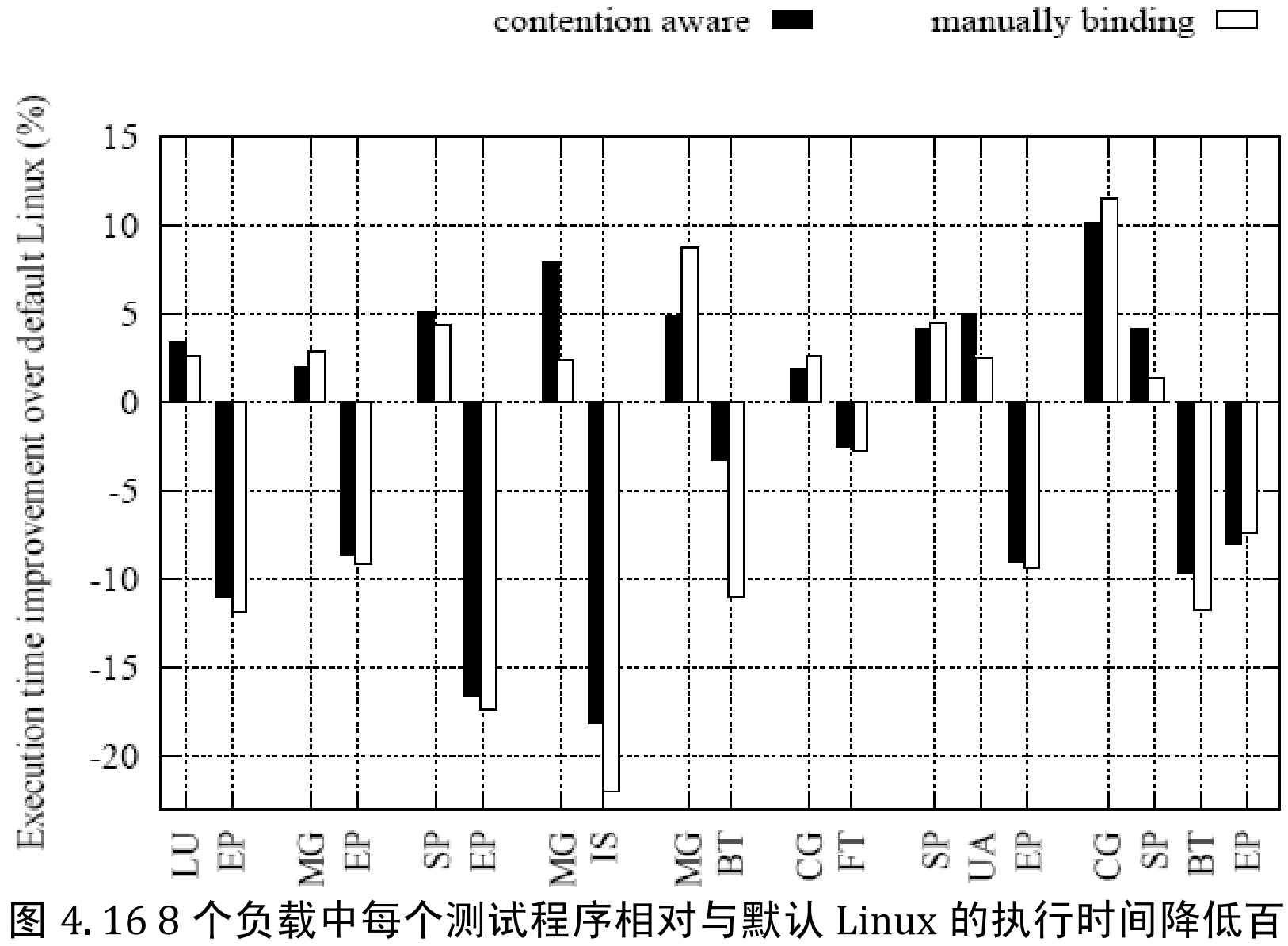

图 4.16 8 个负载中每个测试程序相对与默认 Linux 的执行时间降低百分比

那么，为什么在 8 核平台上的提升不够显著呢？根据本文实验平台的硬件架构图(图 4.7)，我们分析是由缓存竞争，前端总线饱和，地址总线饱和以及内存带宽

饱和等因素之一所导致的。首先，缓存竞争不可能是问题的根源，因为当系统中的核数增加时，最后级缓存的数目也随之增加。而且，本文已经在共享最后级缓存的双核上运行了所有的负载，并且发现大部分负载都有性能提升。另外可以看到，当核的数目增加时，前端总线的压力也随之增大，所以另一个可能的因素是前端总线饱和。本文使用 LU 和 EP 的组合来验证这种可能性。前段总线的利用率利用文献[99]中的公式进行计算，其值仅为 11.72%。这样就排除了前端总线饱和这一因素。第 3 个可能的原因是地址总线饱和[42]。当执行 LU 和 EP 组合时，地址总线的利用率达到了 21.39%。然而，在工业上，地址总线利用率达到 60%以上时，才认为地址总线达到了饱和，所以这一因素也被排除。最后，本文根据文献[100]计算出了当执行 LU 和 EP 组合时,系统的内存带宽占用率。其计算值是 76%，进而表明内存带宽饱和实际上是八核系统上性能提升不够显著的根源。如果该瓶颈被移除，那么，资源竞争感知调度能达到的性能提升将大幅度增加。即使是在带宽受限的系统中，当内存密集任务具有较高优先级时，资源竞争感知调度策略同样有效。

## 4.5 相关工作

共享最后级缓存是多核系统上的一个显著特征。现有工作中有许多是解决多核系统上的资源竞争问题。例如，Fedorova 等[11]提出通过分析每个线程的重用距离来估计所有可能线程组的缓存缺失率，并且只调度组缺失率低于一定阈值的组合。然而，这种方法在收集每个程序的重用距离和选择合适的组合上均会导致较大的开销。在文献[18]中，Fedorova 提出了一个基于进程桶的算法将具有互补资源需求的任务同时执行。该方法解决了文献[11]中涉及到的耗时操作，并且已经实现在 Solaris 内核中。然而，这个方法限制了可以执行内存密集应用的 CPU，从而会导致负载不均衡。尽管总体的系统性能有所提升，每个内存密集应用的性能会受到影响[18]。

Jiang 等[101]提出将局部性信息分析和在线调度相结合。在提出的方法中，每个任务仅需要独立的重用距离信息和每周期不同的数据块信息来计算一个任务是内存密集类任务还是 CPU 密集类任务。利用这种信息，具有互补资源需求的任务被同时调度。然而，重用距离和每周期不同的数据块信息仅能反映任务在最后级缓存上的需求，因此不能建模其它资源竞争的影响。为了支持上述结论， 作者对 Jiang 的方法进行了实现，并且在实际系统上测试它的有效性。本文作者发现该方法的准确性强烈依赖于几个参数和统计函数的准确性。然而, Jiang 等没有给出参数调优的方法和统计函数的构造方法。

Sato 等[17]使用离线测量来获取应用程序使用缓存的需求，并且将使用缓存需

求较大的进程分散在不同的芯片上。然而，调度策略中需要的离线测量将引入过大的开销，而且任务到核的映射是依赖于中心化的排序策略，因此其可扩展性不佳。

Zhuravlev 等[15,90]首先分析了几个应用分类方法，并且发现缓存缺失率是描述程序资源需求的最佳指标。基于这个结论，提出了 DI 和 DIO 调度算法将内存密集类任务均匀地分布到不同的芯片上以避免多个内存密集类应用的同时调度。与这个工作相比，1、本文也提出了一种方法系统地选择启发式指标来代表应用的资源需求。本文的方法不但考虑了指标的准确性，而且考虑到了指标的稳定性。2、本文的方法不仅考虑了如何将任务映射到核上，而且考虑了如何在一个核上调度任务。3、本文提出的调度方法是纯分布式的实现，因此避免了中心化排序方法的瓶颈。最后，本文的竞争感知调度达到的可扩展性在 3 个调度器中均被验证。

Kishore Kumar Pusukuri 等[13]提出了一个调度框架将内存密集类进程分散到不同的芯片上。文中使用了机器学习的方法来指导如何进行进程迁移。这个工作首次同时使用了多个性能指标。然而，其选择指标的方式是基于经验的，而且该文也忽略了如何在一个核上进行调度。

Robert L. McGregor 等[14]基于一系列的可在线获得的运行时指标在处理器内和处理器间将任务进行同时调度。与以前的工作不同，处理器内和处理器之间使用不同的性能指标。然而，所有的指标都是基于经验的选择。而且在每个时间片开始时，该文依赖于中心化的排序策略将任务映射到核上。

Joshua Kihm 等[102]通过调整调度顺序来避免 SMT 平台上线程之间的资源竞争。在该工作中，任务的缓存需求被建模成一个向量。然而，该模型需要特殊的硬件来获取这个向量中的值，并且它不能解决其它资源的竞争问题(例如内存控制器的竞争或者预取硬件的竞争)。

Knauerhase 等[12]提出使用基于性能计数器的调度来解决共享缓存竞争，不公平性和异构的指令集问题。与这个工作相比，本文的工作有几个不同。1、提出一个方法来确定合适的启发式指标用以描述程序的资源使用需求。2、Knauerhase 等的方法仅仅是通过调整调度顺序来达到不同任务之间的同时调度。然而，本文增加了负载均衡时的机制重新进行任务到核的映射。通过将两种方法相结合，本文可以更加有效的解决资源竞争问题。3、竞争感知的调度策略被整合进 3 个最流行的调度器中，这一点在 Knauerhase 等人的工作中没有被涉及。4、本文对资源竞争感知调度的可扩展性和性能提供了详尽的分析。

一些文章[103-105]提出利用页着色技术来降低最后级缓存上的竞争。然而，这种方法有三方面缺陷。1、对内核的改动较大。2、当程序的缓存使用行为频繁发生变化时，这种方法会引入较大开销[104]。3、页着色方法解决的仅仅是最后级缓

存竞争，而忽略了前端总线和地址总线等资源上的竞争。最后,本文提出的方法可以和页着色方法共同使用以进一步地减少资源竞争[106]。

## 4.6 本章小结

本文提出了内核级别的调度器在多核系统上降低共享资源竞争。本文首先提出了一种方法从 5 个候选指标中选择最后级缓存访问率来代表程序的资源需求。基于这种信息，通过任务到核的重新映射和调度顺序的调整实现具有互补资源需求任务的同时调度。提出的调度策略在 3 个最主流的调度器上进行了实现并且在 Intel8 核平台上进行了评测。实验结果表明本文提出的调度算法可以有效降低资源竞争从而提升系统总体可扩展性。而且，当采用了本文提出的调度算法之后，每个程序执行时间的稳定性也有大幅度地提升。

本章提出的研究成果的一个可应用场景是云计算环境。在该场景中，多个应用运行于同一个系统上。而且，最近的研究表明科学计算可以在云上实现，因为云计算环境使得获取强有力的计算资源更加容易和方便[123]。因此，本文提出的调度策略可用于该环境下解决资源竞争问题。

# 第5章 可扩展性瓶颈定位方法

## 5.1 本章引言

芯片制造厂商已经从单核处理器切换到了多核处理器。为了充分的利用多核系统的计算资源，软件需要具有良好的可扩展性。然而，许多重要的实际应用(如网页服务器[42]和 MapReduce[45])的可扩展性均表现不佳。 本章研究了具有较高并行度的应用——在线事务处理(OLTP)应用的可扩展性。在多核处理器平台上的实验表明这些应用也具有有限的可扩展性。本文提出了一种新方法以定位应用的可扩展性瓶颈并且对发现的瓶颈进行解决以验证该方法的有效性。

本文选择 2 个测试程序，即 TPCC-UVa[107]和 Sysbench-OLTP[108]来代表 OLTP 应用。TPCC-UVa 是 TPC-C 标准[109]的一个开源实现。为了有效的使用多核资源，本文对 TPCC-UVa 进行了改进，并且利用内存文件系统来模拟 RAID。Sysbench-OLTP 是一个轻量级的测试程序并且广泛地用于 OLTP 应用的测试上。2 个测试程序的可扩展性由事务的吞吐量随核数的变化来衡量。在 Intel8 核系统上，TPCC-UVa 的可扩展性是 3.68，而 Sysbench-OLTP 的可扩展性是 5.26。

一个核心的挑战是如何找到 2 个应用的可扩展性瓶颈。可扩展性瓶颈可能存在于多进程的应用中，存在于操作系统中，存在于系统库中或者存在于多核硬件中。为了解决这个问题，本文提出了一个基于函数可扩展性值(scalability value)的方法。函数的可扩展性值被定义成一个事务在多核和单核上运行时,该函数的执行时间之差。如果一个函数的可扩展性值为正数，则说明该函数是限制应用可扩展性的。当对所有的函数根据各自的可扩展性值排序之后，具有最大可扩展性值的函数很可能是最大的可扩展性瓶颈。实验结果表明该方法容易使用，而且可以有效的定位程序的可扩展性瓶颈。而且，仅仅需要简单的扩展，该方法即可用于定位所有面向吞吐量的服务器类负载。

使用这个方法，本文发现 TPCC-UVa 中最大的可扩展性瓶颈是数据库缓存池的竞争，这种竞争产生的原因是 TPCC-UVa 测试程序具有较大的工作集；下一个瓶颈是数据库的同步原语竞争；最后是 SystemV IPC 导致的内核锁竞争和调度器开销。对于 Sysbench-OLTP，可扩展性瓶颈是数据库的锁竞争和内核调度器。

本文移除确定出的瓶颈来验证基于函数可扩展性值的瓶颈定位方法的有效性。实验结果表明，MySQL 中使用较为有效的互斥锁实现可以使 Sysbench-OLTP 的执行时间缩短 15.27%，同时提升系统的可扩展性达 10.84%。本文通过使用可扩

展自旋锁[55-56]和基于 RCU 的 IDR API[110]来降低 System V IPC 的锁竞争。

本章在四方面取得了成果。1、在多核平台上, 本文探索了 OLTP 应用的可扩展性问题。2、提出了基于可扩展性值指标的方法以定位服务器应用的可扩展性瓶颈。3、定位出了 2 个测试程序中的可扩展性瓶颈并且通过理解 Linux 内核代码和开源数据库代码对确定出的瓶颈进行分析。4、提出了瓶颈的解决方案并且进行了实际的实现以提升应用可扩展性。

本章的剩余部分进行如下组织: 5.2 节介绍了基于可扩展性值指标的方法以定位服务器类应用的可扩展性瓶颈。5.3 节给出了实验平台信息, 测试程序描述, 数据库配置以及测试程序的运行规则。5.4 节表示了 2 个测试程序的可扩展性。5.5 节分析了可扩展性瓶颈, 提出了解决方案并且对方案的有效性进行了评估。5.6 节讨论了后续工作。5.7 节介绍了相关工作而 5.8 节总结本章。

## 5.2 可扩展性指标

由于 OLTP 应用的重要性, 有许多工作侧重于提升这种应用在多核平台上的可扩展性。然而, 在当前工作中, 还没有系统地分析可扩展性行为的方法。相反地, 软件设计者利用基于经验的代码分析或者 adhoc 的方式来确定可扩展性瓶颈[111-112]。在本节中, 提出了基于可扩展性值指标的方法以确定事务应用的可扩展性瓶颈。然后, 将该方法进行扩展来定位一般的服务器应用中的可扩展性瓶颈。

典型的 OLTP 应用是客户端服务器结构。客户端向服务器发送事务执行请求。当服务器接收请求后, 会创建任务来处理客户端的事务请求(如查询数据库等)。OLTP 应用在服务器端的并行可扩展性是本文的关注点。为了定位可扩展性瓶颈, 本文提出了基于可扩展性值指标的方法。该方法是基于如下分析, 如果 OLTP 应用的可扩展性不佳, 那么, 在多核平台上执行一个事务需要的时间比在单核上的时间长, 否则可扩展性是完美的。而且, 执行时间的差异越大(单核和多核), 整体的可扩展性就越差。

基于上述分析, 执行一个事务的平均时间可以被分解成一系列的<funcID, T>元组。每个 funcID 代表唯一的函数, 而该函数可以是用户态的函数或者内核的函数。T 代表该函数在执行一个事务中的执行时间。

对于一个应用, 可以获得两组函数元组。其中一个是在单核上得到的, 而另外一个是在多核上得到的。这两组元组被重新组织, 使得具有相同 funcID 的函数元组相互合并。每个新的元组被表示成为<funcID, Tm, Ts>。Tm 和 Ts 分别代表一个事务内, 函数 funcID 在多核和单核上的执行时间。Tm 和 Ts 之差被定义成函数 funcID 的可扩展性值。根据上述定义, 如果某个函数影响总体可扩展性, 那么, 该

函数的可扩展性值为正数。而且，函数的可扩展性值越大，该函数对整体可扩展性的影响就越大。当对函数根据可扩展性值进行排序后，可以找到具有较大可扩展性值的函数。对这些函数分析和分类就可以找到可扩展性瓶颈。

尽管基于可扩展性值指标的方法被用于定位事务处理类应用的可扩展性瓶颈，但是对该方法简单扩展之后同样可以用于其它服务器类应用。与科学计算应用和桌面应用相比，服务器类应用具有不同的特征。具体而言，这类应用是多线程或者多进程的，在执行的过程中要大量的使用系统服务。系统性能的衡量是通过总体的吞吐量而非执行时间[113]。基于上述的特征，本文引入计算周期的概念。使用 n 个核时的计算周期被定义为 $n/throughput_n$ ,其中 $throughput_n$ 是当使用 n 核时的总体吞吐量。事务类应用的计算周期即为执行一个事务的时间。

在服务器应用中，如果存在某些可扩展性瓶颈，那么，多核上的平均计算周期比单核环境下的要大。与 OLTP 应用中的事务一样，普通服务器类应用的平均计算周期也可以被分解成一系列函数元组。函数元组需要在单核和多核情况下均进行收集，进而，能够计算出每个函数的可扩展性值。具有较大可扩展性值的函数对整体的可扩展性影响较大，并且很可能成为一个瓶颈。

## 5.3 实验方法

### 5.3.1 测试平台和操作系统

所有的实验均是在 Intel8 核平台上进行的。平台上有 2 个 Xeon 芯片，每个芯片上有 4 个核。每个核有私有的 L1 指令和数据缓存。每个芯片上有 2 个 L2 缓存。每 2 个核共享一个 L2 缓存。芯片上的 4 个核与 4G 内存通过前端总线相连。核和总线的频率分别是 1.6GHZ 和 1.033GHZ。使用 Ubuntu Linux 作为操作系统, Linux 内核的版本是 2.6.25。为了使得结果更加准确，系统中安装了很多内核。大部分的实验结果是从干净内核(没有统计和调试功能)中搜集出来的。其它内核用于搜集锁竞争信息和调度统计信息等。

### 5.3.2 测试程序

本文使用 2 个测试程序来代表 OLTP 应用，后面将会逐一介绍。

TPCC-UVa TPCC-UVa 是 TPC-C 标准的开源实现。该测试程序通过模拟公司活动来代表 OLTP 的运行环境。tpcc 数据库由 9 张表组成，如 warehouse 表，item 表, district 表, customer 表等。测试程序中的每个客户端可以执行 5 种事务，分别是 new order, payment, order-status, delivery 和 stock-level。一次完整的测试将经历 2

个阶段，预热阶段和测量阶段。测试程序的性能是由一定间隔内的 new order 事务数目来决定的，并被定义为 transaction-per-minute-C(tpmC)。

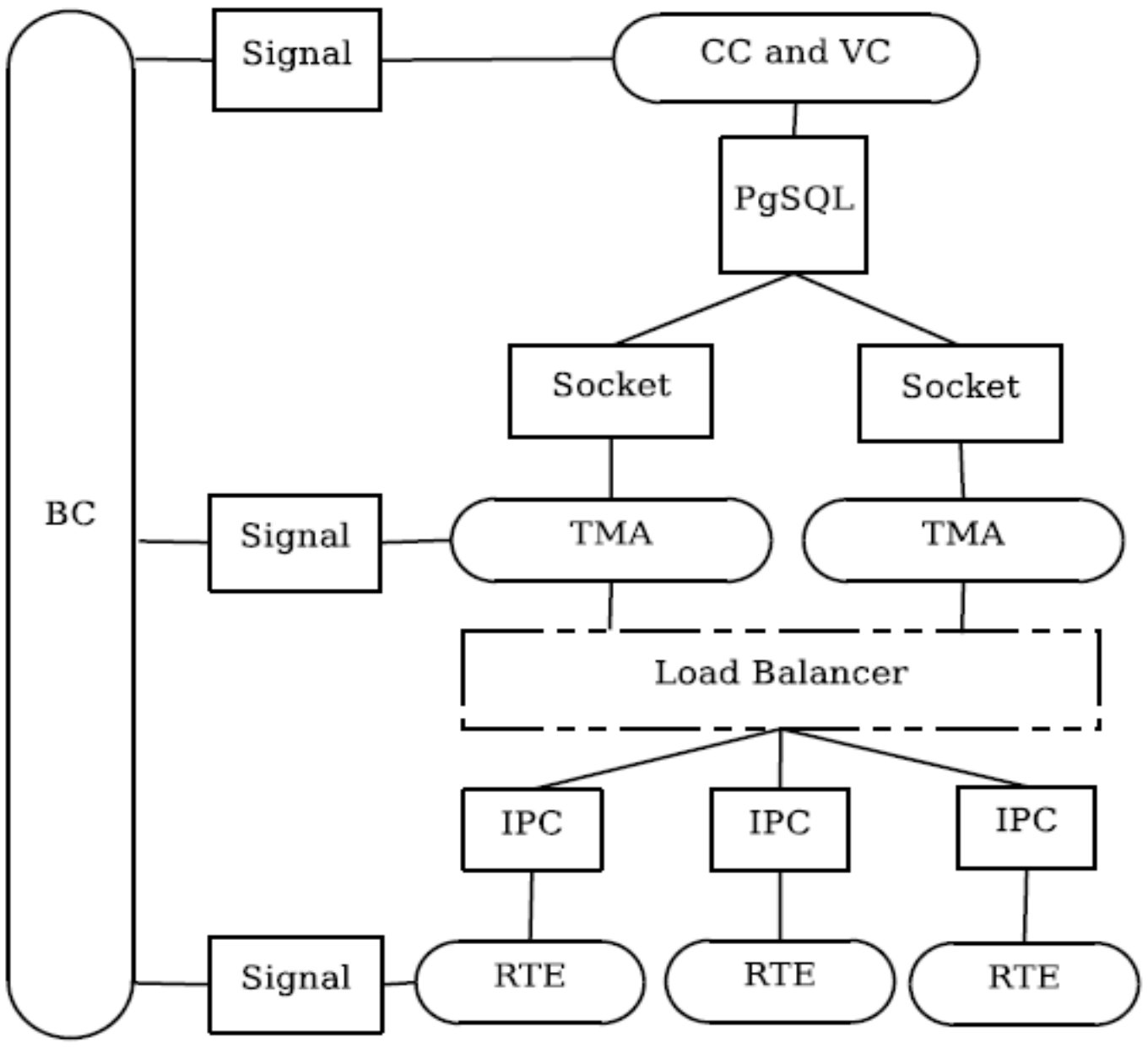

图 5.1 带有 3 个 RTE 和 2 个 TMA 的 TPCC-UVa 配置。2 个 TMA 组成一个 TM。圆角矩形代表测试程序中的功能模块，矩形代表通信方法，方块代表 postgreSQL 数据库，点划线矩形代表负载均衡模块

测试控制(BC)，远程终端模拟(RTE)，事务检测(TM)，检测点控制(CC)，和空间控制(VC)是 TPCC-UVa 的 5 个功能模块。RTE 负责根据 TPC-C 标准创建各种事务。TM 检测 RTE 的请求并且将请求送至数据库进程。检测点控制和空间控制用于控制数据库的相关行为。测试控制控制所有模块的开始和结束，并保证它们正确工作。RTE 和 TM 通过 IPC 进行通信，而 TM 与 PostgreSQL 通过套接字通信。在测试程序中，每个部分均被实现成一个不同的进程。测试程序利用 C 语言进行编写，利用 PostgreSQL 作为底层数据库。

TPCC-UVa 改进 默认的 TPCC-UVa 在利用计算资源方面效率很低，而且，提供的计时控制功能的粒度也较粗。本文将 TPCC-UVa 进行如下改进:

1、多核感知的事务检测 TPCC-UVa 中的事务检测被实现成独立的进程，它与 RTE 通过 SystemV IPC 的消息队列进行通信。当客户端的数目较少时，事务检测可以有效地处理来自客户端的请求。然而，当客户端的数目增加，单一的事务检测模块会变成可扩展性瓶颈，并且使得多核平台上的计算资源不能够被充分利用。为了解决这个问题，本文引入事务检测代理(TMA)模块。每个事务检测代理和原始测试程序的事务检测模块功能相同。在新的设计中，许多 TMA 组成一个 TM (如图 5.1 所示)。在实验中发现，当 TMA 的数目和系统中的核数相等时，测试程序的吞吐量最大。在改进版的 TPCC-UVa 中，sysconf 系统调用被用于获取系统中的核数。而

且，作者在 RTE 和 TMA 之间增加了一个负载均衡模块。该模块根据 RTE 所在的 warehouse ID 将负载以 round-robin 的方式均衡。其它的负载均衡策略将会在以后进行尝试。

2、忙等避免控制 TPCC-UVa 中的测试程序控制，检测点控制和空间控制与其它的进程通过信号(signal)进行通信。在注册了需要的信号之后，控制器将停止运行直到信号到达。这个功能避免了原来设计中的忙等。

3、细粒度的计时控制 在原来的测试程序设计中，计时的粒度是分钟。用户需要指定预热阶段和测量阶段的时间。本文将测试的粒度改为秒。

Sysbench-OLTP Sysbench 是一个模块化，跨平台，多线程的测试工具[108]。在众多的测试模式中，OLTP 测试用于测量数据库服务器的性能。数据库中含有一个表 sbtest。在 Sysbench-OLTP 中仅有一种事务类型。然而，用户可以指定测试是复杂事务型，简单事务型或者非事务型。复杂事务型包含点查询，范围查询，范围之和查询等。简单事务型和非事务型包含的查询是复杂事务型包含查询的子集。准备阶段和运行阶段是复杂测试的 2 个阶段。当产生的临时数据需要被清除时，则需要清除阶段。该测试程序的性能由一段时间内的事务数目即事务吞吐量来衡量。

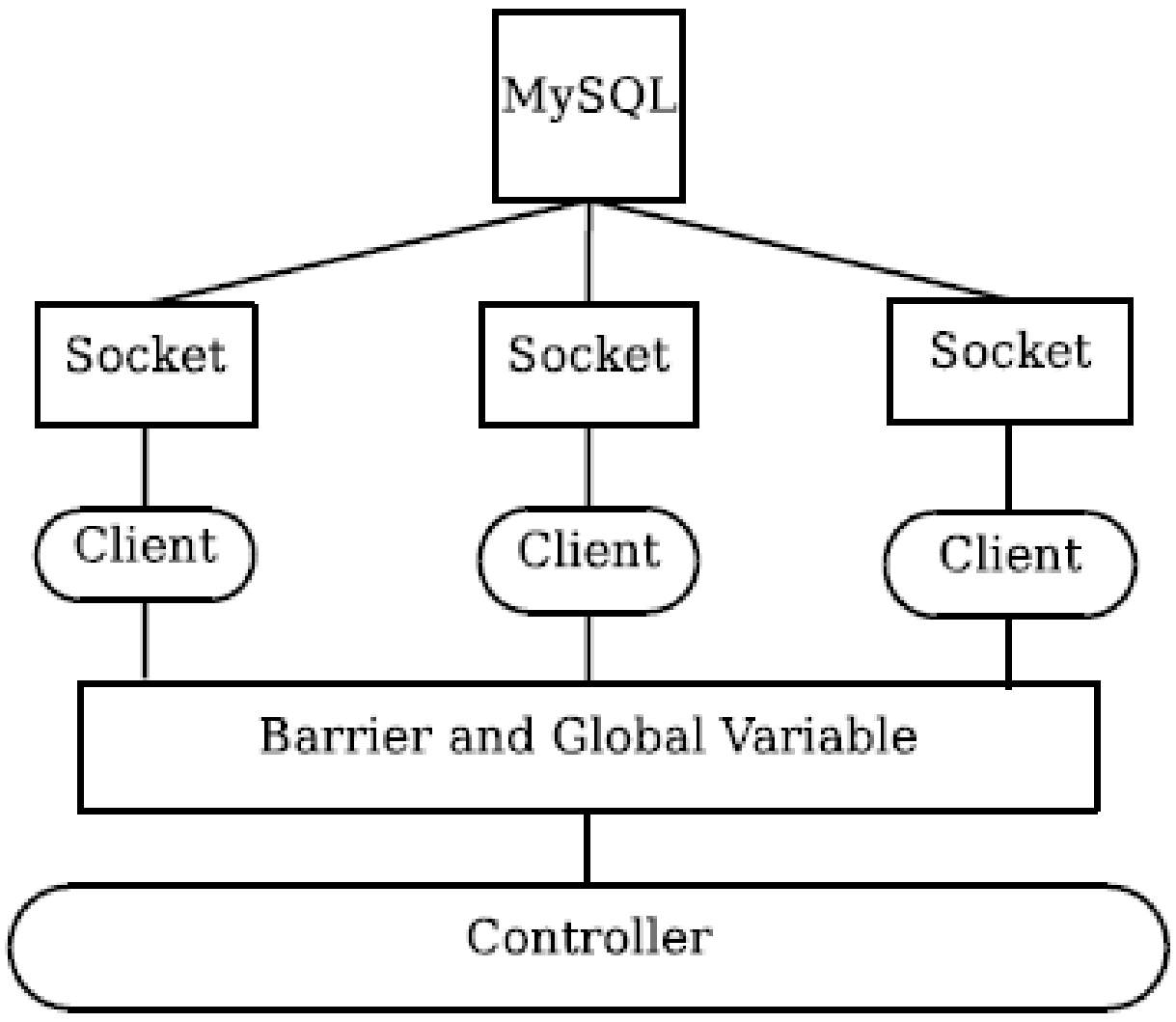

图 5.2 带有 3 个客户端的 Sysbench-OLTP。圆角矩形代表测试程序中的功能模块，矩形代表通信方法，方块代表 MySQL 数据库。

图 5.2 表示了 Sysbench 的软件架构图。控制器和客户端是 Sysbench-OLTP 中的 2 个功能模块。客户端负责根据用户指定创建事务，而控制器控制测试程序的开始和结束。所有的客户端均由线程实现，并且和 MySQL 通过套接字通信。测试程序利用 C 语言编写。

### 5.3.3 数据库配置

在 TPCC-UVa 中，PostgreSQL 的版本号是 8.1.4。PostgreSQL 的配置选项

checkpoint_segments 和 checkpoint_timeout 按照 TPCC-UVa 手册进行设置[114]。而且，增大 shared_buffers, maximum_prepared_transactions 和 effective_cache_size 等参数以提升应用程序的性能并且降低 maximum_connections 以节省共享内存。数据库配置如表 5.1 所示。

表 5. 1 PostgreSQL 和 MySQL 配置, shared_buffers 和 effective_cache_size 的单位是 8K 字节, checkpoint_timeout 的单位是秒

| PostgreSQL | |
|---|---|
| shared_buffers | 3000 |
| effective_cache_size | 375000 |
| checkpoint_segments | 10 |
| checkpoint_timeout | 3600 |
| max_prepared_transactions | 8 |
| max_connections | 8 |
| **MySQL** | |
| innodb_buffer_pool_size | 3G |
| innodb_log_file_size | 750M |
| innodb_log_buffer_size | 500M |
| innodb_flush_log_at_trx_commit | 0 |
| sort_buffer_size | 1M |
| join_buffer_size | 1M |
| max_connections | 3000 |
| thread_concurrency | 16 |

与 PostgreSQL 类似，优化 MySQL 的配置以提升应用性能。使用 InnoDB 作为存储引擎，因为 InnoDB 带有事务处理功能和行级锁策略。同样调整了数据库的缓存池大小和日志文件大小使其可以放入本文使用的多核系统中。具体的配置见表 5.1。MySQL 的版本号是 5.0.51b。

### 5.3.4 测试程序运行规则

为了测量事务型应用的可扩展性而非 I/O，将整个 PostgreSQL 数据库放入内存文件系统中以模拟 RAID 系统。使用 8 个 warehouse，而每个 warehouse 有 10 个 RTE。预热阶段和测量阶段分别设置为 5 秒和 75 秒。使用检测点控制，同时关闭空间控制。为了达到 TPCC-UVa 的上界性能, TPC-C 标准中用于模拟用户行为的参数，按键时间和思考时间，均被设置为 0。

在 Sysbench-OLTP 中，选择复杂事务类型的测试。在一个事务中，允许读操作和写操作。在 300 秒的运行过程中，每个事务在一张表中插入，删除，更新和选择。为了得到更加稳定的结果，运行每个测试之前，对缓存和内存进行预热。

## 5.4 可扩展性

利用上一节描述的实验方法，本节展示 2 个 OLTP 应用的可扩展性。图 5.3 显示了 TPCC-UVa 和 Sysbench-OLTP 的归一化吞吐量随着核数的变化趋势。为了方

便比较，理想的线性加速比也在该图中表示。TPCC-UVa 和 Sysbench-OLTP 的基线性能分别是 2659.2tpmC 和 331.31tps。TPCC-UVa 达到的加速比为 3.67，而 Sysbench-OLTP 为 5.26。可以看出，2 个 OLTP 应用的可扩展性均不佳。

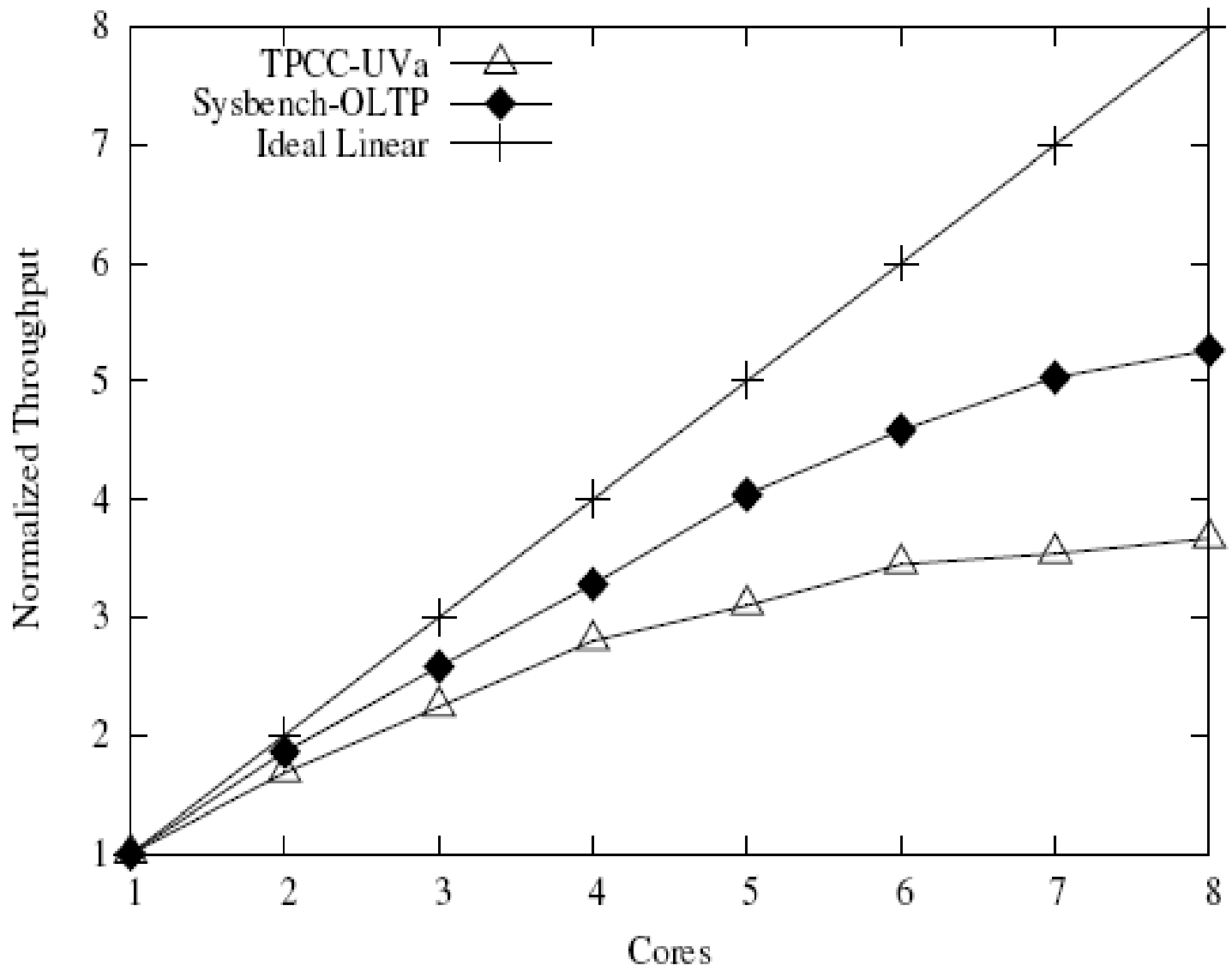

图 5.3 Sysbench-OLTP 和 TPCC-UVa 的归一化吞吐量随着核数的变化

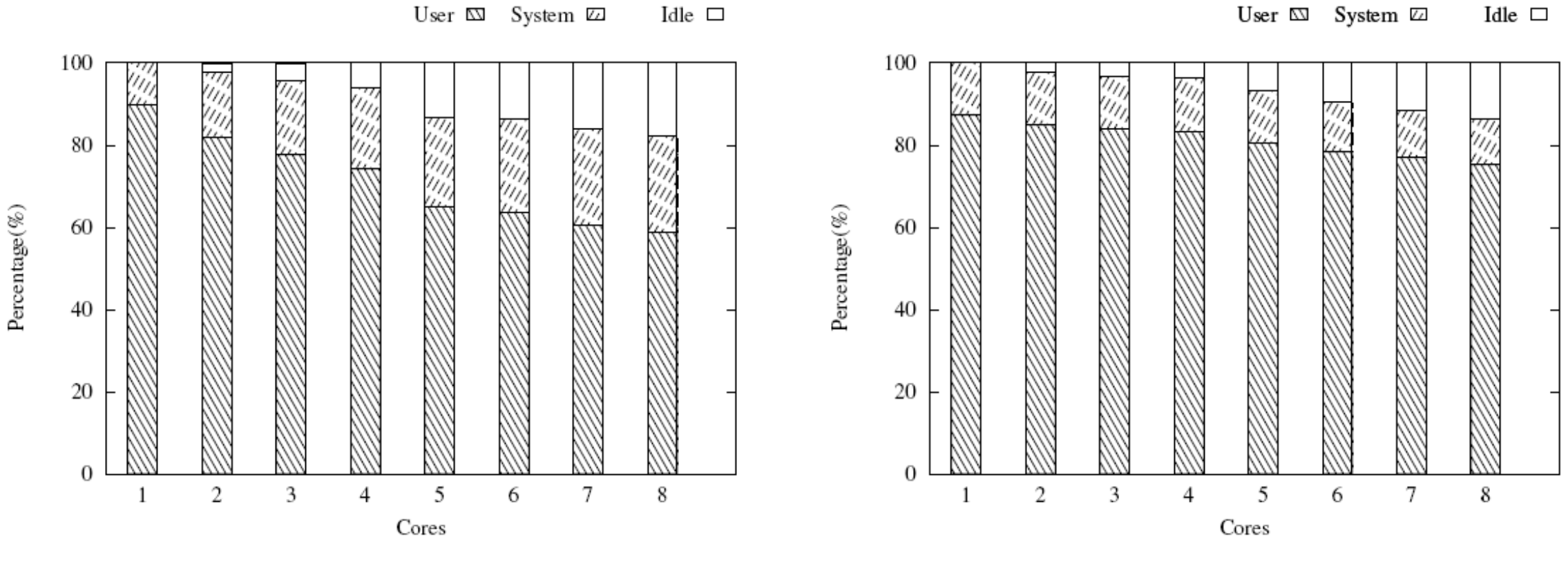

(a) TPCC-UVa (b) Sysbench-OLTP

图 5.4 2 个测试程序的执行时间分解图随着核数的变化

图 5.4 表示了 2 个测试程序的运行时间分解图随核数的变化。对于 TPCC-UVa，当核数由 1 增加到 8 时，每个运行核的平均用户时间从 89.91%降至 58.76%。而系统时间和空闲时间分别由 10.09%涨至 23.58%，由 0%涨至 10.67%。Sysbench-OLTP 中，当增加核数时，用户时间和空闲时间展示出了类似的趋势。增加的空闲时间百分比和降低的用户时间百分比是可扩展性差的有力证明。

## 5.5 瓶颈

本节将展示如何利用可扩展性值指标来定位可扩展性瓶颈。对于每个发现的瓶颈，本文对相关源代码进行分析并且进行实验来验证本文的分析。而且，每个瓶颈的解决方案也被提出。

### 5.5.1 TPCC-Uva中的瓶颈

在 TPCC-UVa 中，选择其可扩展性值为正数的函数并且根据可扩展性值对所有的函数进行排序。符合条件的函数有上千个。然而，排在前 10 位函数的可扩展性值总和占所有函数可扩展性值之和的 58.4%，而排在前 50 位的函数占 86.5%。这样，本文只侧重于排名前 10 位的热点函数以发现可扩展性瓶颈。表 5.2 展示了运行 TPCC-UVa 时，按照可扩展性值排序的前 10 位函数。表中的每一列信息代表的含义如下所示:

**函数** 具有较大可扩展性值的函数。**位置** 该函数存在于软件架构中的位置。位置的取值可能是应用，数据库，系统库和操作系统内核。**Ts** 该函数在单核上的执行时间。**Tm** 该函数在多核上的执行时间。**可扩展性值** 表明可扩展性值的大小，可扩展性值由 Tm-Ts 计算。**权重** 表明该函数影响总体可扩展性的程度。权重实际上是该函数的可扩展性值占所有函数可扩展性值之和的百分比。

表 5.2 TPCC-UVa 中具有较大可扩展性值的函数

| 函数 | 位置 | Ts | Tm | 可扩展性值 | 权重 |
|---|---|---|---|---|---|
| copy_user_generic_string | 内核 | 123.20 | 433.69 | 310.49 | 10.77% |
| ipc_lock | 内核 | 0.21 | 238.14 | 237.93 | 8.26% |
| task_rq_lock | 内核 | 2.13 | 217.82 | 215.69 | 7.49% |
| hrtick_set | 内核 | 3.39 | 161.75 | 158.36 | 5.50% |
| LWLockAcquire | 数据库 | 113.62 | 270.40 | 156.78 | 5.43% |
| hash_search | 数据库 | 82.09 | 178.82 | 96.73 | 3.25% |
| find_busiest_group | 内核 | 0.003 | 88.75 | 88.75 | 3.08% |
| XLogInsert | 数据库 | 62.27 | 149.56 | 87.29 | 3.03% |
| schedule | 内核 | 13.20 | 92.35 | 79.15 | 2.75% |
| LWLockRelease | 数据库 | 82.78 | 161.05 | 78.27 | 2.71% |

1、数据库缓存池竞争 从表 5.2 可以看出，负责将一块内存由内核空间拷贝到用户空间的函数 copy_user_generic_string 在执行 TPCC-UVa 测试时具有最大的可扩展性值。为了理解该函数被哪些函数调用，本文使用 Oprofile 工具。实验结果表明该函数在读取内存文件系统中的文件时被调用。该函数较大的可扩展性值说明数据库的缓存池有频繁的缺失。因为仅有当缓存缺失时，执行路径需要到底层的文件系统中取数据，否则读请求可以通过将缓存池中的数据拷贝到用户态来完成。

数据库进程对缓存池的竞争导致了函数 copy_user_generic_string 具有较大的可扩展性值。竞争的原因是 TPCC-UVa 具有较大的工作集。在实验中，数据库缓存池的大小是固定的，当增加数据库进程的数目时，唯一的缓存池被越来越多的进程所共享。这样，一个进程最近需要的数据可能被另外一个进程替换，因此，读请求在发现缓存池缺失时需要从文件系统中读取数据。

该瓶颈通过持续的增大数据库缓存池的大小来解决。最终，数据库中的

shared_buffers 选项和 Linux 内核中的 shmmax 变量分别被设置为 125000 和 2048000000。图 5.5 表示了当增大数据库缓存池之后的归一化吞吐量。为了方便比较，该图也表示了理想的可扩展性和原来的可扩展性。原来 TPCC-UVa 和增大缓存池之后的 TPCC-UVa 的基线性能分别是 2659.2tpmC 和 2777.6tpmC。从图中可见，增大缓存池大小之后的 TPCC-UVa 比原来的性能好很多。而且，这种提升随着核数的增加而不断变大。在移除缓存池竞争这一瓶颈之后，加速比提升了 46.52%，同时测试程序的执行时间比原来短 53.04%。

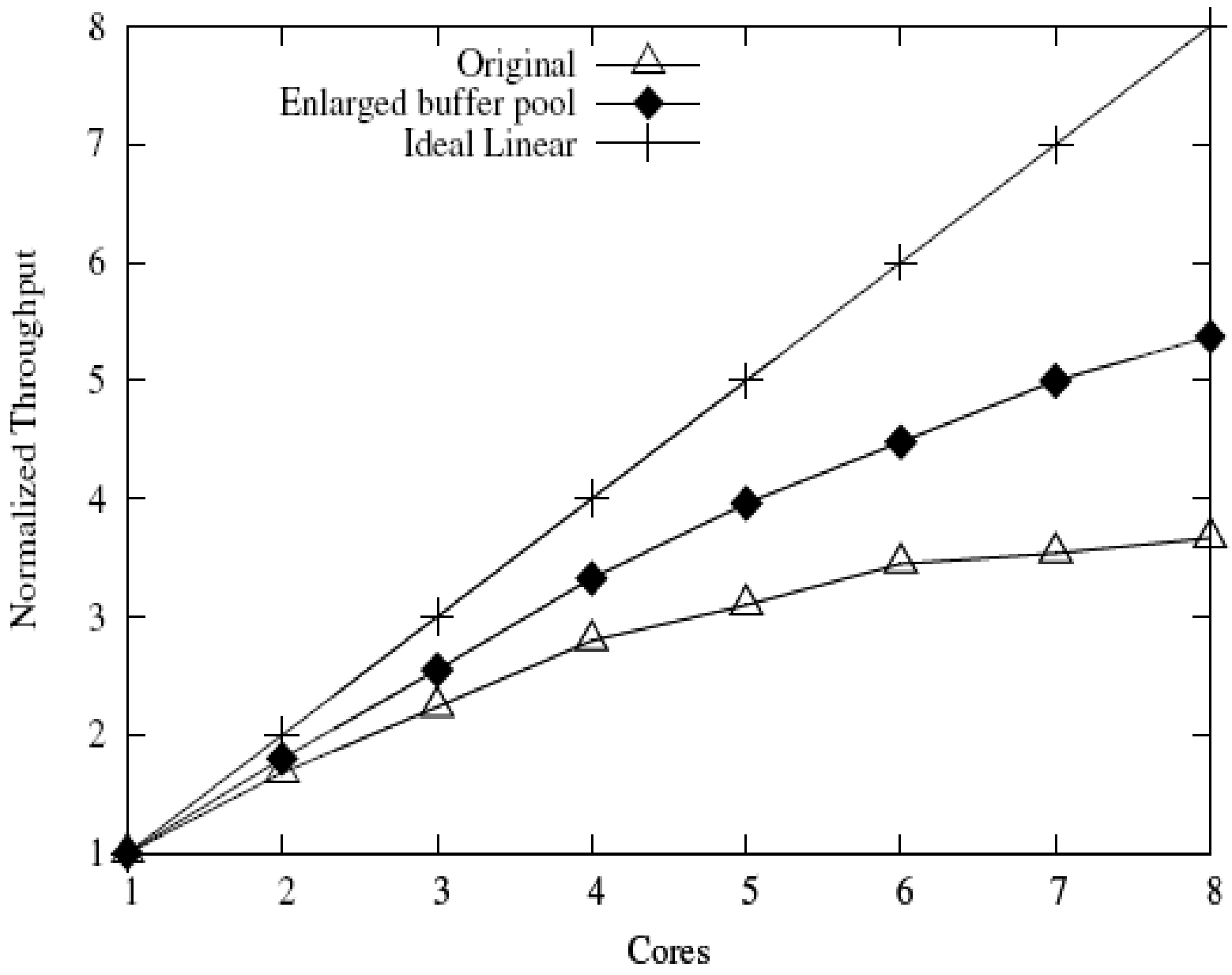

图 5.5 TPCC-UVa 的归一化吞吐量随核数的变化

表 5.3 增大数据库缓存池之后具有最大可扩展性值的前 10 位函数

| 函数 | 位置 | Ts | Tm | 可扩展性值 | 权重 |
|---|---|---|---|---|---|
| LWLockAcquire | 数据库 | 104.93 | 225.81 | 120.88 | 11.14% |
| XLogInsert | 数据库 | 52.76 | 141.16 | 88.40 | 8.15% |
| task_rq_lock | 内核 | 3.65 | 64.40 | 60.95 | 5.62% |
| ipc_lock | 内核 | 0.40 | 51.35 | 50.95 | 4.70% |
| LWLockRelease | 数据库 | 74.35 | 123.99 | 49.64 | 4.58% |
| hash_search | 数据库 | 91.32 | 137.83 | 46.51 | 4.29% |
| hrtick_set | 内核 | 5.93 | 49.86 | 43.93 | 4.05% |
| HeapTupleSatisfiesSnapshot | 数据库 | 69.27 | 105.15 | 35.88 | 3.31% |
| __copy_user_nocache | 内核 | 6.32 | 38.81 | 32.49 | 3.00% |
| memset | 系统库 | 21.55 | 46. 72 | 25.17 | 2.32% |

2、数据库同步原语竞争 表 5.3 表示了增大数据库缓存池之后，根据可扩展性值指标排在前 10 位的函数。这些函数的可扩展性值总和占所有函数可扩展性值的 51.14%。从表中可以看出，函数 copy_generic_string()从排名前十的函数中消失了，而在 PostgreSQL 中用于获取数据库锁的函数 LWLockAcquire 变成了新的热点函数。本文将 LWLockAcquire 和 LWLockReleases 合并，将该瓶颈称为数据库同步原语竞争。当多个进程同时运行时，数据库同步原语会限制应用在多核平台上的可扩

展性，因为获取锁和释放锁的开销会随着核数的增加而增加。本文分析了 PostgreSQL 的源代码以分析哪些锁被激烈地竞争。这种分析是通过在 PostgreSQL 中开启 LWLOCK_STATS 宏来实现的。通过分析发现，预写日志(WAL)和缓存池是竞争的 2 个主要来源。为了减少磁盘写的次数, PostgreSQL 使用 WAL 机制使得在事务提交时仅有日志文件被刷新到磁盘上而不是所有由事务改写的数据[115]。然而，当多个事务同时提交时，用于同步 WAL 行为的锁 WALInsertLock 将会被竞争。同理，当多个进程同时从缓存池中取数据时，缓存池的锁 BufFreelistLock 也会被激烈地竞争。

3、调度开销 调度器的开销会影响 TPCC-UVa 的可扩展性。在测试程序实现上，客户向 TMA 提交事务处理请求。TMA 将请求转化为一系列 SQL 语句并且将这些语句通过套接字提交给数据库进程。在数据库中,语句要通过词法分析器和语法分析器进行检测。而且，查询可能为有效性进行优化。之后，实验结果通过 TMA 返还客户端。

在本文的实验配置中, TMA 进程和数据库进程位于相同的系统上。这样, 2 个系统会以乒乓的方式进行通信。这种通信方式意味着当一个进程运行时，另外一个进程处于等待状态。当运行的进程将要睡眠时，等待的进程需要被唤醒。在 Linux 内核中，函数 try_to_wake_up 用于唤醒睡眠进程。在多核环境下，当运行 try_to_wake_up 函数的 CPU 和运行睡眠进程任务的 CPU 不同时，需要远程唤醒。远程唤醒需要通过函数 task_rq_lock 拿到运行队列的自旋锁。这种行为引入了自旋锁的竞争，因为函数 hrtick_set 和 schedule 同样需要获取该自旋锁。从表 5.3 所示，函数 task_rq_lock 和 hrtick_set 的可扩展性值均很大。

4、System V IPC 中的锁竞争 测试程序 TPCC-UVa 中有两部分使用 SystemV IPC 进行通信。一个是测试程序中，另外一个是 PostgreSQL 数据库中。从表 5.3 中可以看出, System V IPC 相关的函数 ipc_lock 具有较大可扩展性值。函数的调用关系表明函数 ipc_lock 主要由 do_msgsnd, do_msgrcv 和 sys_semtimedop 调用。而函数__down_read 由 ipc_lock 调用。SystemV IPC 对于可扩展性的影响体现在 2 个部分，一个是 ipc_ids 数据结构的读写信号量竞争，另外一个是每个 SystemV IPC 实例的竞争。在 Linux 内核中，每种 SystemV IPC 资源(共享内存，信号量和消息队列)由一个在 ipc_ids 数据结构中的读写信号量保护。在操作一个由 kern_ipc_perm 数据结构表示的 System V IPC 资源之前，进程首先要在函数__down_read 内获取信号量。然而，对于读写信号量的竞争会导致可扩展性有所下降，同时 CPU 会出现空闲时间。另外一部分竞争是 IPC 实例的自旋锁竞争。在 TPCC-UVa 中，属于相同 warehouse 的客户端通过相同的 SystemV IPC 消息队列对相同的 TMA 提出请求。

客户端提出请求和 TMA 的响应均需要获取消息队列的自旋锁。每个 IPC 信号量可以同时保护很多数据结构。然而，在相同信号量集合中对不同信号量同时操作需要获取相同的自旋锁。

表 5.4 TPCC-UVa 中的锁竞争

| 类名 | 竞争次数 | 函数 |
|---|---|---|
| &sem->wait_lock | $8.0\times10^5$ | __down_read __up_write<br>__down_write_trylock<br>__down_write_nested |
| &new->lock | $6.5\times10^5$ | ipc_lock |

利用 Linux 内核中的/proc/lock_stat 工具可以观察到上面描述的现象。实验结果如表 5.4 所示。其中类名代表锁被创建和初始化时的名称，竞争次数列出了某个锁的竞争次数，函数代表了竞争发生的函数。可见，读写信号量和 IPC 实例的竞争次数分别是 $8.0\times10^5$ 和 $6.5\times10^5$ 。在测试程序运行时，这 2 个锁是竞争最为激烈的。

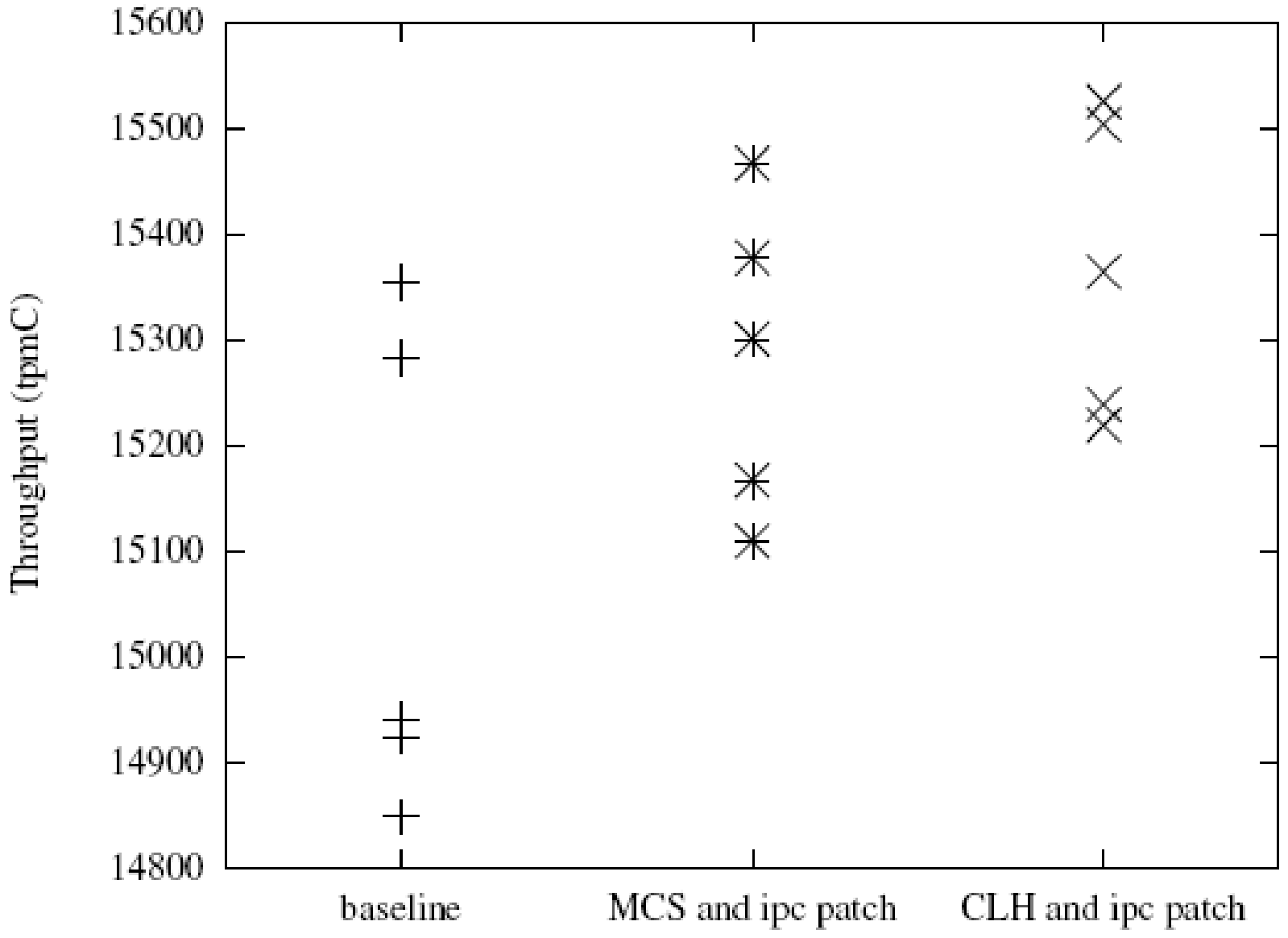

图 5.6 执行 TPCC-UVa 时 SystemV IPC 补丁和可扩展锁带来的提升

自旋锁竞争会导致缓存的局部性差，因为竞争时缓存行的传递很频繁并且会消耗大量的互联资源。而读写信号量的竞争会增加空闲时间和上下文切换次数。Linux 社区已经提出了解决读写锁竞争的内核补丁[110]，该补丁实现了基于 RCU 的 IDR 数据结构 API 并且在 ipc_lock 函数中不需要读写信号量。而可扩展锁可以被用来降低 SystemV IPC 实例的自旋锁竞争。本文同时采用了上述两种实现。图 5.6 表示 8 核系统上针对每种配置的实验结果。从图中可见，采用两方面改进得到的性能均比未采用前达到的性能要好，尽管性能的提升并不大(分别是 2%和 1.4%)。性能提升不大的原因是锁的竞争不够激烈。事实上，使用同一个锁的进程数目越多，改进能得到的性能提升就越大。如果竞争不够激烈，测试程序得到的提升就会不大。

为了验证这一观点，本文进行 2 个实验。首先，当获取一把锁时通过测量锁等待者的数目来衡量 SystemV IPC 的自旋锁竞争。实验的结果表明，大部分时间锁只由 2 个进程相互竞争, 3 个进程同时竞争一把锁的可能性仅为 0.13%。第二，本文利用测试程序 single counter 来探索当竞争一把锁的核数增加时，可扩展锁能够得到的性能提升。在 single counter 中，每个线程在临界区中增加一个全局唯一计数器直到计数器的值超过一定阈值。使用每次操作的平均时间作为性能。

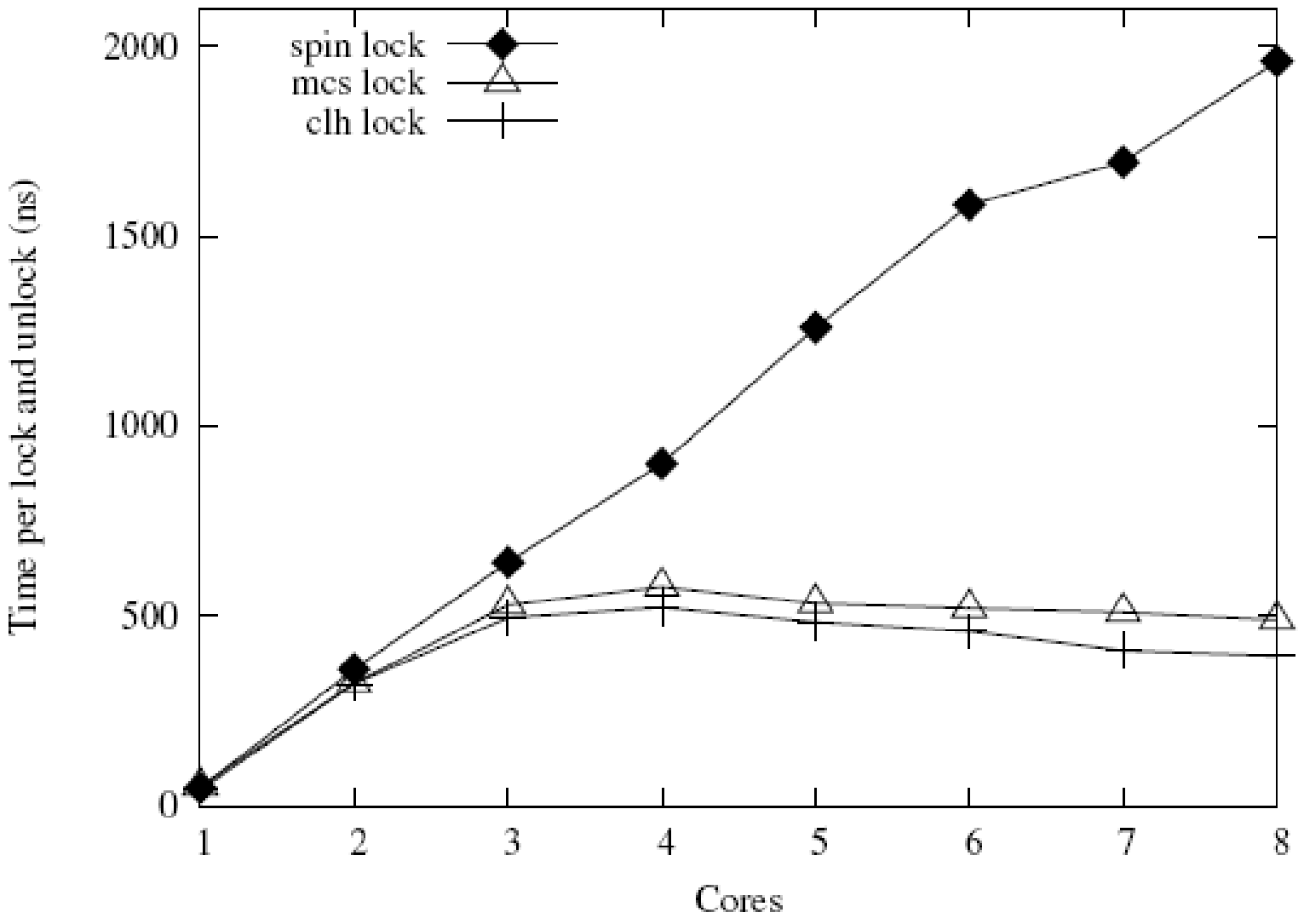

图 5. 7 single counter 中每次操作的平均执行时间随核数的变化

图 5.7 表示了实验结果。从图中可见，随着核数的增加，CLH 锁达到的性能最好。对于 MCS 锁，除了单核(51.59 纳秒 vs 47.03 纳秒)，其性能均优于排号自旋锁。但是，当核的数目较少时，可扩展自旋锁得到的优势不够明显。

### 5. 5. 2 Sysbench-OLTP中的瓶颈

Sysbench-OLTP 中最主要的瓶颈是数据库同步原语和调度器的开销。

1、数据库同步原语 表 5.5 表示了在 Sysbench-OLTP 中具有最大可扩展性值的 10 个函数。可以看到数据库同步原语对该测试程序的可扩展性影响最大。基于 libpthread 中的互斥锁, MySQL 数据库实现了线程之间的同步。这种同步原语被用在很多地方，例如数据库缓存池，锁管理器和日志管理等。在 MySQL 中, libpthread 中的互斥锁被直接使用或者在 InnoDB 互斥锁中被使用。当竞争发生时，等待互斥锁的线程先自旋一段时间，如果没有得到锁，则根据 MySQL 中使用的加锁函数，可能需要进入内核来等待锁释放。

在 Sysbench-OLTP 中，每个线程在事务中获取一系列锁。当多个进程并行执行时，获取一把锁的平均时间会变大。pthread_mutex_lock 和 pthread_mutex_ trylock 是数据库获取锁时需要使用的 2 个函数。区别是执行 pthread_mutex_lock 的函数可

能需要进入内核来获取锁，而 pthread_mutex_trylock 则不需要。从表 5.5 可以看出这 2 个函数具有较大的可扩展性值。另外一个解锁时调用的函数__pthread_mutex_unlock_usercnt 也是在 sysbench-OLTP 执行时的热点函数。实验结果表明其最大的可扩展性值是由最后级缓存缺失导致的。图 5.8 表示了实验结果。可以看到，将核数由 1 核增加至 8 核时，__pthread_mutex_unlock_usercnt 在一个事务之内的缓存缺失大幅度增加。利用 InnoDB 中的互斥锁统计数据可以看到哪些互斥锁被激烈地竞争。在 Sysbench-OLTP 中，热点锁是保护缓存池的读写锁，每个线程私有数据的哈希表锁和内核数据结构的锁。存在两种避免瓶颈的方式。一种是使用细粒度的锁。这样，对一个锁的高度竞争可以被替代成对多个锁的竞争。Yasufumi[116]提出的补丁就将数据库缓存池的锁替换成了多个锁。

表 5.5 执行 Sysbench-OLTP 时具有最大可扩展性值的前 10 个函数

| 函数 | 位置 | Ts | Tm | 可扩展性值 | 权重 |
|---|---|---|---|---|---|
| __pthread_mutex_unlock_usercnt | 系统库 | 0.54 | 2.18 | 1.64 | 9.34% |
| pthread_mutex_trylock | 系统库 | 0.68 | 1.94 | 1.26 | 7.21% |
| pthread_mutex_lock | 系统库 | 0.38 | 1.57 | 1.19 | 6.83% |
| row_search_for_mysql | 数据库 | 0.85 | 1.73 | 0.88 | 5.00% |
| schedule | 内核 | 0.30 | 0.78 | 0.48 | 2.71% |
| free | 系统库 | 0.34 | 0.76 | 0.42 | 2.41% |
| _int_malloc | 系统库 | 0.46 | 0.86 | 0.40 | 2.33% |
| malloc | 系统库 | 0.29 | 0.58 | 0.29 | 1.68% |
| filesort | 数据库 | 0.69 | 0.97 | 0.28 | 1.61% |
| innodb_srv_conc_enter_innodb | 数据库 | 0.02 | 0.30 | 0.28 | 1.57% |

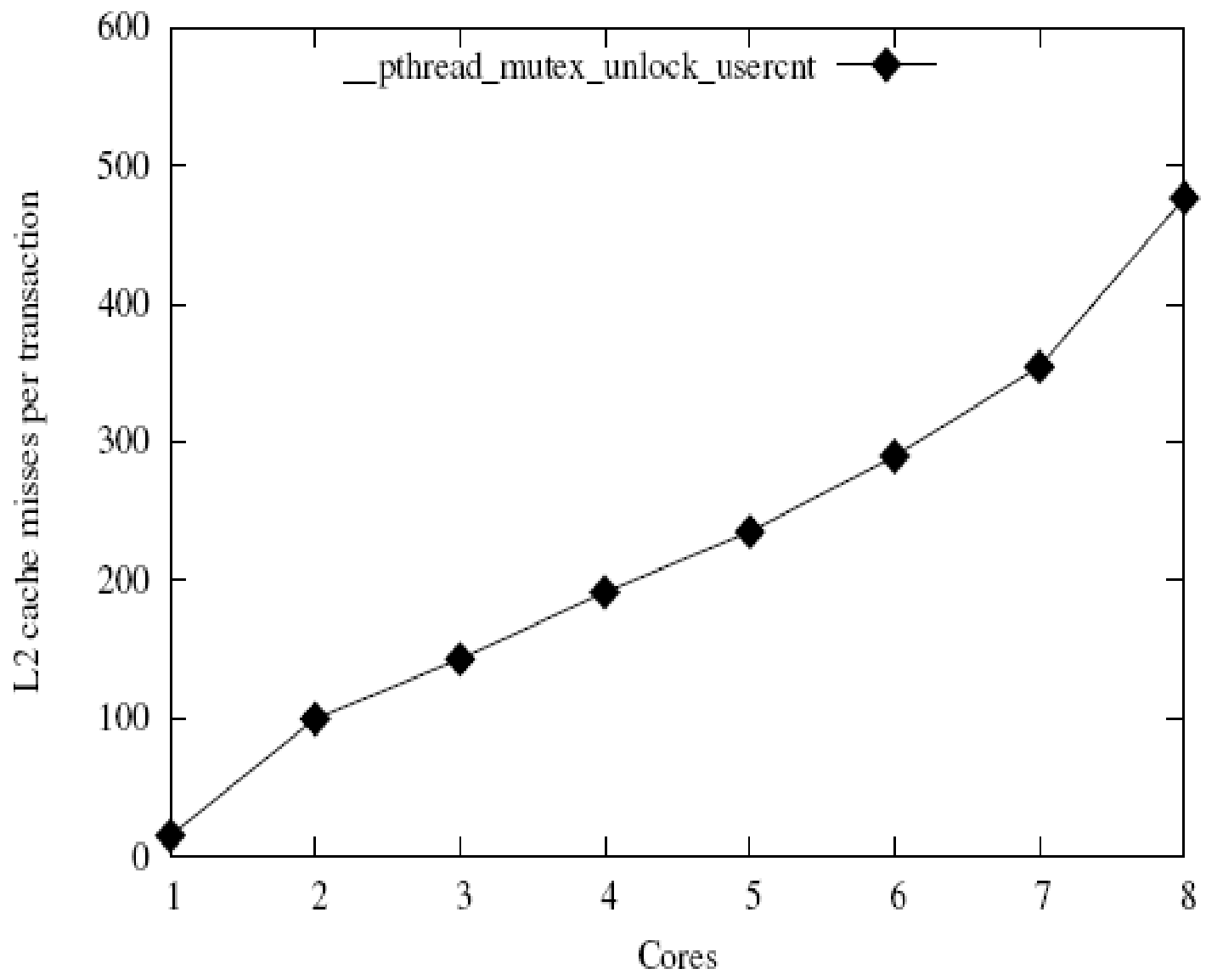

图 5.8 每个事务中__pthread_mutex_unlock_usercnt 的最后级缓存缺失次数随核数的变化

另外一种方式是提高 MySQL 同步原语的有效性。Google 公司开发出一种在 SMP 平台上提升 MySQL 性能的补丁[117]。在补丁中，将 InnoDB 互斥锁和读写互斥锁替换成了原子内存指令。本文将该补丁移植过来并且和原来的系统性能相比

较以观察补丁的有效性。实验结果表明，使用补丁之后的 MySQL 比原来的实现提升了 10.91%的可扩展性。转化成性能的提升是 15.27%。

2、调度开销 从表 5.5 中可见，调度开销同样会影响系统的可扩展性。与 TPCC-UVa 相比较，Sysbench-OLTP 没有事务检测模块。然而，客户端和数据库进程的通信方式和 TPCC-UVa 中事务检测模块与数据库进程的通信方式相同。这样，由远程唤醒导致的自旋锁竞争会增加 schedule 函数的执行时间。但是增加的程度不会太大(2.71%)。

## 5.6 相关工作

由于 OLTP 应用在工业界的重要性，很多工作都是围绕如何提升这类应用在多核平台上的可扩展性展开的。例如，Thakkar 和 Sweiger 等[111]在多核处理器上研究了 TP1 测试程序的可扩展性。进程迁移开销和 I/O 饱和分别被确定为全缓存测试和大压力测试的可扩展性瓶颈。Saylor 和 Khessib[112]在大规模 ccNUMA 平台上部署了 OLTP 应用。Saylor 等首先利用各种性能计数器对 OLTP 应用执行时的特征进行量化，进而发现编译器优化，数据划分和局部性感知的数据库锁均可以提升该应用的可扩展性。然而，Saylor 等在硬件级别确定可扩展性问题(如缓存缺失过多等)，而将硬件级别的数据对应到具体的瓶颈却是基于经验的方式。而本文却是根据基于函数可扩展性值指标的系统级方法来定位瓶颈，根据函数所在的内核模块或者语义可以方便地定位可扩展性问题，因此本文的方法对经验的依赖很少。另外一个和本文相关的工作是在 Intel Itanium2 处理器上利用 OLTP 应用研究操作系统内核的可扩展性[41]。它利用 Itanium 处理器的特殊功能来定位具有较大开销的缓存缺失。因为该工作依赖于特殊处理器的性能计数功能，因此不适合在其它平台上使用。本文的可扩展性值指标同样依赖于性能计数器来获取每个函数的执行时间。然而，大部分的现代处理器均具有基于时间的采样功能。

## 5.7 本章小结

本文研究了 OLTP 应用在多核平台上的可扩展性。使用 2 个测试程序 TPCC-UVa 和 Sysbench-OLTP 来代表 OLTP 应用。TPCC-UVa 是 TPC-C 标准的开源实现而 Sysbench-OLTP 是一个轻量级的 OLTP 测试程序。为了准确而有效的定位事务类应用的可扩展性瓶颈，本文提出了基于可扩展性值指标的瓶颈定位方法。使用该方法，找到了限制应用可扩展性的几个瓶颈。TPCC-UVa 中，最大的瓶颈是数据库进程对缓存池的竞争。数据库锁，调度开销和 System V IPC 的自旋锁竞争

是下一级别的可扩展性瓶颈。在 Sysbench-OLTP 中，数据库同步原语和调度开销影响应用的可扩展性。本文采用可扩展的数据库同步原语，基于 RCU 的 IDR API 和可扩展自旋锁来降低瓶颈带来的影响。

# 第6章　总结和展望

## 6.1　工作总结

如今, 多核芯片已经被广泛应用。相对于传统的对称多处理器, 多核系统具有两方面最为显著的不同:一是核的数目可以变得更多；二是硬件资源共享。然而, 如今的操作系统却是针对小规模的对称多处理器进行设计, 缺乏针对多核系统的硬件特点的设计，可扩展性差。并行可扩展性是衡量软件使用硬件资源能力的有效度量, 提升操作系统在多核平台上的可扩展性, 不但具有较大的研究价值, 而且具有很强的实际需求。

如何分析和定位操作系统的可扩展性瓶颈以及如何针对瓶颈提出有效的解决方案是本领域最为基本和核心的问题, 同时又存在着极大的挑战和进一步解决的空间。本文围绕着如下 4 个具体问题进行深入研究。

1）当前主流的操作系统中, 其服务接口在多核平台上的并行可扩展性如何, 其瓶颈是什么, 如何产生的；2）系统密集类应用中, 排号自旋锁(ticket spinlock)的竞争导致应用程序的吞吐量随着核数的增加而下降(锁颠簸现象)。为什么锁竞争会导致吞吐量下降以及如何对系统进行改进以有效地避免该现象；3）多道程序(multiprogramming)环境中, 共享最后级缓存等共享硬件资源导致程序之间相互竞争, 从而影响系统整体的可扩展性, 如何在操作系统中设计有效的机制来避免这种影响；4）复杂系统操作密集类应用中, 如何定位操作系统的可扩展性瓶颈。瓶颈可能存在于应用, 操作系统或者硬件中。

本文的研究围绕着上述 4 个问题展开, 取得了如下的研究成果:

1、分析和比较了操作系统在多核平台上的可扩展性，设计了针对系统服务接口的微基准测试程序集, 测试结果表明，在 16 核以上的系统上，现有操作系统中保护共享数据的锁竞争会导致可扩展性急剧下降。测试程序集在保证系统之间比较的公平性前提下, 针对最具代表性的服务接口(如进程管理和内存管理等)进行测试。实际系统的测量发现操作系统的系统服务接口均有可扩展性问题。在此基础上的深入分析准确地还原了每种系统服务的可扩展性瓶颈的产生方式。综合测试实验分析,我们得出一条重要结论是, 操作系统中保护共享数据的锁竞争是限制加速比增长的根本因素。

2、针对锁颠簸问题, 设计实现了基于离散事件仿真技术的模拟器, 可准确、快速地重现锁颠簸现象, 进而提出了基于等待者数目的可扩展锁机制和竞争感知

的调度策略，很好地避免了锁颠簸。本文首先深入分析锁颠簸现象，找到导致该现象的主要软、硬件因素。基于排队论模型，利用离散事件仿真技术对所有因素进行整合，实现的模拟器可以准确、快速地重现锁颠簸现象。在此基础上，提出了基于等待者数目的可扩展锁机制以避免锁颠簸现象。该机制利用等待者数目自适应地决定锁等待者的等锁方式(自旋或者进入节能状态)，并且采用了伪共享避免等一系列的可扩展性优化，很好地避免了锁颠簸现象。同时，本文提出的锁竞争感知的调度策略检测每个任务的用锁程度，并利用模型驱动搜索将所有用锁密集任务集中在部分处理器上运行。测试结果表明，采用锁竞争感知的调度策略比相关工作提升加速比达 46%。

3、针对硬件资源竞争问题，提出了资源竞争感知的调度策略，显著提升系统可扩展性。在多核平台上的多道程序执行环境中，共享最后级缓存等硬件资源的竞争导致程序之间相互影响，因而降低了系统可扩展性。对此，本文提出了代表程序共享资源需求的启发式指标的选择方法。该方法不仅考虑了利用指标代表程序资源需求的准确性，而且考虑了在线获取该指标时的稳定性。基于程序的资源需求信息，将任务到核的映射机制和每一个核上的任务选择机制均进行改进，使得具有互补资源需求的任务可以被同时调度。在实际系统上的评测表明，提出的资源竞争感知调度策略可以提升系统的可扩展性达 13%，显著发挥了资源竞争感知调度的潜力。

4、针对操作系统的瓶颈定位问题，提出了基于函数可扩展性值的瓶颈定位方法。函数的可扩展性值被定义为该函数在多核和单核上每单位工作量的执行时间之差。具有较大可扩展性值的函数对整体的可扩展性影响较大。通过对可扩展性值排在前几位的函数进行分析，能够有效地定位可扩展性瓶颈，进而优化相关函数。以在线事务处理应用为例，定位到内存缓冲处理等瓶颈并对其优化改进，提升加速比可达 49%，有效削减了操作系统对加速比的限制。

## 6.2 展望

根据上述的研究内容将来需要进一步研究的工作分别介绍如下：

1、对于开源操作系统的分析和比较，进一步的研究分为两部分。一部分是将当前测试程序集找到的可扩展性瓶颈进行移除。本文已经对部分测试程序的可扩展性瓶颈进行了移除(如 FreeBSD 平台上的 dupbench 和 Linux 平台上的 sembench)，但是还有一些瓶颈仍然没有有效的解决方案。另外一部分是补充微基准测试程序集，使得系统密集类程序的可扩展性可以根据微基准测试程序的结果和简单的程序信息来进行预测。

2、尽管本文提出的机制和策略已经有效地避免了锁颠簸现象，但是用锁密集类应用的可扩展性仍然受到临界区的限制。将来的研究工作包括增加操作系统代码的并行程度，使用无锁操作[40, 44]，设计新型数据结构来避免不必要的锁竞争[118]。

3、本文提出的资源竞争避免策略主要适用于解决共享缓存所导致的硬件资源竞争。后续的工作需要考虑其它的硬件资源竞争，如地址总线饱和和 TLB 竞争等。设计更加通用的资源竞争感知调度将更具挑战性而且更具实用价值。

4、设计更有效的瓶颈定位方法或者工具。目前的多核处理器均带有强有力的性能计数器，例如 AMD 平台上的 IBS[119]和 Intel 平台上的 PEBS[120]。这些性能计数器可以提供更加深层次的程序信息，进而为进一步地提升可扩展性打下基础。后续的工作将利用这些计数器设计一些工具以定位操作系统的可扩展性瓶颈。

# 参考文献

# 致 谢

衷心感谢我的导师史元春教授和陈渝副教授多年来对我的悉心指导和谆谆教诲，他们严谨踏实的治学作风、乐观豁达的人生态度、平易近人的为人之道是我学习的榜样。同时感谢 MIT 的 Frans Kaashoek 教授, Frans 教授虚怀若谷，实事求是。作者在攻读博士学位期间,有幸得到 Frans Kaashoek 教授的指导，受益颇多，在此表示感谢。

感谢普适计算实验室和操作系统与中间件中心的全体同学，与你们在学术和生活上的交流让我获益良多。感谢我的朋友们，一直在我身边关心我、支持我，让我感受到快乐与生活的意义。

最后特别感谢我的父母，他们对我的培养以及给予我的理解、支持与鼓励是我信心和动力的源泉。

# 声　明

本人郑重声明：所呈交的学位论文，是本人在导师指导下，独立进行研究工作所取得的成果。尽我所知，除文中已经注明引用的内容外，本学位论文的研究成果不包含任何他人享有著作权的内容。对本论文所涉及的研究工作做出贡献的其它个人和集体，均已在文中以明确方式标明。

签　名：____________日　期：__________

# 附录A 论文中的图表索引

# 个人简历、在学期间发表的学术论文与研究成果

## 个人简历

1984 年 10 月 5 日出生于吉林省白城市。

2003 年 9 月考入北京邮电大学计算机科学与技术系，2007 年 7 月本科毕业并获得工学学士学位。

2007 年 9 月免试进入清华大学计算机科学与技术系攻读工学博士至今。

## 攻读博士学位期间参加的课题

[1] 2012 年 1 月 1 日——2015 年 12 月 31 日，国家自然科学基金(NSFC)项目”面向众核体系结构的操作系统并行优化关键技术研究”,项目号 61170050。

[2] 2010 年 1 月 29 日——2010 年 7 月 16 日，清华-Intel 国际合作项目”面向众核架构最后一级共享 cache 软件关键优化技术”，项目号 20103000032。

[3] 2011 年 2 月 1 日——2011 年 12 月 20 日，清华-华为科技合作项目”基于 ccNUMA 的性能优化项目”，项目号 20112000284。

(本文作者为上述项目主要承担者)

## 发表的学术论文

第一作者发表的期刊论文：

[1]. **Yan Cui**, Yingxin Wang, Yu Chen, Yuanchun Shi, “Lock-Contention-Aware Scheduler: A Scalable and Energy Efficient Method for Addressing Scalability Collapse on Multicore Systems”, in ACM Transactions on Architecture and Code Optimization (**TACO**), (**CCF Rank A**, SCI &EI, to appear).

[2]. **Yan Cui,** Yu Chen, Yuanchun Shi, Towards Scalability Collapse Behavior, in Concurrency and Computation:Practice and Experience (**CONCURRENCY**), (**CCF Rank B**, SCI &EI, to appear).

[3]. **Yan Cui**, Yu Chen, Yuanchun Shi, Comparing Operating Systems Scalability on Multicore Processors by Microbenchmarking, in IEICE Transactions on Information and Systems (**IEICE Transactions**), Vol:E95-D, NO: 12, pages 2810-2820, 2012 (SCI & EI).

第一作者发表的在审期刊论文:

[1]. **Yan Cui**, Yingxin Wang, Yu Chen, Yuanchun Shi, Mitigating Resource Contention on Multicore Systems via Scheduling, in Computer Journal (**CJ**), (**CCF Rank B**, SCI & EI, major revision).

[2]. Yan Cui, Yingxin Wang, Yu Chen, Yuanchun Shi, Requester-Based Lock: A Scalable and Energy Efficient Locking Scheme on Multicore Systems, in IEEE Transactions on Computers (**TC**), (**CCF rank A**, SCI & EI, under review).

第一作者发表的会议报告论文(oral):

[1]. **Yan Cui**, Yingxin Wang, Yu Chen, Yuanchun Shi, Wei Han, Xin Liao, Fei Wang, Reducing Scalability Collapse via Requester-Based Locking on Multicore Systems, in Proceedings of IEEE 20th International Symposium on Modeling, Analysis and Simulation of Computer and Telecommunication Systems (**MASCOTS 2012**), pages 298-307, (**CCF Rank B**, EI:20124515642164).

[2]. **Yan Cui**, Yingxin Wang, Yu Chen, Yuanchun Shi, Experience on Comparison of Operating Systems Scalability on the Multicore Architecture, in Proceedings of IEEE 13th Cluster Computing (**CLUSTER 2011**), pages 205-215, (**CCF rank B**, EI:20114614519276).

[3]. **Yan Cui**, Yu Chen, Yuanchun Shi, Comparison of Lock Thrashing Avoidance Methods And Its Performance Implications for Lock Design, in Proceedings of workshops on the ACM International Symposium on High Performance Distributed Computing (**HPDC workshops 2011**), pages 3-10, (EI:20113214209165)

[4]. **Yan Cui**, Yu Chen, Yuanchun Shi, Scaling OLTP Applications on Commodity Multi-Core Platforms, in Proceedings of 2010 IEEE 10th International Symposium on Performance Analysis of Systems and Software (**ISPASS 2010**), pages 134-143, (EI:20102112951778).

[5]. **Yan Cui**, Weida Zhang, Yu Chen, Yuanchun Shi, A Scheduling Method for Avoiding Kernel Lock Thrashing on Multicores, in Proceedings of IEEE 16th International Conference on Parallel and Distributed Systems (**ICPADS 2010**), pages 17-26, (**CCF rank C,** EI:20110813687043).

[6]. **Yan Cui**, Weiyi Wu, Yingxin Wang, Xufeng Guo, Yu Chen, Yuanchun Shi, A Discrete Event Simulation Model for Understanding Kernel Lock Thrashing on Multi-core Architectures, in Proceedings of IEEE 16th International Conference on Parallel and Distributed Systems (**ICPADS 2010**), pages 1-8, (**CCF rank C**, EI:20110813687041).

[7]. **Yan Cui**, Yu Chen, Yuanchun Shi, Parallel Scalability Comparison of Commodity Operating Systems on Many Cores, in Proceedings of workshops on 36th International Symposium on Computer Architecture (**ISCA workshops 2009**), pages 1-10.

[8]. **崔岩**, 史元春, 陈渝, 位置传感器网络中移动目标定位精度的视觉测试方法, 第三届和谐人机环境联合学术会议论文集, 2007, pages 854-860.

第一作者发表的会议张贴论文(poster):

[1].**Yan Cui**, Weiyi Wu, Yingxin Wang, Xufeng Guo, Yu Chen, Yuanchun Shi, Reinventing Lock Modeling for Multicore Systems, in Proceedings of IEEE/ACM 18th International Symposium on Modeling, Analysis and Simulation of Computer and Telecommunication Systems (**MASCOTS 2010**), pages 455-457 (EI:20104513365722).

非第一作者发表的会议文章:

## 国家发明专利

[1]. 秦岭, 陈渝, **崔岩**, 吴瑾, 实现自适应锁的方法和装置以及多核处理器系统(申请号: 201110394780, 公开号: CN102566979)

[2]. 刘仪阳, 陈渝, 谭玺, **崔岩**，一种线程调度方法、线程调度装置及多核处理器系统 (申请号:201110362773, 公开号:CN102495762)

[3]. 刘仪阳, 张知缴, 方帆, 陈渝，**崔岩**, 一种内存分配方法、装置及系统 (申请号: 201210176906)